\documentclass[12pt,a4,twoside]{puthesis}
\pdfoutput=1
\usepackage{epsfig}
\usepackage{amsfonts}
\usepackage{amssymb}
\usepackage{amsmath}   
\usepackage{amsthm}
\usepackage{latexsym}
\usepackage{float}
\usepackage{natbib}
\usepackage{lscape}
\usepackage{multicol}

\usepackage{url}
\usepackage{wasysym} 
\usepackage{gensymb} 
\usepackage{color}
\usepackage{pst-node,pst-tree}
\usepackage{mathrsfs} 
\usepackage{bm} 

\usepackage{graphicx}
\title{Detection techniques for the H.E.S.S. II telescope, data modeling of gravitational lensing and emission of blazars in HE-VHE astronomy}

\author{Anna Barnacka}

\abstract{
Recent years have seen a tremendous progress in field of the high energy (HE) astrophysics
and  very high energy (VHE) astronomy.
This progress has been achieved mostly thanks to a new generation of  instruments 
that provide  data of previously unattainable quality. 
This thesis presents the study of four aspects of high energy astronomy.

The first part of my thesis is dedicated to an aspect of instrument development for  imaging 
atmospheric Cherenkov telescopes, namely  the Level 2 trigger system of  
the High Energy Stereoscopic System (H.E.S.S.). 
H.E.S.S. is an array  dedicated to VHE $\gamma$-ray astronomy. 
The array has been in operation since the beginning of 2004. 
Originally it has been composed of four 12 meter diameter telescopes,
which has been completed with a fifth 28 meter diameter telescope in 2012. 
This H.E.S.S. II telescope is designed to 
operate both in stereoscopic mode and in monoscopic mode. 
The Level 2  trigger system is required to suppress spurious triggers of
 the telescope when operating in monoscopic mode.
This dissertation provides  the motivation and principle of the operation of the Level 2  trigger. 
I had the opportunity to work on the Level 2 trigger system for the H.E.S.S. II telescope at IRFU/CEA in France.  
The IRFU is responsible for designing and building this trigger system.
The system consists of both hardware and software solutions. 
My work on the project focused on the algorithm development and the Monte Carlo simulations of 
the trigger system and overall instrument (Moudden, Barnacka, Glicenstein et al. 2011a; Moudden,
Venault, Barnacka et al. 2011b).
The hardware implementation of the system is described and its expected performances are then evaluated.

The H.E.S.S. array has been used to observe the blazar  PKS 1510-089. 
The second part of my thesis deals with the data analysis and modeling of broad-band emission of  this particular blazar.  
PKS 1501-089 is an example of the so-called flat spectrum radio quasars (FSRQs) 
for which no VHE emission is expected due to the Klein-Nishina effects 
and strong absorption in the broad line region (Moderski et al. 2005). 
The recent detection of at least three FSRQs by Cherenkov telescopes has forced a revision of our understanding of these objects. 
In part II of my thesis, I am presenting the analysis of the H.E.S.S. data: the light curve and spectrum of  PKS 1510-089, together with the FERMI data and a collection of multi-wavelength data obtained with various instruments. 
I am presenting the model of  PKS~1510-089 observations carried out during a flare recorded by H.E.S.S..
The model  is based on a single zone internal shock scenario.

The third part of my thesis deals with blazars observed by the FERMI-LAT, 
but from the point of view of other phenomena: a strong gravitational lensing. 
This part of my thesis shows the first evidence for gravitational lensing phenomena in high energy gamma-rays. 
This evidence comes from the observation of a gravitational lens system induced echo in the light curve of the distant blazar PKS 1830-211.
Traditional methods for the estimation of time delays in gravitational lensing
systems rely on the cross-correlation of the light curves from individual images. 
In my thesis, I used 300 MeV-30 GeV photons
detected by the Fermi-LAT instrument.
The FERMI-LAT instrument cannot separate the images of known lenses.
The observed light curve is thus the superposition of individual image light curves.
The FERMI-LAT instrument has the advantage of providing long, evenly spaced, time series with very low photon noise.          
This allows to use directly Fourier transform methods.

A time delay between the two compact images of PKS 1830-211 has been
searched for  both by the autocorrelation method and a new method:
the ``double power spectrum''. The double power spectrum shows a
 4.2~$\sigma$ evidence for a time delay  of  27.1$\pm$0.6 days (Barnacka et al. 2011), consistent
with the results from Lovell et al. (1998) and Wiklind~\&~Combes (2001). 

The last part of my thesis concentrates on another lensing phenomena called "femtolensing". 
The search for femtolensing effects has been used to derive limits on the primordial black holes abundance.
The abundance of primordial black holes is currently significantly
constrained in a wide range of masses.  The weakest limits are
established for the small mass objects, where the small intensity of
the associated physical phenomenon provides a challenge for current
experiments.  
I have used gamma-ray bursts with known redshifts detected by the
FERMI Gamma-ray Burst Monitor (GBM) to search for the femtolensing
effects caused by compact objects.  The lack of femtolensing detection
in the GBM data provides new evidence that primordial black holes in
the mass range $5 \times 10^{17}$ -- $10^{20}\,$g do not constitute
a major fraction of dark matter (Barnacka et al. 2012).  

My Ph.D. studies have been carried out jointly between the Nicolaus Copernicus Astronomical Center 
of the Polish Academy of Sciences, in Warsaw in Poland 
and the IRFU institute of the Commissariat \'a l'\'energie atomique et aux \'energies alternatives (CEA) Saclay in France.
}

\streszczenie{
Ostatnie dziesi\c eciolecie przynios{\l}o niezwyk{\l}y post\c ep w dziedzinie astronomii wysokich i bardzo wysokich energii. 
Post\c ep ten zosta{\l} osi\c agni\c ety g{\l}\'ownie dzi\c eki nowej generacji instrument\'ow, 
kt\'ore umo\.zliwi{\l}y obserwacje z nieosi\c agaln\c a dotychczas precyzj\c a i czu{\l}o\'sci\c a. 
W niniejszej rozprawie przedstawiam cztery problemy z zakresu astronomii wysokich energii, 
w kt\'orych wyniki zosta{\l}y uzyskane dzi\c eki nowej generacji instrument\'ow prowadz\c acych obserwacje gamma w szerokim zakresie.

Pierwsza cz\c e\'s\'c pracy doktorskiej dotyczy rozwoju i budowy teleskop\'ow Czerenkowskich, 
a w szczeg\'olno\'sci systemu wyzwalania kamery zwanego "Level 2 trigger"  dla najwi\c ekszego teleskopu Czerenkowa
 w Wysokoenergetycznym Systemie Stereoskopowym (High Energy Stereoscopic System - H.E.S.S.). 
 H.E.S.S. jest uk{\l}adem teleskop\'ow dedykowanych do obserwacji bardzo wysokoenergetycznego promieniowania gamma. 
 Sie\'c teleskop\'ow pracuje w systemie stereoskopowym od 2004 roku. 
 Przez pierwsz\c a dekad\c e uk{\l}ad sk{\l}ada{\l} si\c e z czterech 12 metrowych teleskop\'ow. 
 W roku 2012 uk{\l}ad zosta{\l} uzupe{\l}niony o pi\c aty 28 metrowy teleskop, 
 tym samym obserwatorium H.E.S.S. wesz{\l}o w kolejn\c a faz\c e swojej dzia{\l}alno\'sci nazwan\c a H.E.S.S. II. 
 Sie\'c sk{\l}adaj\c aca si\c e z pi\c eciu teleskop\'ow zosta{\l}a zaprojektowana, 
 aby m\'oc prowadzi\'c obserwacj\c e zar\'owno w trybie stereoskopowym jaki w systemie monoskopowym (jednoteleskopowym). 
 Pi\c aty teleskop zosta{\l} wyposa\.zony w system wyzwalania kamery wy\.zszego poziomu, 
 w celu zredukowania liczby rejestrowanych zdarze\'n t{\l}a, gdy teleskop pracuje w trybie monoskopowym.
 
 W pracy doktorskiej przedstawiam motywacj\c e i zasady dzia{\l}ania systemu wyzwalania kamery wy\.zszego poziomu teleskopu H.E.S.S. II. 
 Znaczna cz\c e\'s\'c bada\'n w tym zakresie zosta{\l}a wykonana w IRFU/CEA we Francji 
 - w jednostce naukowej odpowiedzialnej za zaprojektowanie i budow\c e tego systemu. 
 W sk{\l}ad uk{\l}adu Level 2 trigger wchodz\c a zar\'owno rozwi\c azania sprz\c etowe (hardware) jaki i programistyczne (software). 
 W prezentowanej rozprawie doktorskiej opisuj\c e r\'owie\.z zastosowane rozwi\c azanie sprz\c etowe wraz z implementacj\c a opracowanego algorytmu, 
 oraz  wyniki symulacji Monte Carlo przedstawiaj\c ace wydajno\'s\'c zaimplementowanego rozwi\c azania 
 (Moudden, Barnacka, Glicenstein et al. 2011a; Moudden, Venault, Barnacka et al. 2011b).  
 
 W roku 2009 w obserwatorium H.E.S.S. przeprowadzono obserwacjie blazara PKS 1510-089. 
 Druga cz\c e\'s\'c rozprawy doktorskiej przedstawia analiz\c e danych jak i wyniki modelowania widma tego blazara. 
 Blazar PKS 1510-089 jest przyk{\l}adem tak zwanego radio-kwazara o p{\l}askim widmie (Flat Spectrum Radio Quasar - FSRQ). 
 Z powodu efektu Klein-Nishiny oraz silnej absorpcji w  obszarze szerokich linii emisyjnych, 
 nie spodziewano si\c e rejestracji emisji bardzo wysokoenergetycznego promieniowania gamma z tego typu obiekt\'ow  (Moderski et al. 2005). 
 Detekcja co najmniej trzech obiekt\'ow tego typu w zakresie VHE spowodowa{\l}a konieczno\'s\'c rewizji 
 dotychczasowej interpretacji wysokoenergetycznego sk{\l}adnika widmowego obiekt\'ow typu FSRQs.
 
 W procesie analizy danych blazara PKS 1510-089 wykorzysta{\l}am dane uzyskane za pomoc\c a obserwatorium H.E.S.S. 
 wraz z danymi uzyskanymi dzi\c eki satelicie FERMI uzupe{\l}nionych o zbi\'or archiwalnych obserwacji w szerokim zakresie widma elektromagnetycznego.
 Obserwacje te umo\.zliwi{\l}y rekonstrukcj\c e widma PKS 1510-089 w zakresie od fal radiowych po najwy\.zsze obserwowane energie 
 ze szczeg\'olnym uwzgl\c ednieniem  rozb{\l}ysku zaobserwowanego w marcu 2009 roku. 
 W rozprawie doktorskiej przedstawiam r\'ownie\.z interpretacj\c e emisji PKS 1510-089 podczas wy\.zej wspomnianego rozb{\l}ysku. 
 Dyskutowana przeze mnie interpretacja bazuje na modelu jednostrefowym,  
 w kt\'orym elektrony s\c a przyspieszane do relatywistycznych  pr\c edko\'sci w procesie Fermiego zachodz\c acym w wewn\c etrznych szokach. 
 
 Trzecia cz\c e\'s\'c rozprawy doktorskiej r\'ownie\.z dotyczy blazar\'ow obserwowanych przez satelit\c e FERMI, 
 jednak przedstawiane wyniki odnosz\c a si\c e do innego zjawiska - silnego soczewkowania grawitacyjnego. 
 W tej cz\c e\'sci przedstawiam pierwszy przypadek soczewkowania grawitacyjnego zarejestrowanego za pomoc\c a obserwacji gamma blazara PKS 1830-211.
 
Tradycyjne metody analizy zjawiska soczewkowania grawitacyjnego polegaj\c a na badaniu wzajemnej korelacji krzywych zmian blasku 
obraz\'ow powsta{\l}ych na skutek soczewkowania grawitacynego w celu wyznaczenia op\'o\'znienia czasowego 
obserwowanych sygna{\l}\'ow. 
W przedstawionej w pracy analizie wykorzystano obserwacje foton\'ow gamma o energii z zakresu 
od 300 MeV do 30 GeV zarejestrowanych detektorem znajduj\c acym si\c e na pok{\l}adzie satelity FERMI. 
Instrument ten nie dysponuje wystarczaj\c ac\c a rozdzielczo\'sci\c a k\c atow\c a, 
aby m\'oc zarejestrowa\'c krzywe zmiany blasku bezpo\'srednio dla ka\.zdego powsta{\l}ego obrazu. 
Obserwowana krzywa zmiany blasku soczewkowanego grawitacyjnie \'zr\'od{\l}a jest wi\c ec sum\c a powsta{\l}ych obraz\'ow. 
Satelita FERMI dostarcza bardzo d{\l}ug\c a pr\'obk\c e danych r\'ownomiernie obserwowanego nieba 
z bardzo niskim poziomem szumu dzi\c eki czemu dane te s\c a niezwykle atrakcyjne dla metod bazuj\c acych na transformacji Fouriera.

Op\'o\'znienie czasowe pomi\c edzy obrazami PKS 1830-211 zosta{\l}o wyznaczone zar\'owno metod\c a auto-korelacji, 
jak r\'ownie\.z opracowan\c a metod\c a podw\'ojnego widma mocy. 
Metoda podw\'ojnego widma mocy pozwoli{\l}a na wyznaczenie op\'o\'znienia  czasowego o warto\'sci 27$\pm$0.6 dnia z istotno\'sci\c a detekcji 4.2 $\sigma$ (Barnacka i in. 2011).
Wynik ten jest zgodny z wynikami uzyskanymi przez Lovell i in. (1998) oraz przez Wiklind i Combes (2001), 
z obserwacji radiowych, jednak zastosowanie danych w zakresie HE pozwoli{\l}o zmniejszy\'c niepewno\'s\'c pomiarow\c a o rz\c ad wielko\'sci. 

W ostatniej cz\c e\'sci rozprawy doktorskiej przedstawiam wyniki bada\'n szczeg\'olnego zjawiska 
soczewkowania grawitacyjnego zwanego femtosoczewkowaniem. 
Efekt femtosoczewkowania grawitacyjnego zosta{\l} wykorzystany do wyznaczenia ogranicze\'n na g\c esto\'s\'c pierwotnych czarnych dziur.  
W badaniach wykorzystano b{\l}yski gamma o znanym przesuni\c eciu ku czerwieni zarejestrowe 
przez detektor b{\l}ysk\'ow gamma znajduj\c acy si\c e na pok{\l}adzie satelity FERMI (FERMI Gamma-ray Burst Monitor - GBM). 
Brak zarejestrowanego efektu femtosoczewkowania w tych b{\l}yskach pozwoli{\l} stwierdzi\'c, 
\.ze pierwotne czarne dziury w zakresie mas od $10^{17}$ do $10^{20}$~g stanowi\c a mniej ni\.z 3\% g\c esto\'sci krytycznej (Barnacka i in. 2012).
 
Zaprezentowane w pracy wyniki zosta{\l}y osi\c agni\c ete w trakcie bada\'n prowadzonych w IRFU/CEA Saclay 
oraz w  Centrum Astronomicznym im. Miko{\l}aja Kopernika Polskiej Akademii Nauk w Warszawie.

}

\resume{
Il y a eu dans les ann\'ees r\'ecentes de nombreuses avanc\'ees 
dans l'astronomie et l'astrophysique des hautes \'energies. 
Ces avanc\'ees sont dues principalement \`a une nouvelle g\'en\'eration d'instruments 
qui fournissent des donn\'ees d'une qualit\'e sans pr\'ec\'edent. 
La pr\'esente th\`ese porte sur quatre aspects diff\'erents de l'astronomie des hautes \'energies. 

La premi\`ere  partie de ma th\`ese est d\'edi\'ee \`a un d\'eveloppement instrumental 
pour les t\'elescopes Cherenkov imageurs, le syst\`eme de d\'eclenchement de niveau 
2 du t\'elescope de 28 m\`etres du r\'eseau H.E.S.S. (High Energy Stereoscopic System).
H.E.S.S. est un r\'eseau d\'edi\'e aux rayons gamma de tr\`es haute \'energie. 
Ce r\'eseau est en op\'eration depuis le d\'ebut 2004, avec quatre t\'elescopes de 12 m\`etres de diam\`etre. 
Le r\'eseau a \'et\'e compl\`et\'e par un cinqui\`eme t\'elescope (LCT) de 28 m\`etres en 2012. 
Le t\'elescope LCT peut fonctionner en mode str\'eoscopique ou monoscopique. 
Le syst\`eme de d\'eclenchement de niveau 2 est n\'ecessaire pour supprimer des d\'eclenchements 
fortuits ou dus aux muons isol\'es lorsque le t\'elescope fonctionne en mode monoscopique. 
J'explique la motivation et le principe de fonctionnement du syst\`eme de d\'eclenchement de niveau 2.  
L'IRFU du CEA, ou j'ai pass\'e une partie de ma co-tutelle est responsable de la conception 
et la construction du syst\`eme de d\'eclenchement de niveau 2 du LCT. 
La conception fait intervenir \`a la fois des d\'eveloppements hardware et software.
 
Mon travail s'est focalis\'e sur l'invention d'algorithmes 
et les simulations Monte-Carlo du syst\`eme de d\'eclenchement, 
ainsi que la comparaison de la reconstruction au niveau de la carte de d\'eclenchement 
 \`a la reconstruction "hors-ligne" 
 (Moudden, Barnacka, Glicenstein et al. 2011a; Moudden, Venault, Barnacka et al. 2011b).  
 Je d\'ecris le syst\`eme et j'\'evalue ses performances. 
 
Le r\'eseau H.E.S.S. a  observ\'e le blazar PKS1510-089.
La deuxi\`eme partie de ma th\`ese traite de l'analyse des donn\'ees et la mod\'elisation 
de l'\'emission large bande de ce blazar. 
C'est un exemple de quasar radio \`a spectre plat (FSRQ), 
pour   lequel il n'est attendu aucune \'emission aux tr\`es hautes \'energies. 
En effet, les photons de haute \'energie ne sont pas produits \'a cause de l'effet Klein-Nishina 
d'un part et sont absorb\`es dans la r\'eion des ÔÕraies largesÔÕ (broad line region) (Moderski et al. 2005). 
La d\'etection r\'ecente d'au moins 3 FSRQ par les t\'elescopes Cherenkov a forc\'e \`a une r\'evision de notre compr\'ehension de ces AGN. 

Dans le deuxi\`eme chapitre de ma th\`ese, je pr\'esente une analyse de 
la courbe de lumi\`ere et du spectre de PKS1510-089, 
obtenus avec les donn\'ees de H.E.S.S.. 
Les donn\'ees du FERMI-LAT, ainsi qu'une collection de donn\'ees multi-longueur d'ondes ont \'egalement \'et\'e exploit\'ees. 
J'ai mod\'elis\'e les  donn\'ees observ\'ees pendant un "flare" de PKS1510-089. 
Ce  mod\`ele est bas\'e sur un sc\'enario de choc interne \`a une zone.  

Le troisi\`eme chapitre de ma th\`ese est une \'etude d'un autre ph\'enom\`ene 
affectant potentiellement les blazars observ\'es par FERMI-LAT: 
l'effet de lentille gravitationnelle fort. 
Cette partie de ma th\`ese montre le premier indice de pr\'esence d'un effet de lentille gravitationnelle 
dans le domaine des photons de haute \'energie.   
Cet indice provient de l'observation dÕun \'echo dans la courbe de lumi\`ere du blazar distant PKS1830-211, 
qui est une lentille gravitationnelle connue. 
Les m\'ethodes d'estimation des retards temporels dans les syst\`emes de lentille gravitationnelles reposent  
sur la corr\'elation crois\'ee des courbes de lumi\`ere individuelles. 
Dans l'analyse pr\'esent\'ee dans cette th\`ese, j'ai utilis\'e des photons de 300 MeV \`a 30 GeV d\'etect\'es par
 l'instrument FERMI-LAT. 
 L'instrument FERMI-LAT ne peut pas s\'eparer spatialement les images des lentilles gravitationnelles fortes connues. 
La courbe de lumi\`ere observ\'ee est donc la superposition des courbes de lumi\`ere des images individuelles. 
Les donn\'ees du FERMI-LAT ont l'avantage d'\^etre des s\'eries temporelles r\'eguli\'erement espac\'ees, 
avec un bruit de photons tr\'es bas. 
Cela permet d'utiliser directement les m\'ethodes de transform\'ees de Fourier. 
Un retard temporel entre les images compactes de PKS1830-211 a \'et\'e recherch\'e par deux m\'ethodes : 
une m\'ethode d'auto-corr\'elation et la m\'ethode du "double spectre". 
La m\'ethode du double spectre fournit un signal de 27 $\pm$ 0.6 jours (statistique) avec une significativit\'e de 4.2 $\sigma.$ 
Ce r\'esultat est compatible avec ceux de Lovell et al (1998) et Wiklind et Combes (2001).

La derni\`ere partie de ma th\`ese es consacr\'ee \`a un effet de lentille diff\'erent, le "femtolensing". 
La recherche d'effets de femtolensing a \'et\'e utlis\'ee pour obtenir des limites sur l'abondance de trous noirs primordiaux. 
Celle-ci a \'et\'e  contrainte de mani\`ere significative dans un large domaine de masses. 
Les limites les moins contraignantes ont \'et\'e \'etablies pour les objets de faible masse, 
pour lesquels la d\'etection repr\'esente un d\'efi exp\'erimental. 
J'ai utilis\'e les sursauts gamma de redshift connus d\'etect\'es par  le Fermi Gamma Ray Burst Monitor (GBM) 
pour rechercher d'\'eventuels effets de femtolensing produits par des objets compacts sur la ligne de vis\'ee. 
L'absence de ces effets de femtolensing montre que des trous noirs primordiaux de masse comprises entre $5\times10^{17}$ et $10^{20}$~g ne constituent pas une fraction importante de la mati\`ere noire. 

J'ai effectu\'e mes \'etudes de th\`ese en co-tutelle entre le Centre Astronomique Nicolaus Copernicus 
de l'acad\'emie des sciences polonaise, \`a Varsovie et le l'Institut de Recherches sur les Lois fondamentales de l'Univers 
du Commissariat \`a l'Energie Atomique et au \'energies alternatives \`a Saclay, en France.
}

\acknowledgements{
It gives me great pleasure in expressing my gratitude to all those people 
who have supported me and had their contributions in making this thesis possible 
and an unforgettable experience as well as big adventure.

Foremost, I would like to express my gratitude to my advisors Jean-Francois Glicenstein and Rafa\l{} Moderski 
for the continuous support of my Ph.D study and research.
I thank Jean-Francois for his patience, motivation, enthusiasm and immense knowledge 
and Rafa\l{} for his valuable comments and many discussions.

Besides my advisors, I would like to thank the rest of my thesis committee: 
Tomasz Bulik, W\l{}odek Bednarek, Stefan Wagner and  Reza  Ansari
for their encouragement, insightful comments, and hard questions. 

I would  especially like to thank  Tomek Bulik who during my internship at CAMK showed me that being a scientist is cool, 
and Bronek Rudak for his endless and priceless support. 

I thank all the Ph.D students at Nicolaus Copernicus Astronomical Center, in particular 
Karolina, Mateusz, Krzysztof, Tomek, Weronika and Torun team. 
I am thankful to my colleagues Pierre, Anais, Emmanuel, Aion, Bernard,  for the hospitality 
and support during my stay in France.

Finally, I take this opportunity to express the profound gratitude to my  parents, my siblings and Jacek
for their continuous support. 

This work was supported by the Polish Ministry of Science and Higher Education under Grants No. DEC-2011/01/N/ST9/06007 and  in part supported by HECOLS.
}

\def\aap{\ifnum\longrefs=1 {Astron.\ Astrophys.}\else 
                           {A\hbox{\rm \&}A}\fi}
\def\aapr{\ifnum\longrefs=1 {Astron.\ Astrophys.\ Rev.}\else 
                            {A\hbox{\rm \&}AR}\fi}
\def\aaps{\ifnum\longrefs=1 {Astron.\ Astrophys.\ Suppl.}\else 
                            {A\hbox{\rm \&}A Suppl.}\fi}
\def\aj{\ifnum\longrefs=1 {Astron.\ J.}\else 
                          {AJ}\fi} 
\def\ao{\ifnum\longrefs=1 {Applied Optics}\else 
                           {Appl.\ Opt.}\fi} 
\def\aspcs{\ifnum\longrefs=1 {Astron.\ Soc.\ Pacific Conf. Series}\else 
                           {ASP Conf.\ Ser.}\fi} 
\def\apj{\ifnum\longrefs=1 {Astrophys.\ J.}\else 
                           {ApJ}\fi} 
\def\apjl{\ifnum\longrefs=1 {Astrophys.\ J. Lett.}\else 
                            {ApJ}\fi} 
\def\aplett{\ifnum\longrefs=1 {Astrophys.\ J. Lett.}\else 
                            {ApJ}\fi} 
\def\apjs{\ifnum\longrefs=1 {Astrophys.\ J. Suppl.}\else 
                            {ApJS}\fi}
\def\apss{\ifnum\longrefs=1 {Astrophys.\ and Space Science}\else 
                            {Astrophys.\ Space Sci.}\fi}
\def\araa{\ifnum\longrefs=1 {Ann.\ Rev.\ Astron.\ Astrophys.}\else 
                            {ARA\hbox{\rm \&}A}\fi}
\def\azh{\ifnum\longrefs=1 {Astronomicheskii Zhurnal}\else 
                            {Astron.\ Zhur.}\fi}
\def\baas{\ifnum\longrefs=1 {Bull.\ Am.\ Astron.\ Soc.}\else 
                            {BAAS}\fi}
\def\bain{\ifnum\longrefs=1 {Bull.\ Astronom.\ Institutes Netherlands}\else
                            {Bull.\ Astr.\ Inst.\ Neth.}\fi}
\def\gca{\ifnum\longrefs=1 {Geochim.\ Cosmochim.\ Acta}\else 
                           {Geochim.\ Cosmochim.\ Acta}\fi}
\def\grl{\ifnum\longrefs=1 {Geophys.\ Res.\ Lett.}\else 
                           {Geoph.\ Res.\ Lett.}\fi}
\def\iaucirc{\ifnum\longrefs=1 {IAU Circulars}\else 
                          {IAU Circ.}\fi}
\def\ip{\ifnum\longrefs=1 {in press}\else 
                          {in press}\fi}
\def\jgr{\ifnum\longrefs=1 {J.\ Geophys.\ Res.}\else 
                           {J.\ Geophys.\ Res.}\fi}  
\def\jrasc{\ifnum\longrefs=1 {J.\ Royal Astron.\ Soc.\ Canada}\else 
                           {JRAS Can.}\fi}  
\def\memsai{\ifnum\longrefs=1 {Mem.~Soc.~Astron.~Italiana}\else
                              {MemSAI}\fi}
\def\mnras{\ifnum\longrefs=1 {Mon.\ Not.\ Roy.\ Astron.\ Soc.}\else 
                             {MNRAS}\fi} 
\def\nat{\ifnum\longrefs=1 {Nature}\else 
                           {Nat}\fi}
\def\pasj{\ifnum\longrefs=1 {Pub.\ Astron.\ Soc.\ Japan}\else 
                            {PASJ}\fi} 
\def\pasp{\ifnum\longrefs=1 {Pub.\ Astron.\ Soc.\ Pacific}\else 
                            {PASP}\fi} 
\def\physscr{\ifnum\longrefs=1 {Physica Scripta}\else 
                            {Phys.\ Scrip.}\fi} 
\def\planss{\ifnum\longrefs=1 {Planetary \& Space Science}\else 
                            {Plan. \& Space Sci.}\fi} 
\def\procspie{\ifnum\longrefs=1 {Proc.\ SPIE}\else 
                            {Proc.\ SPIE}\fi} 
\def\qjras{\ifnum\longrefs=1 {Quarterly J.\ Royal Astron.\ Soc.}\else 
                            {QJRAS}\fi} 
\def\sa{\ifnum\longrefs=1 {Soviet Astron..}\else 
                               {Sov.\ Astron.}\fi}
\def\skytel{\ifnum\longrefs=1 {Sky \& Telescope}\else 
                            {Sky \& Tel.}\fi} 
\def\solphys{\ifnum\longrefs=1 {Solar Phys.}\else 
                               {Sol.\ Phys.}\fi}
\def\ssr{\ifnum\longrefs=1 {Space Science Rev.}\else 
                               {Space\ Sci.\ Rev.}\fi}
\def\zap{\ifnum\longrefs=1 {Zeitschr.\ f.\ Astrophysik}\else
                               {Z.\ Astrophys.}\fi}

\begin{document}



\setcounter{chapter}{1}
\part{Level 2 trigger system for H.E.S.S.~II telescope}  

\section{Introduction}
\label{sec:intro}

The High Energy Stereoscopic System (H.E.S.S.) is an observatory of very high energy gamma rays ($>$100~GeV). 
It is located on Khomas Highland in Namibia and became operational in 2004.
The H.E.S.S. array consists of four imaging atmospheric Cherenkov telescopes (IACTs) 
working as a stereoscopic system \citep{2004NewAR..48..331H}.
 The stereoscopic technique is used to achieve a better background rejection power,
 especially to reject events  triggered by single muons and  night sky background photons (NSB), 
 and also it is used  to improve image reconstruction. 
 These single muons come from hadronic showers and 
 became a dominant source of spurious triggers for a single telescope  \citep{2005AIPC..745..753F}. 

Recently, the H.E.S.S. observatory has been completed with a 
 28 meter diameter telescope. 
The large Cherenkov telescope (LCT) saw its first light on 26 of July 2012.
In the low energy range (below 50~GeV), the LCT shall work detached from the rest of the array, 
in the so-called "mono mode". 
Therefore, the LCT camera is equipped with a Level 2 trigger board \citep*{2011APh....34..568M,2011ITNS...58.1685M} to improve the rejection of accidental NSB 
 and single muon triggers. 
Such Level 2 trigger systems have been used also by other
Cherenkov instruments, as well. For example, the MAGIC collaboration 
\citep{2001NIMPA.461..521B} is using a Level 2 trigger to perform a rough 
analysis of data and apply topological cuts to the obtained images.    

In the first part of this thesis, I introduce the phase I and II of the  H.E.S.S. project
(sections \ref{sec:hessI} and \ref{sec:hessII}).
Then, in section \ref{sec:cherprinc}, I describe the principle of the Cherenkov technique.
Section \ref{sec:AnalyticalThreshold} contains a discussion about the energy threshold of the array, 
which I  obtained using Monte Carlo simulations.  
Next, in section \ref{sec:trigger}, I present the trigger system,
 and then in sections \ref{sec:algosoft} and \ref{sec:TriggerSimulations}, 
 the results of my work on the algorithm for the Level~2 trigger. 
 Section \ref{sec:hardware} describes the hardware solution of the Level 2 trigger developed at IRFU/CEA. 
The results are summarize in section~\ref{sec:triggerconclusions}.
    
\newpage
\section{ The H.E.S.S. I phase } 
\label{sec:hessI}
Originally the H.E.S.S.  observatory was design to observe  high energy photons
with energy in the 100~GeV to 100~TeV range. 
 The instrument consisted  of four  Cherenkov telescopes,
 located at the vertices of a square with side 120~m.
 This configuration was selected
 to provide multiple stereoscopic views of air showers.
 The telescopes are made of steel, with altitude/azimuth mounts.
 The dishes have a Davis-Cotton design  with  an hexagonal arrangement, 
 composed of 382~round mirrors, each 60~cm in diameter \citep{2003APh....20..111B}.

\begin{figure}[ht!]
  \centering
  \includegraphics[angle=0,width=145mm,bb=0 0 1500 426]{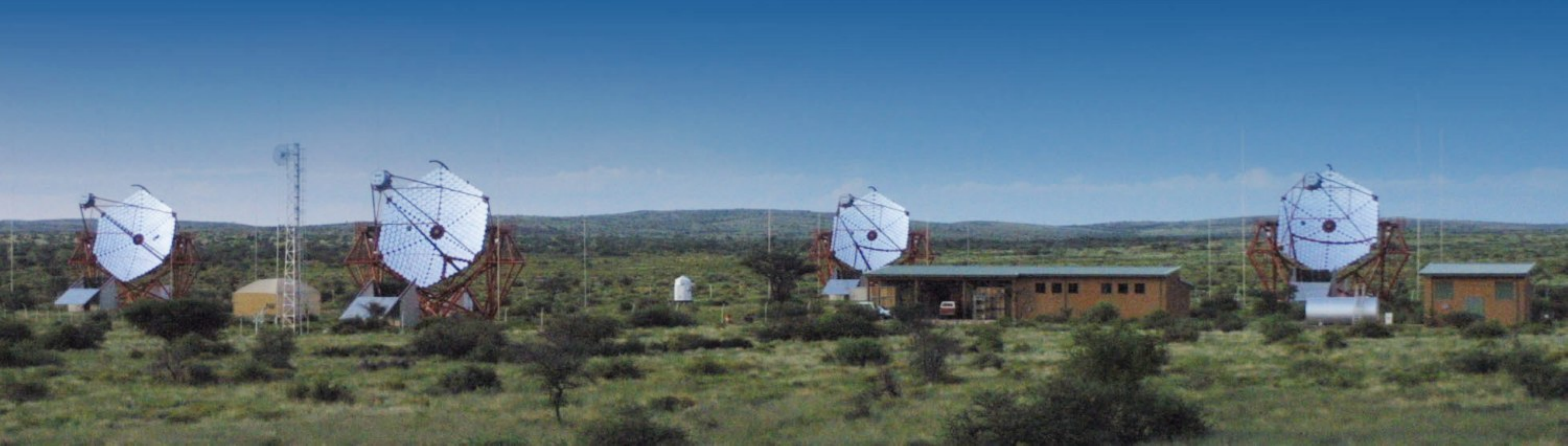}
  \caption{The H.E.S.S. array of four telescopes. Image credit www.unam.na/research/hess/ }
  \label{fig:hess}
\end{figure}

Each of the present small Cherenkov telescopes (SCTs)  has a 12~m diameter mirror 
and is equipped with a camera consisting of 960~Photonis XP2960 photo-multiplier tubes (PMTs). 
Each tube corresponds to an area of 0.16$^{\circ}$ in diameter on the sky, 
and is equipped with a Winston cone. 
The Winston cones  capture the light which would fall in between the PMTs, 
and simultaneously   reduce the background light. 
The camera  design groups the PMTs  in 60~drawers 
of 16~tubes each \citep{2003ICRC....5.2887V}. 
Each drawer contains the trigger and readout electronics for the tubes, 
as well as the high voltage supply, control and monitoring electronics. 
The field of view (FoV) of the detector is 5$^{\circ}$ in diameter. 
The camera is placed  at the focus of the dish, 15~m above the mirrors.
The H.E.S.S. system of four telescopes is presented in figure~\ref{fig:hess}.

\newpage
\section{H.E.S.S. II phase }
\label{sec:hessII}
\begin{figure}
  \centering
  \includegraphics[angle=0,width=140mm,bb=0 0 914 608]{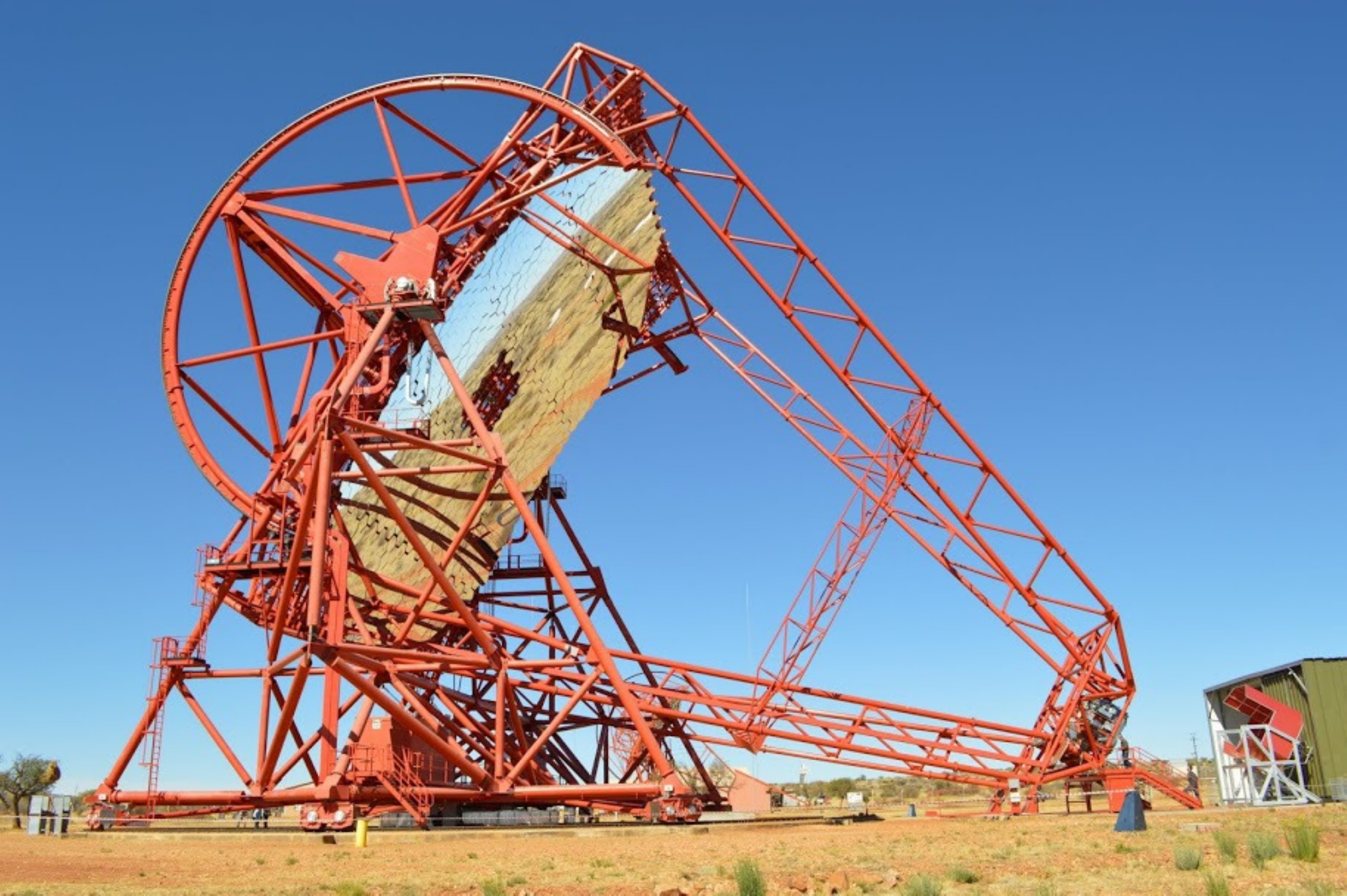}
  \caption{The overall picture of the large Cherenkov telescope of the H.E.S.S. array.
                  Image taken during the H.E.S.S. II inauguration ceremony on 28 September 2012.}
  \label{fig:hess2}
\end{figure}

Recently (July 2012), a fifth telescope has been added to the array, 
what greatly enlarges the observatory capabilities.
The  28~meter diameter telescope uses a parabolic-shaped mirror
to minimize time dispersion.  
The dish is composed of 875~hexagonal faces of 90~cm size, 
with a focal length 36~m.
The overall picture of the LCT is shown on figure \ref{fig:hess2}.

The LCT camera follows the design of the H.E.S.S. I cameras, but is much larger.
It is equipped with 2048~PMTs in~128 drawers.
The physical pixel size is 42~mm, which is equivalent to a~0.067$^{\circ}$ pixel FoV.
The LCT pixels have the same physical size as of the SCT, 
but, due to the larger focal length, shower images are much better resolved. 

The LCT is sensitive to photons down to 10~GeV.
In the normal operations, any of the SCTs  will be triggered only in 
case of a coincidence  with another telescope (LCT or SCTs). 
Low energy photons will not trigger the SCTs. 
To increase the acceptance of low energy photons, 
standalone LCT triggers will have to be accepted.

\section{The principle of the Cherenkov technique}
\label{sec:cherprinc}
\subsection{Cherenkov light emission} 

The Cherenkov light is emitted by a charged particle 
passing a dielectric medium with a velocity $v=\beta c$ greater than the phase velocity of light in that medium.
The particle threshold velocity for the Cherenkov light  production is:

\begin{equation}
\beta_{min}=\frac{1}{n},
\end{equation}

where $n$ is a refraction index of the medium.

This can be translated into an energy threshold, $E_{th}$, for the particle: 

\begin{equation}
E_{th}=\frac{m_0c^2}{\sqrt{1-\beta^2_{min}}}=\frac{m_0c^2}{\sqrt{1-n^{-2}}},
\end{equation}

where $m_0$ is a particle rest mass.

\begin{figure}[h]
\vspace{2mm}
\begin{center}
\hspace{3mm}\psfig{figure=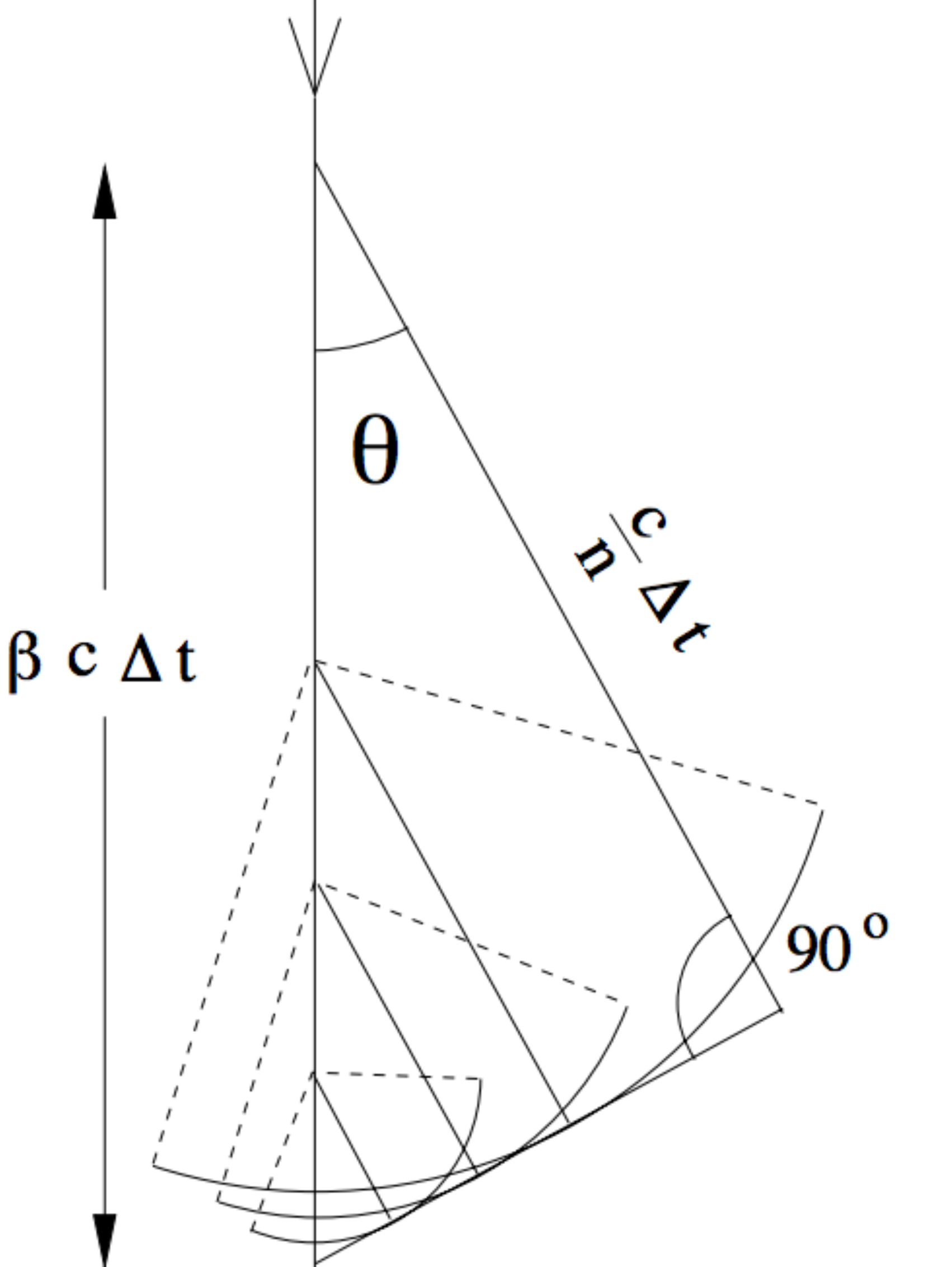,height=80mm,angle=0.0,bb=0 0 490 652}
\caption{Cherenkov light wavefront.}
\label{fig:wavefront}
\end{center}
\end{figure}

Cherenkov photons are emitted with a fixed angle $\Theta$ to the particle trajectory.
The angle  can be calculated using  the relation between the distance traveled by 
the particle and by the emitted radiation $t$ (see figure \ref{fig:wavefront}).

\begin{equation}
\cos \Theta = \frac{(c/n)/t}{\beta c t}=\frac{1}{\beta n}.
\end{equation} 

In the case of IACTs, the relativistic particle o interest  comes from 
the cascades initiated by cosmic rays  (CR) particles  in the Earth's atmosphere. 

\newpage
\subsection{Cosmic rays}
\label{sec:CR}

\begin{figure}[h]
\vspace{2mm}
\begin{center}
\hspace{3mm}\psfig{figure=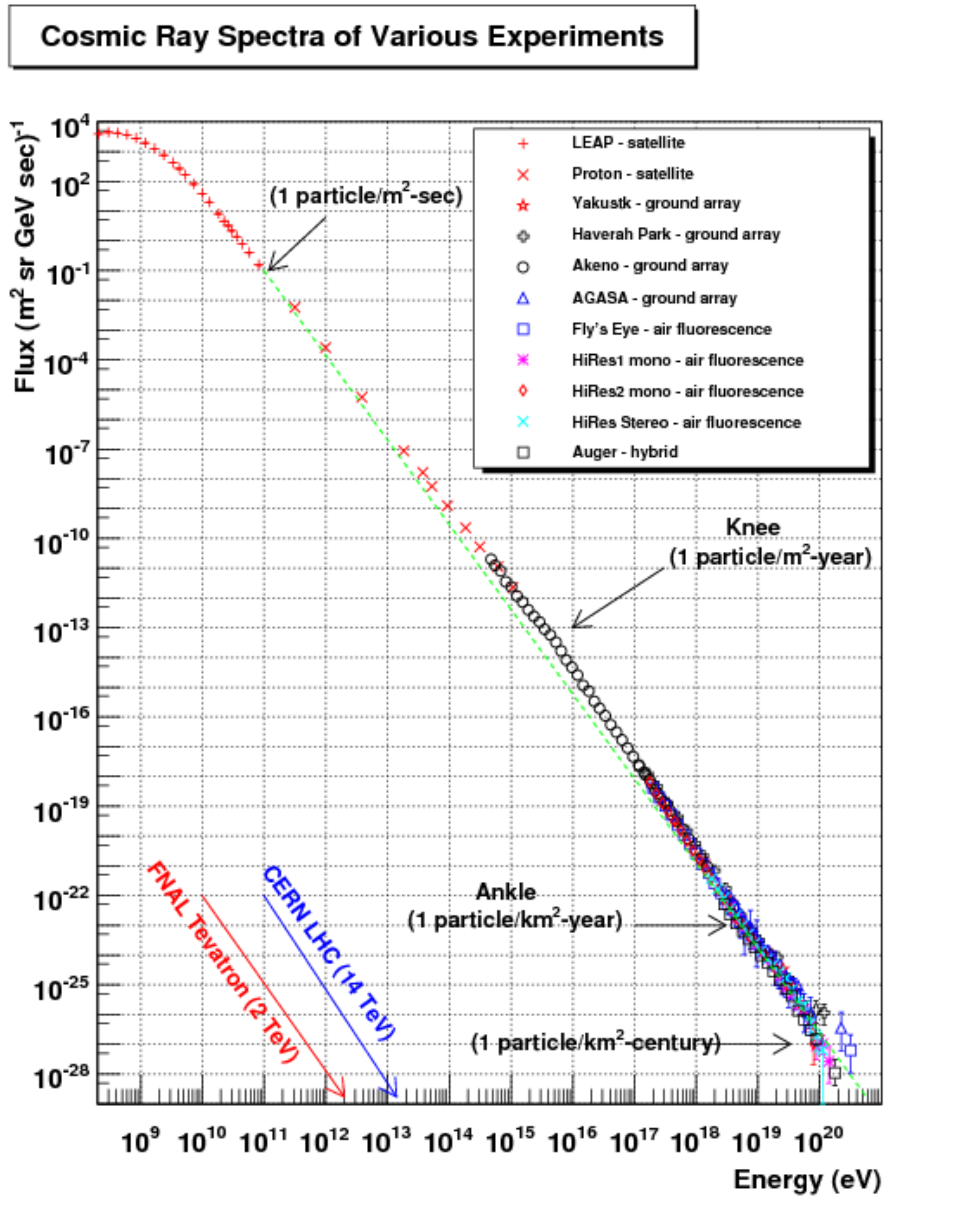,height=120mm,angle=0.0,bb=0 0 567 710}
\caption{Differential cosmic ray particle flux spectrum  as a function of energy. 
Figure shows the spectrum reported by different experiments. 
(Figure from http://www.physics.utah.edu/~whanlon/spectrum.html)}
\label{fig:CR_spectrum}
\end{center}
\end{figure}

Cosmic rays (CRs) are charged particles and atomic nuclei  arriving at the Earth location from the all directions.
They were discovered by the Austrian physicist Viktor Hess 
in a series of balloon experiments in the first decade of the XX$^{th}$ century. 
Viktor Hess was awarded the Nobel prize in 1936 for this discovery.  
The CR spectrum  is shown on figure \ref{fig:CR_spectrum}. 
The spectrum, measured at the top of the Earth's atmosphere, has the following composition: 98\% of the particle are protons and nuclei, the remaining 2\% are electrons. 
The protons and nuclei part is composed of 87\% protons, 12\% helium nuclei and 1\% heavier nuclei. 

\subsection{The atmospheric  air showers}
When the high energy CRs enter the Earth's atmosphere, 
they collide with O$_2$ and N$_2$ molecules and produce new energetic particles.
The generation of secondary particles 
starts at the  height of about 20~km above the sea level,
and continues until the depletion of the energy of the primary particle.
The set of generated particles is called an atmospheric air shower.
The~showers initiated by hadronic or leptonic primary particles or photons have different compositions due 
to the nature of the physical processes involved.  

Atmospheric showers caused by electrons or $\gamma$-ray photons develop as a pure electromagnetic cascades
through electron positron pair production and bremsstrahlung radiation. 
When gamma rays interact with nuclei they produce  $e^{\pm}$ pairs. 
In the next step, the electrons and positrons regenerate gamma rays by bremsstrahlung radiation.

The growth of hadronic cascade involves more types of possible interactions, 
thus results  in the production of 
a greater  variety of  secondary particles like pions, kaons, nuclei, etc. 
The vast majority of the secondary particles produced after the first interaction are pions ($\pi^+,\,\pi^-,\,\,\pi^0$). 
The charged pions decay into muons $\mu$ and neutrinos $\nu$:

\begin{equation}
\pi^{\pm} \rightarrow \mu^{\pm} + \nu_{\mu} (\bar{\nu}_{\mu}).
\end{equation} 

The muons have a life-time of $\Gamma\times \tau_{\mu}$ with  $\tau_{\mu} \sim 2.2 \times 10^{-6}$~s and Lorentz factor $\Gamma$,
which implies $c\Gamma \tau_{\mu}$=~1$\times\Gamma$~km.
Therefore, they can travel through the atmosphere and reach the Earth's surface.
The Cherenkov light of such muons can trigger the cameras of IACT 
if muons reach the distance of a few hundred meters above the telescope.
 In such a case, typical arc shaped images or  ring images  are observed.  

The neutral pions  decay with 99\% probability into  photons $\gamma$:
\begin{equation}
\pi^{0} \rightarrow \gamma + \gamma,
\end{equation} 
and initiate a pure electromagnetic sub-shower. 

\begin{landscape}
 \begin{figure}
  \centering
  \includegraphics[angle=0,width=210mm,bb=0 0 720 504]{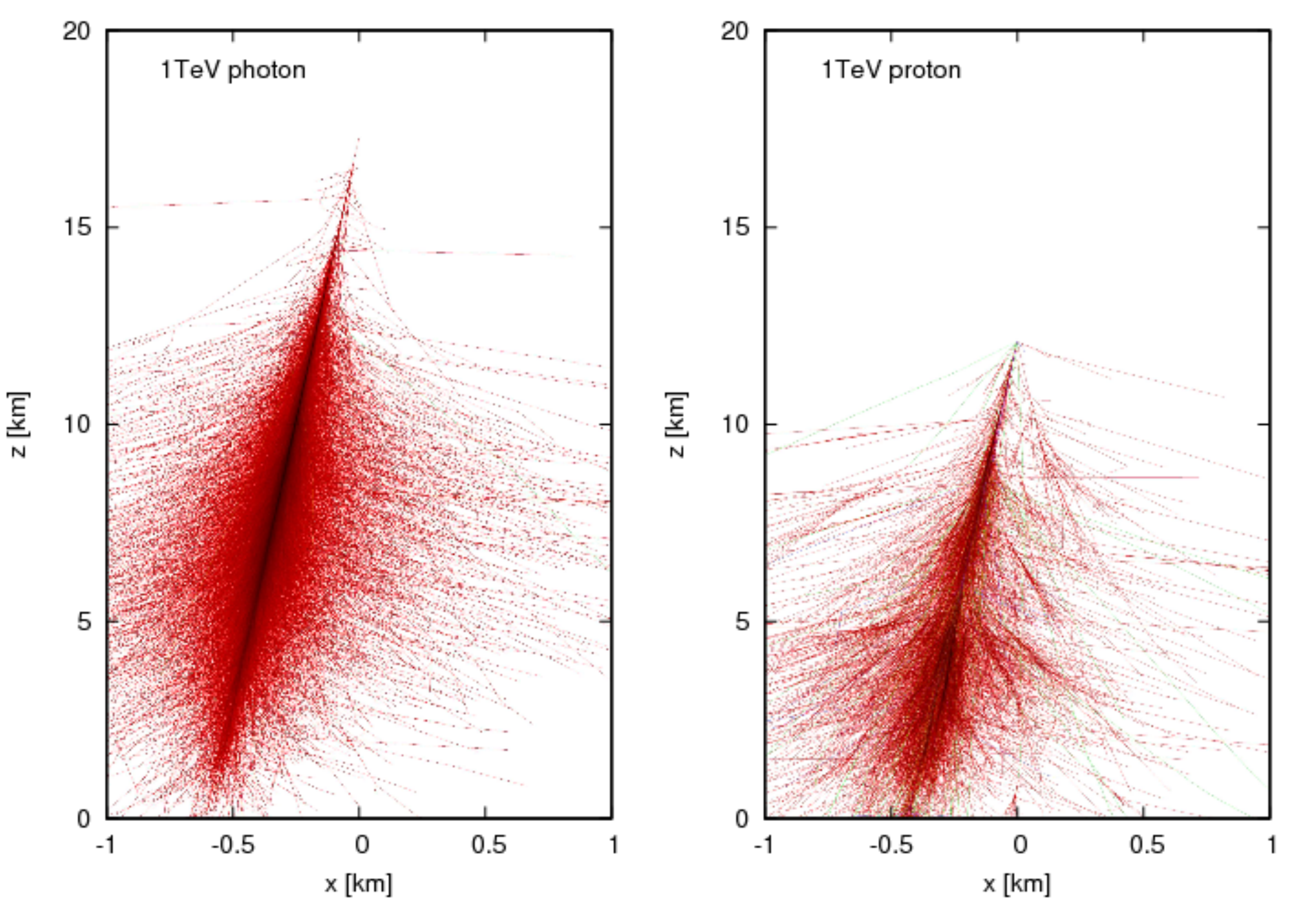}
  \caption{Two examples of extended air showers simulated with {\tt CORSIKA},
  developing in the Earth's atmosphere.  Each line represents a path of
    elementary particle created during shower evolution. The left panel
    presents the shower generated by a single photon of energy 1~TeV,
    while the right panel shows the shower originating from a single
    proton of the same energy.}
  \label{fig:showerexample}
\end{figure}
\end{landscape}

\begin{figure}[h]
\vspace{2mm}
\begin{center}
\hspace{3mm}\psfig{figure=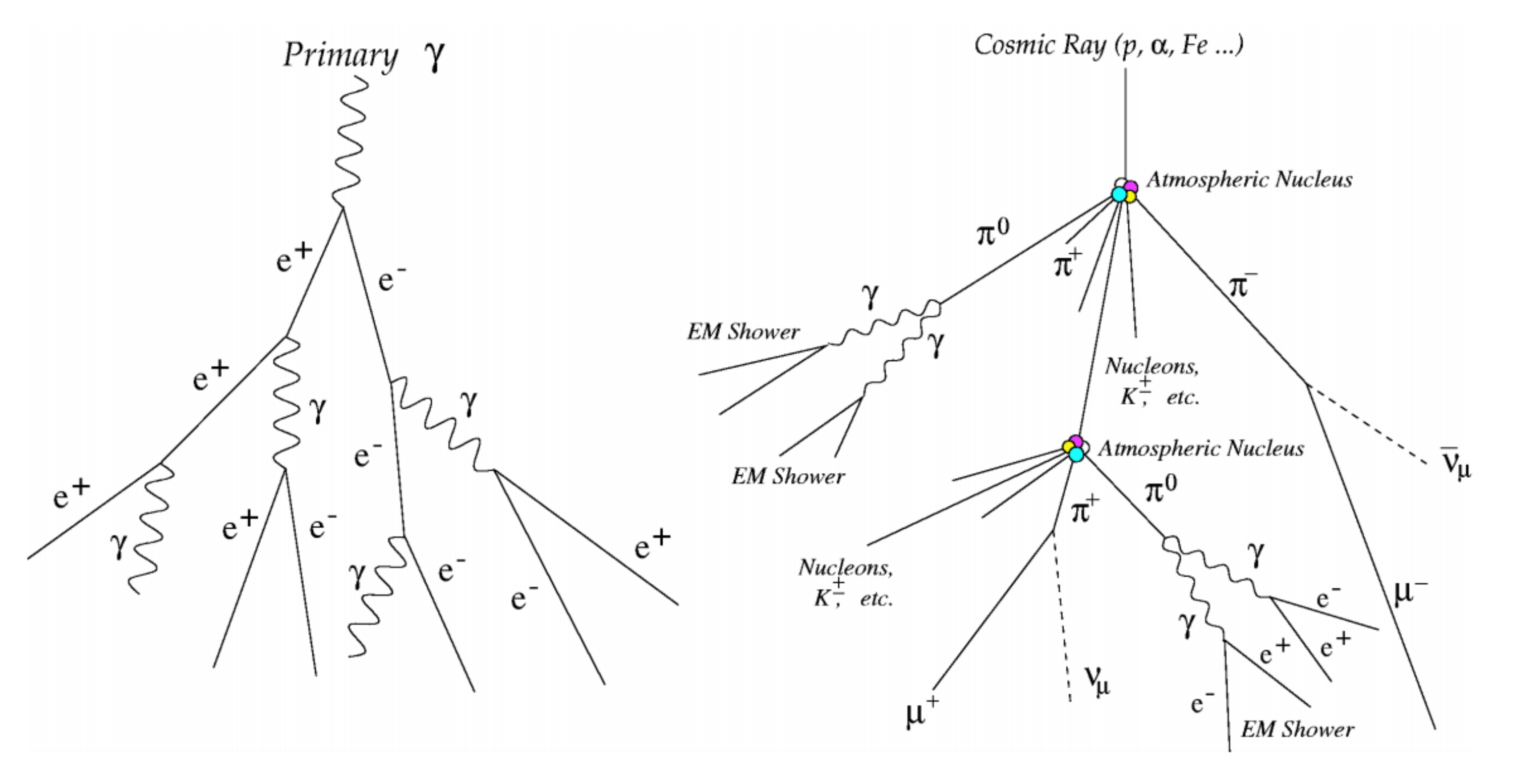,height=80mm,angle=0.0,bb=0 0 1211 622}
\caption{Schematic view of an electromagnetic (left) and a hadronic (right) air shower.}
\label{fig:cascade}
\end{center}
\end{figure}

In the case of electromagnetic showers, the vast majority of generated electrons 
 is well collimated with the shower axis.
 This make  the images of gamma showers very compact and regular. 
Hadronic showers are much less regular  and less compact, 
because the nature of secondary particle is hadronic as well as leptonic,
 and the secondary particles are less collimated with respect to the shower axis. 
Figures \ref{fig:cascade} and \ref{fig:showerexample} show  
two examples of the atmospheric air showers generated by 
the photon and proton, respectively.
 
\subsection{Air shower development in the atmosphere}
In the electromagnetic cascade the number of secondary particles 
is nearly proportional to the energy, $E$, of the primary particle.
After each radiation length $X_{r}$, the number of secondary particles increase by a factor of 2. 
After $n=X/X_{r}$ radiation lengths, the number of secondary particles is $N(X)=2^{X/X_r}$,
where X is the slant depth along the shower axis \citep{1991crpp.book.....G}.

The showers stop to develop when energy losses of secondary particles due to the pair production  
 and bremsstrahlung emission are smaller than their losses by ionization.
After this happens, secondary particles  are absorbed by the atmosphere.
The ionization energy loss, $E_{io}$, is about 2.2~MeV${\rm \,g^{-1}cm^{2}}$. 
The critical energy $E_{cr}$, below which the shower stops expanding is $E_{io}\times X_r$=~81~MeV,
where radiation length $X_r$ in air is equal to 37~${\rm g\,cm^{-2}}$.

\begin{figure}
\vspace{2mm}
\begin{center}
\hspace{3mm}\psfig{figure=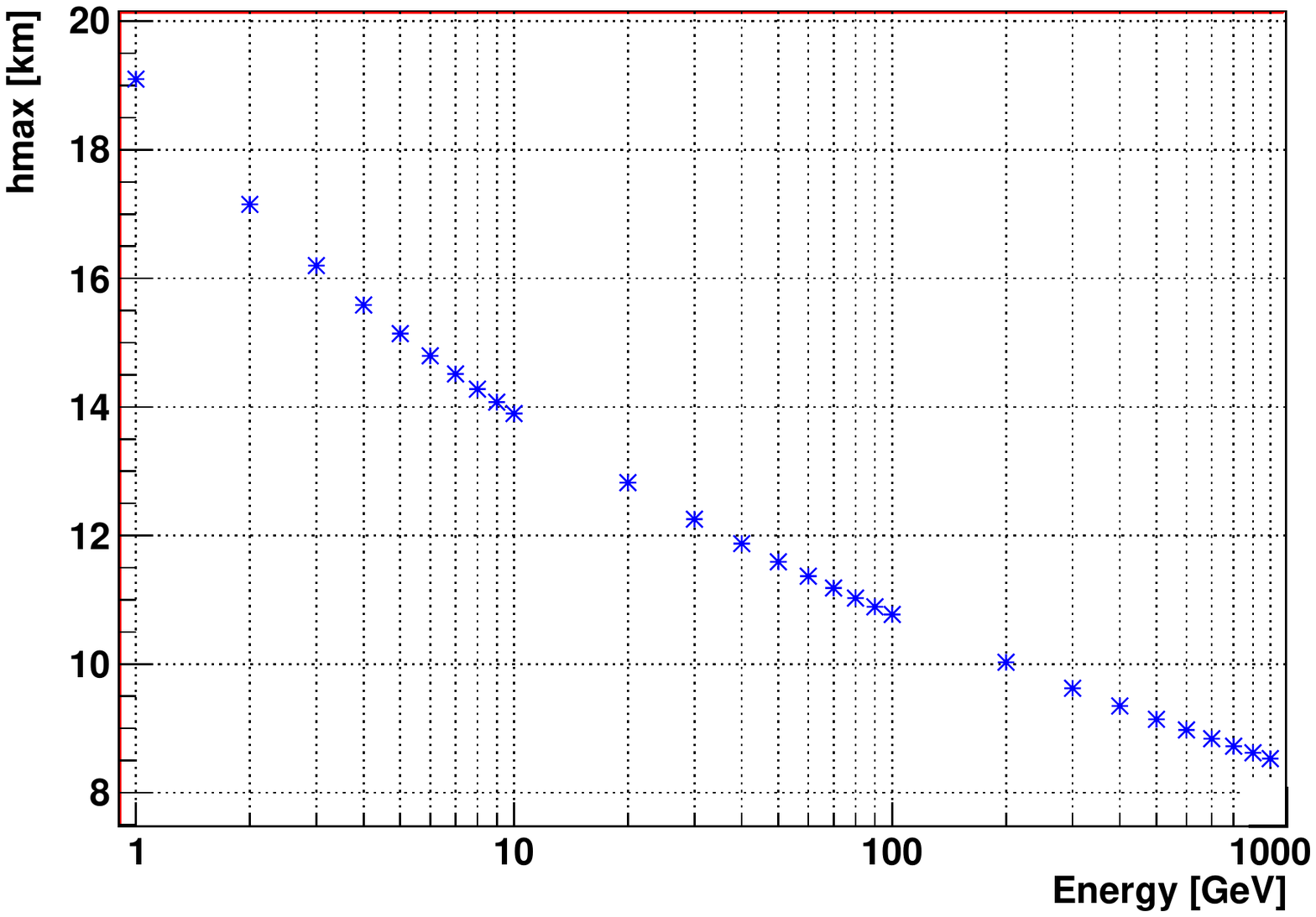,width=150mm,angle=0.0,bb=0 0 595 842}
\caption{Shower maximum height as a function of primary particle energy.}
\label{fig:hmax}
\end{center}
\end{figure}

The maximum number of particles in the shower, is reduced at the  shower maximum $X_{MAX}$:
\begin{equation}
X_{MAX}=n  X_0  {\rm ln}(2)= X_0  {\rm ln}\left(\frac{E}{E_{cr}} \right)\,.
\label{eq:xmax}
\end{equation}
The atmospheric depth, given in units of [${\rm g\,cm^{-2}}$], 
corresponds to an atmospheric heigh, $h$, in km:
\begin{equation}
X= X_1 e^{-\frac{h}{h_1}}   \,,
\label{eq:xh}
\end{equation}
where $X_1=1013\,{\rm g\,cm^{-2}}$ and $h_1=8\,$km. 
Figure \ref{fig:hmax} shows the shower maximum height as a function of energy, 
assuming that the maximum of Cherenkov emission corresponds to the shower maximum $X_{MAX}$. 

\subsection{Cherenkov light distribution}
\begin{figure}[h]
\vspace{2mm}
\begin{center}
\hspace{3mm}\psfig{figure=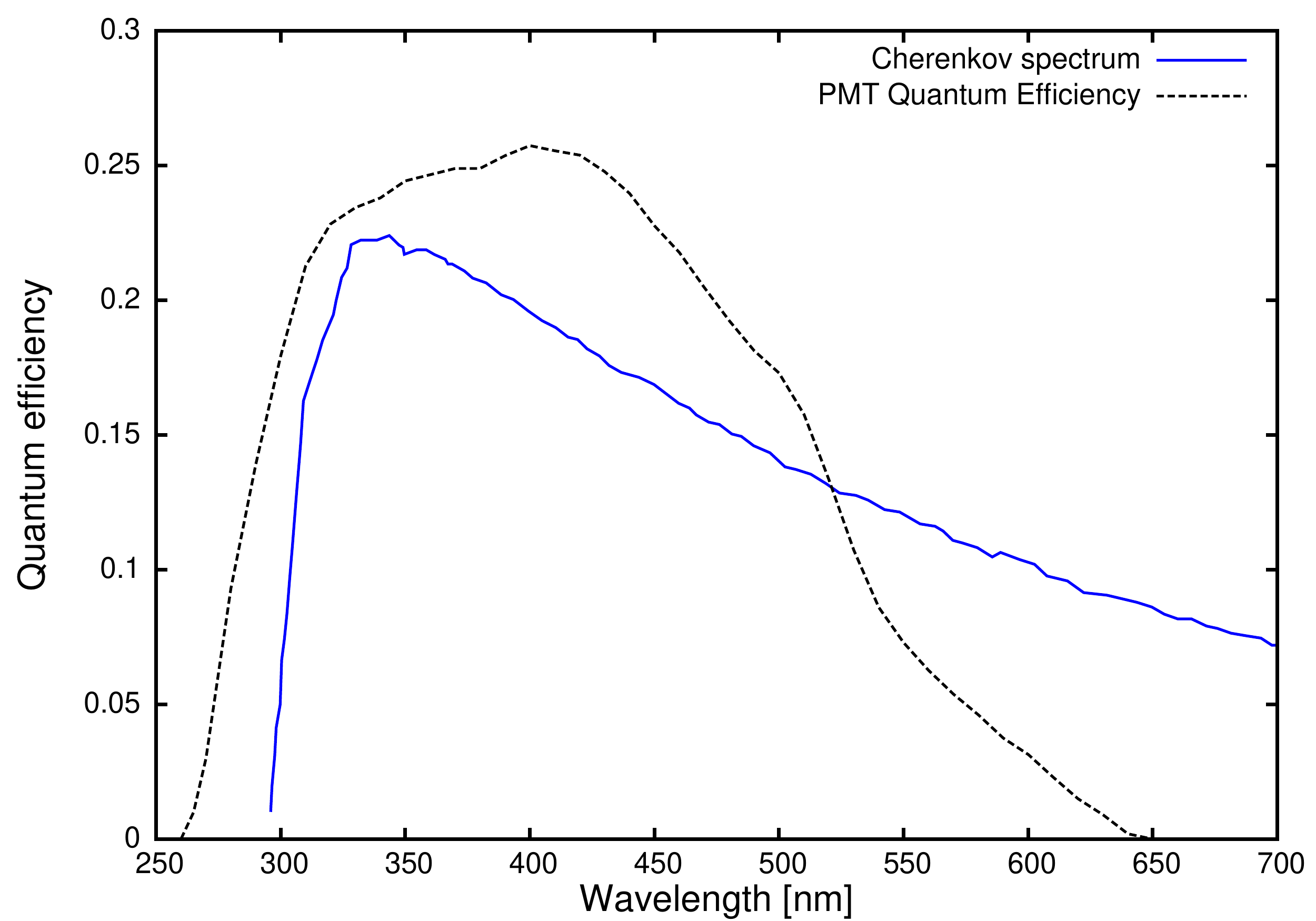,width=100mm,angle=0.0}
\caption{The blue line represents the Cherenkov light spectrum for a $0^o$~zenith 
                angle detected at 2000~m above see level.
	       The spectrum is shown after atmospheric absorption.
                The black dotted line shows the quantum efficiency of PMTs used in the H.E.S.S. camera.}
\label{fig:qu}
\end{center}
\end{figure}

The number of Cherenkov photons per unit of track length of the particle and per 
unit of wavelength is given by the Frank-Tamm formula:
\begin{equation}
\frac{d^2N}{dl\,d\lambda}=\frac{2\pi \alpha Z^2}{\lambda^2}\sin^2\Theta = \frac{2\pi \alpha Z^2}{\lambda^2}\left(1-\frac{1}{\beta^2n^2(\lambda)}\right) \,,
\end{equation}
where $\alpha=e^2/\hbar c$ is the fine structure constant, and $Z$ is the charge of the particle in units of the elementary charge.  
For  atmospheric air showers
the maximum intensity of the Cherenkov light emission corresponds to UV and blue light (300-700~nm).
For shorter wavelengths it is cut off by the decrease of $n(\lambda)$. 
The cut off appears before X-rays because $n($X-rays$)<1$ in all materials. 
In addition, the Cherenkov light  is strongly absorbed in the atmosphere before reaching the ground.
The atmospheric absorption is more efficient toward the short wavelengths.
This effect modifies  significantly the observed spectrum. 
Figure \ref{fig:qu} shows the spectrum of Cherenkov light emitted by a particle  at 0$^\circ$ zenith angle. 
The spectrum includes  atmospheric absorption and for the comparison it is shown together with the quantum efficiency of PMT used by IACTs.

The electromagnetic cascades of atmospherical showers 
are initiated at $\sim 22 \pm 4$\,km above the see level. 
Then, depending on the primary particle energy, the shower reaches its maximum between 15~km and 5~km. 
The charged secondary particles traversing the atmosphere emit  Cherenkov photons 
with the cone angle $\Theta \sim1^\circ$.  
Cherenkov photons emitted at 10~km produce a ring  of 
radiation at the ground level with a radius of $\sim$120~m, centered on the particle trajectory.  
The shape of the ring depends also on the shower axis angle.
The Cherenkov light distribution become more diffuse if the initial particle had a larger zenith angle.  

The majority of Cherenkov photons, emitted between the first interaction 
and shower maximum, will arrive approximately within 120~m of the shower core. 
However, Cherenkov photons may have a significant flux many hundreds of meters 
from the shower axis.
This is a consequence of the angular distribution of particles and the scattering of Cherenkov light.

Two examples of the Cherenkov light density distributions as a function of distance from the shower axis (impact parameter) 
are presented in figures \ref{fig:CLD1} and \ref{fig:CLD2}, 
for $\gamma$-ray initiated showers and proton initiated showers,  respectively. 
The distributions have been obtained using the  \texttt{CORSIKA} package \citep{1993AIPC..276..545C}.
The light density  of   Cherenkov photons in the wavelength range  250-700~nm were  simulated for gamma and proton showers. 
The magnetic field has been set in simulations  for the H.E.S.S. site, and zenith angles $0^\circ$, azimuth angle $90^\circ$, and an altitude of 2000~m. 

The Cherenkov light distribution of gamma showers shows a very regular structure with a characteristic bump at 120~m. 
The average photon density of proton showers is 3 times smaller than for the gamma showers of the same energy,
since, on average  only third part of energy of hadronic shower goes to electromagnetic sub-showers. 

\begin{landscape}

  \begin{figure}[h]
\vspace{2mm}
\begin{center}
\hspace{3mm}\psfig{figure=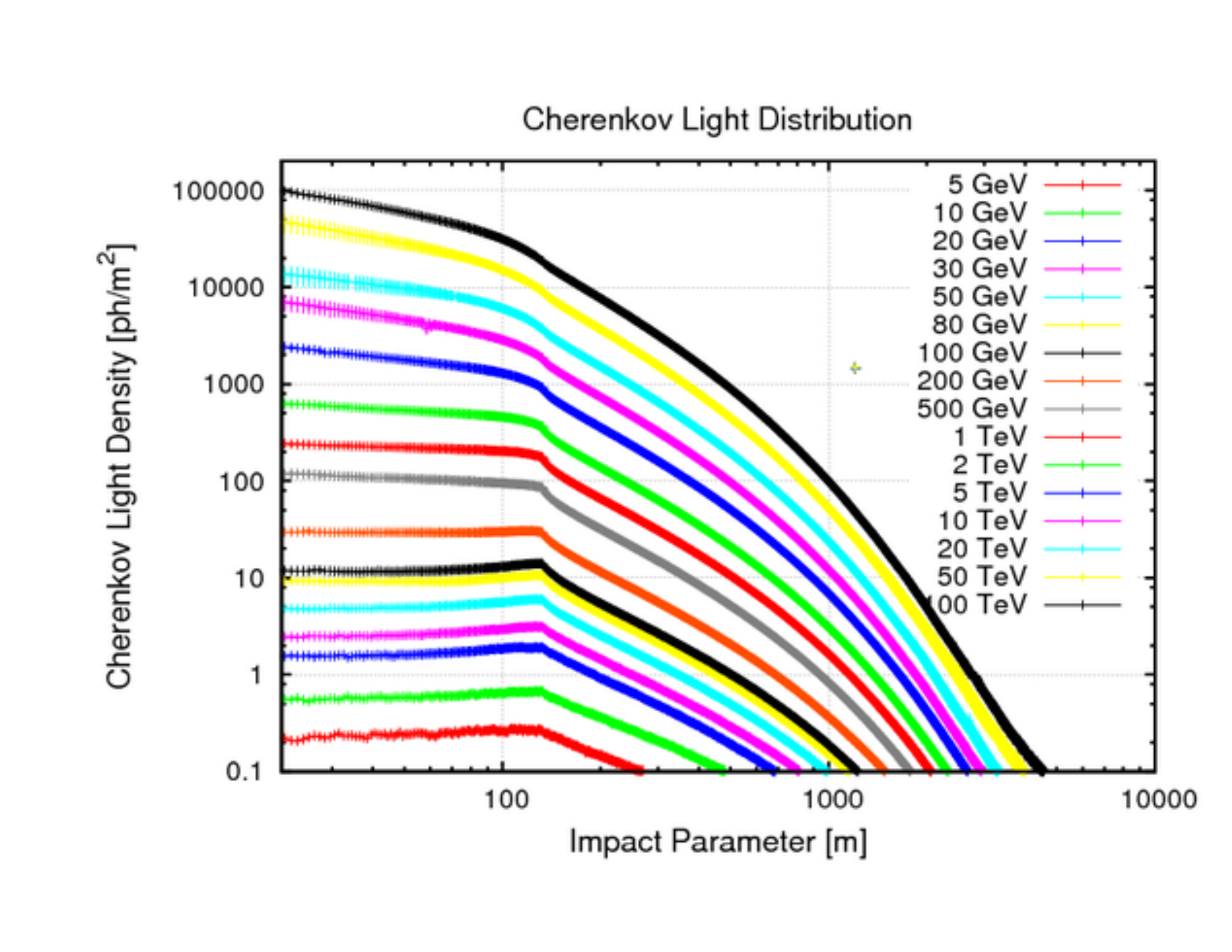,height=150mm,angle=0.0,bb=0 0 634 490}
\caption{Cherenkov light distribution: $\gamma$-ray initiated showers.} \label{fig:CLD1}
\end{center}
\end{figure}

  \begin{figure}[h]
\vspace{2mm}
\begin{center}
\hspace{3mm}\psfig{figure=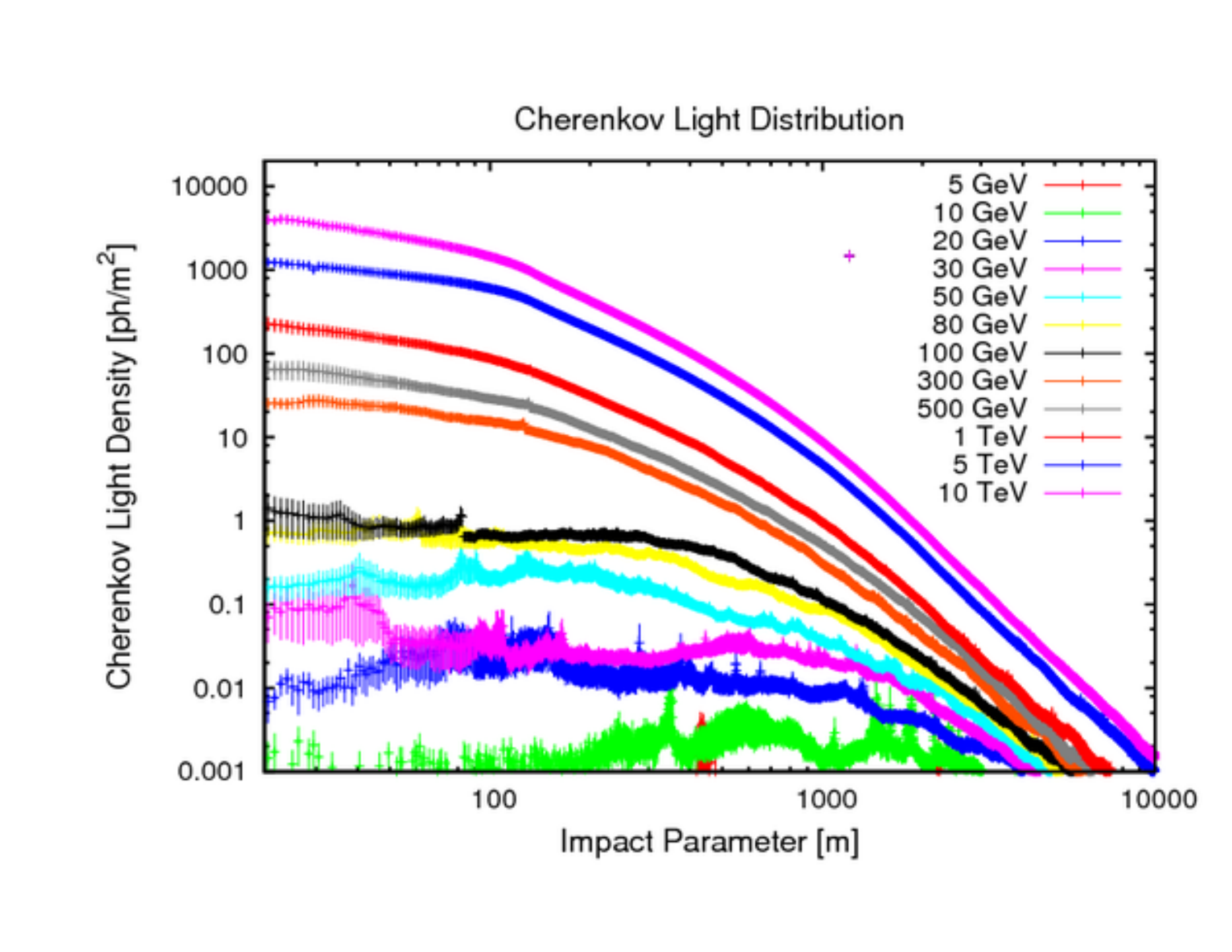,height=150mm,angle=0.0,bb=0 0 634 490}
\caption{Cherenkov light distribution: proton initiated showers.}
\label{fig:CLD2}
\end{center}
\end{figure}

\end{landscape}

\subsection{Shower geometry}
\label{sec:ShowerGeometry}

The Cherenkov light emitted by the atmospheric shower is observed in the focal plane of  ground based  instruments. 
The image of gamma showers  in the camera has an elliptical shape which can be characterized 
by  Hillas parameters \citep{1985ICRC....3..445H}.  
The Hillas parameters are obtained by calculating  moments of 
the photo-electron (phe) distribution in the camera. 
The   most comonly used parameters in the image analysis are 
\texttt{Size}, \texttt{Length}, \texttt{Width}, \texttt{Alpha} and \texttt{Dist}. 
These parameters are shown on figure \ref{fig:shower_geometry}. 

    \begin{figure}
\vspace{2mm}
\begin{center}
\hspace{3mm}\psfig{figure=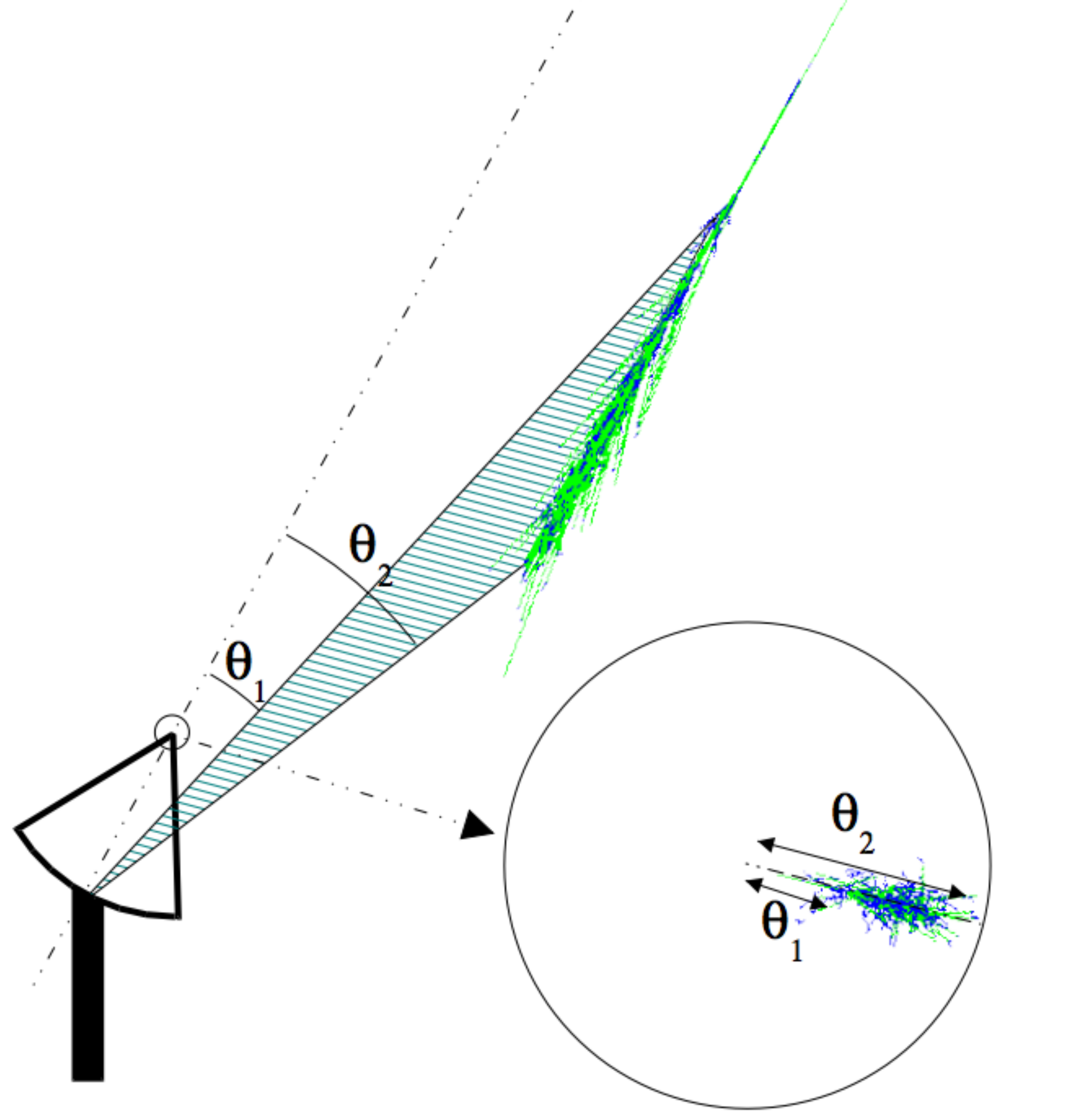,height=80mm,angle=0.0,bb=0 0 670 682}
\caption{The principle of the Cherenkov imaging technique  \citep{GuyThesis}.}
\label{fig:shower_geometry}
\end{center}
\end{figure}

  \begin{figure}
\vspace{2mm}
\begin{center}
\hspace{3mm}\psfig{figure=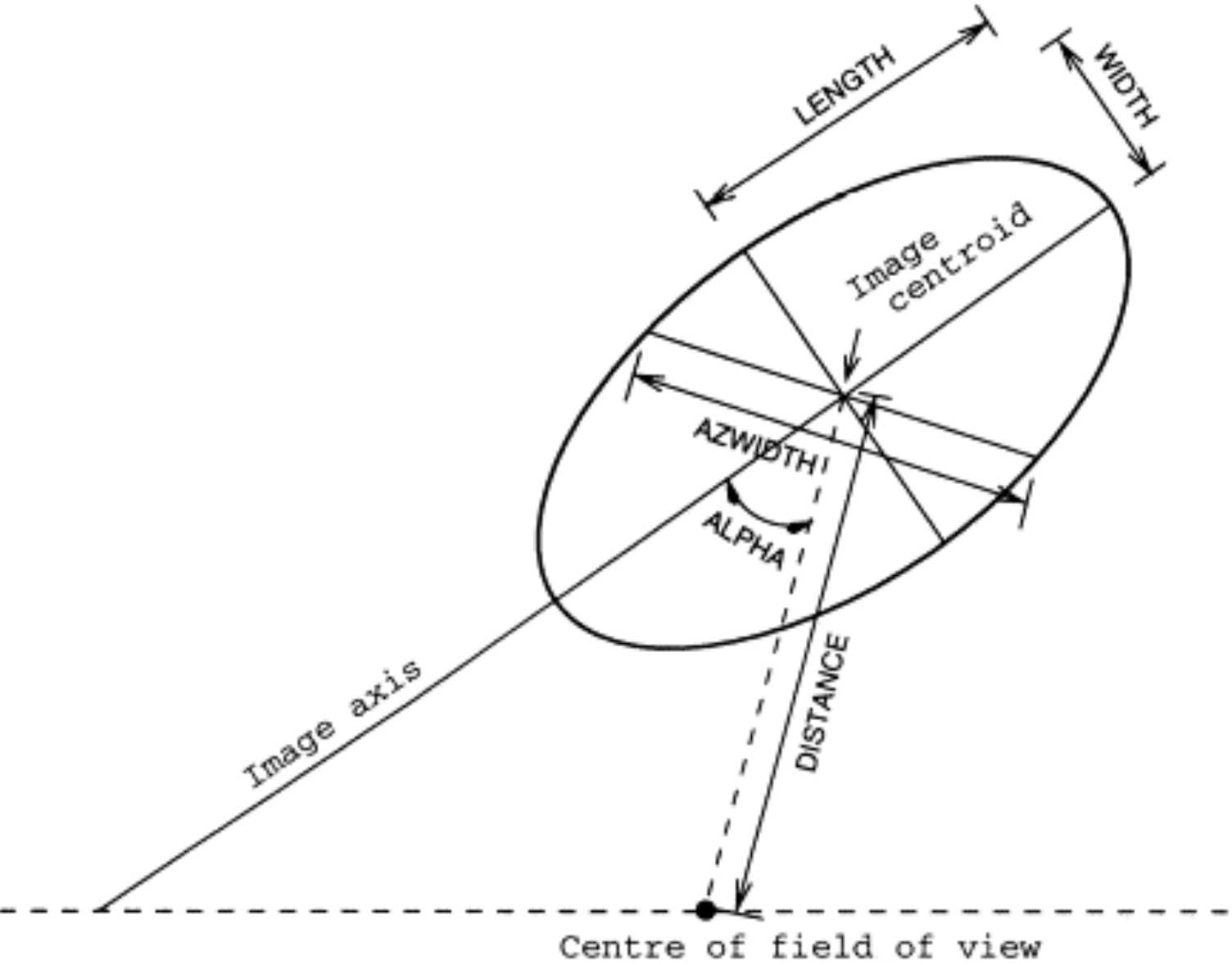,height=80mm,angle=0.0,bb=0 0 433 339}
\caption{The elliptical contour of a typical Cherenkov light image as seen in the focal plane of camera of IACTs.}
\label{fig:shower_geometry}
\end{center}
\end{figure}

The \texttt{Size} parameter  is the total number of photo-electrons in the image and 
is roughly proportional to the energy of the primary particle, it is also called the {\tt Amplitude}. 
 The second moment  of the phe distribution  with substracted  {\tt COG} (the image center of gravity) 
 along the major image axis is the \texttt{Length}, 
and along the minor axis is the \texttt{Width}.
The \texttt{Alpha} is the angle between the direction of the major axis and the line joining 
the image centroid with the source position.

\begin{figure}
\vspace{2mm}
\begin{center}
\hspace{3mm}\psfig{figure=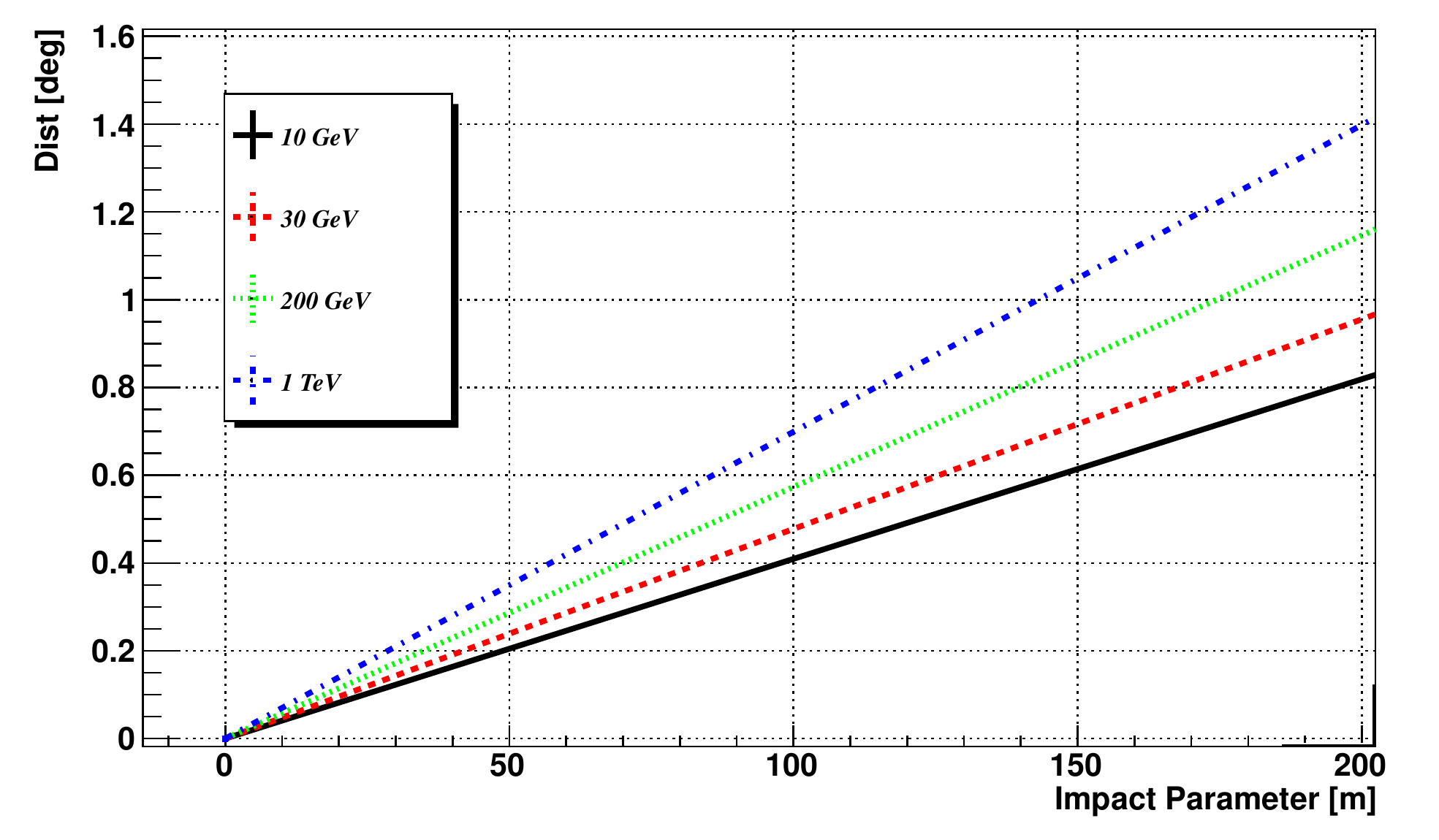,height=80mm,angle=0.0}
\caption{ \texttt{Dist} parameter as a function of the impact parameter 
                 for different energies of the primary gamma photon.}
\label{fig:Disp1}
\end{center}
\end{figure}

The \texttt{Dist} is the distance between the image centroid and the source position in the camera plane.
The  \texttt{Dist}  parameter is correlated with the distance of the shower to the telescope axis (impact parameter). 
The correlation between \texttt{Dist} and impact parameter (IP)  is presented in figure \ref{fig:Disp1}. 
The correlation is energy dependent because the image maximum 
for different energies of primary particle appears at different heights (see figure \ref{fig:Disp1}). 
 The relation between the energy and shower maximum height has been calculated from equations (\ref{eq:xmax}) and (\ref{eq:xh}).

\section{Analytical estimation of the energy threshold of the H.E.S.S. LCT}
\label{sec:AnalyticalThreshold}
The theoretical energy threshold of the H.E.S.S. telescope can be 
calculated using the~Cherenkov light distribution 
and the overall performance parameters of the telescope. 
The minimum photon density, $\rho_{ph}$, required to obtain the signal of at least $N_{phe}$  is defined as: 
\begin{equation}
\rho_{ph}= \frac{N_{phe}}{(A-A_C)\times R \times QE }
\label{eq:th}
\end{equation}
where $A$ is a mirror effective area, $A_C$ is the camera and masts shadowing,
$R$  is  system reflectivity, $QE$ 
is  integrated detector efficiency weighted by the Cherenkov light spectrum.
\begin{table}
\center
\begin{tabular}{|l|c|c|}
\hline
\hline
                                                            & H.E.S.S. SCT & H.E.S.S. LCT \\
\hline
Mirror effective area $A[$m$^2]$      & 113 & 600  \\
Camera shadowing $A_C[$m$^2]$    & 1 & 1  \\
System reflectivity $R$                   & 0.8  & 0.8 \\
Average photodetector QE              & 0.1  & 0.1 \\
\hline
\hline
\end{tabular}
\caption{The parameters of the H.E.S.S. telescopes.}
\label{tab:hess}
\end{table}

Typically 50~--~100 phe are necessary to perform an image reconstruction. 
Using equation (\ref{eq:th}) and the telescope specifications listed in table \ref{tab:hess}, 
one can derive the photon density required to detect certain number of photo-electrons. 
For the H.E.S.S.~I telescopes, a density of $\sim$5.5--11$\,{\rm ph\,m^{-2}}$  
is required to detect 50~--~100 photo-electrons.
This density corresponds to an energy of 60~--~100~GeV,  
according to figure \ref{fig:CLD1}.
This is the energy threshold of  the H.E.S.S.~I detector. 

The H.E.S.S.~II telescope has a reflection area more than 5~times larger than the H.E.S.S.~I telescopes.
This allows to collect enough photons from a much smaller  signal.
In the case of the H.E.S.S.~II, a minimum photon density of $\sim$1--2\,${\rm ph\,m^{-2}}$ 
is required to detect 50~--~100~photo-electrons. 
This photon density correspond to gamma shower energies of 10~--~20 GeV. 

  \begin{figure}
\vspace{2mm}
\begin{center}
\hspace{3mm}\psfig{figure=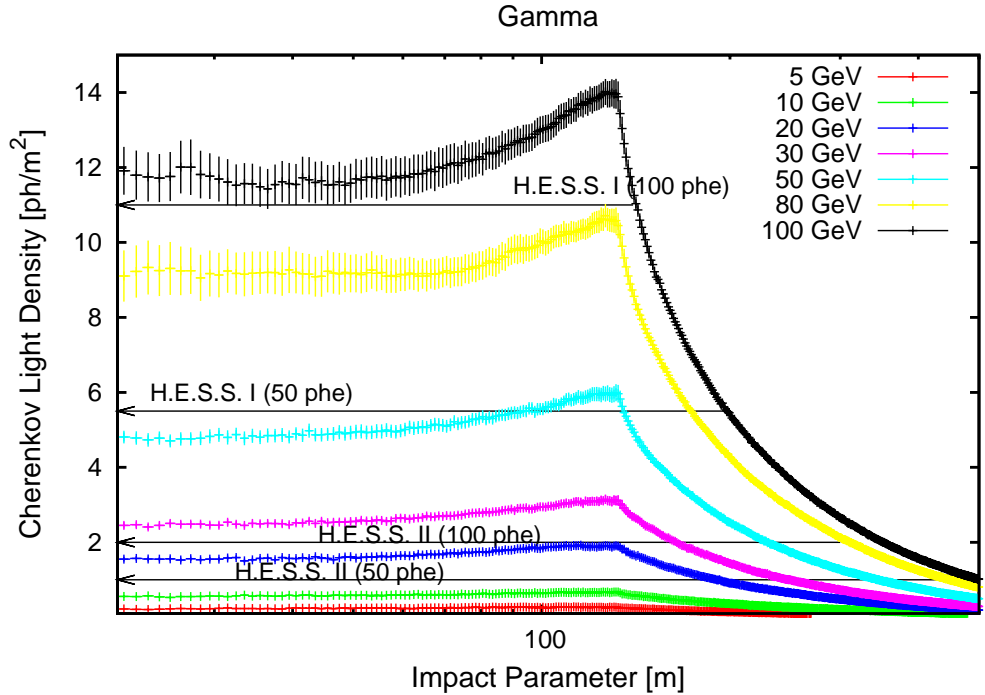,width=150mm,angle=0.0}
\caption{Cherenkov light distribution for low energy gamma showers. 
                The arrows indicate the minimum photon density required 
                to detect 50~--~100 phe by the H.E.S.S.~I and the H.E.S.S.~II systems.}
\label{fig:CLD_low}
\end{center}
\end{figure}

The numbers quoted above refers to the trigger threshold, and are estimated 
for  observations at zenith angle~$0^o$.
The numbers give a rough estimate of the system performances. 
The LCT will work alone in the energy range from 10~GeV to 60~GeV. 
The telescopes should work efficiently  together in stereoscopic mode for energies  above 60~GeV. 
In the case of observations at larger zenith angle, the energy threshold for the stereoscopic trigger  is larger. 
 
\section{Trigger system of the H.E.S.S. II telescope}
\label{sec:trigger}
The trigger system of  the H.E.S.S. II will operate at three levels: 
the Level 1 trigger (camera level), the Level 2 trigger (LCT level) and the stereoscopy (array level). 
In addition to time coincidences between SCT Level 1 triggers, 
the central trigger system will check  for time coincidences of LCT and SCTs triggers. 
The result of the latter coincidence test (monoscopic or stereoscopic event)
 will be  sent back to the LCT trigger management. 
 As in H.E.S.S. I, stereoscopic events will always be accepted. 

\subsection{Level 1 trigger }
\label{sec:l1trig}
The LCT has a Level 1 trigger similar to the Level 1 trigger of the four small  telescopes. It is a local camera trigger
described in  details by \cite{2005AIPC..745..753F}.
A camera Level 1 trigger occurs if the signals in M pixels (pixel multiplicity) of a camera sector,
 exceed a threshold of N photoelectrons (pixel threshold). 
 Each  sector consists of 64~pixels. 
 The LCT camera was divided to   96~overlapping sectors  to ensure trigger  homogeneity.
The effective time window for coincidence is~1.3 ns. 
 
 \subsection{Level 2 trigger}
The small Cherenkov telescopes are not equipped with a Level 2 trigger, 
since they do not operate in mono mode.
The LCT was build to lower the energy threshold of triggered gamma events. 
Normally, the background rejection is achieved in the stereoscopic mode 
when more than one small telescope is triggered at the same time as the large telescope. 
The stereoscopy  with the large telescope 
will allow to lower the  energy threshold down to 50~--~60~GeV 
(as was discussed in section \ref{sec:AnalyticalThreshold}). 
The LCT has to work in mono mode below this energy range.
The mono mode does suffer from high trigger rates caused by single muons.
The solution with Level 2 trigger has been proposed for LCT to reduce the trigger rate.

 \subsection{Stereoscopy}
 
\begin{figure}[h]
\vspace{2mm}
\begin{center}
\hspace{3mm}\psfig{figure=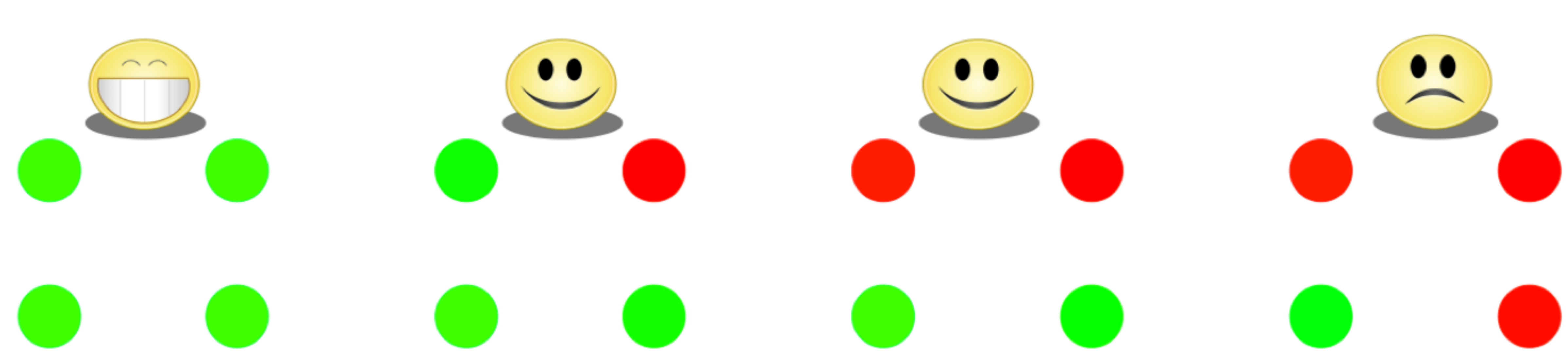,height=30mm,angle=0.0, bb=0 0 1379 315}
\caption{The array of 4 telescopes requires a coincidence of at least two SCTs. 
		Red color indicates  not triggered telescope, white green color indicates triggered telescope. }
\label{fig:stereo1}
\end{center}
\end{figure}

\begin{figure}[h]
\vspace{2mm}
\begin{center}
\hspace{3mm}\psfig{figure=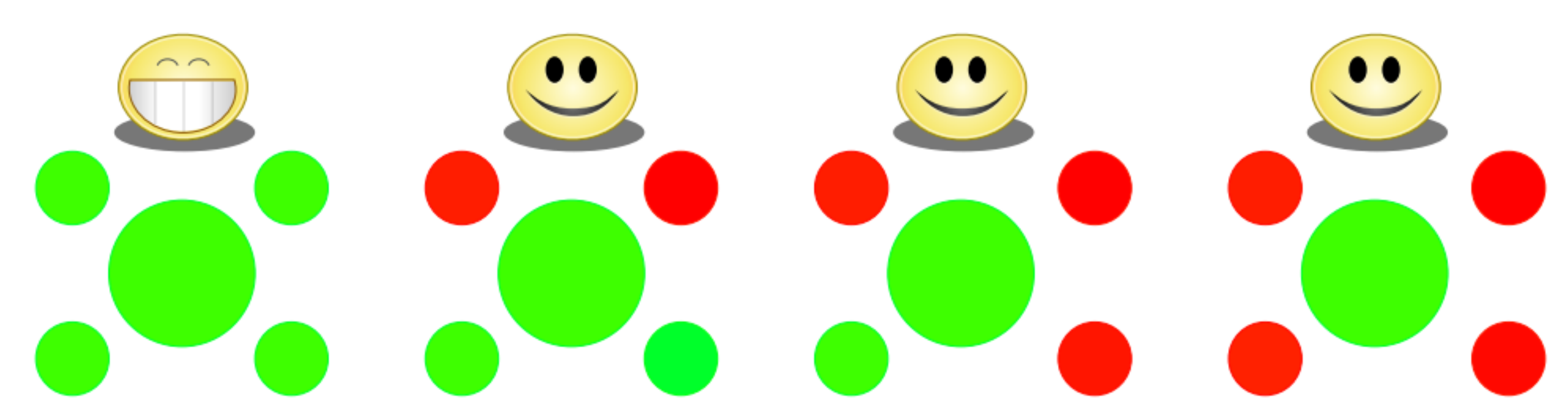,height=40mm,angle=0.0,bb=0 0 995 261}
\caption{The configuration of the full array (4 SCT + 1 LCT) required to fire the central trigger.
		Red color indicates  not triggered telescope, white green color indicates triggered telescope. }
\label{fig:stereo2}
\end{center}
\end{figure}

The step after the camera trigger level (Level 1 trigger) is the so-called central trigger. 
The central trigger  system looks for coincidences of telescope triggers inside a 40~ns
time window. 
A coincidence of at least 2 telescopes is required in the central trigger time window. 
LCT monoscopic events are accepted or rejected depending on the result of Level 2 system evaluation. 
The possible configurations of stereoscopy are shown on figures~\ref{fig:stereo1} and~\ref{fig:stereo2}.


\section{Algorithms for the Level 2 trigger}
\label{sec:algosoft}

\subsection{Requirements for the Level 2 trigger}
The input rate to the Level 2 trigger is limited to less than roughly 100 kHz by 
the dead-time of the front-end readout board. 
In turn, the output rate is limited to a few kHz by the capacity of the ethernet
connection to the acquisition system.  

Table \ref{tab:nsb} shows the trigger rates caused by the NSB for different pixel multiplicities and thresholds. 
The input rate  gives a strong constraint on possible Level 1 trigger conditions: pixel multiplicity and pixel threshold. 
The total particle and NSB rate is  at the level of a few kHz. 
A further reduction of this rate by a factor of 2 or 3 allows to fulfill 
the output rate condition even in very noisy environments. 

\subsection{Principle of the Level 2 trigger}
The idea of the Level  2 trigger is to have the whole trigger information
at the pixel level, instead of the sector level as in the Level 1 trigger.
A reduced, 3-level image of the camera,  so-called ``combined map'', 
is sent to the Level 2 trigger system
whenever the LCT has an confirmation of  Level 1 trigger. 

The combined map consists of two black and white images of the camera, 
with 3 possible pixel intensities, 
which are 0 (when a pixel is below the threshold), the Level~1 pixel threshold and another higher pixel threshold. 
The black and white image obtained by taking only the Level~1 threshold information
 (resp. the Level~2 threshold information)
is called ``Level 1 map'' (resp. ``Level 2 map'').

The background rejection is performed by a dedicated  algorithm,
described in detail in subsections \ref{sec:denoising} or \ref{sec:topological}. 
Since stereoscopic events should always be accepted, 
the Level 2 trigger operates differently on stereoscopic and monoscopic events. 
When the Level 1 trigger of the LCT occurs, the central trigger checks, 
if another telescope was triggered. 
If this is the case, then the event is accepted by  the Level 2 system. 
If on the contrary the event is monoscopic, the decision depends on the Level 2 trigger algorithm.  

\subsection{Approach to the algorithm }

For monoscopic events the trigger rate can be reduced with a  two step procedure.
The first step rejects NSB events, which have been accepted by Level 1 system
in the procedure called clustering/denoising. 
The second step  lowers the rate caused by the particle background events (protons, muons, electrons),
through the topological algorithms.
The crucial requirement  is to keep as many gamma events as possible during each of the above steps. 

\subsection{Clustering/denoising}
\label{sec:denoising}

The NSB consist of photons from stars and a diffuse light. 
Therefore, no correlation is expected between the pixels illuminated by NSB.
These events can be rejected requiring pixels with signals above the pixel threshold 
from a cluster of neighboring pixels,
the so-called "clustering" condition. 
Most of the pixels fired by NSB photons are isolated, and they can be removed by a step called "denoising".
 
The denoising algorithm removes all the isolated pixels from the Level 1 map. 
If the resultant map is empty then the event is rejected. 
There are several possible clustering algorithms. 
One variant simply demands  a group of 2 or 3 neighboring pixels above a trigger threshold. 
The effect of the denoising/clustering on the trigger rate cased by NSB for a cluster of 
at least 2 pixels above the threshold is illustrated in table \ref{tab:nsb}.
The NSB trigger rates decrease by large factors, in some cases by several order of magnitude 
(see e.g. the trigger rates  in table \ref{tab:nsb} for a pixel threshold of 3 photoelectrons and multiplicity of 3 pixels). 
The efficiency of the clustering/denoising algorithm allows to decrease     
the Level 1 trigger threshold and thus to reach a smaller photon energy threshold.

Protons, electrons, and total particle rates as a function of trigger condition are displayed on figures 
\ref{fig:protonM4}, \ref{fig:electronM4} and \ref{fig:allM4}, respectively. 
These rates are little affected by the clustering cut. The electron rate is dominated by low energy events, 
so that most electron events will trigger only the LCT. 

\subsection{Topological algorithms} 
\label{sec:topological}
The  topological algorithms  rely on the
fact that the images of showers observed in the camera plane have a characteristic shape.
The images of gamma-like events are well defined  by the Hillas parameters described in section \ref{sec:ShowerGeometry}. 
The images created by hadrons have much less regular shape, and thus Hillas parameters 
 can be used to separate the electromagnetic from hadron-like showers. 
The single muons created in the hadronic cascade produce a very characteristic ring or arc shapes. 

Therefore, it is worth investigating, which of the Hillas parameters 
can be used in the trigger to reject hadron like events, without losing too many gamma events. 
The time duration of the shower depends on the primary particle energy and the  impact parameter. 
The shower event on the ground can  lasts from a few to dozen of  ns in the case of very energetic events. 
The maps processed by the Level 2 trigger contain the signal integrated in $\sim$~1~ns,
so it contains only a fraction of the shower image. 

Figures \ref{fig:dst30GeVC13} and \ref{fig:dst100GeV}   
show shower parameters ({\tt Width, Length, Amplitude, COG, Length/Size, Width/Length})  computed for the signal  integrated over 16~ns  
 compared to the one calculated from 1~ns Level 2 combined maps. 
The comparison is presented for 30~GeV and 100~GeV simulated gamma showers.
The figures show barely any correlation for the \texttt{Width} and \texttt{Length} parameters,
but there are strong correlations for the \texttt{COG} and the \texttt{Amplitude} parameters.
Therefore, the \texttt{COG} or the \texttt{Amplitude} cuts can be used to further reduce the hadron rate.

\begin{landscape}

  \begin{figure}[h]
\vspace{2mm}
\begin{center}
\hspace{3mm}\psfig{figure=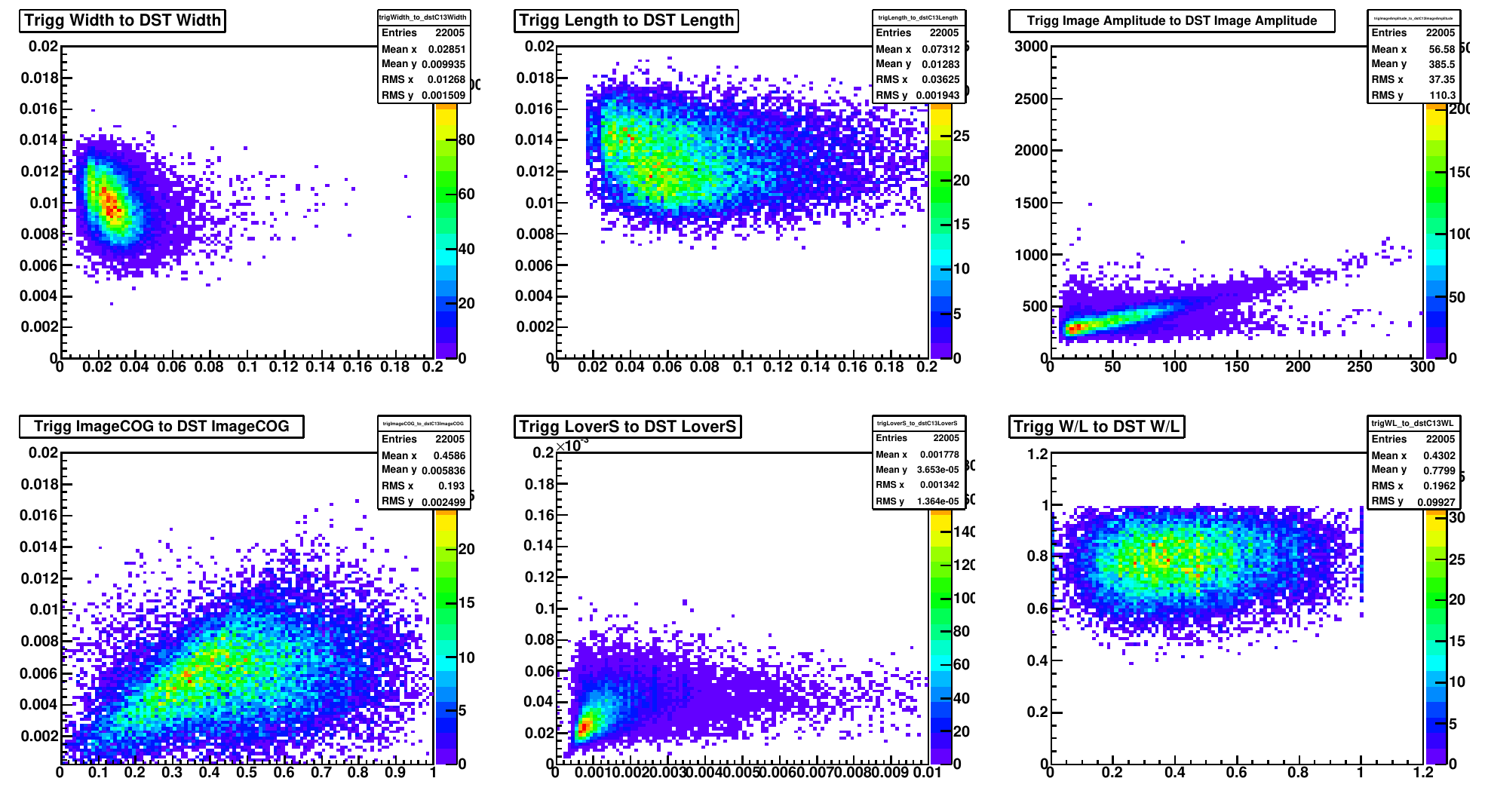,height=120mm,angle=0.0,bb=0 0 567 310}
\caption{Comparison of the showers parameters calculated using the Level 2 maps 
	       and the processed images after cleaning (DST). 
	       The comparison has been performed for 30 GeV gamma showers. 
	       The Hillas parameters have been calculated using a 1/3 image cleaning.}
\label{fig:dst30GeVC13}
\end{center}
\end{figure}

  \begin{figure}[h]
\vspace{2mm}
\begin{center}
\hspace{3mm}\psfig{figure=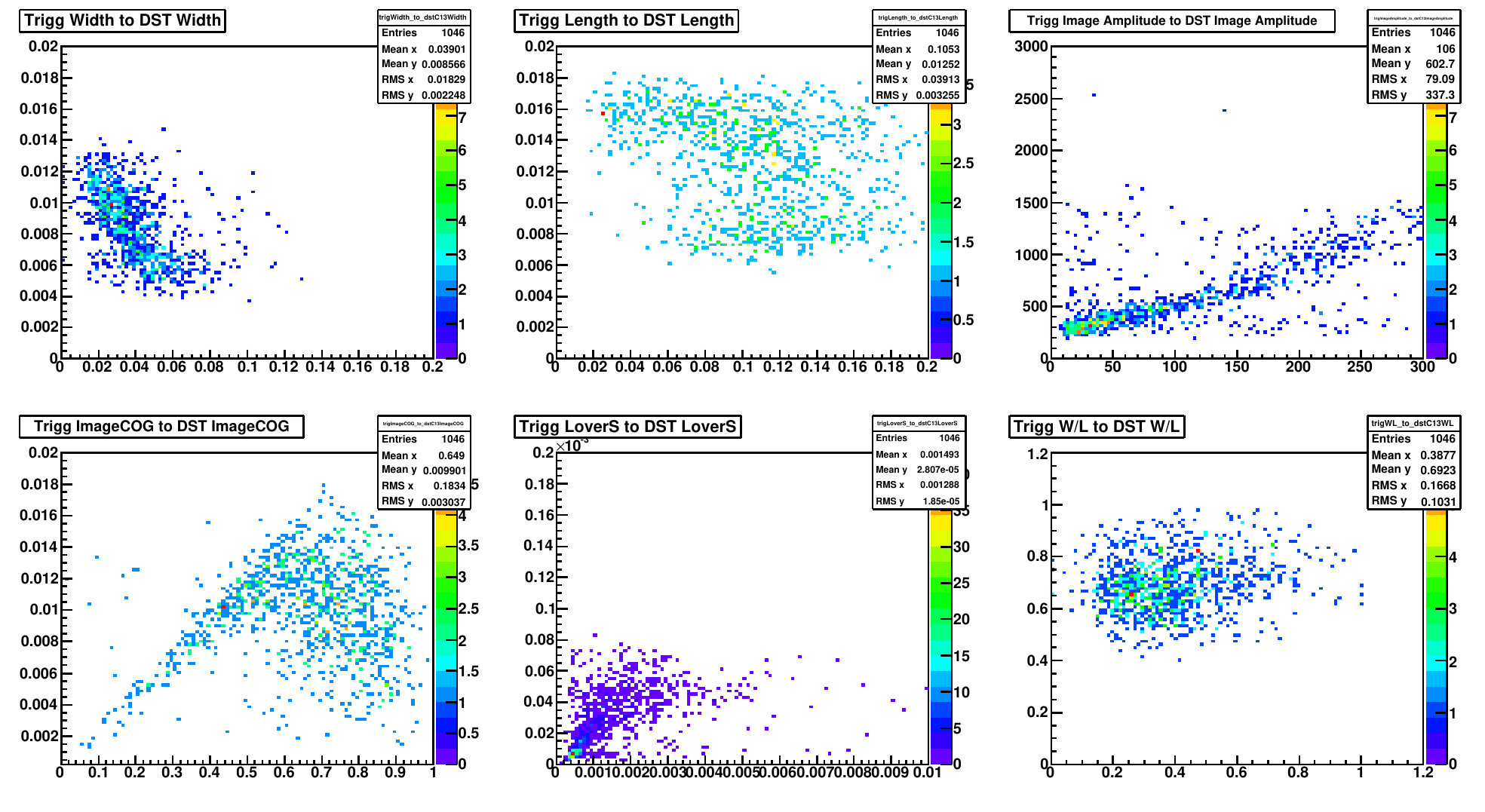,height=120mm,angle=0.0,bb=0 0 567 310}
\caption{  Comparison of the showers parameters calculated using the Level 2 maps 
		and the processed images after cleaning (DST). 
		The comparison has been performed for 100 GeV gamma showers. 
		The Hillas parameters have been calculated using the H.E.S.S.  DST  with a 1/3 image cleaning.}
\label{fig:dst100GeV}
\end{center}
\end{figure}

\end{landscape}

\subsection{The center of gravity cut}

The algorithm that can be used to reject a part of the background particles 
is based on the center of gravity ({\tt COG}) cut. 
The {\tt COG} parameter was chosen, 
because for low energy gamma showers, the {\tt COG} shows a clear correlation 
between the one calculated from instantaneous Level~2 
maps and that calculated from the whole image integrated over 16~ns. 

The other reason for using the {\tt COG} parameter comes from the shower geometry. 
The LCT is design to detect low energy gamma events. 
The showers initiated in the atmosphere by low energy $\gamma$-ray photons 
have their maximum higher in the atmosphere (see. figure \ref{fig:hmax}). 
The Cherenkov light distribution at low energy (figure \ref{fig:CLD_low}) shows that low energy gammas  
have  a photon density large enough to be detected up to $\sim$ 200 m. 
The lateral distribution reaches its maximum at  $\sim$120 m, and then decreases rapidly.
Figure \ref{fig:Disp1} shows the \texttt{DIST} parameter as a function of the impact parameter calculated 
for different energy of primary gammas. 
From figure \ref{fig:Disp1} its clear that at low energies ($\sim 10-30$ GeV) the majority of showers will have their 
{\tt COG} positions within 1$^\circ$ radius from the source position.
We thus demand that  the {\tt COG} of accepted showers be located at less than $1.75^{\circ}/\sqrt{3} = 1^{\circ}$ 
from the expected position of the source.

The higher energy $\gamma$-rays produce air showers with enough 
Cherenkov photons to trigger more than one telescope. 
Thus they will be accepted by the stereoscopic trigger. 
This algorithm is valid for point sources or weakly 
extended sources of photons. 

The direction of the charged  primary particle is changed by the Galactic and the Earth magnetic field.  The observed distribution of the background events is then isotropic. 
The {\tt COG} of such particles are uniformly distributed in the focal plane of the telescope. 
The fraction of the background events rejected with the {\tt COG} cut is proportional to the excluded area. 
The {\tt COG} cut set at 1$^\circ$ should reject 1~--~{\tt COG}$_{cut}^2/(\mbox{FoV}/2)^2\approx$ 70\% of the background events.

\section{Trigger simulations}
\label{sec:TriggerSimulations}
The trigger simulations have been performed using \texttt{KASKADE} and \texttt{SMASH} tools. 
\texttt{KASCADE} \citep{1994NIMPA.343..629K}  package is a  computer software  that simulates in three dimensions  the Cherenkov photons produced 
by VHE gamma-rays and hadronic air showers.

{\tt SMASH} is a package dedicated to the H.E.S.S. detector simulation. 
The package is used to simulate the response of the detector to the Monte Carlo photon data
produced with \texttt{KASCADE}. 
 
 \texttt{SMASH} reproduces  the camera, dish and telescope structure geometry. 
 The package simulates the whole electronics as well as  the background and noise contributions.
The different Level 2 schemes have been implemented by the  author to the {\tt SMASH} software. 
The electronic channel outputs were simulated with realistic photon signal shapes and a realistic electronics readout. 
The results of the simulations have been used to estimate
the various trigger rates with the method described by \cite{GuyThesis}. 

\subsection{Background rates}
The largest contributions to the trigger rate of a single telescope in $\gamma$-ray astronomy are background events.
The largest fraction of the events triggering the camera are photons from the NSB.
The other major source of background are cosmic ray showers. 
These showers have either hadron (proton, helium, etc.) or electron/positron primaries.
The typical proton flux is  larger than 100~particles~m$^{-2}$sr$^{-1}$s$^{-1}$ taking into account protons with energies above 10 GeV. 
The expected muon flux is about $\sim$10~particles~m$^{-2}$sr$^{-1}$s$^{-1}$,
while the electron flux above 7~GeV is $\sim$3~particles~m$^{-2}$sr$^{-1}$s$^{-1}$.  
The background rates are calculated according to the formula:
\begin{equation}
\mbox{Background\,rate} = \frac{N_{trigg}}{N_{sim}}\times \Omega \times S \times \int_{E_{min}}^{E_{nax}}\frac{dN}{dE}
\label{eq:backgroundrate}
\end{equation}
where $\Omega =2\pi[1-\cos(\mbox{viewcone})]$ is the solid angle of the viewcone in steradians, 
$S=\pi\times IP_{max}^2$ is the area with radius which corresponds to 
the maximum impact parameter $IP_{max}$, of simulated events.
$N_{trigg}$ is a number of triggered events and $N_{sim}$ is a number of simulated events.
The particle flux, $dN/dE$, is known from observations of many instruments and differ for each particle type.

\subsection{Proton rate}
Protons were simulated in the energy range from 0.005~TeV to 500~TeV, with the maximum
 impact parameter of 550~m and the viewcone of 5$^\circ$.  
The proton trigger rates were calculated using the particle flux given by
\cite{GuyThesis} [Chapter 13, p.135] and using equation (\ref{eq:backgroundrate}):
  \begin{equation}
   \frac{dN}{dE}=1.49\times10^4(E+2.15e^{-0.21\sqrt{E}})^{-2.74} {\rm m^{-2}s^{-1}sr^{-1}GeV^{-1} }.
  \end{equation}
\begin{figure}
\vspace{2mm}
\begin{center}
\hspace{3mm}
\psfig{figure=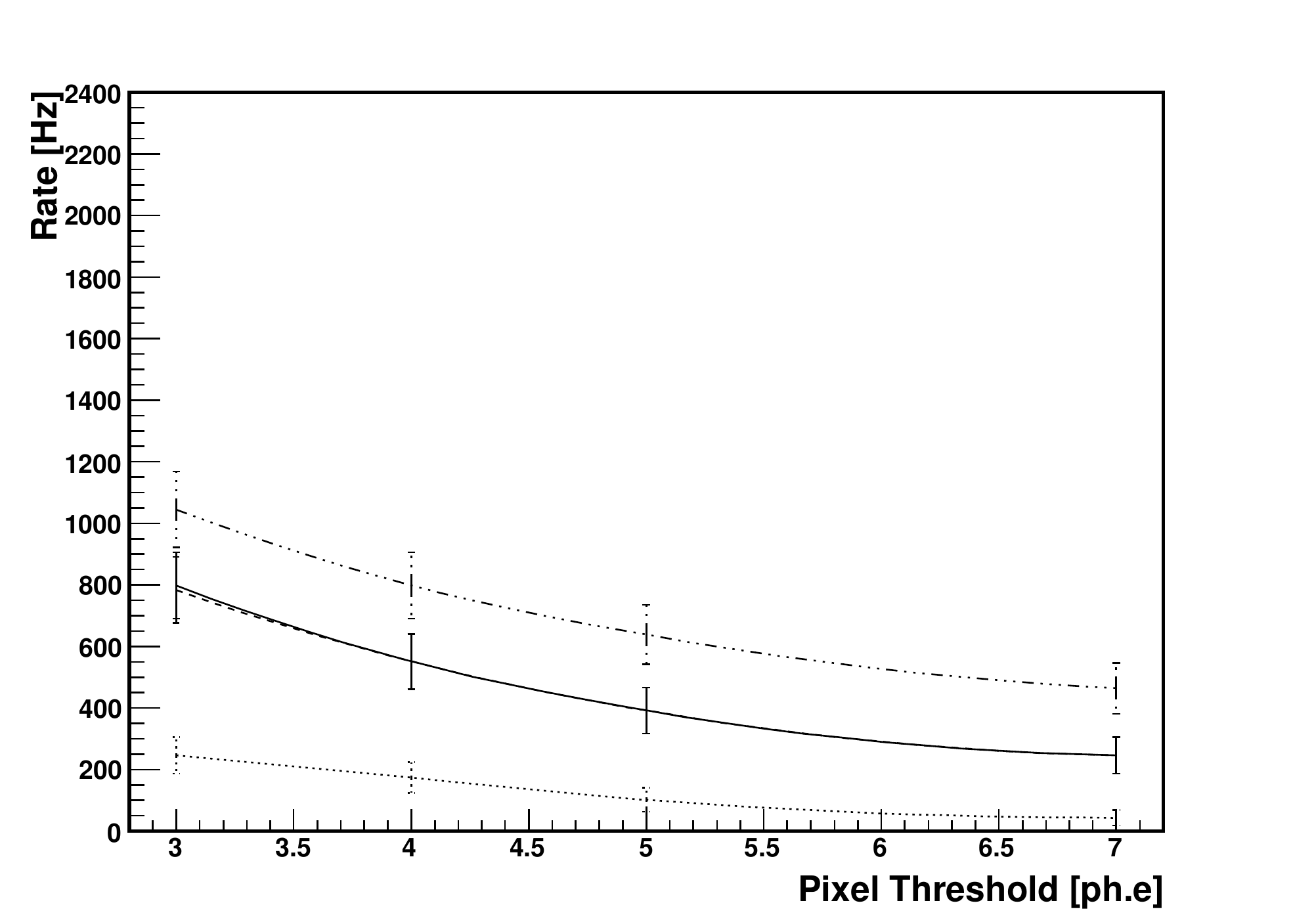,width=130mm,angle=0.0,bb=0 0 567 405} 
\psfig{figure=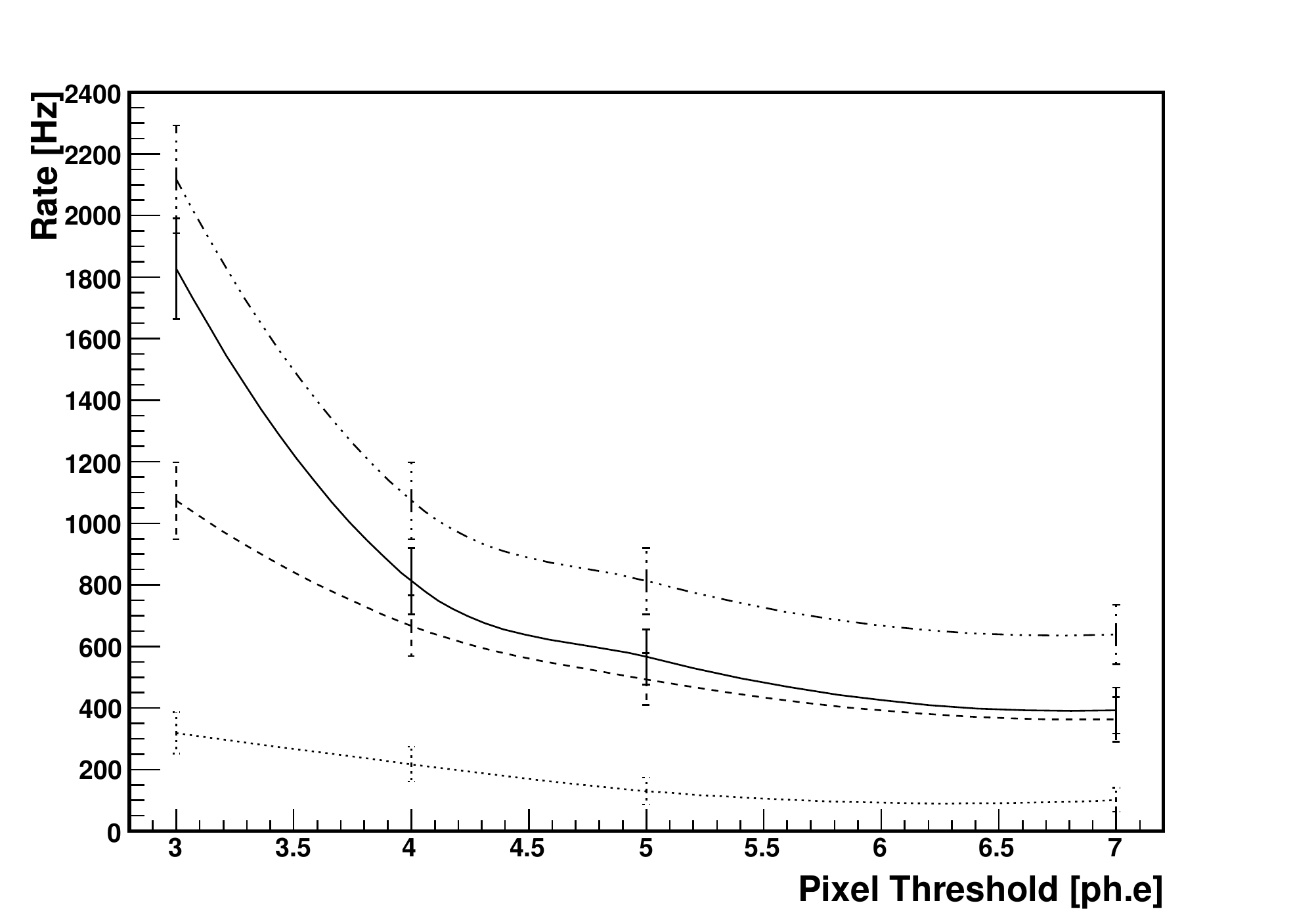,width=130mm,angle=0.0,bb=0 0 567 405}
\caption{ Proton trigger rate versus pixel threshold (in photoelectrons).
	       On the top figure the pixel multiplicity is 4 and on the bottom figure is 3. 
	       A  Level 2 pixel threshold of 7 have been assumed. 
	       The dash-dotted line gives the raw level 1 rate. 
	       The solid line shows the rate of monoscopic events. 
	       The dashed line gives the rate of events passing the cleaning/denoising pixel cut. 
	       Finally, the dotted line is the rate of events passing the {\tt COG} cut. 
	       Note that the {\tt COG} cut reduces the proton rate by a factor of 3.}
\label{fig:protonM4}
\end{center}
\end{figure}
The proton trigger rate is shown on figure \ref{fig:protonM4} as a function of the pixel threshold in photoelectrons. 

\subsection{Muon rate}
\begin{figure}
\vspace{2mm}
\begin{center}
\hspace{3mm}
\psfig{figure=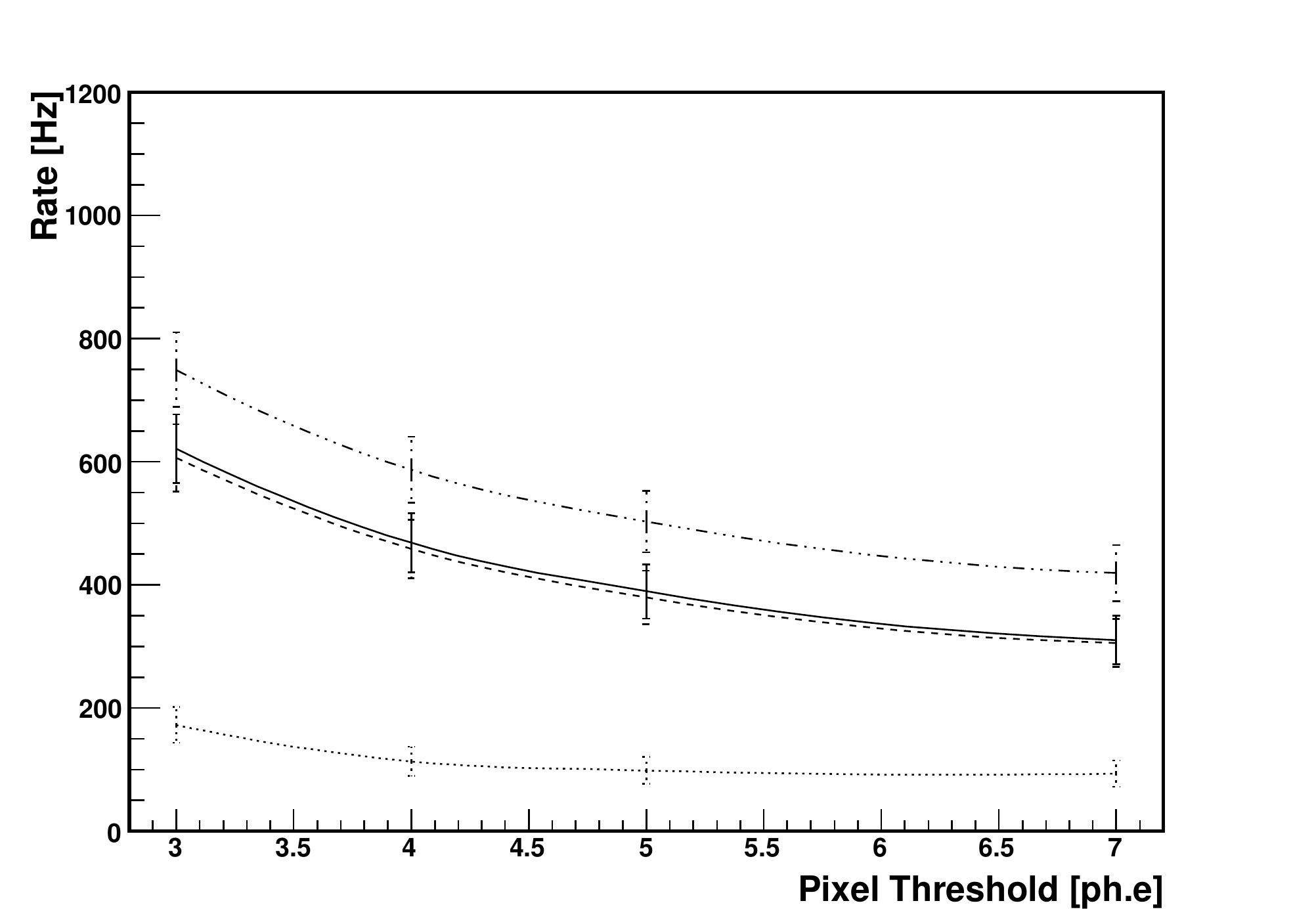, width=130mm, angle=0.0,bb=0 0 567 405}
\psfig{figure=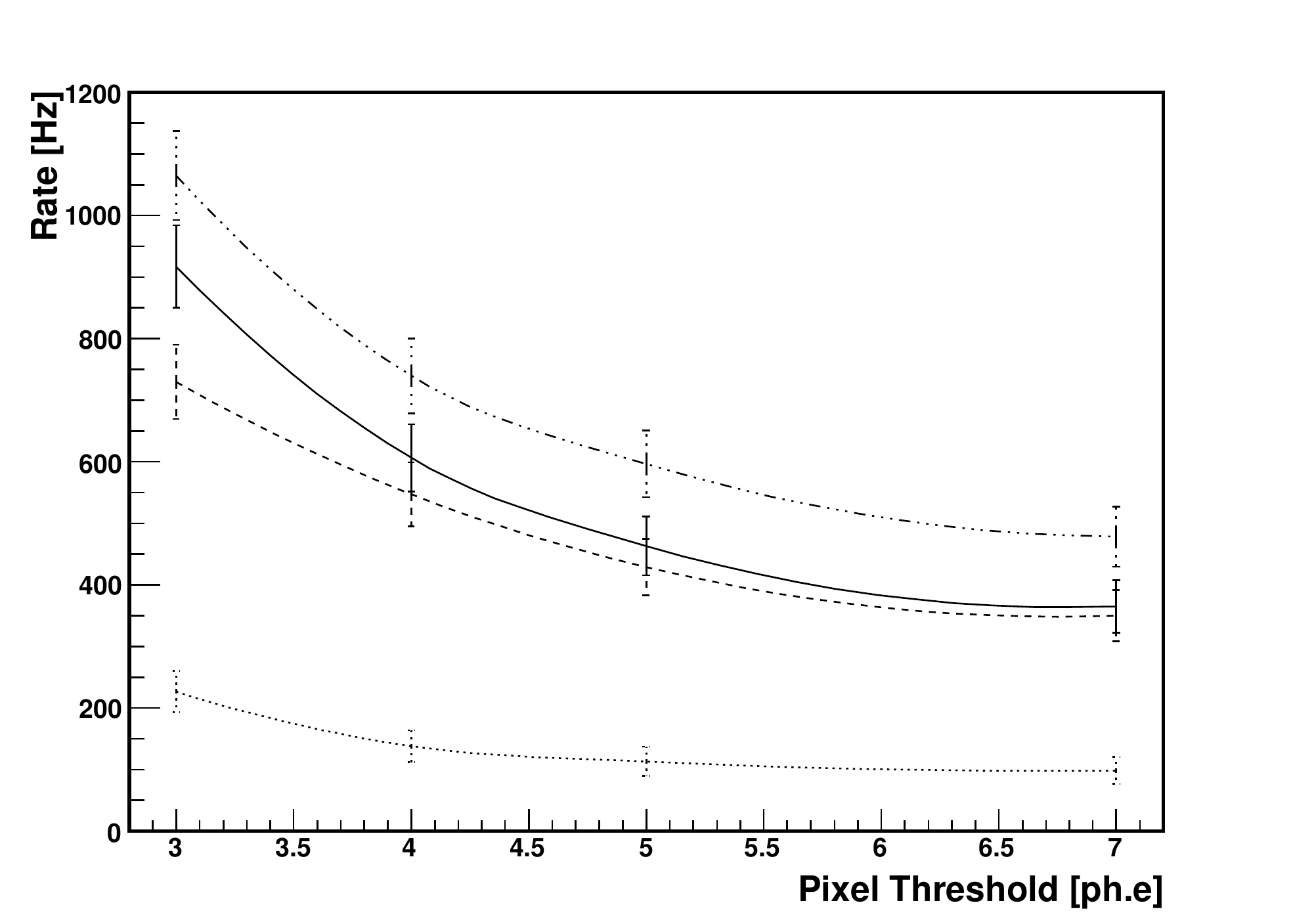, width=130mm, angle=0.0,bb=0 0 567 405}
\caption{ Part of the trigger rate due to the isolated muon component of the shower.
	        Level 1 pixel multiplicity of 4 (top figure) and 3 (bottom figure), 
	        and Level 2 pixel threshold of 7 have been assumed. 
	        The dash-dotted line gives the raw Level 1 rate. 
		The solid line shows the rate of monoscopic events. 
	       The dashed line gives the rate of events passing the cleaning/denoising pixel cut. 
	       Finally, the dotted line is the rate of events passing the {\tt COG} cut. 
                Note that the {\tt COG} cut reduces the single muon rate by a factor of 3.}
\label{fig:muonM4}
\end{center}
\end{figure}

Isolated muons from distant hadronic showers can trigger  Cherenkov telescopes. 
These muon triggers dominate the single telescope  triggers \citep{2005AIPC..745..753F}
and can be rejected by demanding a multi-telescope trigger (stereoscopy). 
The  muons flux  were calculated using:
 \begin{equation}
   \frac{dN}{dE}= 431 E^{-2.3} e^{-0.38/E}(1-e^{\frac{-33}{E}})  {\rm m^{-2}s^{-1}sr^{-1}GeV^{-1} }\,.
    \label{eq:muon}
 \end{equation}
The energy range of simulated muons was 10~--~100~GeV. 
The trigger rate contributed by single muons is shown on figure \ref{fig:muonM4}. 

\subsection{Electron rate}
 \begin{figure}
\vspace{2mm}
\begin{center}
\hspace{3mm}\psfig{figure=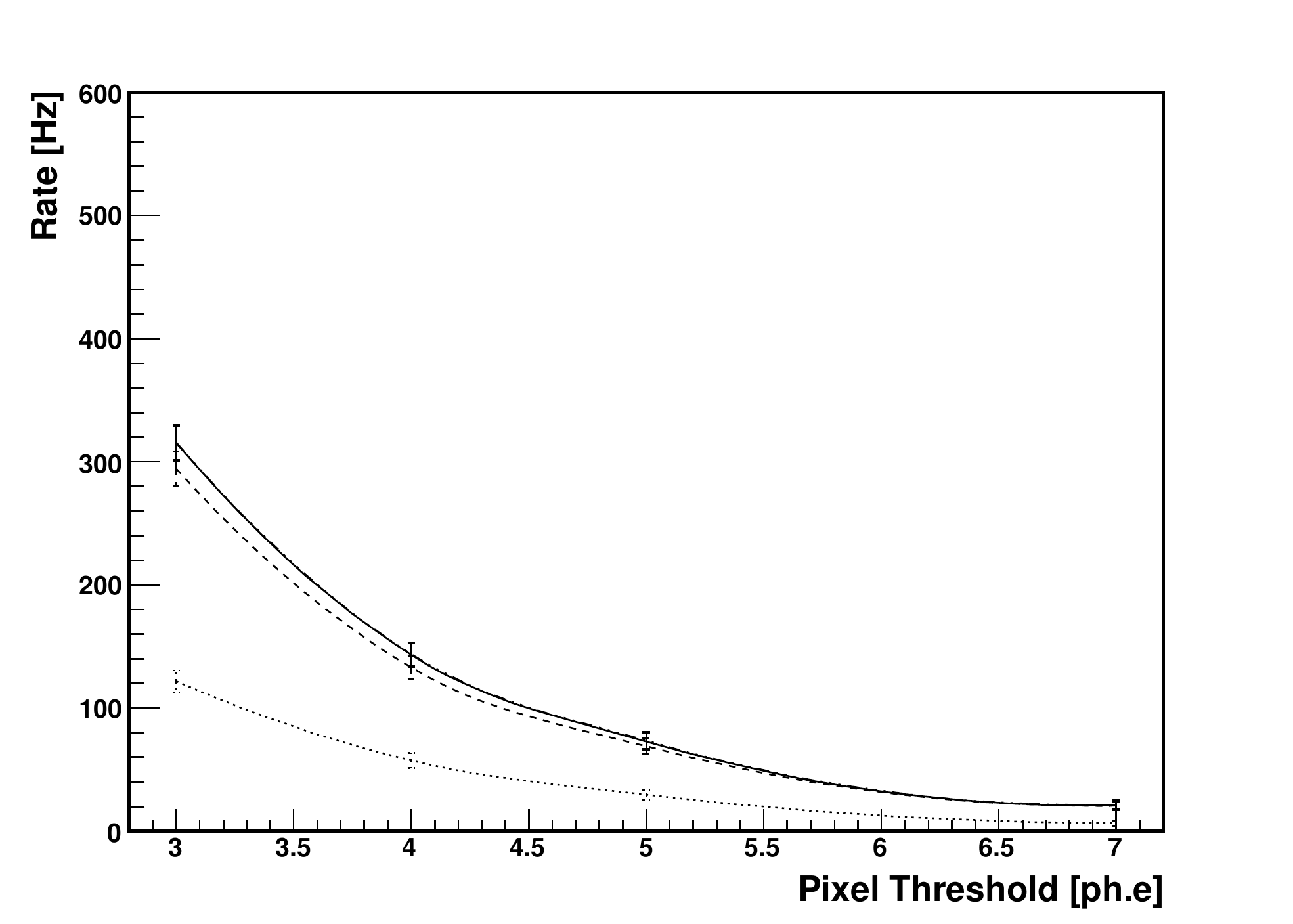, width=130mm, angle=0.0,bb=0 0 567 405}
\hspace{3mm}\psfig{figure=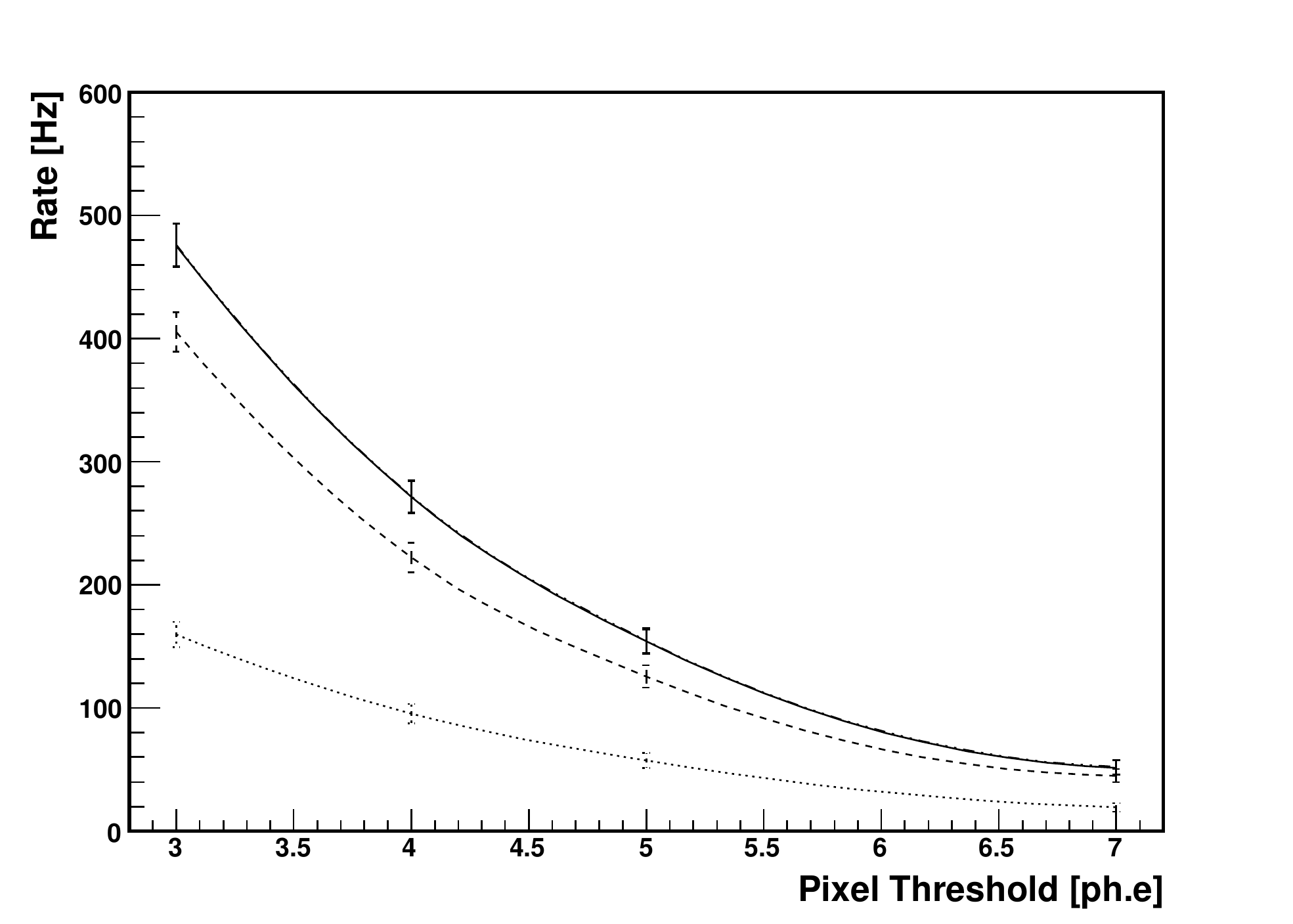, width=130mm, angle=0.0,bb=0 0 567 405}
\caption{  Electron rate as a function of the pixel threshold.
		Level 1 pixel multiplicity of 4 (top panel) and 3 (bottom panel), 
		and Level 2 pixel threshold of 7 have been assumed. 
		The dash-dotted line gives the raw Level 1 rate. 
		The solid line shows the rate of monoscopic events. 
		These 2 lines are almost superimposed since the electron rate is dominated by low energy events.
		The dashed line gives the rate of events passing the cleaning/denoising pixel cut. 
		Finally, the dotted line is the rate of events passing the {\tt COG} cut.}
\label{fig:electronM4}
\end{center}
\end{figure}

Cosmic ray electrons give a Cherenkov signal very similar to the signal of high energy gamma rays. 
It is thus not possible to eliminate electrons from the analysis. 
However, the electron background, which is a diffuse source, can be reduced in point source studies. 
The trigger rates were
calculated using the particle flux:  
  \begin{equation}
   \frac{dN}{dE}= 0.95\times10^{-4} \left(\frac{E}{1TeV}\right)^{-3.26}{\rm  m^{-2}s^{-1}sr^{-1}TeV^{-1} }.
  \end{equation}
The minimum energy of electron entering the atmosphere  depends on the rigidity cut-off \citep{2001APh....15..203C}.
The geomagnetic field bends the cosmic ray trajectories preventing low rigidity particles from reaching the Earth's surface. 
The rigidity of a particle is defined as $pc/Z$, where $c$ is the speed of light, $p$ is the particle momentum and $Z$ is the charge of the particle. The minimum allowed rigidity is known as rigidity cut-off, $R_c$.
The electron rigidity cut-off can by estimated from
\begin{equation}
 R_c=\frac{59.4\cos^4\lambda}{r^2(1+\sqrt{1-cos^3\lambda \sin\theta \sin\phi})^2} \,,
 \label{eq:Rc}
\end{equation}
where:\\
$\theta$ - is the zenith angle ($\theta=0^o)$ \\
$\phi$ - is the azimuth angle ($\phi=90^o)$ \\
r - is the distance from the dipole center \\
$\lambda$ - is the magnetic altitude\\
A simple manipulation of the rigidity definition gives an expression for the minimum energy of the particle which are able to penetrate into the Earth's atmosphere:
\begin{equation}
E_{min}=\sqrt{(Z\,R_c)^2+m_0^2c^4}\,.
\end{equation}
For the H.E.S.S. site, the geographic longitude is 18$^\circ$~E, the geographic latitude 22$^\circ$~S, 
and the corrected magnetic latitude is 33$^\circ$.  
The rigidity cut-off for H.E.S.S. site is then $R_c$=7~GV.
The cut-off energy for the electrons is thus  $E_{min} \simeq 7$~GeV.

Monte Carlo samples of cosmic electrons were simulated using the
following parameters: energy range from 0.007~TeV to 300~TeV,
viewcone 5$^\circ$.  
The electron trigger rate is typically a few hundred Hz, and is plotted on figure \ref{fig:electronM4}. 

\subsection{Total particle trigger rate}
\begin{figure}
\vspace{2mm}
\begin{center}
\hspace{3mm}
\psfig{figure=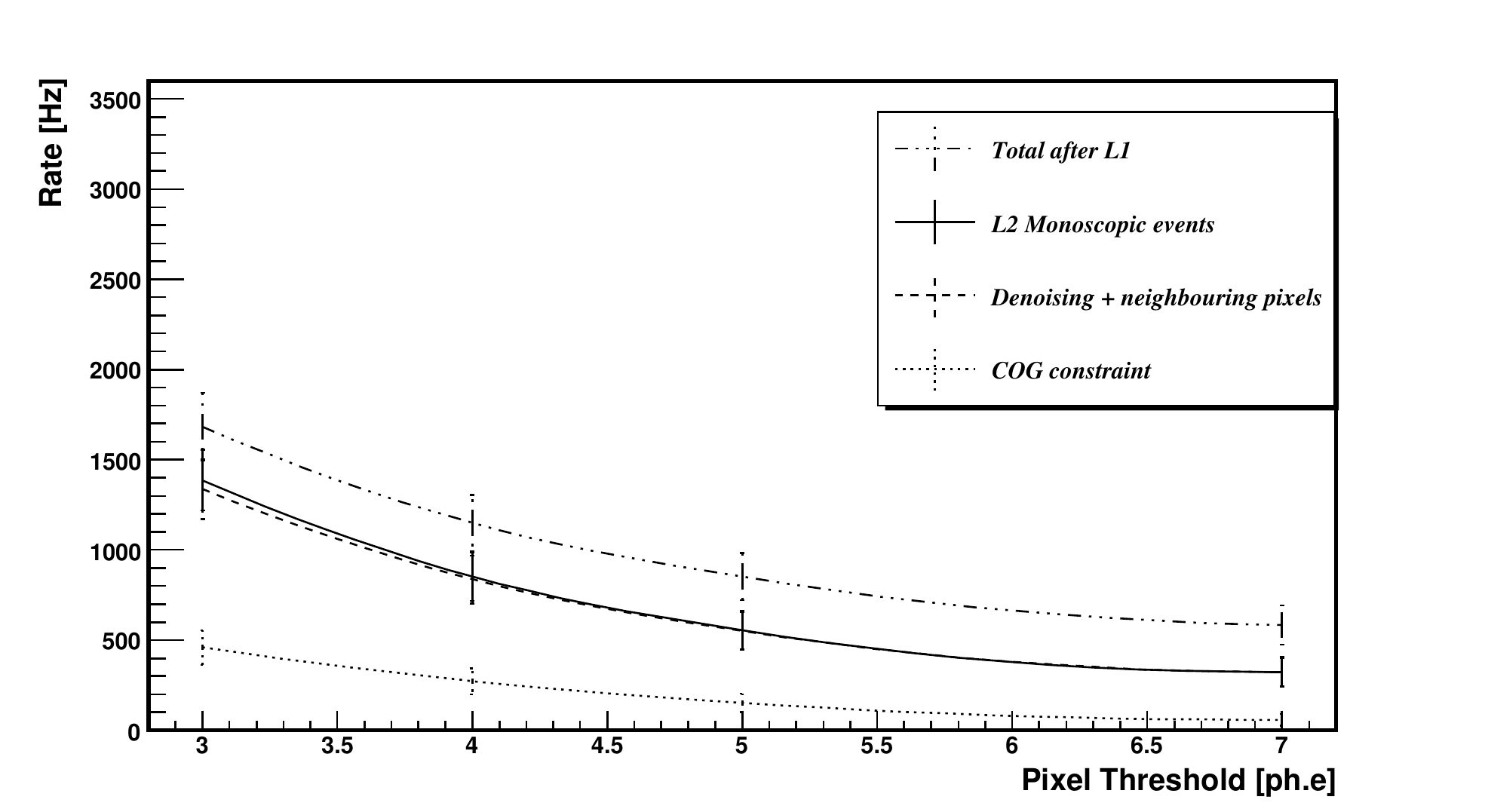,width=150mm,angle=0.0,bb=0 0 567 405}
\psfig{figure=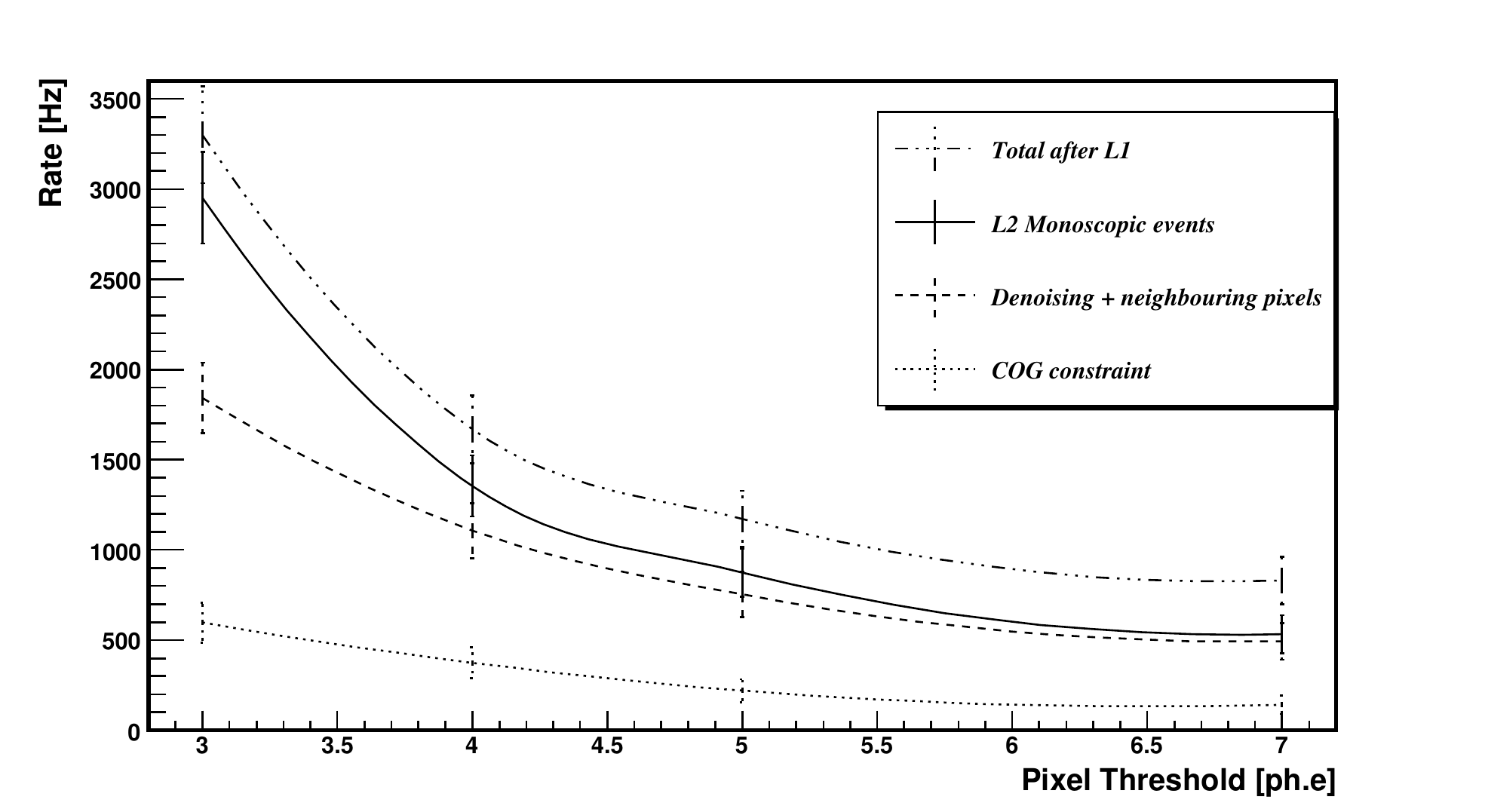,width=150mm,angle=0.0,bb=0 0 567 405}
\caption{  Total hadronic+electron rate as function of the pixel threshold.
		Level 1 pixel multiplicity is 4 (top panel) and 3 (bottom panel).
		Level 2 pixel threshold of 7 have been assumed. 
		The dash-dotted line gives the raw Level 1 rate. 
		The solid line shows the rate of monoscopic events. 
		The dashed line gives the rate of events passing the cleaning/neighbouring pixel cut. 
		Finally, the dotted line is the rate of events passing the {\tt COG} cut.}
\label{fig:allM4}
\end{center}
\end{figure}

The particle trigger rate is shown as a 
function of the pixel threshold on figure~\ref{fig:allM4}. 
The particle trigger rate is the sum of 
the proton, the helium and the electron rate. The helium rate is taken 
into account by multiplying the proton rate by 1.2 \citep{GuyThesis}. 
The total particle trigger rate is of the order of 1 kHz.

\subsection{Night Sky Background  rate}
The NSB comes from diffuse sources, such as the zodiacal light and the galactic plane, and light from bright stars. 
The NSB flux has been measured at the H.E.S.S. site and NSB photoelectron rates were derived
for the 12 meter telescopes \citep{2002NIMPA.481..229P}. 
The calculated NSB photoelectron rate is $100\pm13$ MHz per pixel at zenith in extragalactic fields. 
In galactic fields, the single pixel rate is higher and reaches 200-300 MHz per pixel. 
The 28 meter telescope has a larger collection area (596 m$^{2}$  as compared to 108 m$^{2}$), 
but with more pixels (2048 instead of 960) and a smaller angular acceptance ($3\times10^{-3}$ sr instead of $6\times10^{-3}$ sr), 
the expected NSB rate per pixel of the LCT is only a factor of 1.3 higher as compared to SCT. 

{\tt KASKADE} simulations have been used to generate gamma particles with very
low energies not producing detectable Cherenkov light 
to reproduce the response of the detector to the NSB events.
The gamma energy has been set arbitrarily to  $5\,$MeV.  
The NSB trigger  was then simulated by adding  random photoelectrons to every readout channel. 
NSB single pixel rates of 100, 200 and 300 MHz were studied. 
The different NSB levels have been set to reproduce different observation conditions. 
The low NSB level $\sim$~100 MHz is relevant for an extragalactic observation. 
The high NSB level $\sim$~300 MHz corresponds to the photon background for the Galactic plane observations.

The NSB rate has been calculated by looking for a trigger in a 40~ns coincidence window: 
\begin{equation}
\mbox{NSB\,rate} = \frac{N_{trigg}}{N_{sim}}\times \mbox{Window\, Duration}\,. 
\end{equation}

The LCT trigger rates due to the NSB are shown on table~\ref{tab:nsb} for several Level 1 trigger conditions. 
Depending on the conditions, the estimated rates range from several MHz to less than a few tens of Hz. 
Since the dead-time per event of the LCT acquisition is of the order of a few microseconds, the acquisition rate should be less than roughly 100 kHz. Table~\ref{tab:nsb} shows that some Level 1 trigger condition (e.g. a~pixel multiplicity of 3 and  a~pixel threshold of 3) lead to unmanageably high trigger rates. 

\begin{table}   
\begin{tabular}{cccc}
\hline
(Multiplicity,   & L1 rate      & L1 rate  & L1 rate       \\
Pixel Threshold) & 100 MHz  &   200 MHz               &   300 MHz         \\     
\hline
\hline
(4,3)           & $<$ 63 Hz       & 655 $\pm$ 182 Hz & 183 $\pm$ 3.6 kHz \\
(4,4)           & $<$ 63 Hz       & $<$ 120 Hz      & 142 $\pm$ 51 Hz   \\
(4,5)           & $<$ 63 Hz       & $<$ 120 Hz      & $<$ 160 Hz       \\
(4,5)           & $<$ 63 Hz       & $<$ 120 Hz      & $<$ 162 Hz       \\
(3,3)           & 803 $\pm$ 80 Hz  & 125 $\pm$ 2.3 kHz& 7 $\pm$ 0.18 MHz  \\
(3,4)           & 84 $\pm$ 40 Hz   & 1 $\pm$ 0.2 kHz  & 16 $\pm$ 1 kHz    \\
(3,5)           & 21 $\pm$ 20 Hz   & 63 $\pm$ 37 Hz   & 1 $\pm$ 0.3 kHz   \\
(3,7)           & $<$ 63 Hz       &$<$ 120 Hz       & 320 $\pm$ 156 Hz  \\
\hline
\hline
\end{tabular}
\begin{tabular}{cccc}
\hline
(Multiplicity,   & clustering& clustering    & clustering  \\
Pixel Threshold) & 100 MHz              &  200MHz                 &   300 MHz\\     
\hline
\hline
(4,3)           & $<$ 63 Hz    & 230 $\pm$ 112 Hz  & 171 $\pm$ 3.5 kHz \\
(4,4)           & $<$ 63 Hz    & $<$ 120 Hz       & $<$ 160 Hz  \\
(4,5)           & $<$ 63 Hz    & $<$ 120 Hz       & $<$ 160 Hz  \\
(4,5)           & $<$ 63 Hz    & $<$ 120 Hz       & $<$ 160 Hz  \\
(3,3)           & $<$ 63 Hz    & 13 $\pm$ 0.24 kHz & 8.7 $\pm$ 0.17 kHz\\
(3,4)           & $<$ 63 Hz    & $<$ 120 Hz       & 510 $\pm$ 212\\
(3,5)           & $<$ 63 Hz    & $<$ 120 Hz       & $<$ 160 Hz \\
(3,7)           & $<$ 63 Hz    & $<$ 120 Hz       & $<$ 160 Hz \\
\hline
\hline
\end{tabular}
\caption{Night sky background rates for NSB levels of 100 MHz, 200 MHz and 300 MHz. 
Upper limits are given at the 95\% C.L.
Upper table: Night Sky Background rates for various trigger conditions. 
Lower table: effect of denoising and clustering. 
The clustering condition asks for at least 2 neighbors around at least one triggered pixel.}
\label{tab:nsb}
\end{table}

\subsection{Effective area}
  
The advantage of the Cherenkov imaging technique is its large collection area. 
 The gamma efficiency (number of triggered events divided by number of simulated events) 
alone does not gives even a rough estimate of the telescope performance. 
To check the performance more accurately it is much better to calculate the effective area, 
which include also the information about the trigger efficiency as a function of the  impact parameter. 
The effective collection area, $A_{{\rm eff},\gamma}$, of a single telescope is determined by 
the lateral and angular distribution of the Cherenkov light.
 For gamma-rays from a point source:
\begin{equation}
A_{{\rm eff},\gamma}(E) = 2 \pi \int_0^\infty P_\gamma(E,r) r {\rm d} r\,, 
\end{equation}

where $P_g(E,r)$ is the detection probability for a gamma-ray shower 
induced by a primary photon with energy E and impact parameter r.

Figure~\ref{fig:DetectionProbability} shows the detection probability 
as a function of the impact parameter for 20~GeV gamma showers at different trigger conditions. 
It is worth pointing out that at low energy the trigger efficiency depends 
very strongly on the Level 1 trigger conditions. 
However, at very low trigger thresholds the trigger efficiency is increased by random NSB hints. 
The simulations of gamma showers has been performed with additional 
photon noise at level of 100~MHz. 
Figure~\ref{fig:EffectiveArea} shows the comparison of effective areas 
at  different trigger conditions and trigger algorithm stages. 

\begin{landscape}
\begin{figure}
\vspace{2mm}
\begin{center}
\hspace{3mm}\psfig{figure=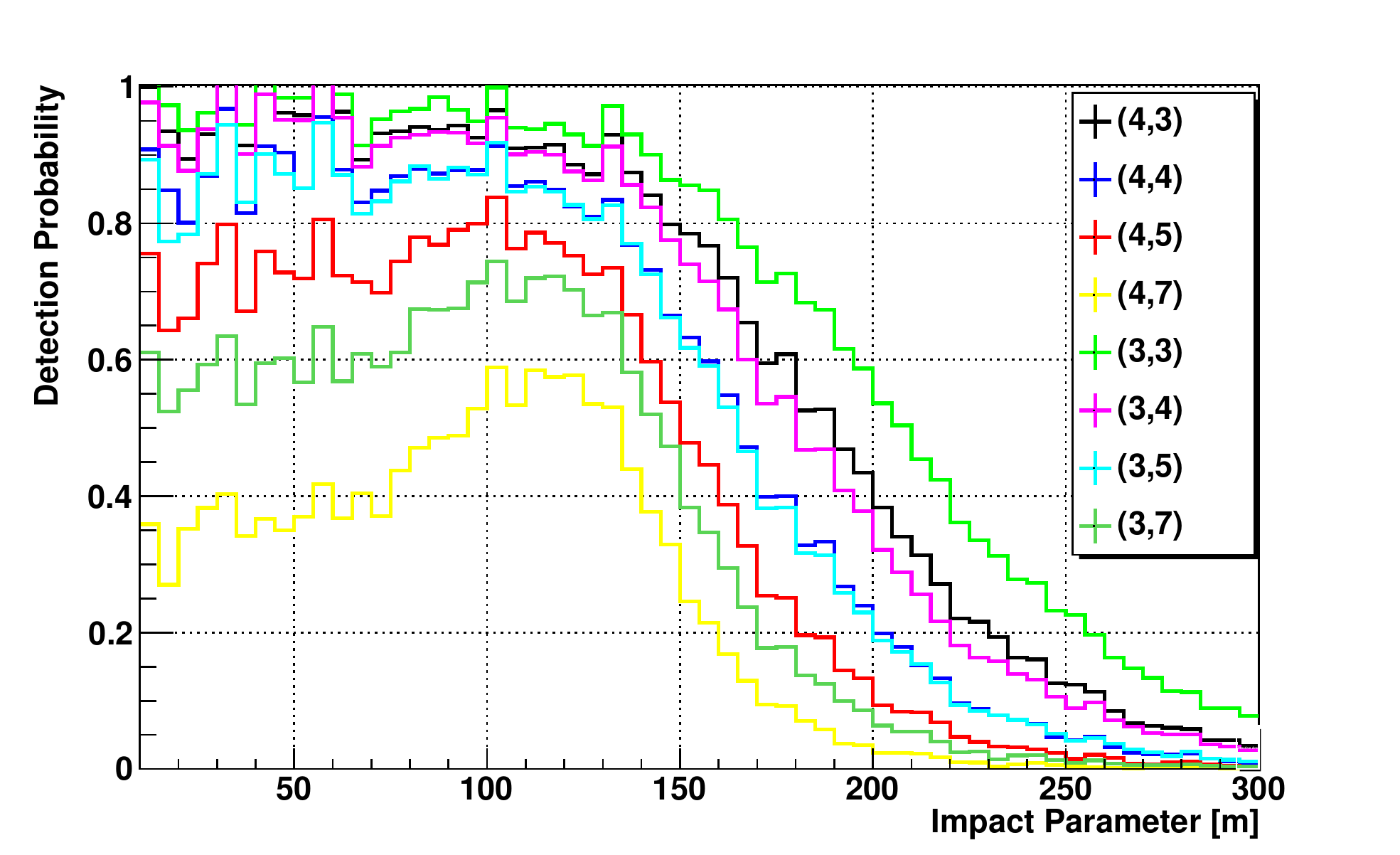,width=180mm,angle=0.0}
\caption{  The detection probability as a function of impact parameters for gamma showers of energy 20 GeV. 
		The legend indicate the Level 1 trigger condition (multiplicity and pixel threshold). 
                   The gamma detection probability is presented for events accepted by the Level 1 trigger.}
\label{fig:DetectionProbability}
\end{center}
\end{figure}

\begin{figure}
\begin{center}
\psfig{figure=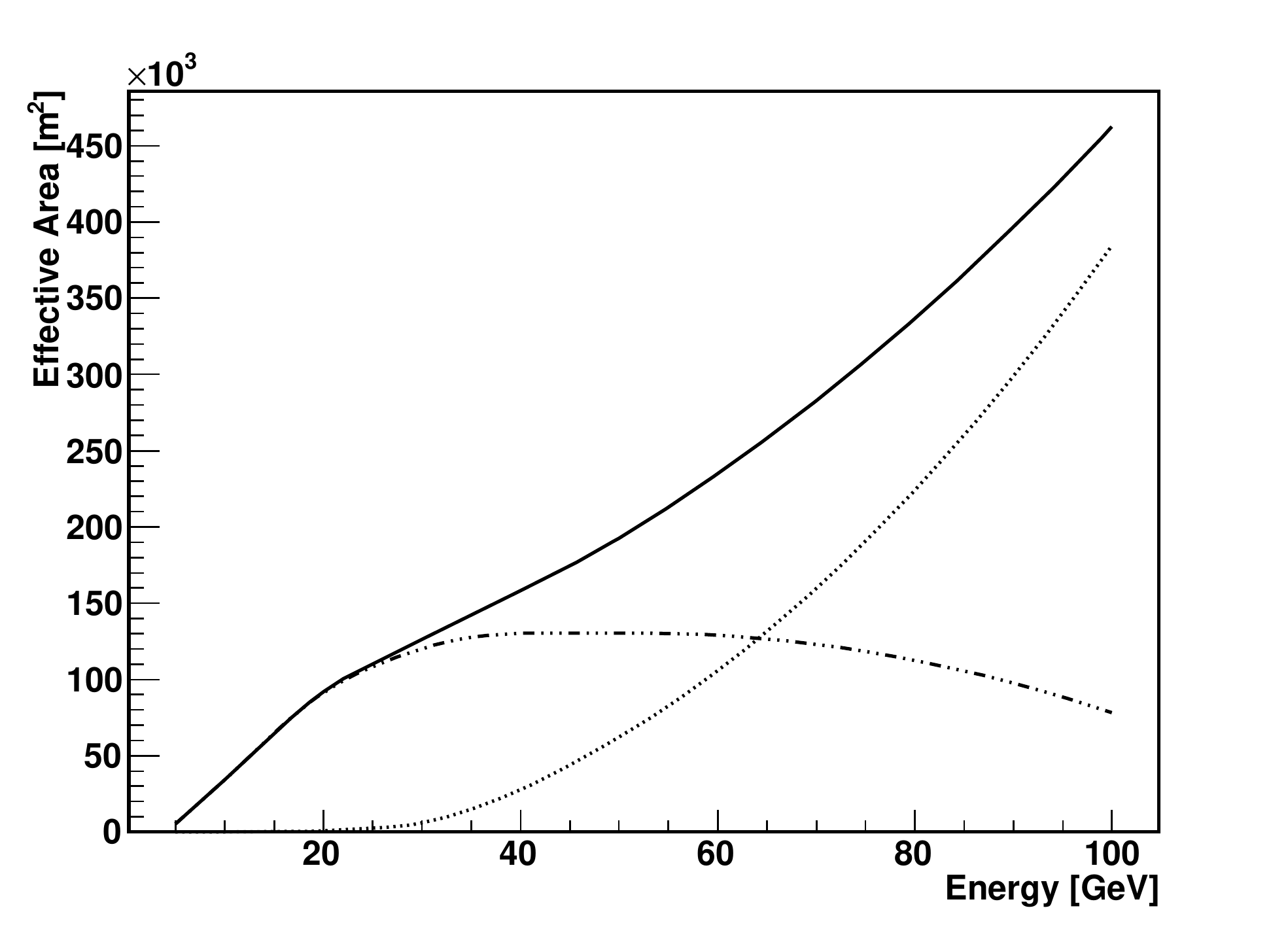,width=90mm,angle=0.0,bb=0 0 567 407}
\psfig{figure=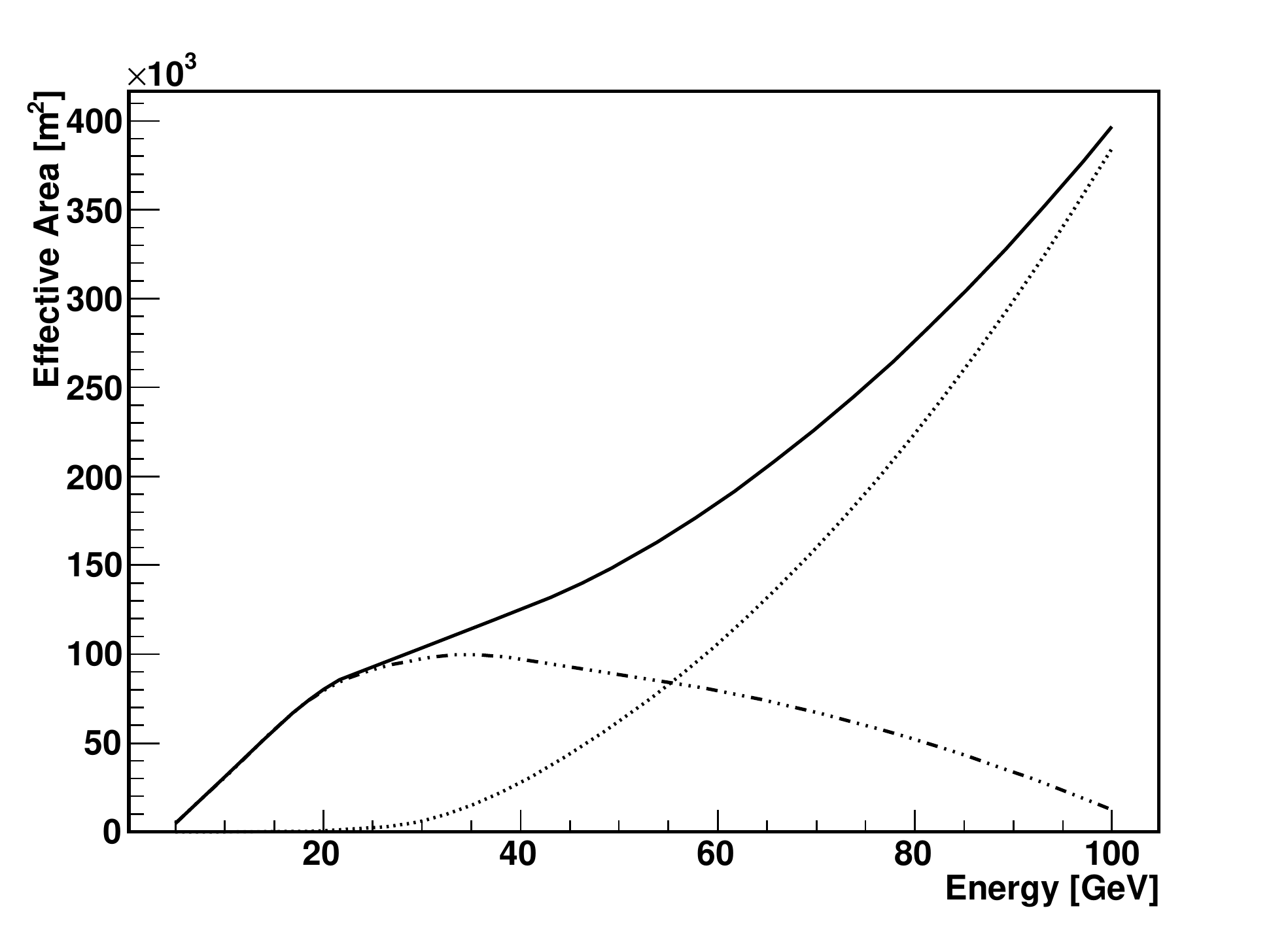,width=90mm,angle=0.0,bb=0 0 567 407}
\psfig{figure=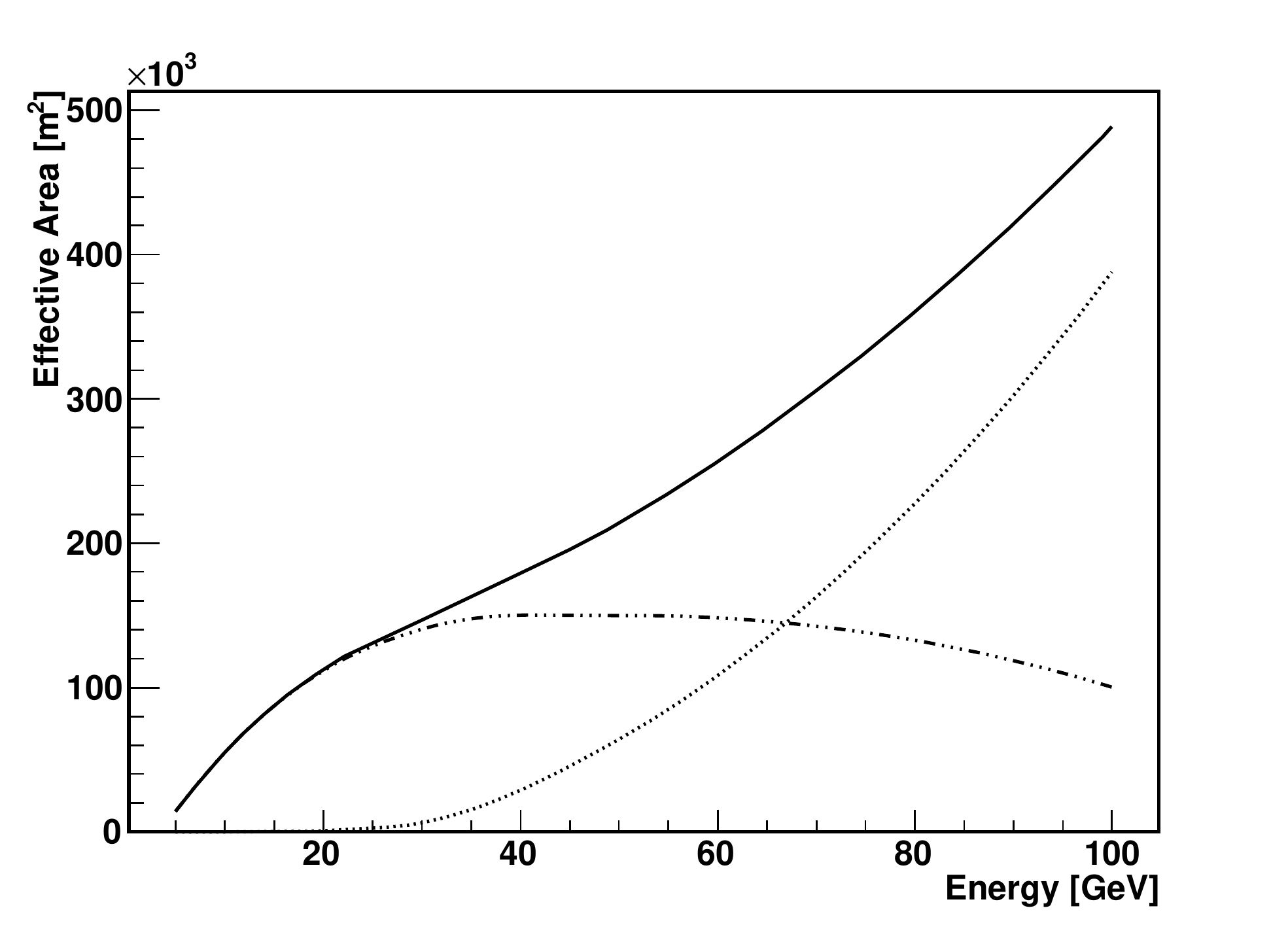,width=90mm,angle=0.0,bb=0 0 567 407}
\psfig{figure=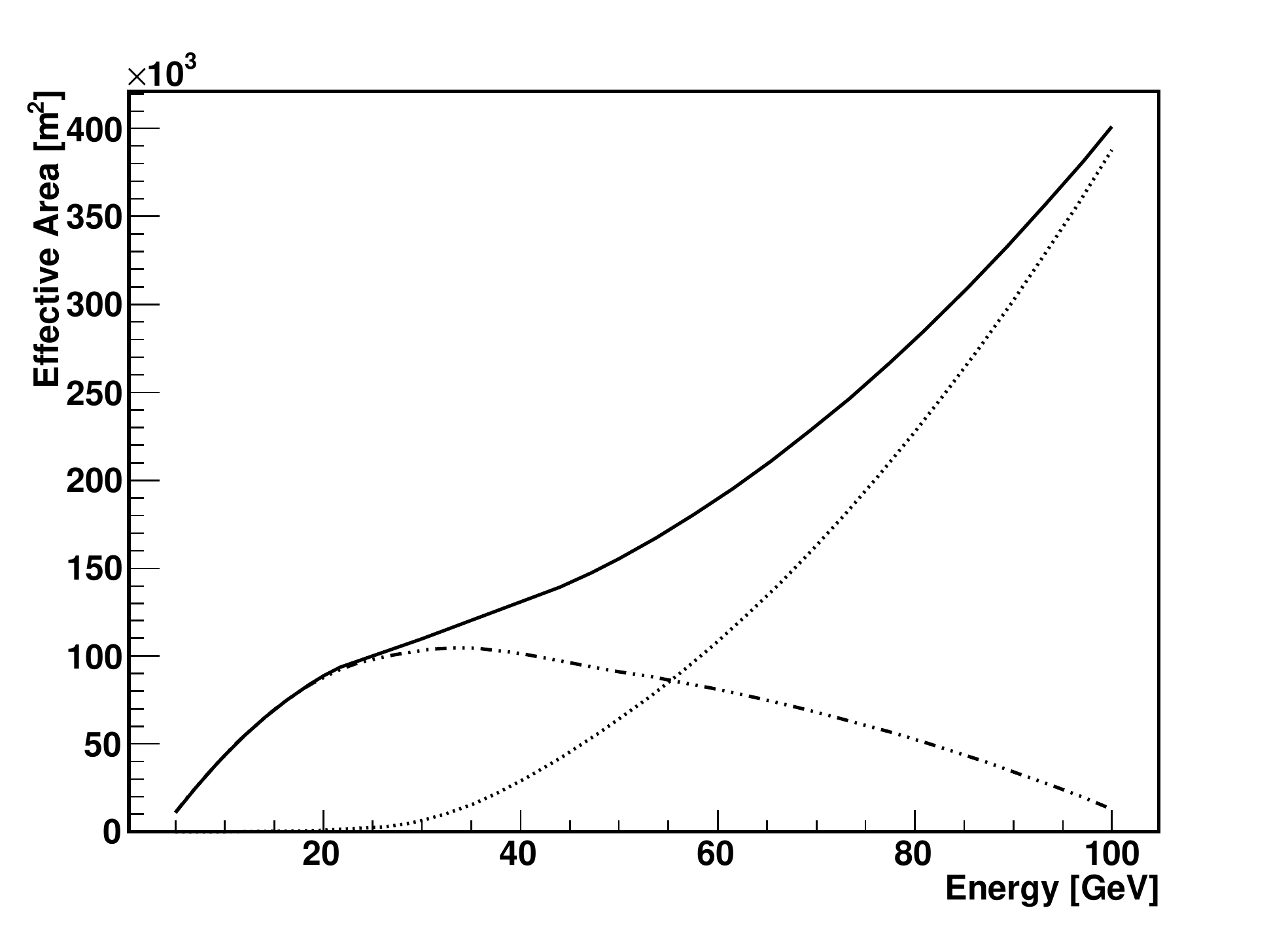,width=90mm,angle=0.0,bb=0 0 567 407}
\caption{The effective area calculated for Level 1 trigger multiplicity 3 (up panel)  and 4 (down panel).
	      The effective area is presented at two different stages of the trigger.
                The left panel shows the effective are after Level 1 trigger 
                and the right panel the effective area after {\tt COG} cut. 
                The dashed dotted lines indicate the effective area of monoscopic events. 
	       The dotted lines indicate the stereoscopic events. 
	       The solid line is the sum of monoscopic and stereoscopic events.}
\label{fig:EffectiveArea}
\end{center}
\end{figure}

\end{landscape}

\subsection{Level 2 trigger efficiency}
\begin{figure}
\vspace{2mm}
\begin{center}
\hspace{3mm}
\psfig{figure=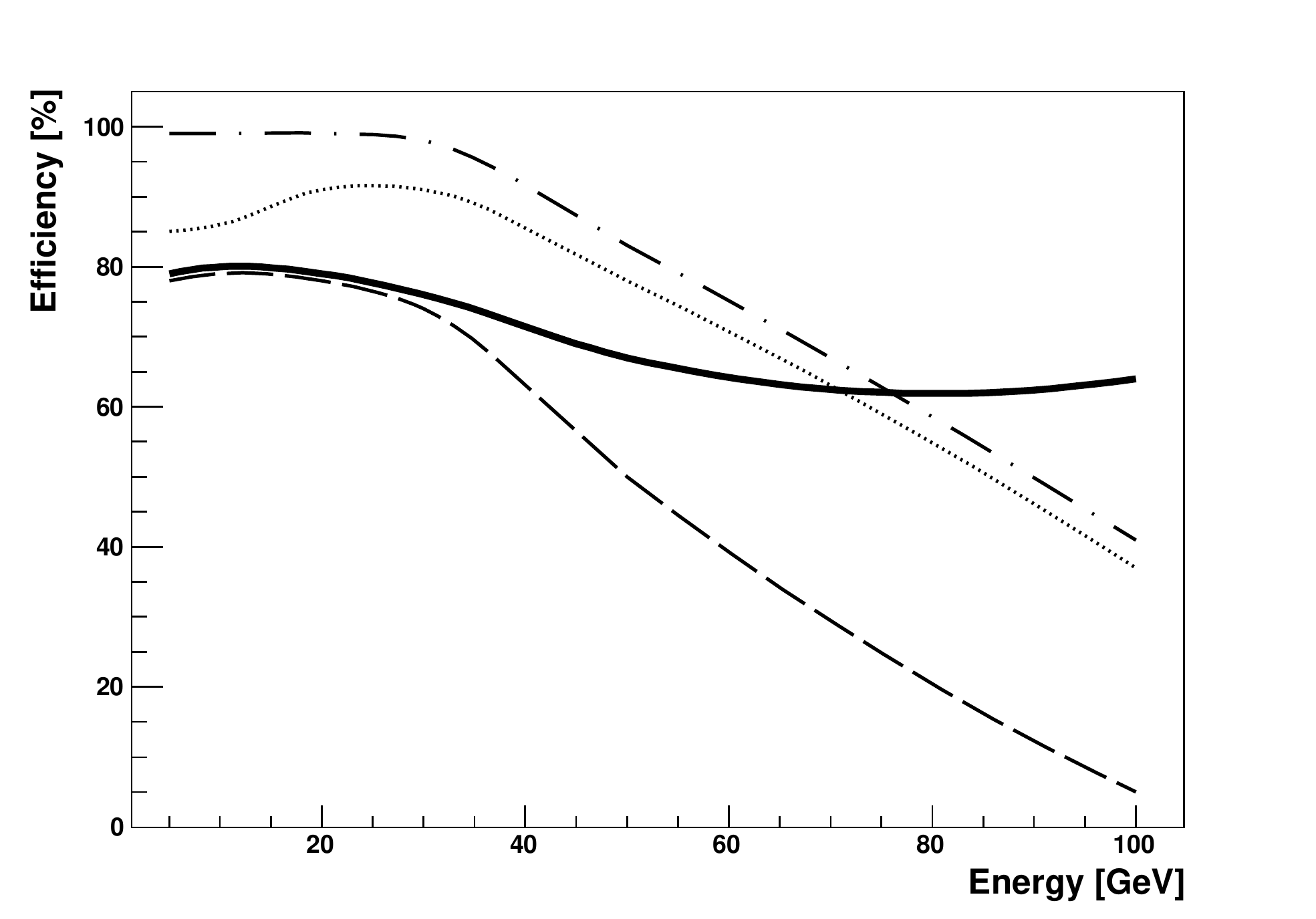,width=150mm,angle=0.0}
\caption{Level 2 trigger efficiency as a function of shower energy in the case of filtering 
	      by the simple {\tt COG} algorithm described in the text. 
	      The efficiency is normalized to the Level~1 trigger photon efficiency. 
	      The dot-dashed line shows the fraction of monoscopic events. 
	      The dotted line and the dashed line show respectively 
	      the effect of the clustering/denoising algorithm and the combined effect 
	      of the nearest neighbor and {\tt COG} algorithms. 
	      The Level 2 trigger efficiency is the sum of the dashed line contribution 
	      and of the stereoscopic events. 
	      The fraction of events accepted by the Level 2 trigger is shown by the solid line.}
\label{fig:efficiecyvsenergy}
\end{center}
\end{figure}

The efficiency of the different algorithm steps has been tested with the Monte Carlo simulations. 
Figure~\ref{fig:efficiecyvsenergy} shows the results of the simulations for each step of the algorithm. 
The solid line indicate the total efficiency of the algorithm as a function of $\gamma$-ray event energies. 
At  energies below 30~GeV all events are monoscopic (the dot-dashed line).
Above 30~GeV, a small fraction of the events start to be stereoscopic, 
then stereoscopy is starting to be efficient above an energy of 60~GeV. 
All stereoscopic events are accepted. 
The fraction of the stereoscopic events is represented by the area above the dot-dashed line. 

The algorithm based on the {\tt COG} cut is very efficient at low energies (the dotted line).
At higher energies the stereoscopy is starting to work very efficiently  
and the majority of gamma events are thus accepted. 
The solid line indicate that for energies below 40~GeV 
the efficiency of the algorithm  for accepting gamma events reach 80\%. 

\subsection{Summary of {\tt COG} algorithm}

Figure~\ref{fig:protonM4} shows that the proton and single muon  rates are reduced by a factor of 3  when the {\tt COG} cut is applied. 
The same applies to the electron background, as shown on figure~\ref{fig:electronM4}. 
The total background rate is summarized on figure~\ref{fig:allM4}.

The {\tt COG} cut also affects the photon efficiency. 
The photon efficiency, shown on figure~\ref{fig:efficiecyvsenergy}, 
has been normalized to the efficiency of the Level 1 trigger. 
As the photon energy increases, the fraction of monoscopic events (solid line) decreases. 
Note that stereoscopic events are automatically accepted by the Level 2 trigger. 
The clustering/denoising  algorithms (dot-dashed line) remove a fraction ($\sim 15\%$) 
of the low energy ( $\le 20$ GeV) photons. 
After the {\tt COG} cut (dotted line), around $80\%$ of the low energy photons pass the Level 2 trigger. 
This fraction decreases with energy, and reaches a minimum of roughly $60 \%$ around 50~GeV
then  raises again because of the increasing fraction of stereoscopic events. 

As has been shown, it is possible to efficiently remove the NSB background 
with a clustering/denoising algorithms. 
The  background rate can be reduced by algorithms which based on the statistic sum  
similar to {\tt COG} cut.

\section{The Level 2 trigger hardware}
\label{sec:hardware}

The H.E.S.S. II telescope is going to observe in a standalone mode a variety of 
different sources with different background conditions. 
The hardware solution of the Level 2 trigger  has to be then 
reconfigurable depending on the inset of a given observation run. 
The reconfiguration of the system should be possible without affecting the observation schedule.

The reconfiguration condition can be achieved by using FPGA (Field Programmable Gate Array) chip. 
The algorithm described in section \ref{sec:algosoft} has been implemented and tested
using dedicated hardware board on Xilinx Virtex4 FPGA.
The details of the hardware solution has been described by  \cite*{2011ITNS...58.1685M} and \cite*{2011APh....34..568M}.

\subsection{The Level 2 trigger board}

The Level 2 trigger hardware is based on an FPGA with an embedded 32-bit PowerPC (PPC) processor, 
which runs  up to a frequency of 300~MHz,  namely a Xilinx's Virtex4-FX12\footnote{http://www.xilinx.com/support/documentation/user\_guides} (V4FX12).  
The PPC in the V4FX12 is equipped with an  auxiliary processor controller unit (APU).
The unit allows   the processor to externalize the execution of custom instructions to the hardware FPGA fabric,  while still using simple function calls in the software. 

\begin{figure}[h!]
\begin{center}
\includegraphics[width=10cm, clip = true,bb=0 0 2592 1944]{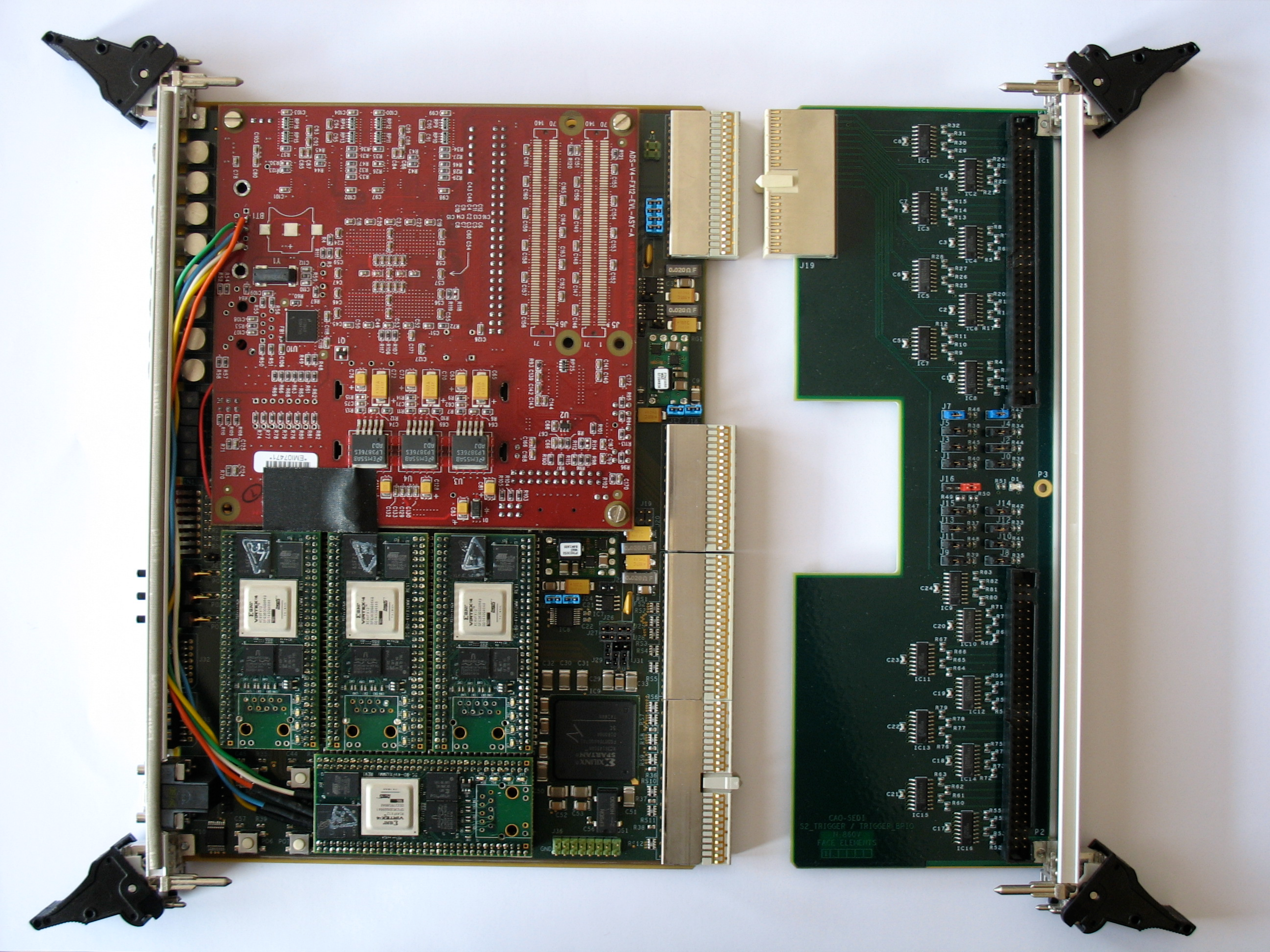}
\caption{
		The Level 2 trigger board equipped with an Avent evolution board and 4 Mini-Modules (on the left) 
		and a rear I/O board for the conversion of 64 LVDS links from the Front End electronics (on the right).} 
\label{fig:finalhardware}
\end{center}
\end{figure}

The evoluation board (EB) distributed by 
Avent\footnote{www.em.avent.com} has been used to ensure optimal   
combination of the sequential and parallel processing capabilities for real time execution. 
 V4FX12 Evaluation 
Kit\footnote{http://www.silica.com/services/engineering/design-tools/ads-xlx-v4fx-evl12-g.htm}
 and  V4FX12 Mini-Module
 \footnote{http://www.files.em.avnet.com/files/177/fx12\_mini\_module\_user\_guide\_1\_1.pdf} (MM) are used in the proposed Level 2 trigger design. 
The large number of accessible user I/O's on the FPGA was the decisive feature of the EB.
The view of the final Level 2 trigger board equipped with the Avent EB is  presented in figure~\ref{fig:finalhardware}.

\begin{figure}[t!]
\begin{tabular}{cc}
\begin{minipage}[b]{0.13\linewidth}
\centering{\includegraphics[bb=630 0 800 450, scale=0.4, clip = true]{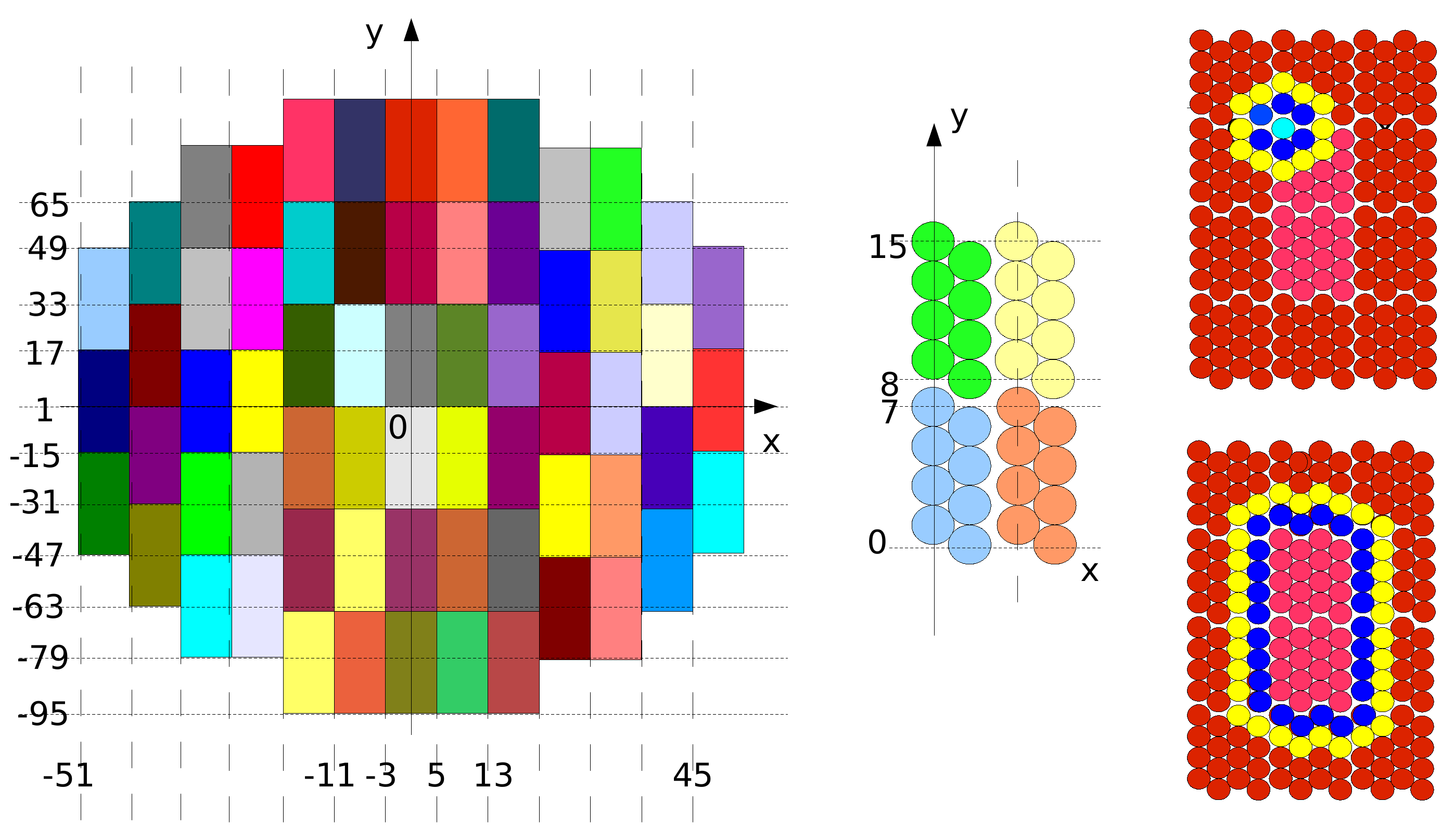}}
\end{minipage}
&
\begin{minipage}[b]{0.87\linewidth}
\centering{\includegraphics[bb=5 100 830 520, scale=0.4, clip = true]{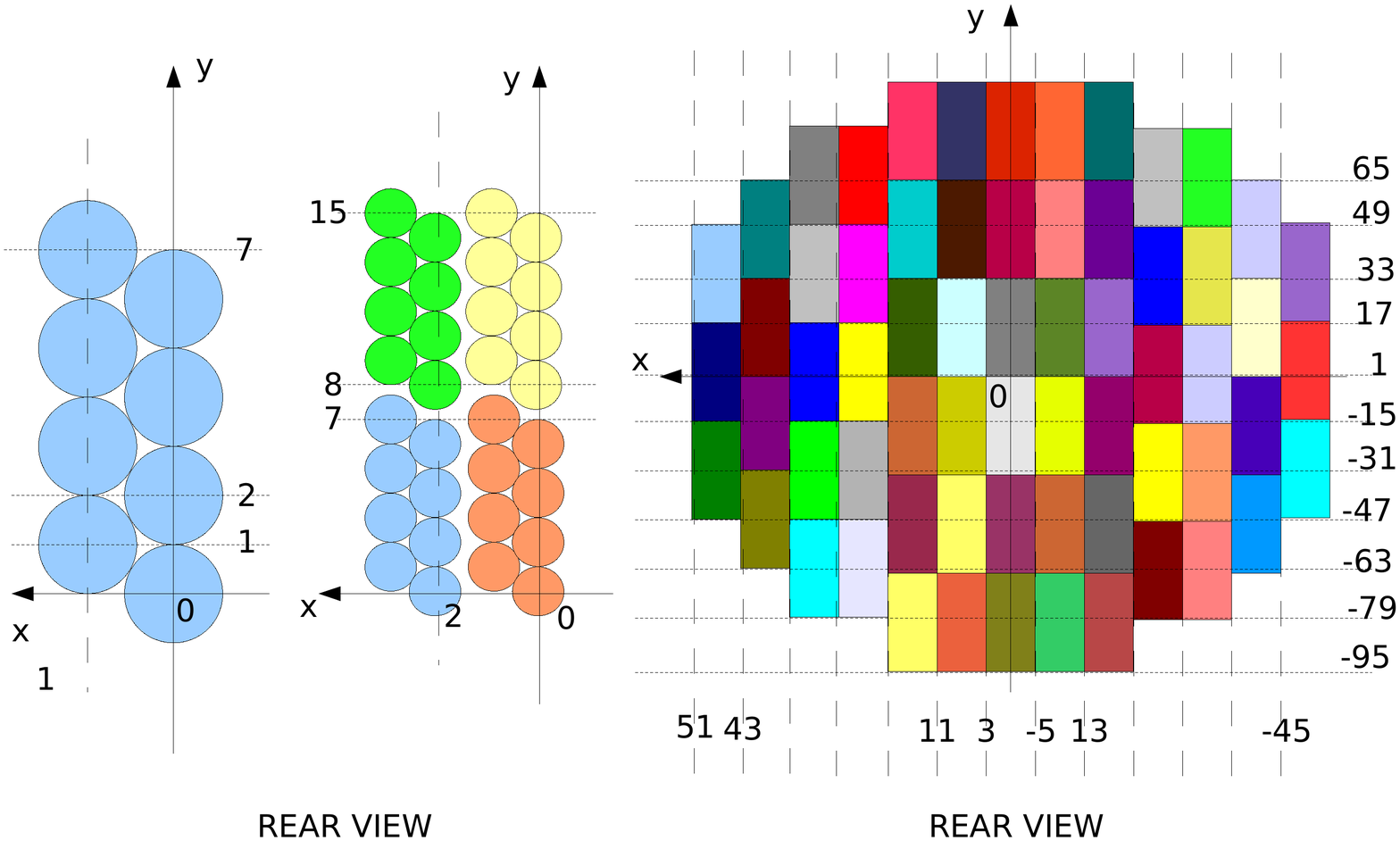}}
\end{minipage}
\\
\end{tabular}
\caption{
\textbf{right~:} The camera of the LCT is composed of 64 drawers. 
Each drawer contains 2 Front End (FE) boards and each FE board carries 8 pixels. 
The 2048 pixels of the camera are on an equilateral triangular grid. 
\textbf{middle~:} Local integer coordinate frame used for Hillas parameter estimation. 
\textbf{left~:} First and second neighbors to the pixels on 4 FE boards in a pair of drawers.}
\label{fig:neighbors}
\end{figure}

The Level 2 trigger system is receiving the data from the Front End (FE) electronics 
on 64 LVDS\footnote{Low-voltage differential signaling (LVDS).} links.
The received data contain two binary images of the camera. 
Each image is made up of 2048 pixels on an equilateral triangular grid (see figure \ref{fig:neighbors}).   

The EB is equipped with the Micron 32 MB DDR SDRAM memory.
It can be used by the PPC  to hold code and data.
The Virtex-4 FPGA is accessible through 76 user I/Os and  is connected to 32M x 16 of DDR memory. 
Both boards hold a 100~MHz oscillator for clocking purposes. 

The EB is in charge of receiving the data from the FE boards, through the backplane. 
Additional information about the stereoscopic nature of the incoming event reaches 
the EB through the front panel. 
The Level 2 system sends the algorithm decision (L2A\footnote{Level 2 Accepted} 
or L2R\footnote{Level 2 Rejected})  as an output
to the Data Acquisition System (DAQ).
The data acquisition FIFOs\footnote{First In, First Out} buffer have a capacity of 50~events, 
and is used to hold the camera data 
while awaiting for the Level 2 trigger response. 
The capacity of the buffer sets an upper bound on the latency of the Level 2 system, 
constraining the real-time implementation of the Level 2 trigger algorithms.\\

\subsection{The algorithm implementation} 
If the processed event is tagged as \textit{stereo}, the Level 2 trigger uses its selection algorithm 
to decide if the event will be issued as \textit{accepted} (L2A) or \textit{rejected} (L2R).
The algorithm proceeds as follow:
\begin{tabbing}
\quad \=\quad \=\quad \kill
1. Set to 0 all pixels in $map_1$  that are not in clusters of 3 at least $\rightarrow \widetilde{map}_1$\\
2. {\bf IF}     $\widetilde {map}_1 = 0$ {\bf THEN} Reject {\bf ELSE} \\
\>3. Set to 0 all isolated pixels in $map_1$ $\rightarrow \widehat{map}_1$\\ 
\>4. Compute Hillas parameterd of $ \delta_1 \widehat{map}_1  +( \delta_2 -\delta_1 ) map_2 $\\
\>5. Compute distance $\Delta$  from center of gravity ({\tt COG}) to target $(x_c, y_c)$\\
\>6. {\bf IF}     $\Delta \ge  \tau_{\textrm{{\tt COG}}}$  {\bf THEN} Reject  {\bf ELSE}  Accept
\end{tabbing}
where  ${map}_1$ and ${map}_2$ are the two input binary maps associated 
with threshold values of $\delta_1$ and $\delta_2$,  
 $(x_c, y_c)$ are the pointed target's coordinates in the camera plane and $\tau_{\textrm{{\tt COG}}}$ is the decision threshold on the nominal distance $\Delta$ between the {\tt COG} of the event  and the target position.

The Level 2  trigger is build as a pipeline system. 
The first step of the Level 2 pipeline is a  transposition of the input 64$\times$64 binary data matrix.
The step is performed before the matrix is available to the PPC in a dedicated cacheable memory block. 
The first half of the data represents the binary camera image with the true 
values for pixels above threshold $\delta_1$ ($map_1$).
The second half is a $map_2$ obtained for the pixel threshold $\delta_2$, respectively.  

The 32 bit words are then read by the PPC from the block.

Each byte corresponds to the 8 pixels from one FE board. 
The geographical position of pixels  follows a constant logical pattern. 
The auxiliary processor of the V4FX12 is used to achieve 
a parallel implementation of the non-linear filters in step 1 and 3. 
This has been built using logic AND and OR gates. 
In this way denoising (Step 1) was implemented by convolving $map_1$ with the filter:  
\begin{equation}
\widehat{map}_1(i) = map_1(i)\wedge \left(\bigcup_{j=1..6} map_1(i_j) \right)\,,
\end{equation}
where $i_j$ is used to index the 6 nearest neighbors of  a pixel.
The similar filter is used to detect the cluster of at least 3 or 4 pixels. 

A fast implementation of step 4 is obtained by reorganizing 
the wighted sums that define the $1^{st}$ and $2^{nd}$ order moments of the input data.
Computing the $1^{st}$ and $2^{nd}$ order moments of the denoised combined map is  
common for the estimation of Hillas parameters and other parameters of interest~\citep{1985ICRC....3..445H}. 
It can be defined as:
\begin{eqnarray}
m_x =  \sum_i m_i x_i\,,  \quad & \quad m_y =  \sum_i m_i y_i\,,\\  
m_{xx} =  \sum_i m_i x_i^2\,, \quad & \quad m_{yy} = \sum_i m_i y_i^2\,, \\
m_{xy} = \sum_i m_i x_i y_i\,, \quad & \quad m = \sum_i m_i\,,
 \end{eqnarray}
where $i$ indexes the 2048 pixels in the processed data maps, and $m_i$ is the weight assigned to pixel $i$. 
The binary maps  $\widehat{map}_1$ and  ${map}_2$ can be processed separately and the sums are profitably rearranged for an efficient hierarchical computation of the moments. 

First, a byte-addressable \textit{look-up table} (LUT)  is used to compute the $1^{st}$ and $2^{nd}$ order moments on each FE board. These are combined locally to compute these statistics on each of the 64 pairs of drawers.  
This local summation requires the LUT outputs to be properly \textit{translated} 
depending on the position of a given FE board in the current pair of drawers. 
Summation over the 64 pairs of drawers requires an additional transformation of these statistics 
to account for the translation and scaling of the local frame.
This  requires  a move of the current drawer to its correct position within the global coordinate frame of the camera. 
In the end, the contributions of the two binary maps are linearly combined with weights $\delta_1$ and $ \delta_2 -\delta_1$ providing final 32 bit integer statistics $m_{yy}$, $m_{xy}$, $m_{xx}$, $m_y$, $m_x$ and $m$ for the combined map. 

With this fast implementation, the PPC computes the first and second order moments of the input data in a maximum of 18000 clock cycles. 
For an even faster execution time, given that these statistics will most often be estimated for low energy events when only very few pixels are high in $\widehat{map}_1$ and even less in  ${map}_2$, it is worth checking if a byte is zero before using the LUT. 
As a result the computation time will vary almost linearly with the number of \textit{active} bytes in the data. 

The algorithm proposed in section \ref{sec:algosoft} uses only the first-order statistics to compute the nominal distance  in finite precision:
\begin{equation}
\Delta = \sqrt{  (\frac{ m_y}{ m} - y_c ) ^2 +  3( \frac{m_x}{ m } - x_c ) ^2}\,,
\end{equation}
where the factor $3$  is due to the equilateral triangular grid and the accompanying $\sqrt 3 $ left out in the moment computation for simplicity. The specified precision on the target coordinates is $1/32^\textrm{th}$ of the unit length, giving the precision to which the 
{\tt COG} coordinates have to be computed. 
%
\subsection{Experimental timing results }\label{sect:results}
%
%
\begin{figure}
\begin{center}
\includegraphics[width=140mm,angle=0.0,bb=0 0 567 405]{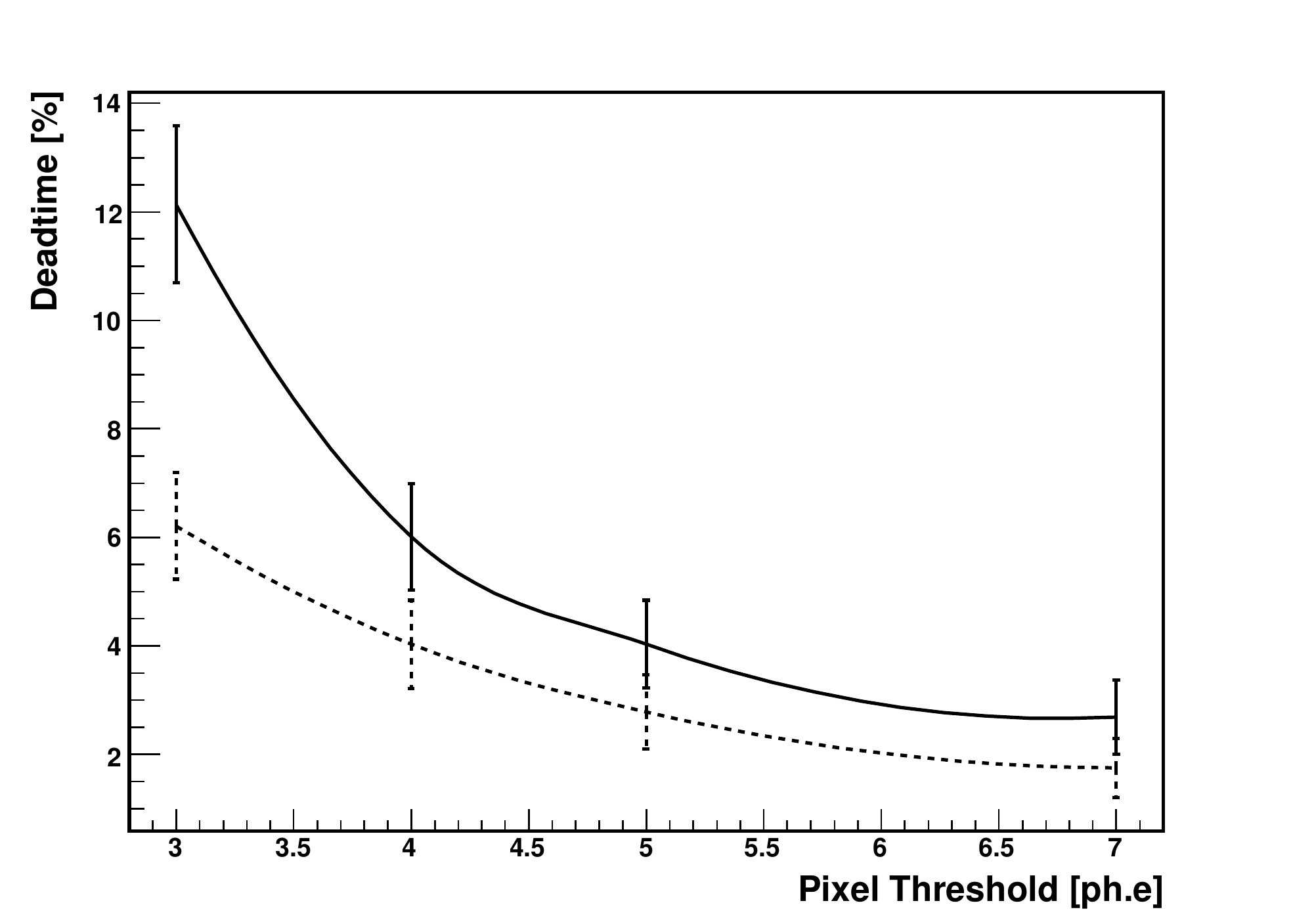}
\caption{
Estimated deadtime of the PPC in the single FPGA implementation of the Level~2 trigger system. 
The continuous and dashed lines correspond to  Level~1
pixel mulitiplicities equal to 3 and 4 respectively.}
\label{fig:maxmeanrate}
\end{center}
\end{figure}
The design and real time implementation of the Level 2~algorithm 
is constrained by the maximum latency.
If the Level 1~rate is of the order of 100~kHz, 
the maximum latency of the Level 2 trigger is $\sim$500~$\mu$s.
The minimum time between two events is then 10~$\mu$s. 
If the Level 1 rate is reduced to 5~kHz, the maximum latency of the Level~2 trigger increases to 
10~ms and the minimum mean time between two events is 200~$\mu$s. 

The Level~2~architecture was tested with timing experiments in order to benchmark the performance of the proposed hardware, firmware and software solution for the Level 2 trigger system. 
With this setup, a stable behavior of the system was observed up to a maximum Level 1 rate slightly above 10~kHz. The multi-FPGA system could sustain a maximum mean rate close to 30~kHz.
A more realistic estimation of the maximum acceptable  Level~1~rate is plotted on figure~\ref{fig:maxmeanrate}.
This results were obtained for the first implementation of the Level 2 trigger system. 
The dead-time estimate is based on the simulated rates  
reported in sections~\ref{sec:TriggerSimulations} and on time measurements of the different elementary steps in the Level 2 trigger pipeline. 
For typical Level 1 trigger conditions (multiplicity 3 or 4 and pixel threshold between 3 and 7) 
the estimated average processing time is  
$\sim$ 37~$\mu$s, which corresponds to a maximum Level 1 trigger rate of 27~kHz. 
Actually, for these trigger conditions, the system occupancy is estimated to be $\le 20$ \% as shown on figure \ref{fig:maxmeanrate}.
The proposed multi-FPGA system will provide thus a safe margin. 
However, the real Level 1 and Level 2 trigger rate have to be determined on site.
%

%
%
\section{Conclusions}
\label{sec:triggerconclusions}
The Level 2 trigger is going to be used  to reduce the trigger rate of the LCT at low energy. 
The principle of the Level 2 trigger is to build a 2-bit (``combined'') map of the camera pixels at the time of trigger. 
The NSB events can then be rejected by demanding clusters of pixels on the combined map. 
Further rejection of the hadronic background can be obtained by
 using quantities such as the {\tt COG} of the pixels above a pixel threshold. 
 A possible, illustrative, algorithm for the Level 2 trigger system has been described in section \ref{sec:algosoft}. 
 This algorithm shows that the required rejection
of the NSB and isolated muon triggers is achievable.

The hardware and software integration into the LCT camera of the previously 
described system based on a single V4FX12 has been achieved. 
The Level 2 system is already  fully integrated in the H.E.S.S. II acquisition system and is currently undergoing tests with real data.


\setcounter{chapter}{2}
\setcounter{section}{0}
\setcounter{equation}{0}
\setcounter{figure}{0}

\part{Data analysis and modeling of PKS 1510-089}
\section{Introduction}
The observations of the FERMI satellite  in the high energy (HE) range resulted in the identification of 1873 sources, 
according to the second FERMI catalog \citep{2012ApJS..199...31N}.
Among these 1873 sources majority are blazars. 
Blazars are very luminous active galactic nuclei (AGNs)
with a relativistic jet pointing toward the observer.

The broadband spectrum of blazars is dominated by  non-thermal  emission 
produced in the jet \citep{1978bllo.conf..328B}. 
The spectral energy distribution (SED)  is characterized by two broad spectral components.
One component, which extends from the radio to optical/UV/X-rays, peaks at low energy, 
and is produced by the synchrotron radiation of relativistic electrons. 
The second one, from X-rays to $\gamma$-rays, peaks in the HE range and in most current interpretations  
is produced by inverse Compton (IC) radiation with as possible source of seed photons either the synchrotron radiation, 
or the broad line region (BLR) or the dusty torus (DT).   

Blazars can be divided into two classes: Flat Spectrum Radio Quasars (FSRQs) and BL Lac  objects.
FSRQs are distinguished from BL Lac objects by the presence of broad emission lines,
which are not found in BL Lac objects.
The FSRQs have HE components much more luminous than low energy ones.
The seed photons for IC  radiation most probably come from BLR. 
The seed photons, for BL Lacs objects, probably come from the synchrotron radiation,
and both spectral components have comparable luminosities. 

According to the prediction of  \cite{2005MNRAS.363..954M}, 
the spectra of blazars should have a cut-off at a few GeV due to the Klein-Nishina  effect, 
if the high energy component is produced by IC of photons reemitted in BLR. 
Spectral breaks at a few GeV have been found in the $\gamma$-ray spectra of many 
FSRQs and BL Lacs \citep{2011ApJ...733L..26A}. 
The most prominent example is 3C 454.3 \citep{2009ApJ...699..817A}.
In addition, the luminous IR-UV photon field, from the BLR and the DT,
can cause a strong absorption of HE and VHE photons by electron-positron pair production \citep{2003APh....18..377D,2006ApJ...653.1089L}.

The other possible explanation of the spectral break involves  photo-absorption by 
He II and  Ly$\alpha$ \citep{2010ApJ...717L.118P}.   
Gamma rays photo-absorbed by He II recombination (54.5 eV) and Ly $\alpha$ (40.8 eV) photons 
from the BLR would create a break at $\sim$5~GeV.   

The number of known blazars in HE range is larger than 1000, 
but in VHE range only about 50 blazars have been detected so far \citep*{2012arXiv1205.0068E}.
Up to now,  only BL Lacs objects, from the blazar class, were clearly detected as VHE sources.

Recently however, Cherenkov telescopes have detected  3 FSRQ in sub-TeV range. 
The first detected object was 3C 279, observed with the MAGIC telescope \citep{2011A&A...530A...4A}.
Two additional FSRQs are 4C 21.35 detected by the MAGIC telescope \citep{2011ApJ...730L...8A} 
and PKS 1510-089 detected with H.E.S.S. \citep{2011ATel.3509....1H}. 
The detection of these objects shows that FSRQs can also emit photons in the VHE range.
The emission in this energy range  is very difficult  to explain  by inverse Compton of photons from BLR.
%

\section{Unification schemes of  active galaxies}

\begin{figure}[h]
\vspace{2mm}
\begin{center}
\hspace{3mm}\psfig{figure=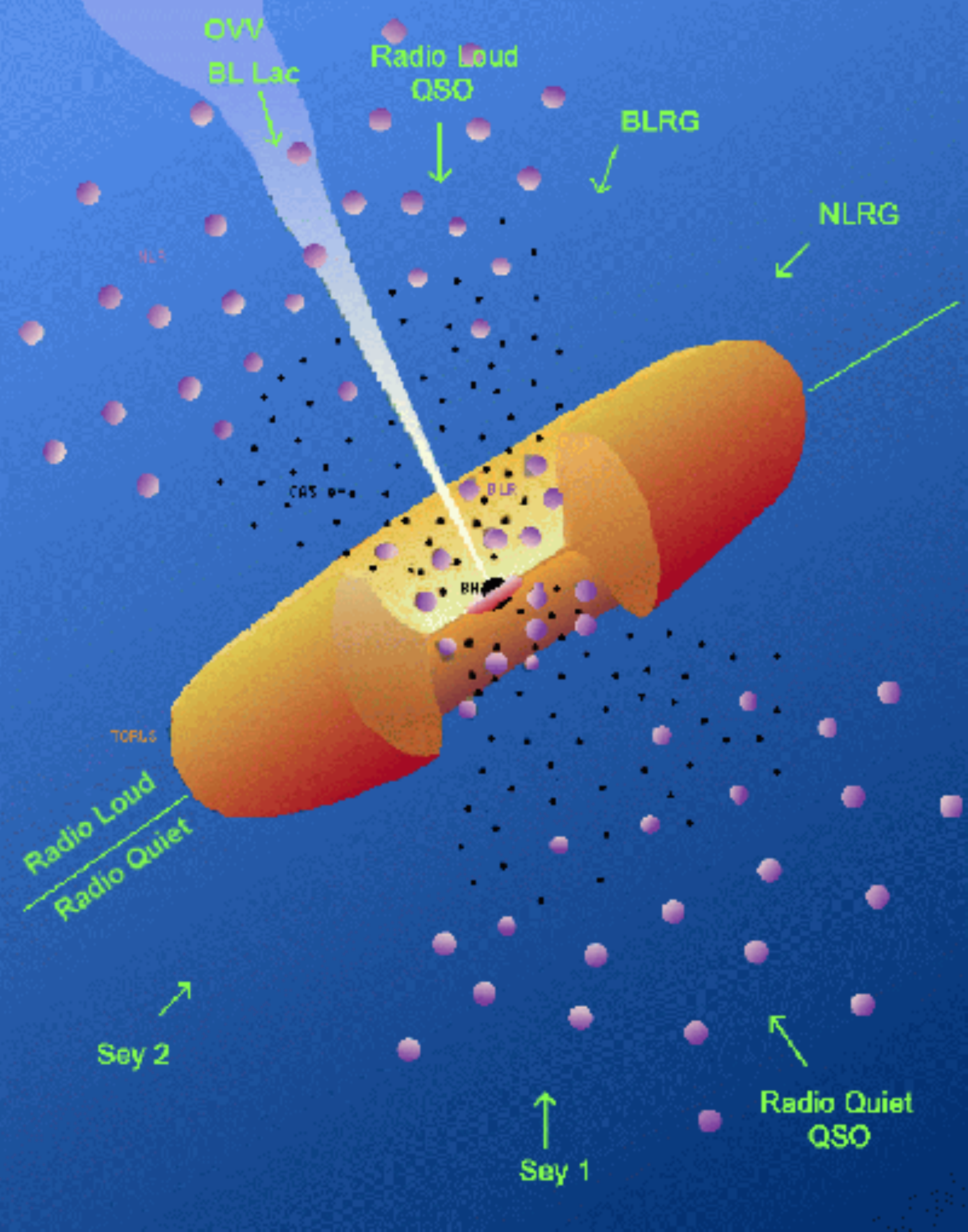,height=100mm,angle=0.0,bb=0 0 400 509}
\caption{A schematic diagram of AGN.}
\label{fig:agn_unification}
\end{center}
\end{figure}

The unified theory of AGNs has been developed since the 70s \citep{1993ARA&A..31..473A}. 
The~basic idea of the unification assumes  that all of the active galaxies have 
similar internal structure of  their nuclei, 
but their appearances depend on their  orientations. 
Figure \ref{fig:agn_unification} shows the model proposed by \cite{1995PASP..107..803U}.
This model assumes that all the AGNs are powered by accretion of surrounding matter 
onto the supermassive black hole located in the center of the host galaxy. 
The accreting  matter forms a geometrically thin accretion disk and corona 
heated by magnetic or viscous processes.   
Farther out there is a geometrically thick DT.
The emission from the accretion disk is reprocessed in DT and BLR.
The fraction of reprocessed emission by BLR and DT ($\xi_{BLR,DT}$) can not be larger than 1. 
The typical values of  $\xi_{BLR,DT}$ are in the range of $\sim$~0.1--0.3 \cite{2012ApJ...760...69N}.

The AGNs are divided to many subclasses. 
The most common classifications based on the properties like an appearance (if the source is observed as a point-like or a clear galaxy host), an presence or an absence of the broad  or narrow line regions, a variability or a polarization. 
The most popular groups are 
\begin{itemize}
	\item {\it radio-load} active galaxies (10\%)
	\item {\it radio-quiet} active galaxies (90\%)
	\item {\it Seyfert galaxies} named after Carl Seyfert, who pointed out the first six {\it Seyfert galaxies}. This group was later subdivided into two types (according to presence or absence of the broad or narrow emissions lines)
	\item {\it Optically Violently Variable} (OVV), this class is marked by exceptionally rapid and large amplitude variability in the optical band
\end{itemize}

The further classification distinguishes  also quasars group, which consists of objects found at large distances with very bright emission from the jet.
Quasars include radiogalaxies and blazars.

\section{Blazar sequence}
\label{sec:BlazarSequence}
Blazars are the most luminous  AGNs. 
Their emission is dominated by the boosted radiation from the jet,
and their spectra consist of two broad components.
The low-frequency component (LFC) has a peak in the IR-X-ray range,
while the high-frequency component  (HFC) has a peak in MeV to TeV range.
Both components are highly variable, with time scale ranging from years to the fraction of a day.
Blazars are also characterized by high radio and optical polarization,
and in many cases strong $\gamma$-ray emission. 
  \begin{figure}[h]
\vspace{2mm}
\begin{center}
\hspace{3mm}\psfig{figure=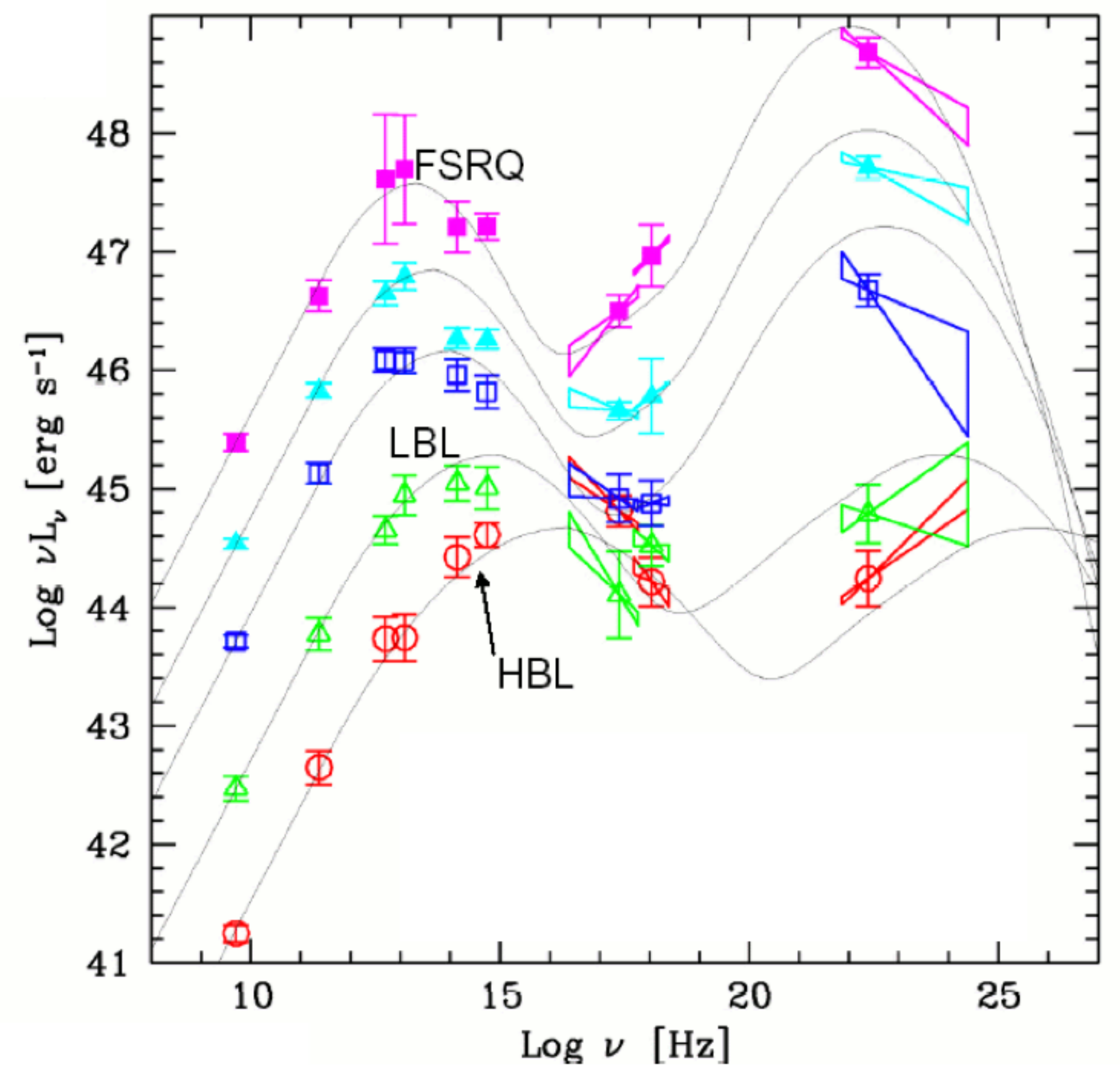,height=100mm,angle=0.0,bb=0 0 731 696}
\caption{Blazar Sequence. Figure credited from Donato et al. (2002) and Fossati et al. (1998).}
\label{fig:blazar_seq}
\end{center}
\end{figure}

The superimposed blazar  spectral energy distributions (SEDs) from figure \ref{fig:blazar_seq} 
suggest the correlation between the luminosities and the peaks positions. 
The sequence is characterized by an increasing synchrotron peak frequency,
a decreasing overall luminosity and a decreasing dominance of the $\gamma$-ray emission
over the synchrotron component. 
\cite{1998MNRAS.299..433F}  based on this behavior  elaborated an unified view of the SEDs called the blazar sequence (see figure \ref{fig:blazar_seq}). 

FSRQs and BL Lacs occupy the opposite sides of the blazar sequence.
In the case of FSRQs, the peaks of low-frequency ($\nu_{syn}$) components
are shifted toward lower frequencies as  compared to the BL Lacs.
 The broad-band spectrum of FSRQs is characterized by a large luminosity ratio
 of their high-frequency and low-frequency spectral components.
 This luminosity ratio can reach values up to 100.
In the case of BL Lac objects, the luminosities of high and low frequency components are comparable.

Since the location of the low-frequency peak is quite broad, 
blazars are sometimes further divided into  sub-groups based on the peak
position ($\nu_{syn}$). 
Blazars with $\nu_{syn}\le10^{14}$~Hz are called low-synchrotron-peaked (LSP).
LSP group contains both FSRQs and LSP BL Lac objects (LBLs).
Blazars with a low-frequency peak located in frequency range 
$10^{14}$~Hz~$\le \nu_{syn} \le 10^{15}$~Hz are called 
intermediate-synchrotron-peaked (ISP).
The ISP group primarily consists of intermediate BL Lac objects (IBLs).
Finally, the last group, high-frequency peaked (HSP) BL Lac objects, is characterized by $\nu_{syn}>10^{15}$~Hz 
\citep{2010ApJ...715..429A}.
The sequence then appears as follows:  FSRQ$\rightarrow$LBL$\rightarrow$IBL$\rightarrow$HBL.   
The FERMI satellite provided a large sample of sources with spectra measured over almost 4 years,
the blazar classification seems to be much more complicated \citep{2012MNRAS.420.2899G,2011ApJ...740...98M}.
However, here I use this simplified approach to illustrate the basic properties of the blazar class. 

\section{Accretion disk}
\label{sec:AD}

Lets assume a "standard" \citep{1973A&A....24..337S} accretion disk.
The accretion disk is optically thick and emits a large amount of thermal radiation from infrared to ultraviolet.  
The thermal emission of accretion disks peak in the UV  ("big blue bump").
In blazars the UV observations during the low state of synchrotron radiation  can be used to estimate the upper limit of the accretion disk luminosity $L_d$.
Following the prescription given by 
\cite{2008NewAR..52..253K} and  \cite{2009MNRAS.399.2041G}, the temperature of the disk is given by:
\begin{equation}
T^4_{disk}(r) = \frac{3R_SL_d}{16\pi \epsilon \sigma_{SB} r^3}\left[1-\left(\frac{3R_S}{r}\right)^{\frac{1}{2}}\right]\,,
\end{equation}
where $\sigma_{SB}$ is the Stefan Boltzman constant, $R_S$ is the Schwarzschild radius of a black hole, 
$3R_S$ refers to the last stable circular orbit for the Schwarzschild black hole and the disk extends up to 500 $R_S$, 
 $\epsilon \, \sim \,0.05\,-\,0.42$ is the efficiency of rest-mass conversion, 
which depends on the inner  boundary conditions, and the black hole spin.  
The accretion efficiency, $\epsilon$, is  linked to the bolometric disk luminosity, $L_d$, and to the accretion rate $\dot{M}$ as
$L_d = \epsilon \dot{M} c^2$.
%
The radiation region of the disk extends from $\simeq 3R_S$ to 500~$R_S$.
The disk temperature peaks  at $R \simeq 4R_S$.
\section{Broad line region}
\label{sec:BLR}

The clouds surrounding the central part of AGN 
are ionized by ultraviolet radiation from the accretion disk. 
The clouds reprocessed the disk radiation and produce emission lines,
which are broadened due to high velocity of the cloud around the central black hole. 
This part of the AGN model is called the broad lines region (BLR). 
The luminosity of the BLR is proportional to the $L_d$ as
\begin{equation}
L_{BLR}=\xi_{BLR} L_d\,,
\end{equation}
where $\xi_{BLR}\simeq0.1-0.3$ is the fraction of the disk radiation reprocessed in BLR.
The radius of the BLR can be derived using the method called "emission-line reverberation mapping" \citep{2004ApJ...613..682P}.
This technique uses the time-lag of the emission line light curve with
respect to the continuum light curve to determine the light crossing size of the BLR in AGNs. 
The reverberation studies resulted in
the empirical relationship  between $R_{BLR}$ 
and the optical continuum luminosity at 1350$\mbox{\AA}$ \citep{2005MNRAS.361..919P}:
\begin{equation}
R_{BLR}=(22.4\pm0.8) \left [ \frac{\lambda L_{\lambda}(1350\mbox{\AA})}{10^{44}\mbox{erg\, s}^{-1}} \right ]^{0.61\pm0.02} \, \mbox{light days}\,.
\label{eq:RBLR}
\end{equation}

The energy density of BLR radiation fields $u_{BLR}$ is constant within $R_{BLR}$ and declines as $r^{-2}$ outside: 
\begin{equation}
u_{BLR}\simeq \frac{\xi_{BLR}L_d}{4\pi c R_{BLR}^2}\frac{1}{1+(r/R_{BLR})^2}.
\label{eq:ubel}
\end{equation}

\section{Dusty torus}
\label{sec:DT}
The radiation from accretion disk is also reprocessed by the dust, which forms a torus located outside the accretion disk. 
The luminosity of the DT can be expressed as: 
\begin{equation}
L_{DT}=\xi_{DT} L_d\,,
\end{equation}
where $\xi_{DT}$ is the fraction of reprocessed emission. 
The reflectivity, in this case, is of the order of $\sim$ 0.1 (Ghisellini et al. 2009).

The temperature of the DT \citep{2000ApJ...545..107B} is expressed as
\begin{equation}
T_{DT}^4=\left(\frac{L_d}{4\pi \sigma_{SB} R_{DT}^2}\right)\,. 
\end{equation}
The temperature of the dust gives the characteristic frequency of the DT radiation field:
\begin{equation}
h\nu_{DT} \simeq 3.92 \,k\,T_{DT}\,.
\end{equation}
The size of DT is approximated by  $R_{DT} \simeq 2.9\,$pc$\times T^{-2.6}_3$ 
\citep{2008MNRAS.386..945T,2008ApJ...685..160N,2009ApJ...704...38S}, 
where the dust temperature $T_3=T\times1000$.

Within the $R_{DT}$, the energy density, $u_{DT}$ of external
radiation fields, is roughly independent of the radius, and  outside decreases with distance:
\begin{equation}
u_{DT}\simeq \frac{\xi_{DT}L_d}{4\pi c R_{DT}^2}\frac{1}{1+(r/R_{DT})^2}\,.
\label{eq:udt}
\end{equation}

%
\section{Jets} 
\label{sec:jet}
Jets are streams of hot plasma, 
which moves with relativistic speeds.
Jets transport the energy up to distances of many kpc.
Around 10\% of the jet energy is dissipated in the very first parsecs. 
When the jet collide with the intergalactic matter it produces luminous  lobes. 

The exact mechanism of the jet formation and collision is unknown.
It has been suggested that it has to be mediated by the magnetic field in the inner part of AGN \citep{1977MNRAS.179..433B}.
Its strong collimation suggest a large density of magnetic energy. 

Jets at small distances (subparsec) 
are dominated by magnetic field \citep{1983AJ.....88..245B}. 
At larger scales the energy of the jets is dominated by matter \citep{2005ApJ...625...72S}. 
The majority of emission is produces at the parsec scale.  

The radio observations of superluminal motion impliy that  matter in the jet reaches relativistic speeds. 
Other observations suggest that jets may have a Lorentz factors $\Gamma$ of the order of 10 to 20. 
The observed  luminosity  of  objects moving with a large Lorentz factor 
is  boosted by the Doppler effect, where the
Doppler factor  is defined as
\begin{equation}
{\cal D} = \frac{1}{\Gamma \left( 1 - \beta_\Gamma \cos \theta_{obs} \right)}\,,
\end{equation}
where $\theta$  is the angle between the direction of the source motion and the line of sight between the source and the observer, and $\beta_\Gamma = \sqrt{\Gamma^2-1}/\Gamma$.
The observed radiation flux is
\begin{equation}
\nu_{obs} F_{\nu_{obs}} 
= {\cal D}^4 \frac{\nu L_\nu}{4 \pi d_L^2} \,,
\end{equation}
where $\nu = \nu_{obs} (1+z)/{\cal D}$, $L_\nu$ is the source luminosity 
and $d_L$ is the luminosity distance.

\section{Models of jet emission}

The spectrum energy distribution (SED) is composed of two broadband components. 
There is an agreement that the lower component, which peaks at infrared to X-rays, 
is  produced by the synchrotron emission of relativistic electrons within the jet. 
The non-thermal character of emission is confirmed by the observations of  rapid variability on time scales of days or less throughout the entire wavelength range \citep{1995ARA&A..33..163W} and high polarization (even 40\%) in the radio and the optical range  \citep{1990A&AS...83..183M}.

The origin of the high energy component is far more debated.
The most common interpretation suggests that the origin is the inverse Compton (IC) emission
of relativistic electrons, or pairs -- the so-called leptonic models. 
The obvious choice of seed photons would be the synchrotron radiation 
from the same population of electrons.
The family of such scenarios is called the synchrotron self-Compton (SSC) models,
and seems to explain well the spectra of BL Lac objects, 
where the lack of any emission and absorption lines suggests the absence 
of any external radiation fields.  
For more details see  \cite{1981ApJ...243..700K,1985ApJ...298..114M,1989ApJ...340..181G}.

For FSRQs, SSC models cannot easily explain the large difference of luminosities  
between the low-frequency component and high-frequency component peaks observed \citep{2009ApJ...704...38S}. 
The luminosity of the high energy component exceeds the luminosity of the low energy component  by a factor 10 to 100 times. 
An external source of seed photons for IC was proposed to explain 
such a large difference of the peak luminosities  \citep{1992A&A...256L..27D,1993ApJ...416..458D,1995ApJ...441...79B,1994ApJ...421..153S}.
  
In addition to the electrons,  protons are an inevitable component of the jet.   
It is worth noting that even if the number of electrons largely exceed the number of protons, 
because of the large mass difference $m_p/m_e\sim 1836$, the jet power may still be dominated 
by ultra-relativistic  protons. 
In some scenarios, the so-called hadronic models, 
protons are also involved  in the radiation mechanisms of jet emission. 
Such hadronic models were discussed for example by 
\cite{2003APh....18..593M,2010arXiv1006.5048B,2012arXiv1210.5024C}.

The direct proton synchrotron emission or proton IC process 
are much less efficient in emitting radiation  than the same processes  involving electrons ($\sigma_T \sim m_{e/p}^{-2}$).
The proton cooling time scale is too long to explain the emission of blazars. 
To overcome this difficulty  some hadronic models
involve pion production. 
The high energy protons interacting with low energy photons may produce pions: 
\begin{equation}
\begin{array}{ccc}
 & p+\gamma  \rightarrow n + \pi^{+} \,,& \\
 & p+\gamma  \rightarrow p + \pi^{0} \,.& \\
\end{array}
\end{equation}
  
The pions initiate cascades of high energy particles.
The neutral pions can decay to TeV photons. 
The hadronic models require an extreme environments to explain blazars emission. 
They  require in particular  a magnetic field of the order of 100 G, 
or an extreme number of protons with  large Lorentz factors
(see e.g. \cite{2009ApJ...704...38S} for a current review).
 
In the present thesis the leptonic model is used to explain the emission of blazar PKS 1510-089.

\section{ Internal shock scenario}

The observations of blazars show a correlated variability of the low and high peak components
with  a time scale as low as hours  suggesting a compact region of emission \citep{2005A&A...430..865A}.
This may suggest that a significant fraction of the jet energy is dissipated
at a sub-parsec scale from the center of the region of emission. 
The energy dissipation  can be explained by an internal shock scenario 
\citep{1994ApJ...421..153S,2001MNRAS.325.1559S},
or the  reconnection  of magnetic field \citep{1992A&A...262...26R,1996ApJ...456L..87B}.
Both approaches are sufficient to produce an efficient acceleration of particles. 

The internal shock scenario assumes  some instability in the central part of the active galactic nuclei, 
 which results in the ejection of  blobs of matter. 
The jet consists of blobs  of different velocity, masses, and densities \citep{1978PhyS...17..193R}. The blobs with the larger velocity catch up with those with smaller velocities, and a nonelastic collision occur. 
Consider two blobs of matter with velocities $\Gamma_1 < \Gamma_2$ and masses $m_1$ and $m_2$. 
Conservation of energy and momentum implies that the dissipation "efficiency" $\eta$ of the total bulk 
kinetic energy during the collision is \citep{2004ApJ...611..770M}:
\begin{equation}
\eta = 1 - \Gamma_{sh}\frac{m_1+m_2}{\Gamma_1 m_1 + \Gamma_2 m_2}\,,
\end{equation}
where $\Gamma_{sh}=(1-\beta_{sh}^2)^{-1/2}$ is the bulk Lorentz factor of the forward shell after the blobs collision, 
and $\beta_{sh}$ the shell velocity is:
\begin{equation}
\beta_{sh}=\frac{\beta_1\Gamma_1 m_1 + \beta_2\Gamma_2 m_2}{\Gamma_1 m_1 + \Gamma_2 m_2}\,.
\end{equation}

Typically $\eta$ is  of the order of 5\% - 10\% \citep{2001MNRAS.325.1559S}.
The efficiency of dissipation increases when the velocity difference is large. 
The largest efficiency is obtained during the first collisions, close to the jet base \citep{2001MNRAS.325.1559S}.  
Assuming  that the blobs velocities before the collisions are $v_1$ and $v_2$, respectively, 
ejected with initial separation $\Delta r_{ej}$, the blobs will collide at a distance: 

\begin{equation}
R_0 = \frac{v_2}{v_2-v_1}\Delta r_{ej} \simeq \frac{2\Gamma^2_1 \Gamma^2_2}{\Gamma^2_2-\Gamma^2_1} \,.
\end{equation}

This region is called the blazar zone, where the majority of the non-thermal emission is produced. 
Collisions of blobs produce a shock structure, which accelerates particles to relativistic energies.
In the absence of a detail model
of particle acceleration
the resulting distribution of particles is assumed to be a pawer law or a broken power law.  

In this thesis the electron  injection 
function is assumed to take  the broken power law form  
\begin{equation} 
Q_{\gamma} = K_e \frac{1}{\gamma^p + \gamma_{\rm br}^{p-q}~\gamma^q} \, ,
\label{eq:Qgamma}
\end{equation} 
where $K_e$ is the normalization factor, $p$ and $q$ are  
spectral indices  of the injection function at the low and high energy 
limits, respectively, and $\gamma_{\rm br}$ is the break energy.

\section{Synchrotron radiation}
When a relativistic electron with a Lorentz factor $\gamma$
is moving in a magnetic field $B$, it emits non-thermal radiation with 
$\nu_{syn}=4/3\gamma^2 \nu_L$, where $\nu_L=eB/(2\pi m_ec)=2.8\times10^6 \frac{B}{1G}\,$Hz. 
The average photon energy is thus:  
\begin{equation}
\langle \varepsilon \rangle \simeq \frac{4}{3} \gamma^2
\frac{B}{B_{cr}} ~,
\end{equation} 
where
\begin{equation}
 B_{cr} \equiv \frac{m_e^2 c^3}{e \hbar} \simeq 4.414 \times
10^{13}{\rm G} ~.
\end{equation}
and the energy is expressed as $\varepsilon=h\nu/m_ec^2$.

The rate of synchrotron cooling 
is calculated according to the formula:
\begin{equation}
\dot{\gamma} =\frac{4c\sigma_T}{3m_ec^2}\gamma^2 u_B \,,
\end{equation}
where $u_B=B^2/8\pi$ is the magnetic field energy density.

\section{Inverse Compton radiation}
Consider a collision between a soft  (low energy) photon and a relativistic electron of velocity $\beta c$.
If energy of the photon is $h\nu$, the incidence  angle $\theta$, and the electron velocity $\beta$
then the photon is boosted in energy to: 

\begin{equation}
\frac{\nu}{\nu_0}=\frac{1+\beta \cos{\theta_0}}{1-\beta \cos{\theta}+\frac{h\nu_0}{\gamma m_e c^2}(1+\cos(\theta-\theta_0))} \,.
\end{equation}
  
When the ratio between the energy of the photon and the energy of the electron before the collision 
$\gamma\frac{h\nu_0}{(m_e c^2)}$ is $\ll$ 1,  the interaction proceeds in the, so-called, Thomson regime.
 The electron energy loss in a single collision is negligible and the photon collision 
 is elastic in the center-of-momentum frame.
 
The maximum energy gain of a photon occurs in $\theta=\theta_0=0$ and $\beta \rightarrow 1$.
Then
$\frac{\nu}{\nu_0}\simeq 4 \gamma^2$.
The average energy gain at given $\theta_0$ over all possible interaction angles is 
\begin{equation}
\left\langle \frac{\nu}{\nu_0} \right\rangle=\gamma^2(1+\cos{\theta_0}).
\end{equation}

The rate of IC energy losses of relativistic electron, isotropically distributed photons is%
\begin{equation}
|\dot \gamma_T| = \frac{4 c \sigma_T}{3 m_e c^2} \gamma^2 u_0, 
\label{eq:ICcool}
\end{equation}
where $u_0$ is the total energy density of the radiation field ($u_0=\int_{\epsilon_{min}}^{\epsilon_{max}} u_{\epsilon} d\epsilon$), 
 $u_{\epsilon}$ is the energy density distribution of the ambient photons.

\section{The Klein-Nishina effect}
\label{sec:KN}
When the photon energy becomes   comparable to the electron energy or larger ($\gamma \frac{h\nu_0}{m_e c^2}>1$),
 the electron may loose most of its energy during a single collision. 
 In that case the IC cross-section has to be expressed by the full Klein-Nishina (KN) formula \citep{2005MNRAS.363..954M}.
The cross-section is reduced as compared to the Thomson regime when the photon energy becomes larger.
The main effect is a reduction of the electron energy loss rate:
\begin{equation}
\dot \gamma=\dot \gamma_T \, F_{KN} \,,
\end{equation}
where $\dot \gamma_T$ is given by equation (\ref{eq:ICcool}) and a factor $F_{KN}$ is given by
\begin{equation}
 F_{KN}= {1 \over u_0} 
\int_{\epsilon_{min}}^{\epsilon_{max}}
 f_{KN}( b) u_{\epsilon} d \epsilon  \, ,\label{eq:FKN} 
 \end{equation}
where $b=4\gamma \epsilon$.   
The $ b=1$ corresponds to 
the transition between the Thomson and KN scattering regimes.
The function $f_{KN}(b)$ is given by \cite{2005MNRAS.363..954M}:
\begin{equation}
f_{KN}(b) = \frac{9g(b)}{b^3} ~, 
\label{eq:fKN}
\end{equation}
\begin{equation}
g(b) = \left( \frac{1}{2}b + 6 + \frac{6}{b} \right) \ln (1+b) -
\left( \frac{11}{12} b^3 + 6 b^2 + 9 b + 4 \right) \frac{1}{(1+b)^2} -
2 + 2 {\rm Li}_2 (-b)
\end{equation}
where $Li$ is dilogarithm and $b = 4 \varepsilon_0 \gamma$.

For $b \ll 1$ (Thomson limit), $f_{KN} \simeq 1$; for $b\gg1$ 
(KN limit), $f_{KN} \simeq [9/(2 b^2)](\ln b - 11/6)$.
For $ b \le 10^4,$   
$f_{KN}( b)$ can be approximated by 
 $f_{KN} \simeq {1 \over (1+ b)^{1.5}}$.

Electrons, for which cooling by Comptonisation is inefficient, 
are loosing more energy throughout the synchrotron emission. 
Usually, the effect of Klein-Nishina is seen as an excess of synchrotron luminosity 
and the hardening of synchrotron spectrum, while the Compton component shows a cut-off. 

%
\section{Absorption}
The influence of low energy photons present in the Universe on 
the propagation of the HE gamma-rays was  pointed out by \cite{1967PhRv..155.1404G}.
The fundamental process responsible of the HE $\gamma$-ray absorption is the
electron-positron pair production: 
\begin{equation}
\gamma_{HE} + \gamma_{LE} \rightarrow e^+ + e^- \,.
\end{equation}
The observed HE $\gamma$-ray spectrum after attenuation is 
\begin{equation}
F_{obs}= F_{int} e^{-\tau} \,,
\end{equation}
where $e^{-\tau}$ is the  attenuation, $\tau$ is the optical depth,  
and $F_{int}$ is the intrinsic spectrum of the source. 
The optical depth given by  \cite{1967PhRv..155.1404G} is 
\begin{equation}
\tau(E) = \int dl \int_{\cos\theta_{min}}^{\cos\theta_{max}} d\cos{\theta} \frac{1-\cos{\theta}}{2} \int_{E_{th}}^{\infty} d\epsilon n(\epsilon)\sigma(E,\epsilon,\cos{\theta}) \,,
\label{eq:taue}
\end{equation}

where $dl$ is the differential path travelled by the HE photon, 
$\theta$ is the angle between the momenta of HE and LE photons.
The energy of HE photon is (1 + z)E and the energy of LE photon is  (1 + z)$\epsilon$, 
where E and $\epsilon$ are the observed photon energies at z = 0.
The low energy photons density number is $n(\epsilon)d\epsilon \, \mbox{cm}^{-3}$.
The cross-section of pair production is given by \cite{2008arXiv0809.5124B}:

\begin{equation}
\sigma(E,\epsilon,\cos{\theta}) = \sigma_T \frac{3m_e^2}{2s} 
\left\{ -\frac{p_e}{E_e} 1+\frac{4m_e^2}{s}+\left[1+\frac{4m_e^2}{s}\left(1-\frac{2m^2_e}{s}\right) \right]
\log\frac{(E_e+p_e)^2}{m^2_e} \right\} \,,
\end{equation}

where $s=2E\epsilon (1-\cos\theta)(1+z)^2$,   $E_e=\sqrt{s}/2$, 
and  $p_e=\sqrt{E_e^2-m_e^2}$. 

The condition for pair production is  $E\epsilon (1+z)^2(1-\cos\theta) > 2(m_ec^2)^2$,
which corresponds to a threshold energy of
\begin{equation}
E_{th} = \frac{2(m_ec^2)^2}{\epsilon(1+z)^2 (1-\cos{\theta})} \,.
\end{equation} 

HE photons can be absorbed by several backgrounds of LE photons 
during their travel to the observer.
The first source  of LE photons is an extragalactic background light (EBL). 
The EBL is the IR/UV radiation  generated by  stars  (UV) 
or radiation emitted through the absorption and re-emission of star light by dust in galaxies (IR).
The EBL models have been reviewed by, e.g.  \cite{2001ARA&A..39..249H}
and recently new constraints have been provided by \cite{2012A&A...542A..59M} using the FERMI data
and \cite{2012arXiv1212.3409H} using the H.E.S.S. observations of the brightest blazars.

For characteristic frequency of EBL photons of $\epsilon_{EBL} \simeq$0.1 eV 
the photons with energy above 5 TeV will be affected by absorption: 

\begin{equation}
\frac{E_{th}}{0.1\,\mbox{eV}} \sim 5.2 \frac{E}{1\, \mbox{TeV}} \,. 
\end{equation} 

The distribution of angles $\theta$, at which background photons can collide with the HE photons  is flat
 when the photon is traveling over cosmological distances, 
therefore, the  $\cos\theta$ of the scattering angle (see equation \ref{eq:taue})  changes from -1 to 1.
The differential path $dl$ travelled by the HE photon can be calculated 
using equations (\ref{eq:ds}) or (\ref{eq:dDRdistance}).
 
 The second possibility of absorption arises from the photon fields present in the blazar itself. 
 When the HE photon is produced in the jet, 
 it has to travels through two fields of LE photons: the broad line region (BLR)
  and IR radiation from the dusty torus (DT), depending on where it was emitted.
 
The characteristic frequency of the BLR radiation is of the order of 10 eV. 
Since, the energy of BLR  photons is larger than that of DT photons, 
the threshold energy of pair production is smaller 
and HE photons can be absorbed above energies of a few GeV.  
 To avoid significant absorption by BLR photons,
 the blazar zone where HE photons are emitted, has to be located outside of BLR,
 otherwise the HE emission is significantly absorbed.  
 The photon number  density of external radiation (from BLR and DT) decreases
  with distance as in equation (\ref{eq:ubel}) and  (\ref{eq:udt}), respectively. 
 When HE photons propagates inside the BLR or DT then the $\theta$ distribution is flat, 
 therefore the $\cos\theta$ in the equation (\ref{eq:taue}) ranges from -1 to 1. 
 However, outside the BLR or DT, 
 the $ \theta$ can have values ranging
 from $\theta_{min}=(\pi+{\rm atan}(R_{BLR,DT}/r))$ to $\theta_{max}=(\pi-{\rm atan}(R_{BLR,DT}/r))$.
 
 The distance $dl$ is measured from the place where the gamma photons are emitted  up to the distance where the energy densities of BLR or DT are negligible.

\section{The PKS 1510-089 blazar}
%
PKS~1510-089 ($\alpha_{\rm J2000} = 15^{\rm h}12^{\rm m}50.5^{\rm s}$, 
$\delta_{\rm J2000} = -09^{\rm d}06^{\rm m}00^{\rm s}$) at
redshift $z=0.361$ is the FSRQ detected in the MeV-GeV
band by EGRET\,\citep{1999ApJS..123...79H}.  
It is characterized by a highly relativistic jet that makes a $3^\circ$ 
angle relative to the line of sight\,\citep{2005ASPC..340...67W}.
The radio jet of PKS 1510-089 is  curved and shows an apparent superluminal motion 
as high as 45 times the speed of light \citep{2001ApJ...549..840H,2002ApJ...580..742H}. 
The first large multi-wavelength campaign on PKS~1510-089 took place in August
2006\,\citep{2008ApJ...672..787K} and involved Suzaku, Swift and  ground-based optical and radio instruments.  
The campaign resulted in a broadband spectrum
ranging from $10^{9}$ to $10^{19}\,$Hz, 
which was succesfully modeled within the internal shock scenario. 
\cite{2008ApJ...672..787K}  focused their work on the explanation of the soft X-ray part of the SED,
where an excess of emission has been observed,
which if interpreted as bulk-Compton radiation,  allowed to
obtain the e$^+$e$^-$ to proton ratio.
In the case of PKS~1510-089, the ratio was estimated to be of the order of $10$.  
This implies that although the number of e$^+$e$^-$ pairs is larger then the number
of protons, the power of the jet is dominated by the latter.

As an alternative interpretation, \cite{2008ApJ...672..787K} showed that
the observed soft X-ray excess was explained
by a synchrotron self-Compton (SSC) component, which, although energetically inefficient, 
shows its presence in the X-ray range.  

\cite{2010ApJ...721.1425A} have reported multi-wavelength observations
during a high activity period of PKS 1510-089 between 2008 September and 2010 July. 
These observations revealed a complex variability at optical, UV, X-ray and $\gamma$-ray bands
on time scale down to 6-12 hours. 
The study of the correlation of variability in different passbands, 
performed by \cite{2010ApJ...721.1425A}, shows 
no correlation between the $\gamma$-rays and the X-rays, 
a weak correlation between the $\gamma$-rays and  the UV(R) band, 
and a significant correlation of $\gamma$-rays with the optical band.  

\cite{2010ApJ...721.1425A} attempted to model three flares with simultaneous data 
from radio to $\gamma$-ray energies.  
They adopted the inverse Compton scenario with seed photons originating from the BLR 
to explain the HE emission (IC/BLR).
The blazar zone in their model was assumed to be at sub-parsec scale. 
The IC/BLR in their model occurred under the KN regime leading to the curved MeV/GeV 
spectral shape that matches the observed spectrum in the HE range. 

Recently in 2011, a multi-wavelength campaign,  which included Herschel data combined with 
the publicly available multi-wavelength data from FERMI, Swift, SMARTS and 
the Submillimeter Array (SMA),  covered the SED  of PKS 1510-089 in a quiet state \citep{2012ApJ...760...69N}. 

\cite{2012ApJ...760...69N} consider a two-zone blazar model to interpret the entire dataset. 
They suggest that the observed infrared emission is associated with the synchrotron component 
produced in the hot-dust region at the sub-pc scale. 
To explain  the gamma-ray emission, they proposed 
an External-Compton component produced in the BLR at the sub-pc scale.
In such a scenario, the optical/UV emission would be associated with the accretion disk thermal emission, 
with the accretion disk corona likely would be  contributing to the X-ray emission.
\cite{2012ApJ...760...69N} showed that to explain the ratio of the maximum luminosity peaks, 
and the peaks frequency ratio within single zone scenario would require an unrealistically high energy density of the external radiation. 

\section{Spectral energy distribution of PKS 1510-089}
\begin{figure}[h!]
\vspace{2mm}
\begin{center}
\hspace{3mm}\psfig{figure=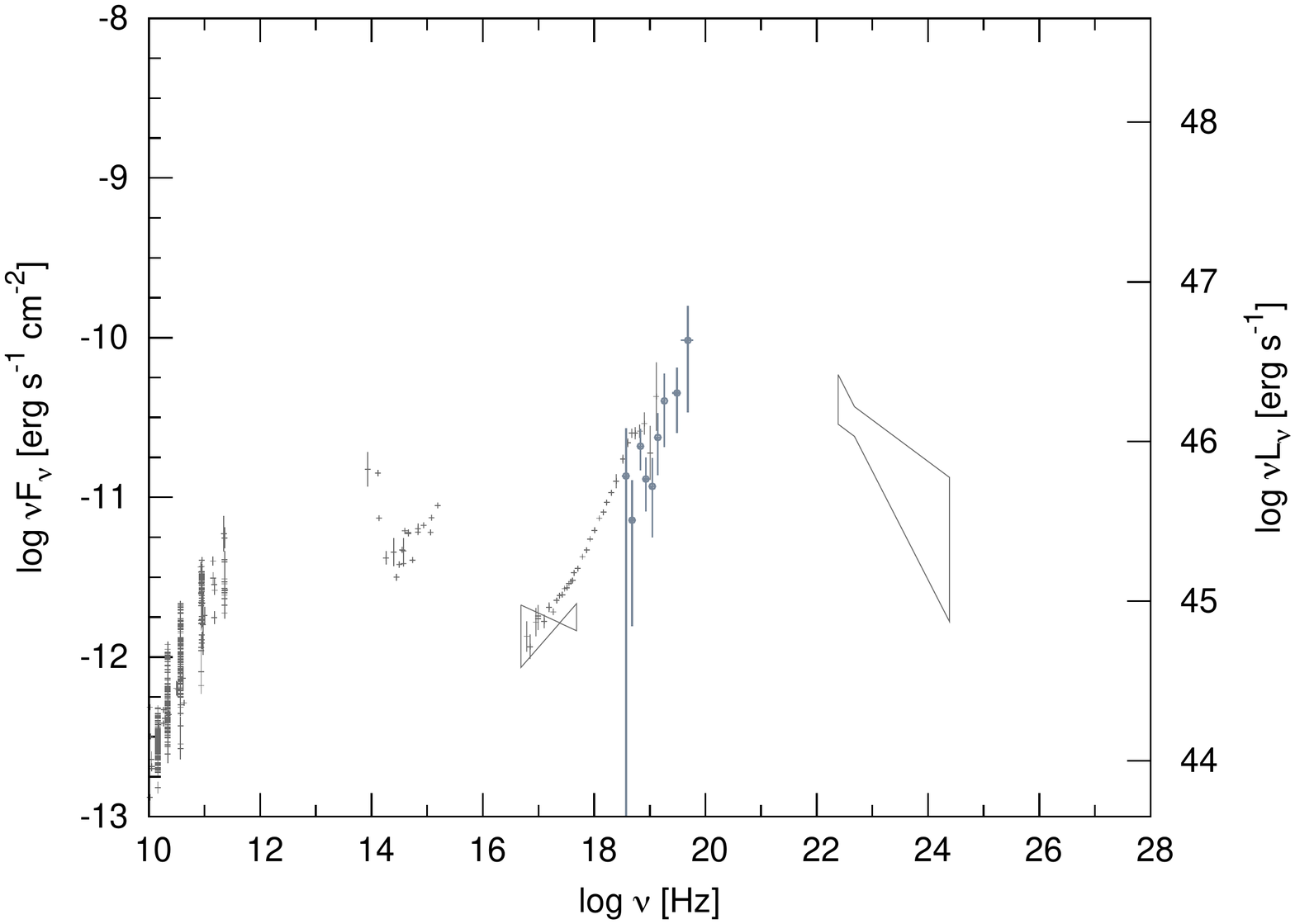,height=120mm,angle=0.0}
\caption{The archival  data of PKS 1510-089 published by \cite{2008ApJ...672..787K},
                and the {\em INTEGRAL}  observations \citep{ABIntegral}.}
\label{fig:his_sed}
\end{center}
\end{figure}

The archival spectrum of  PKS 1510-089 is presented in figure \ref{fig:his_sed}.
The grey points plotted on figure \ref{fig:his_sed} are  
the data published by \cite{2008ApJ...672..787K}. 
The dark grey points are {\em INTEGRAL} data analyzed by \cite{ABIntegral}. 
The {\em INTEGRAL}  observations were carried out in January 2008 with an exposure of 600~ks. 
Data were analyzed using the {\em INTEGRAL} Data Software package OSA 7.0. 
The observed spectral index was \ $1.2 \pm0.3$, and 
the flux, obtained extrapolating ISGRI results to
lower energies, was $F_{10-50\,{\rm keV}} = 15.5 \times
10^{-12}\,$erg\,s$^{-1}$\,cm$^{-2}$. 

\begin{figure}
\vspace{2mm}
\begin{center}
\hspace{3mm}\psfig{figure=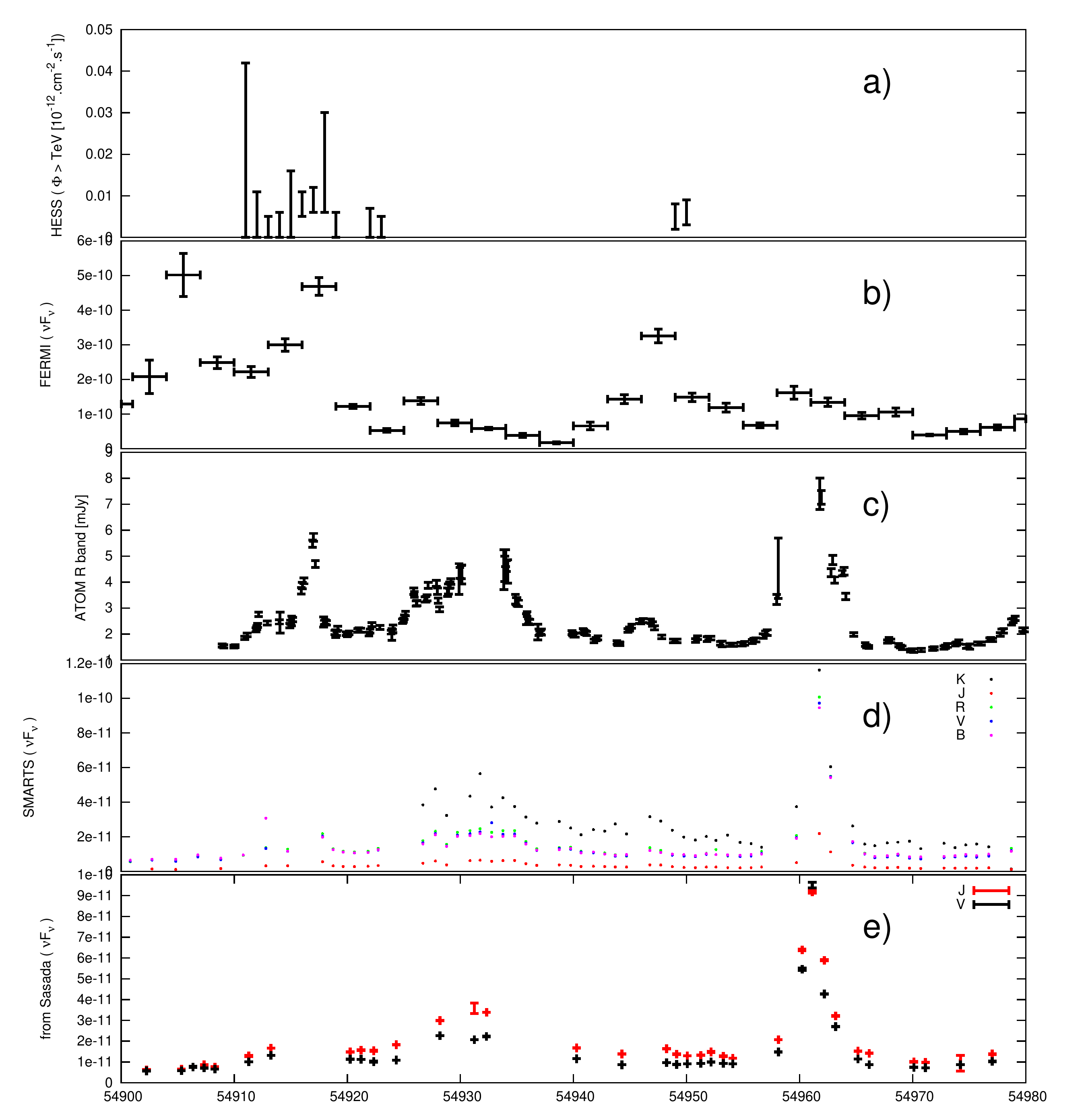,height=160mm,angle=0.0,bb=0 0 801 777}
\caption{The multi-wavelength light curves of PKS 1510-211 during March - April 2009. 
	       \textbf{a)} The VHE light curve recorded by  H.E.S.S.
	       \textbf{b)} The HE energy observations carried out by FERMI LAT, shown in three-days bins.
	       \textbf{c)} The ATOM R-band fluxes after correction for the Galactic absorption.
	       \textbf{d)} Data taken with SMARTS:  K,J,R,V,B filters (data were downloaded from 
	       http://www.astro.yale.edu/smarts/fermi/).
	       \textbf{e)} The optical observations obtained using filters J and V (data taken from \cite{2011PASJ...63..489S}).}
\label{fig:pks1510lc}
\end{center}
\end{figure}

The overall spectrum of  PKS 1510-089 is presented in figure \ref{fig:sed}.
The blue points on the SED are the simultaneous observations taken around the VHE flare on March 2009.
The Swift XRT spectrum as well as the radio data are from \cite{2010arXiv1002.0806M}.
The radio observation at 14.5 GHz were taken at the Michigan Radio Astronomy observatory,
observations with 37 GHz  were recorded with the Metsahovi Radio Observatory 
and 230 GHz at the sub-millimeter Array.
The optical ATOM data  are from Abramowski et al (paper in preparation). 
The FERMI and the H.E.S.S. observations were analyzed by the author 
following the procedure described in sections \ref{sec:FERMIdata} and \ref{sec:HESSdata}.
The multi-wavelength light curve is presented in figure \ref{fig:pks1510lc}.

\subsection{H.E.S.S. data analysis}
\label{sec:HESSdata}
The observations with the H.E.S.S. telescope followed the  report of 
flaring  activities in HE \citep{2009ATel.1957....1D} and in the optical band observed by ATOM. 
The H.E.S.S. data  were simultaneous with the peak of the HE flare recorded by FERMI.
The H.E.S.S. telescopes carried  out observations  of PKS 1510-089 at two periods.
The first observations were taken between 23 March 2009 (MJD 54910) and 2  April 2009 (MJD 54923). 
The second follow-up of the HE activity triggered the H.E.S.S. observations 
between 27 April 2009 (MJD 54948) and 29 April 2009 (MJD 54950). 
The H.E.S.S. observations resulted in 15.8 hours of good quality data.

The data quality selection is based on various variables like the trigger rate, 
the telescope tracking or the fraction of the PMTs turned off.
Runs with the mean trigger rate less than 70\% of the predicted value \citep{2005AIPC..745..753F} are rejected. 
The mean system rate is 240 Hz  for the four telescope data,
and 180 Hz for the three telescope data. 
In addition, if the rms variation in the trigger rate is above 10\%,  the runs also are rejected.
The instability of trigger rate can be caused by the presence of clouds or excessive dust in the atmosphere, 
which leads to the Cherenkov light  absorption, and thus to the fluctuation 
in the trigger system efficiency. 

The problems with telescope tracking can lead to errors in the reconstructed position of the source
and thus can affect the flux.
Runs with the tracking error problems reported by the DAQ are then rejected,
when the rms deviations are greater than 10 arcseconds in altitude or azimuth. 

An alternative check  of the tracking system is performed by producing 
an intensity map of light of bright stars in the telescope field of view.
The positions of known stars are then correlated with this map,
giving a measure of the pointing position of each telescope.
Runs are rejected, if the pointing deviation is greater than 0.1$^\circ$.

Runs with more than 10\% of the PMTs missing are rejected from the analysis.
PMTs can be turned off, if any bright transient light source passes through it. 
Such a source of light can be  bright stars,  meteorites, lightening, airplanes  or even satellites. 
Detail criteria of quality cuts are described in \cite{2006A&A...457..899A}.

  \begin{figure}
\vspace{2mm}
\begin{center}
\hspace{3mm}\psfig{figure=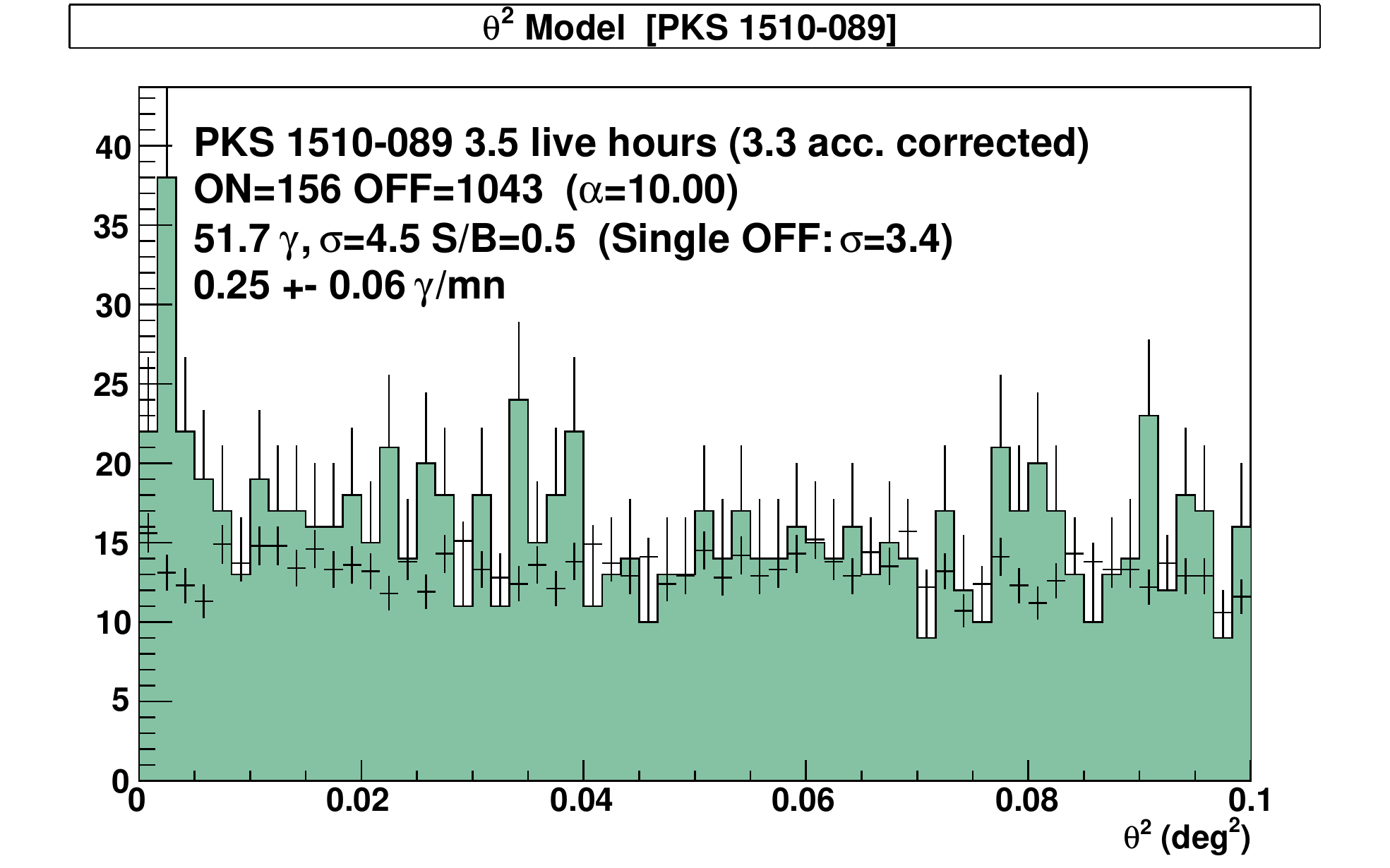,height=90mm,angle=0.0}
\caption{The angular distribution of events around the nominal position of PKS 1510-089. 
                 $\theta^2$ is the square of the angular distance between the nominal source position 
                  and the reconstructed arrival direction of an event. }
\label{fig:theta2}
\end{center}
\end{figure}

The H.E.S.S. data  analysis has been performed using 
the \texttt{Model Analysis} developped by \cite{2009APh....32..231D}.
Events were reconstructed using \texttt{loose cuts} appropriate for sources with steep spectra.
PKS 1510-089 is a high redshift source, therefore its spectrum is expected to be soft 
due to the EBL absorption. 

  \begin{figure}
\vspace{2mm}
\begin{center}
\hspace{3mm}\psfig{figure=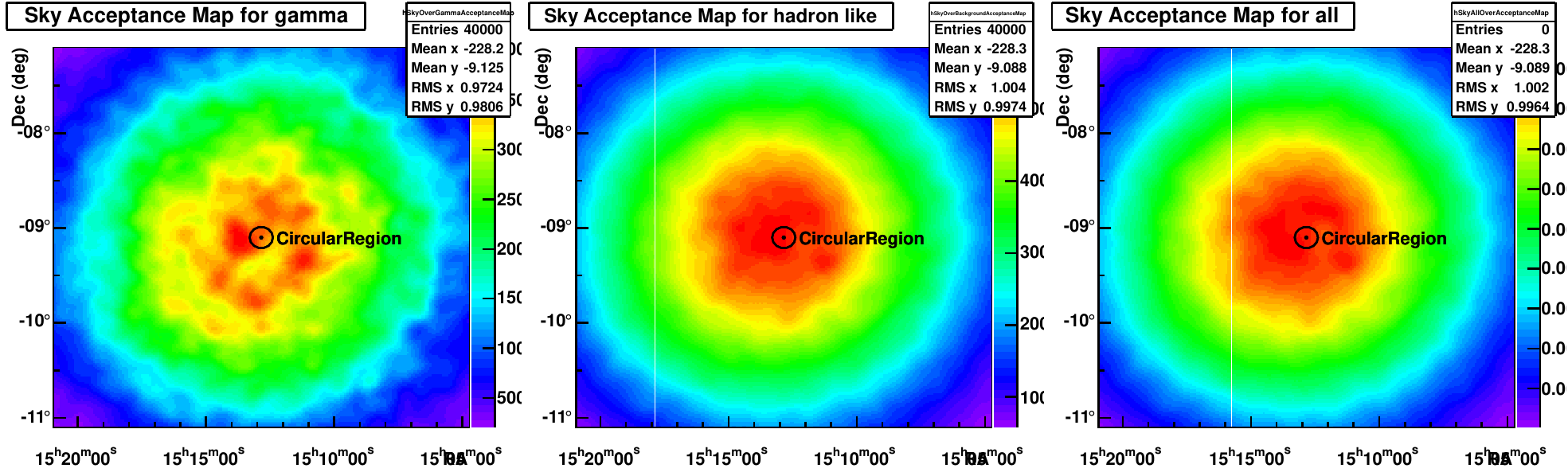,height=45mm,angle=0.0}
\caption{The camera response for $\gamma$-like (left panel), hadron-like (middle panel) 
                and all (right panel) events, over the field of view (angular acceptance).}
\label{fig:acceptance}
\end{center}
\end{figure}

The analysis of the first flare (MJD 54916-54917), based on  3.5 hours of observations,
recorded 51.7 photons from the source direction. 
This corresponds to a detection significance of 4.5 $\sigma$, following the method of \cite{1983ApJ...272..317L}.
The angular distribution of events around 
the position of PKS 1510-089 (figure \ref{fig:theta2}) shows 
the excess in the source region. 
The acceptance distribution suffers from the low statistic of $\gamma$-like events, 
leading to an apparent inhomogeneity. 
The observed off-set in the $\theta^2$ distribution (figure \ref{fig:theta2}) is due to  problems with the pointing model.
A new DST production with a better pointing correction is undergoing, 
however the pointing problem does not influence significantly the detection significance or the energy spectrum, 
since the whole excess  in the $\theta^2$ distribution is  within 0.1$^\circ$ from the target.

The spectrum was derived using a forward-folding technique.
The threshold energy, $E_{thr}$, is given by the 
energy at which the effective area falls to 10\% of its maximum value. 
For these observations the energy threshold was estimated to be $E_{thr} \sim$ 0.15 TeV.					

\begin{figure}[h]
\vspace{2mm}
\begin{center}
\hspace{3mm}\psfig{figure=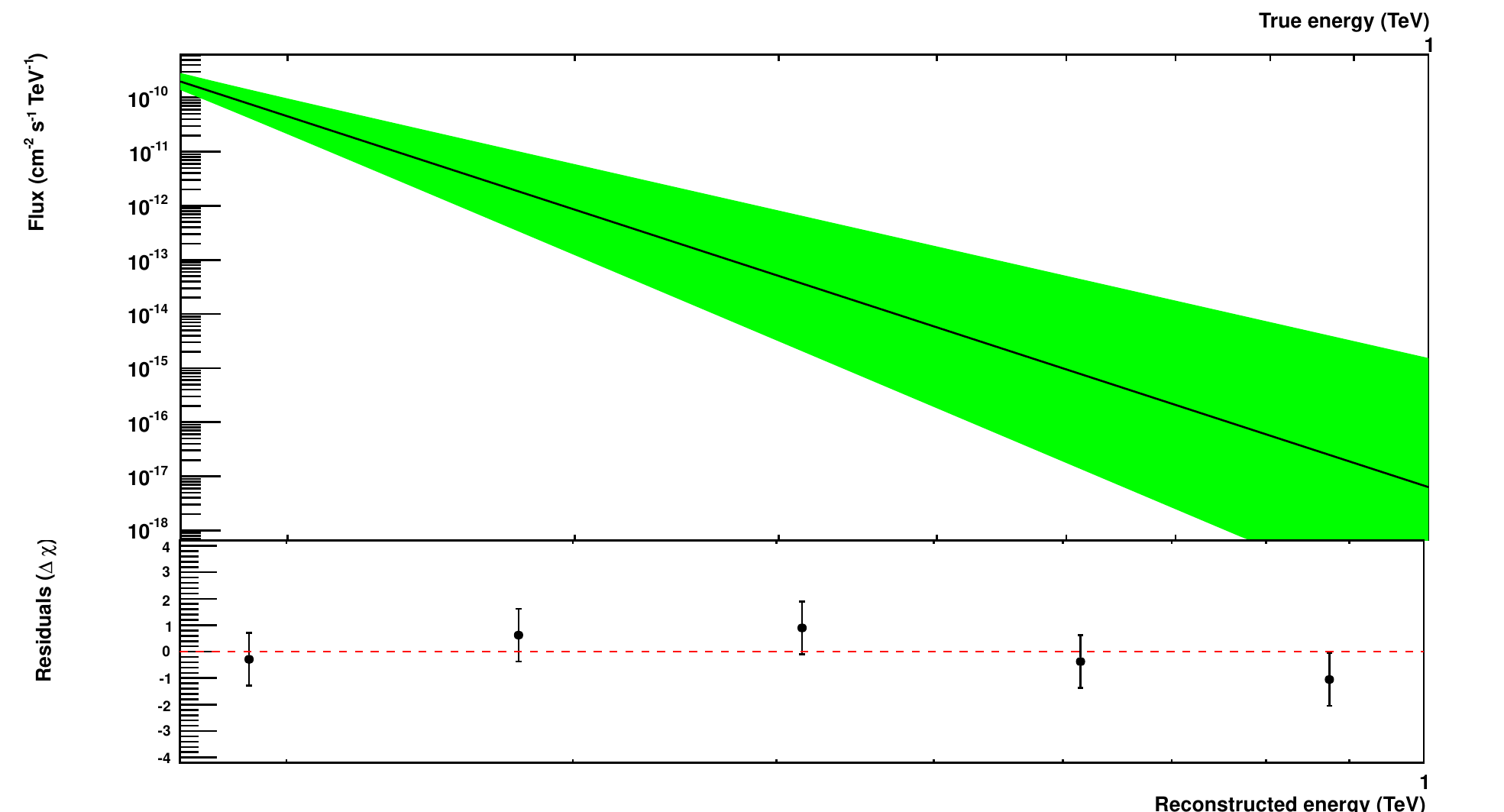,height=80mm,angle=0.0}
\caption{The VHE spectrum of PKS 1510-089 measured with the
                H.E.S.S. instrument, during March 2009. 
                The solid line is the best-fit Power Law obtained using forward folding. 
                The "butterfly" is the 68\% confidence band, while the points with errorbars (1$\sigma$ statistical errors)
                 are the residuals in binned energy. Vectors denote the 99\% C.L. upper limits.}
\label{fig:spectrum}
\end{center}
\end{figure}

Most of the VHE events were detected below $\sim$ 400 GeV. 
The spectrum (figure \ref{fig:spectrum}) is fitted with a Power-Law:
$dN/dE=N_0(E/E_0)^{-\Gamma}$, with an index $\Gamma=9.8\pm2.9$ 
and a normalization $N_0=(6.32\pm 0.35)\times 10^{-14}\,$m$^{-2}\,$s$^{-1}\,$TeV$^{-1}$
at $E_0$=157 GeV. The equivalent $\chi^2 is 7.8/13$ndf. 

\subsection{FERMI data analysis}
\label{sec:FERMIdata}
The Fermi-LAT \citep{2009Atwood} data, simultaneous
with the H.E.S.S. observations period, were analyzed using the
publicly available Fermi Science Tools (version v9r15p2) and
the P6\_V3\_DIFFUSE instrument response functions. 
The light curve (figure \ref{fig:pks1510lc}, b) is produced by an unbinned
likelihood analysis taking into account photons (the diffuse class events)
with energies between 200 MeV and 100 GeV from a region of
interest (ROI) with a radius of 20$^\circ$ around the position of PKS 1510-089. 
All sources, from the Fermi-LAT First Source Catalog \citep{2010ApJ...715..429A}, 
within an angular distance of 25$^\circ$  PKS 1510-089, were modeled simultaneously. 
\texttt{Model v02} of the Galactic and extragalactic backgrounds were used. 
Two flares are evident on the light curve (figure \ref{fig:pks1510lc},b), one centered around MJD 54916, 
and the second centered around MJD 54948. 

\begin{figure}[ht!]
\vspace{2mm}
\begin{center}
\hspace{3mm}\psfig{figure=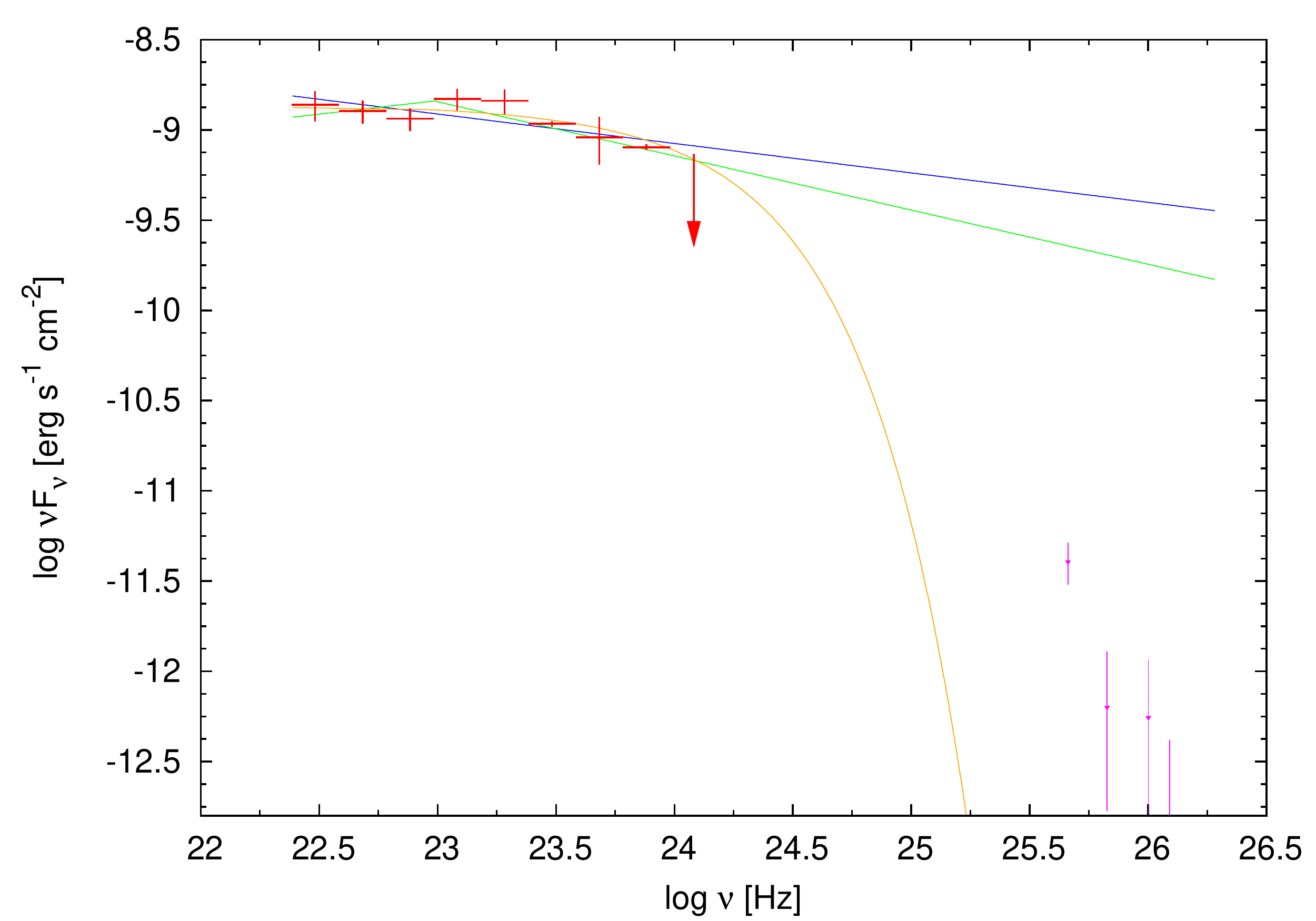,height=90mm,angle=0.0}
\caption{The HE spectrum of PKS 1510-089 observed with the
                FERMI LAT, during VHE flare on March 2009. 
                The red points are the FERMI data and the magenta points are the H.E.S.S. data for comparison.
                The blue line represents the Power Law fit to the FERMI data, the green line is the Broken Power Law fit
                and the orange line is the exponentially cut-off power law (Exp Cut-off) fit. 
                The values of the parameters obtained by fitting the models are summarized  in table \ref{tab:FERMIfit}.  }
\label{fig:spectrumFERMI}
\end{center}
\end{figure}

Figure \ref{fig:spectrumFERMI} shows the LAT SED of PKS 1510-089 extracted for the flare observed 
around MJD 54916.  Three different models have been fitted to the data: 
a Power Law, a Broken Power Law and an exponentially cut-off Power Law (ExpCut-off) model. 
Table \ref{tab:FERMIfit} summarizes the values of the fitted parameters for all 3 models. 
The spectrum is best fitted by the ExpCut-off model, 
with an index of $-2.02\pm0.1$, a break energy of 400 MeV and  P1=$7.9\pm 4.5$. 
The flux obtained using the ExpCut-off model is $5.04\times10^{-6}$ph~cm$^{-2}$s$^{-1}$.
The result of the fit is compatible with the one obtained by \cite{2010ApJ...721.1425A}.

\begin{table}
      \begin{tabular}{|l||c|}
          \hline
          Model & Parameters  \\
          \hline
          \hline
          Power Law & Integral:~$828.2 \pm 60.7 $,  
                    Index:  $-2.16 \pm 0.06$, \\
                    & Lower~Limit: 100,
                     Upper~Limit: 10000,  \\
                    & Npred: 305.551,
                     ROI distance: 5.76807e-05, \\
                    & TS value: 1290.37           \\
         \hline
         Broken Power Law & Prefactor: $5.64  \pm 4.45 (1e-09)$,
                       Index1:    $-1.85 \pm 0.16$,  \\
                    & Index2:    $-2.3  \pm 0.11$,
                        Break~Value:$399.7 \pm 156 $, [MeV]  \\
                    & Npred:              304.527, 
                     ROI distance:   5.76807e-05,  \\
                    & TS value:           1292.73,  \\
         \hline
         Exp Cut-Off  & Prefactor: $3.09 \pm 2.4 (1e-08)  $,
                     Index:     $-2.02 \pm 0.1$,  \\
                    & Scale:     $ 163 \pm 62   $, 
                     Ebreak:    $400.4 \pm 65   $,  \\
                    & P1:        $7.9 \pm 4.5 (*1000)  $,
                     Npred: 304.878, \\
                    & ROI distance: 5.76807e-05,    \\
                    & TS value: 1293.67,            \\
        \hline
      \end{tabular}
 \caption{  Results of the unbinned likelihood spectral fit of the flare of PKS 1510-089 (FERMI LAT data).
		The description of the spectral models can be found on page
		http://fermi.gsfc.nasa.gov/ssc/data/analysis/scitools/source\_models.html. 
		$Npred$ is the number of photons used in the fit.
		TS value is the Test Statistic resulting from the fit \citep{1996ApJ...461..396M}.
		The region of interest  (ROI) distance indicates the angular separation between 
		the center of the ROI and the location of the fit for that source. }     
\label{tab:FERMIfit}
 \end{table}
\section{Modeling }

I have aimed to reproduce the spectrum energy distribution (SED) of PKS 1510-089 during 
the HE and VHE flare recorded on March 2009.
The approach was to use the one zone leptonic 
model implemented in the \texttt{BLAZAR} code \citep{2003A&A...406..855M}.
The \texttt{BLAZAR} code  requires some input parameters, notably 
the value of the energy density of an external diffuse radiation field, 
the injected electron energy distribution, the value of the  magnetic field 
and the description of the overall geometry of the source. 

In this work I consider two sources of the external photons: DT and BLR.  
The luminosity of the accretion disk has been reported by  
\cite{2012ApJ...760...69N} to be $L_d=5\times 10^{45} \,\mbox{erg\, s}^{-1}$. 
Following  equation (\ref{eq:RBLR}), the size of the BLR, $r_{BLR}$,  is estimated to be $0.12\times 10^{18}\,$cm.
The size of the DT is approximatively  by  $r_{DT} \simeq1.94\times 10^{18}\, $cm (see section \ref{sec:DT}).

The energy density of external radiation fields have been calculated using equations (\ref{eq:udt}) and (\ref{eq:ubel}).
To satisfy the physical boundaries on the fraction of reprocessed emission from the accretion disk, 
I have used  $\xi_{BEL}\sim0.1$ and $\xi_{BEL}\sim0.2$. 
The values of $\xi_{BEL,DT}$ together with values of $L_d,\,r_{BEL,DT}$ listed above, and 
taking into account the dust temperature $T_3=1.8$ from \cite{2012ApJ...760...69N},
one can get the energy density of external radiation fields: $u_{BEL}=0.09 \,{\rm erg\,cm^{-3}}$ and $u_{DT}=0.0005 \,{\rm erg\,cm^{-3}}$.
The jet opening angle, $\theta_{\rm jet}$ is assumed to be $1/\Gamma$.

\begin{table}
  \caption{The input parameters for modeling of the non-thermal emission of PKS~1510$-$089.}\label{tab:modelfit}
  \begin{center}
    \begin{tabular}{lc}
    \hline
    Parameter & Model  \\
    \hline
    \hline
    minimum electron Lorentz factor $\gamma_{\rm min}$  & $1$  \\
    break electron Lorentz factor $\gamma_{\rm br}$            & $900$  \\
    maximum electron Lorentz factor $\gamma_{\rm max}$ & $10^5$ \\
    low-energy electron spectral index $p$                             & $1.2$ \\
    high-energy electron spectral index $q$                           & $3.5$ \\
    normalization of the injection function $K_{e}$                & $2.0 \times 10^{46}\,{\rm s^{-1}}$ \\
    bulk Lorentz factor of the emitting plasma $\Gamma_{\rm jet}$ & $22$ \\
    jet opening angle $\theta_{\rm jet}$ & $0.045$\,rad  \\
    jet viewing angle $\theta_{\rm obs}$ & $0.045$\,rad  \\
    location of the blazar zone $R_{\rm 0}$ & $0.7 \times10^{18}$\,cm  \\
    jet magnetic field intensity $B$ & $0.75$\,G  \\
    scale of the BEL external photon field $r_{\rm BEL}$ & $0.12 \times 10^{18}$\,cm \\
      energy density of the external photon field $u_{\rm BEL}$ & $0.09\,{\rm erg\,cm^{-3}}$ \\
    photon energy of the external photon field $h \nu_{\rm BEL}$ & $10$\,eV  \\
    scale of the DT external photon field $r_{\rm DT}$ & $1.94 \times 10^{18}$\,cm \\
    energy density of the external photon field $u_{\rm DT}$ & $0.0005 \,{\rm erg\,cm^{-3}}$ \\
    photon energy of the external photon field $h \nu_{\rm DT}$ & $0.15$\,eV  \\
    \hline
   \end{tabular}
  \end{center}
  \label{tab:model}
\end{table}

The other parameters are estimated to best reproduce the observed multi-wavelength spectrum of PKS 1510-089.
The parameters are summarized  in table \ref{tab:model}, while the  spectrum obtained is presented in figure~\ref{fig:sed}. 
 
\begin{figure}[h]
\vspace{2mm}
\begin{center}
\hspace{3mm}\psfig{figure=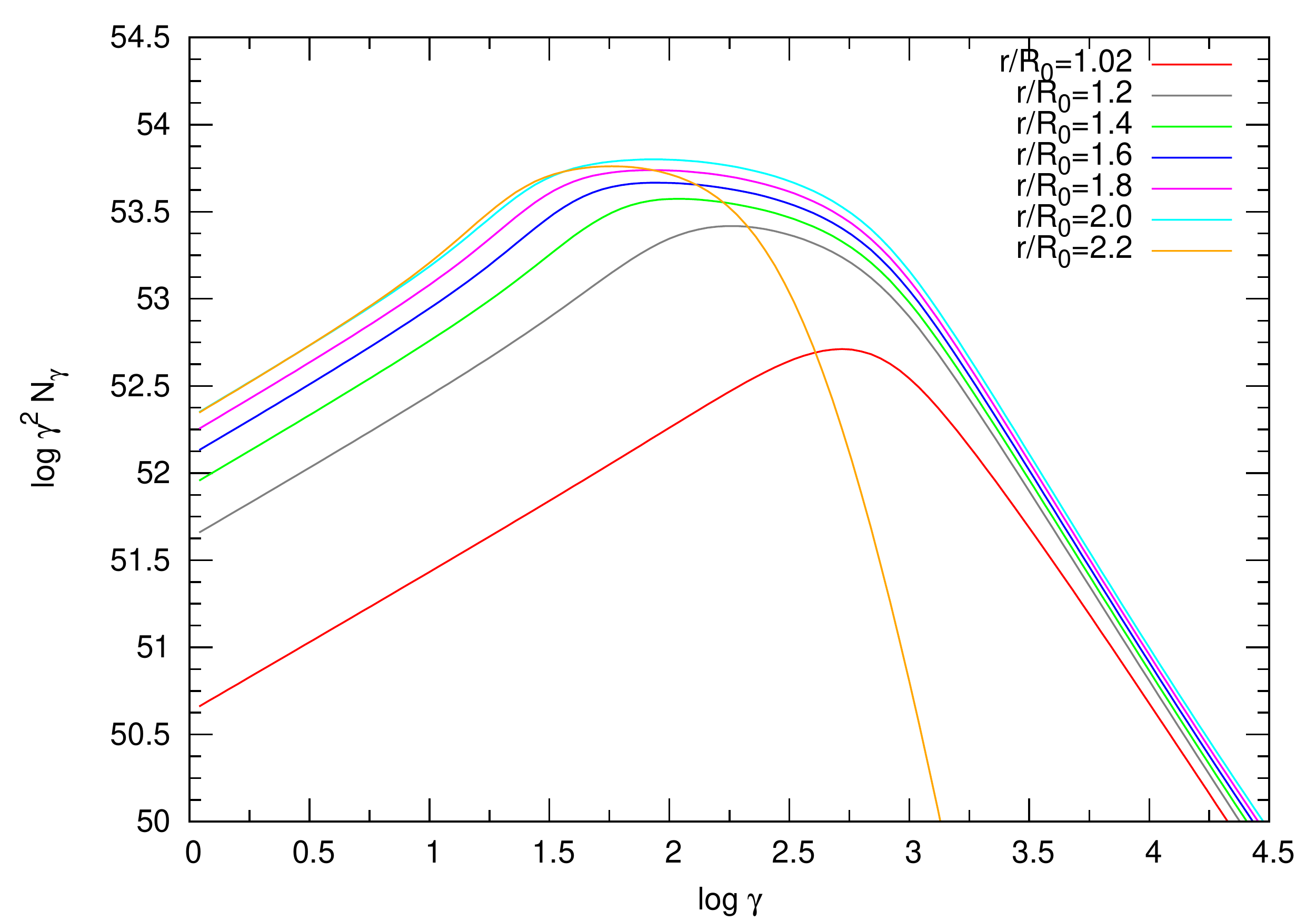,width=150mm,angle=0.0}
\caption{Evolution of the electron energy distribution during the March 2009 flare of the blazar PKS 1510-089.
                The evolution is calculated using the \texttt{BLAZAR} code and model parameters taken from  table \ref{tab:model}.
                 }
\label{fig:electron_evolution}
\end{center}
\end{figure}

The \texttt{BLAZAR} code calculates the evolution of electrons injected along the jet.   
Figure \ref{fig:electron_evolution} shows the electron evolution during the flare of PKS 1510-089.
The injected electrons follow a broken power law distribution 
(see equation \ref{eq:Qgamma}).
Parameters of the electron injection function are listed on table \ref{tab:model}.
The electron injection starts at $R_0$ and continues until $R=2R_0$, 
while the electron evolution is followed up to 3$R_0$. 
The proper choice of $R_0$ is crucial for the SED modelling. 
The next section is dedicated to the discussion of the location of the blazar zone in  PKS 1510-089.

\subsection{Location of the $\gamma$-ray emitting region in PKS 1510-089}
\begin{figure}
\centering
\begin{tabular}{cc}
\psfig{file=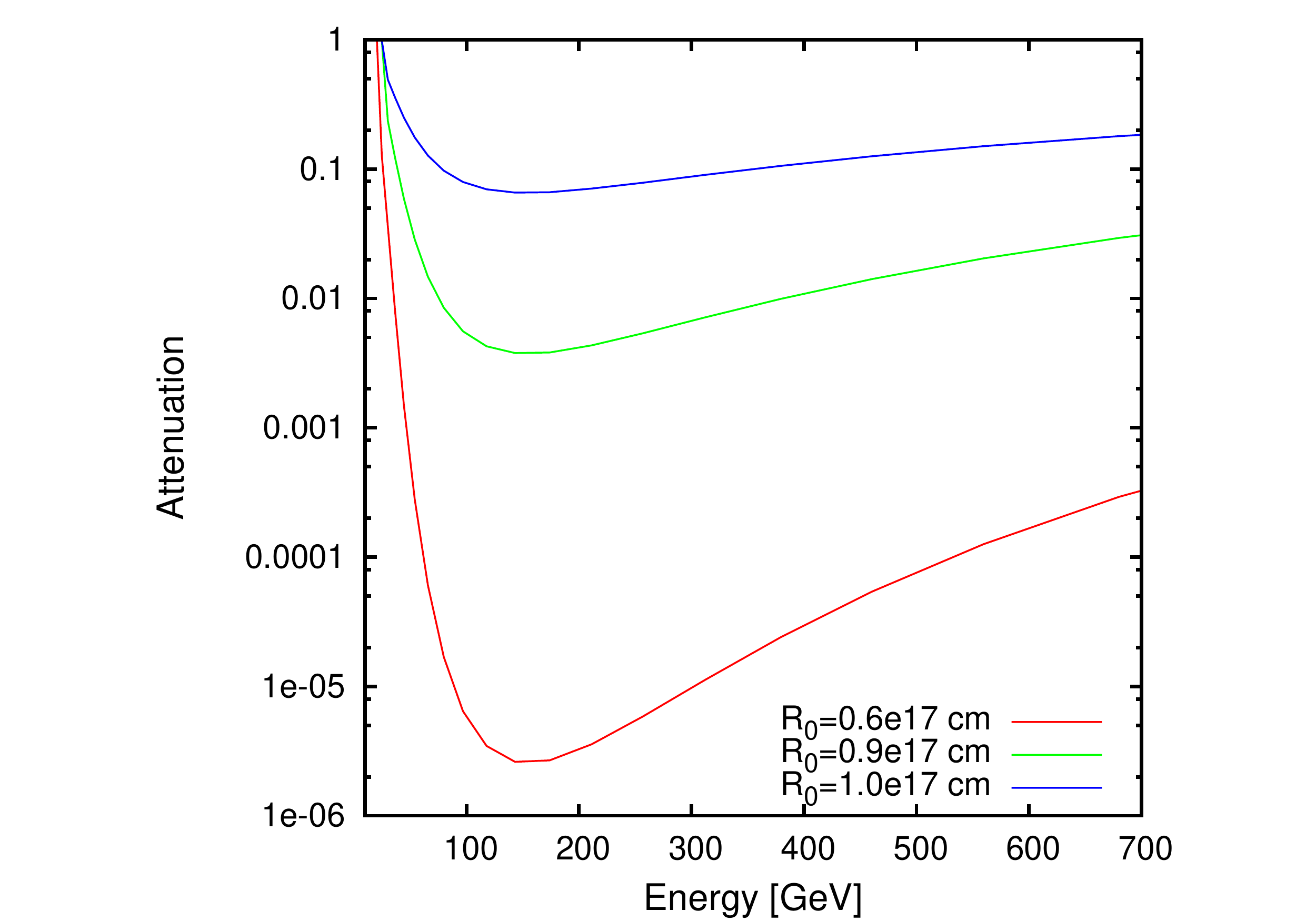,width=0.5\linewidth,angle=0.0} & 
\psfig{file=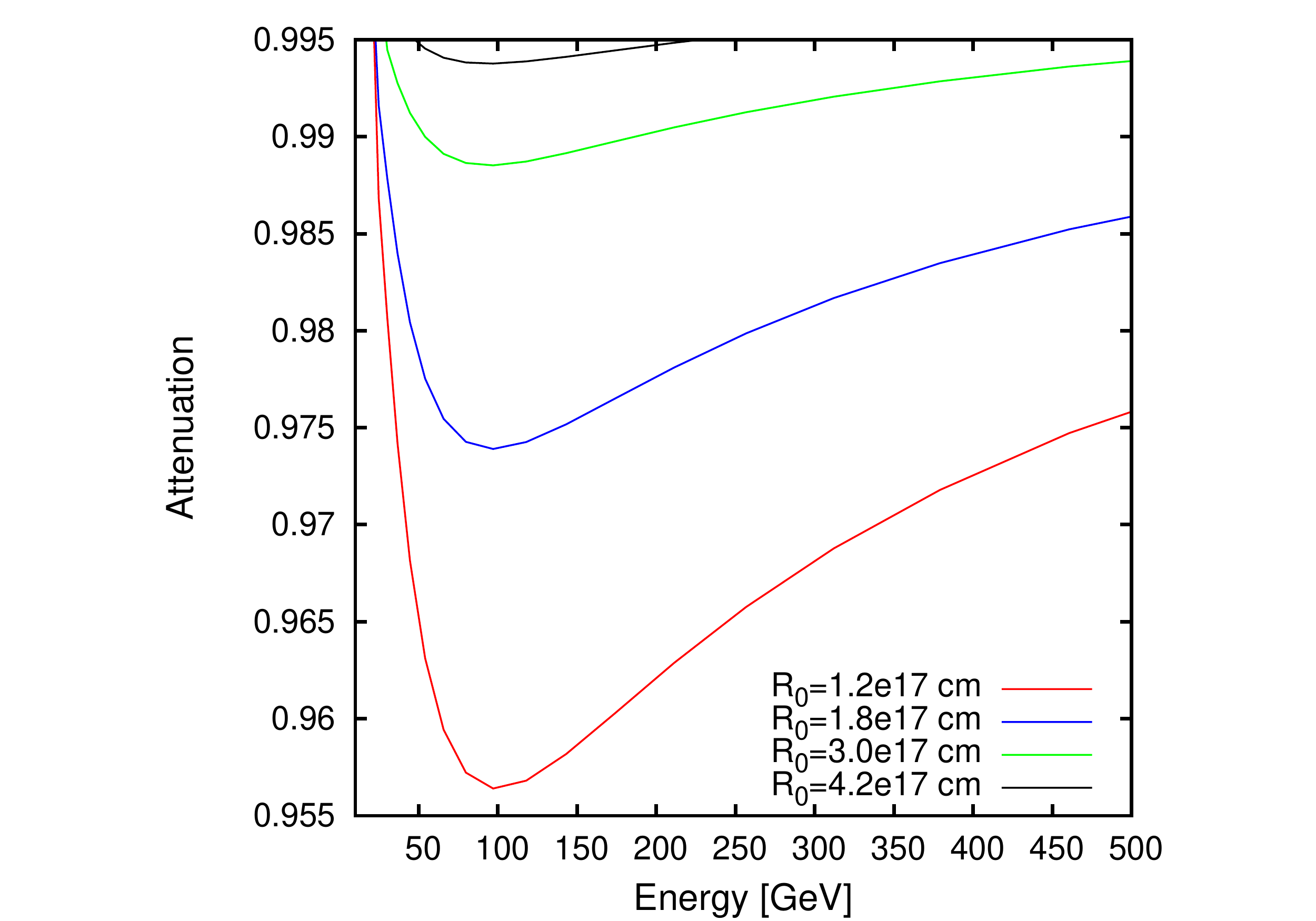,width=0.5\linewidth,angle=0.0} \\
\end{tabular}
\caption{The internal absorption as a function of photon energy emitted in the blazar zone. 
                The values of attenuation, $e^{-\tau}$, is calculated for different  distances from the center 
                and assuming PKS 1510-089 model parameters listed in table \ref{tab:model}.
                {\bf Left panel}: $R_0$ below $r_{BEL}$. 
                {\bf Right panel}: $R_0$ above $r_{BEL}$.}
\label{fig:attenuation}
\end{figure}

The choice of the distance of the shock formation from the central source, $R_0$, is constrained  
from one side by  the internal absorption of gamma rays in BLR, and from the other side by the IC efficiency. 
Figure~\ref{fig:attenuation} shows the internal absorption, $e^{-\tau}$, as 
a~function of the photon energy emitted at different $R_0$. 
When $R_0$ is below $r_{BLR}$ then a significant fraction of the HE radiation 
(from a several dozen to several hundred GeV)  is absorbed.  
If $R_0$ is greater than $r_{BLR}$ then only a few percent of the HE emission is absorbed. 

\begin{figure}[h]
\vspace{2mm}
\begin{center}
\hspace{3mm}\psfig{figure=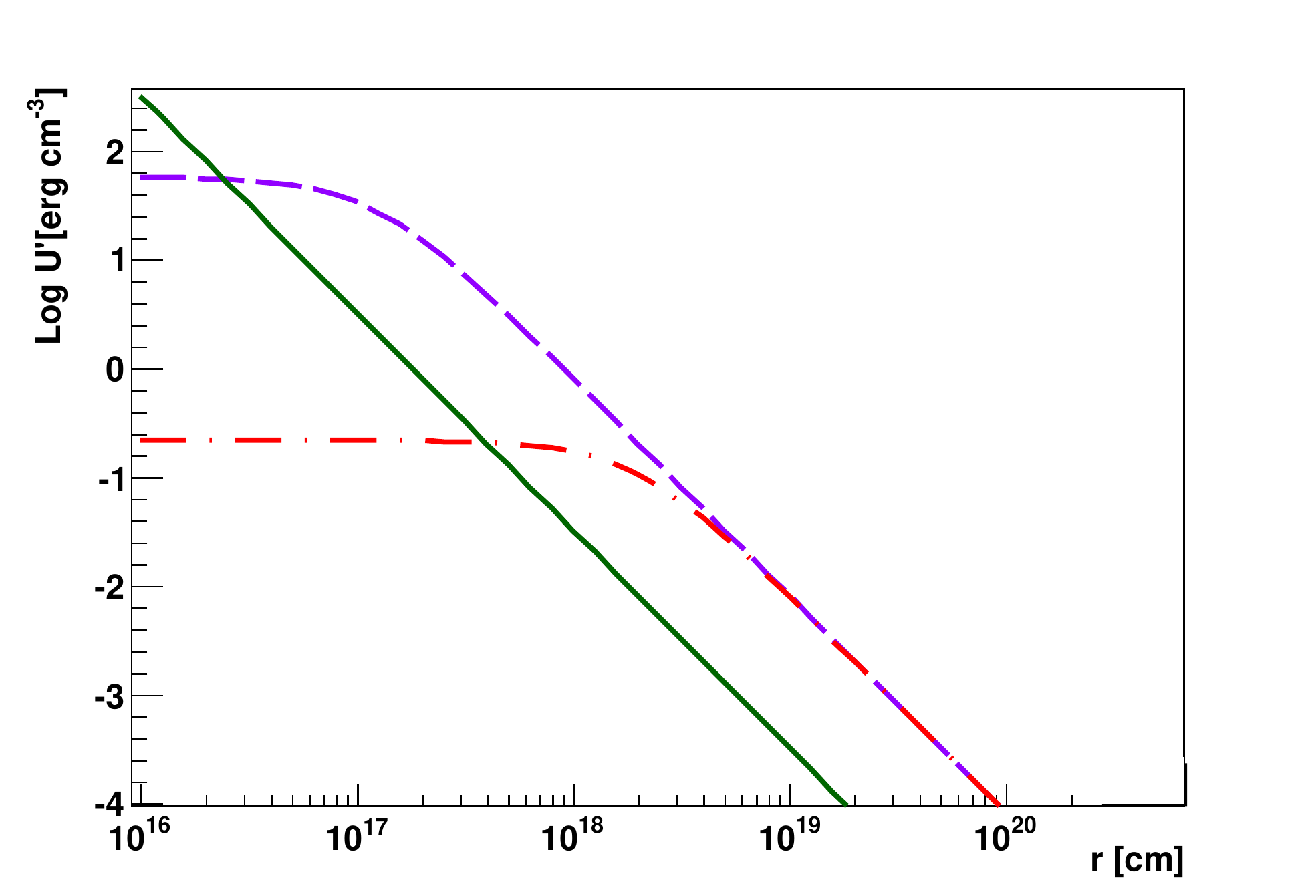,height=90mm,angle=0.0}
\caption{The energy density as a function of the distance from the central source,
                calculated using parameters listed in the table \ref{tab:model}. 
                The violet dashed line represents the energy density of BLR, 
                the red dashed-dotted line shows the energy density of DT, 
                and the green line is the energy density of magnetic field. }
\label{fig:energy_density}
\end{center}
\end{figure}

However, if the blazar zone is too far away from the BLR then the photon energy density is too small 
to produce  a sufficient $\gamma$-ray emission by IC. 
The photon energy densities of different radiation fields are presented in figure \ref{fig:energy_density}.

There exists some  observational evidence that the blazar zone may be located outside the BLR. 
The radio observations of PKS 1510-089 between 2011 September 9 and 2011 October 17
 \citep{2010arXiv1002.0806M}  show a $\sim$ 40 days ($\Delta t_{obs}$) increase in radio flux. 
If the $\gamma$-ray flare is indeed  associated with the same region as the slower radio flare
 \citep{2010arXiv1002.0806M,2012arXiv1210.4319O}, the projected distance 
 between the regions where the shock formed and the site responsible for the $\gamma$-ray emission  is: 
\begin{equation} 
D_{projected} = \frac{\beta_{app} c \Delta t_{obs}}{(1+z)} = \frac{(25c\times40\,\mbox{days})}{1.36}=0.6\, \mbox{pc} = 1.8 \times 10^{18} \,\mbox{cm} 
\end{equation}
where $\beta_{app}$ is the apparent velocity. 
One  then finds that the $\gamma$-ray flare is produced at the projected distance of $\sim\,$0.6~pc 
from the site where the shock  detected in the radio band was formed.
Assuming a jet inclination angle of $\theta=3^\circ$ \citep{2012arXiv1210.4319O,2010arXiv1002.0806M},
the de-projected distance ($D_{de-projected}=D_{projected}/\sin{\theta}$) is about $\simeq$~10~pc.
This shows that at least for some $\gamma$-ray flares, the blazar zone may be located far outside the BLR.

In the presented modeling of PKS~1510-089, I adopted $R_0=0.7\times 10^{18}\,$cm. 
At that distance, the absorption by the low energy photons originating from the BLR, is very small, of the order of 1\%,
while the energy density (see figure \ref{fig:energy_density}) in the blazar zone of PKS 1510-089 
($0.7-1.4 \times 10^{18}\,$cm) is still dominated by radiation from the BLR. 
Outside  $r_{BLR}$, the external radiation field is dominated by $u_{BEL}$
up to a distance $r_{DT}$, where $u_{BEL}$ became comparable to
the energy density of the radiation from DT ($u_{DT}$).

\subsection{Spectral energy distribution}
\begin{figure}[h!]
\vspace{2mm}
\begin{center}
\hspace{3mm}\psfig{figure=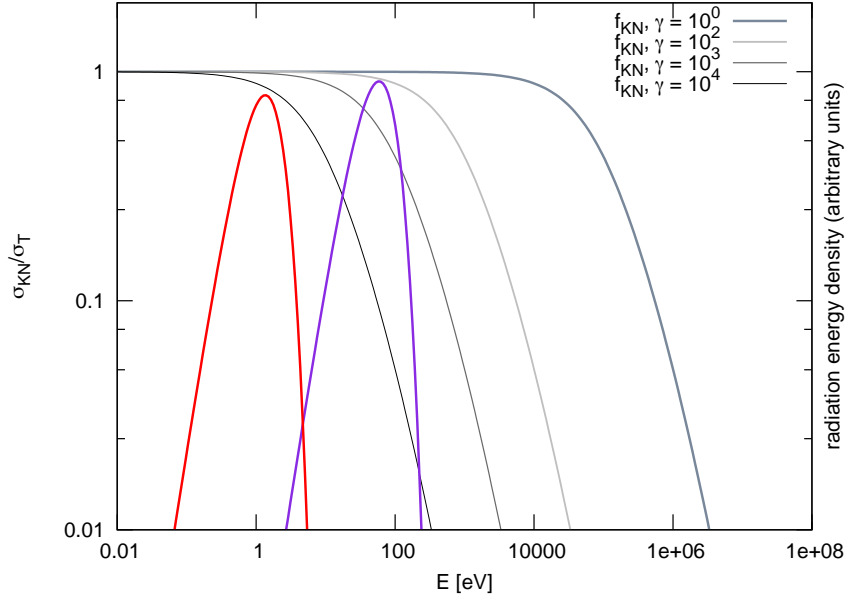,height=100mm,angle=0.0}
\caption{The $f_{KN}$ correction (see equation (\ref{eq:fKN})) at different electron 
                Lorentz factors $\gamma$ (gray lines).  $f_{KN}$ is expressed 
                as a  fraction of Thomson cross-secton  ($\sigma_{KN}/\sigma_T$).  
                The violet and red lines represent the energy densities of two external photon fields: BLR and DT, respectively.
                The energy density is approximated by a black body distribution with characteristic frequency 
                in the comoving frame $\nu'=\Gamma \nu$, and  the corresponding energy density seen 
                in comoving frame is approximated as $u'=4/3\Gamma^2 u_{ext}$. 
                The photon densities for corresponding external photon fields were calculated using 
                values listed in table \ref{tab:model}.}
\label{fig:fKN}
\end{center}
\end{figure}

Figure \ref{fig:fKN} presents the Klein-Nishina (KN) correction ($\sigma_{KN}/\sigma_T$) 
as a function of energy for different electron Lorentz factors $\gamma$. 
The KN correction is presented together with the radiation of BLR and DT approximated as black body. 
For small $\gamma$, electrons cool in the Thomson regime.
The electrons with Lorentz factors above $10^3$ are responsible for the HE and VHE emissions. 

\begin{figure}[h!]
\vspace{2mm}
\begin{center}
\hspace{3mm}\psfig{figure=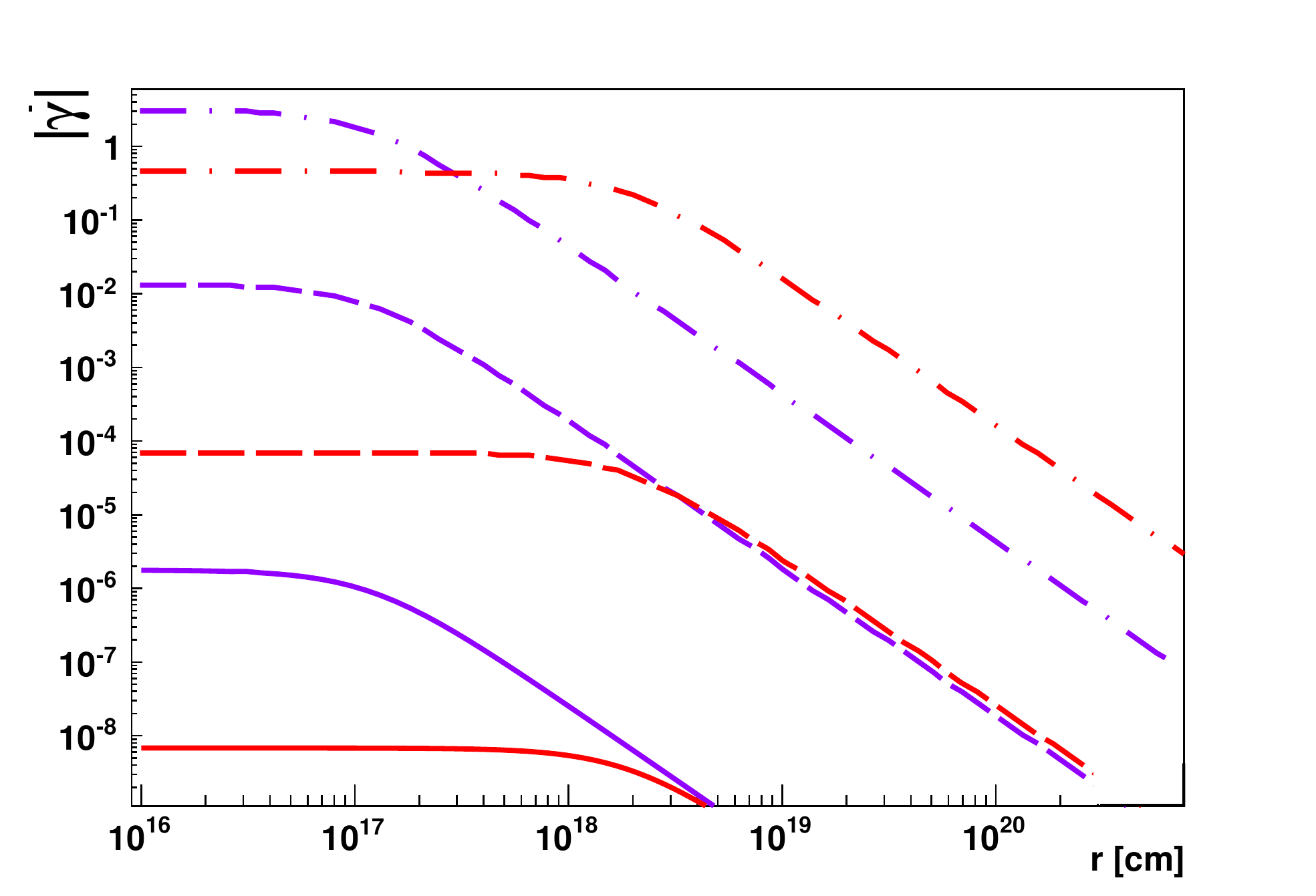,height=85mm,angle=0.0}
\caption{The rate of IC energy losses of relativistic electrons calculated using equation~(\ref{eq:ICcool}).
                The IC cooling is presented for two sources of seed photons: BLR (violet lines) and DT (red lines).
                The cooling rate was calculated for different electron Lorentz factors $\gamma$:
                1 (solid lines), 100 (dashed lines), 10000 (dashed-dotted lines). }
\label{fig:electron_cooling}
\end{center}
\end{figure}

The rate of IC energy losses of relativistic electrons, calculated using equation   (\ref{eq:ICcool})
is shown on figure \ref{fig:electron_cooling}. 
The external  photon energy densities  in the region located at $R_0$ to $2R_0$ 
shows that the energy density of the BLR exceeds a few times the energy density of DT (see figure \ref{fig:electron_cooling}).
However, the electron cooling, in this region is faster for seed photons  originating from DT. 
Therefore, despite of the greater energy density of BLR over DT, 
the last one  is responsible of the VHE photon production via IC radiation.

All above arguments suggest the following scenario for the PKS 1510-089 flaring  activity recorded on March 2010:
\begin{itemize}
	\item The low energy component is produced by the synchrotron radiation.
	\item The high energy component  (from X-rays to VHE) is produced by two components: 
	\item	The first component is the IC radiation with  seed photons from the BLR. 
	This component dominates the emission in the FERMI range. 
	Due to the KN effect this component alone cannot explain the highest part of the spectrum ($>$100 GeV).
	\item The VHE emission is produced via IC scattering of the seed photons originating from DT. The same component is also responsible for the X-ray part of the spectrum, as in the previous modeling attempts of this object \citep{2008ApJ...672..787K}. 
	\item The modeled emission of PKS 1510-089 convolved with the EBL attenuation, calculated using Spitzer constraints on the EBL \citep{2006A&A...451..417D}, fits well all  the observations.
\end{itemize}

\begin{figure}[h!]
\vspace{2mm}
\begin{center}
\hspace{3mm}\psfig{figure=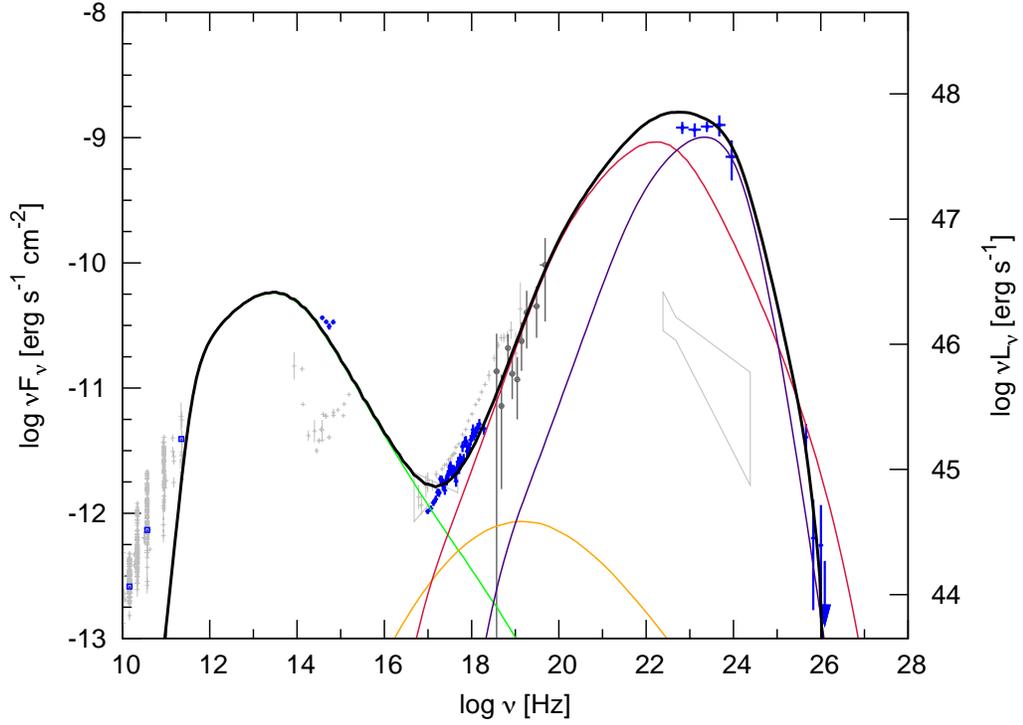,height=120mm,angle=0.0}
\caption{The overall spectrum of the PKS 1510-089. 
                The green line represents the synchrotron component, 
                the red line is IC component with seed photons originating from DT before absorption, 
                and the violet line is IC radiation with seed photons originating from BLR.
                The orange line is the SSC component. 
                The black line represents the sum of the all components, corrected 
                for EBL absorption (Spitzer model) and absorption on the low energy photons 
                originating from BLR and DT. 
                }
\label{fig:sed}
\end{center}
\end{figure}

Figure \ref{fig:sed} shows the result of the modelling of the PKS 1510-089   during the flare.
  
\section{Discussion}
%

The most recent attempt of~PKS 1510-089 modeling was undertaken by 
\cite{2012ApJ...760...69N}, who analyzed the 2011 low state of of the object.
They used the data obtained with the Herschel satellite to constrain the theoretical models.
\cite{2012ApJ...760...69N} concluded that a multi-zone emission model is necessary to explain the spectral properties of PKS 1510-089.
The model of \cite{2012ApJ...760...69N} differs from the model proposed by \cite{2010ApJ...721.1425A}, who analyzed the object in 2009 during an active state. 
The fast optical flares observed in 2009 were significantly brighter and strongly polarized as
compared to those observed in 2011.  

The optical flaring activity in 2009 was maybe accompanied by 
an increase of the magnetic field. 
This is confirmed by the observation of the significant increase in the optical 
degree of polarization  \citep{2011PASJ...63..489S} and observations  of the superluminal knot with VLBA at 7 mm \citep{2010arXiv1002.0806M}.
This behavior of the low energy component was not  observed in 2011 when Herschel data were taken.

PKS 1510-089 was detected with the H.E.S.S. system on March -- April 2009 
during the high state in the HE and  optical domains.
\cite{2010ApJ...721.1425A}  model the emission for energies below 100 GeV, without predictions for the VHE emission. 
In this thesis, I developed  a single zone model to explain the emission of PKS 1510-089 during the flare 
observed on March 2009, where the emission of   the low energy component is produced
by the synchrotron radiation and the high energy component is  produced by the same population of 
ultra-relativistic electrons via an IC process.

\section{Conclusions}
I have successfully modeled PKS~1510-089 with the single zone internal shock scenario.
It has been confirmed that the IC BLR cannot explain the VHE emission due to the KN effect,
as was anticipated by \cite{2005MNRAS.363..954M}.

The observations of PKS 1510-089 with the H.E.S.S. array show a VHE emission
up to 400 GeV.
The observations during the flare provide a 4.5~$\sigma$ detection of this emission. 

The absorption of the HE and the VHE photons in the blazar itself 
has been also investigated.  
The absorption in the BLR  is avoided by locating  the blazar zone outside the BLR. 
The absorption by photons from DT and EBL absorption become  significant only for photons 
with energies above 400 GeV. 
This emission was not observed in the case of PKS~1510-089.  

\cite{2010arXiv1002.0806M}  demonstrated that the HE emission from the jet of PKS~1510-089 is quit complex.  
The emission arises from different regions and probably multiple emission mechanisms are involved. 
I have tried to explain the emission of PKS 1510-089 during the flare on March 2009. 
The peak luminosity of the low energy component during the flare was much higher 
than that of observations shown by \cite{2012ApJ...760...69N}.

The VHE emission can be a very common feature  of FSRQs.  
The H.E.S.S. II telescope, with its energy range from tens of GeV, will provide a great opportunity 
to search for emission from other objects of this class.

\setcounter{chapter}{3}
\setcounter{section}{0}
\setcounter{equation}{0}
\setcounter{figure}{0}

\part{Theory of Gravitational Lensing \label{chapter:TheoryGL}}

\section{Introduction}
The lensing phenomena is an important topic in the cosmology. 
Parts \ref{chapter:helens} and \ref{chap:femto} of my thesis shows two examples of 
the lensing phenomena studied at high energy range. 
The first study (part \ref{chapter:helens}) provides the method of the time delay estimation 
when the lens images are spatially unresolved.   
Then, in the last section of part \ref{chapter:helens}, I  have showed the application of the obtained time delay 
 to the Hubble constant estimation. 
The second study (part \ref{chap:femto}) presents the limits on the abundance of compact objects. 
The limits have been obtained searching for femtolensing effect in the spectra of GRBs. 
The follow part of the thesis provide a brief theoretical introduction concerning 
parts \ref{chapter:helens} and \ref{chap:femto}.

\section{Gravitational Lensing} 
Gravitational lensing observations can be divided  into strong, weak and micro-lenses.
These lenses have different masses and image characteristics. 

Strong lensing events have multiple resolved images and arcs or arclets.
They are produced by the lensing of a distant object by galaxies or clusters of galaxies. 

The weak lensing is a regime where background galaxies are slightly distorted by foreground masses.
Weak gravitational lensing can thus be detected by studying the morphology of a large number of galaxies. 
It is therefore, an intrinsically statistical measurement. 
 
A large part of the lensing  observations  is dedicated to microlensing. 
In the case of microlensing,  the lenses have stellar masses. 
The images separation and time delay are too small to be detected. 
However, the characteristic time dependent magnification pattern  helps to  distinguish  
the microlensing event from the intrinsic variability of lensed sources.
The analysis of the light curve of a microlensed source can provide  informations on the nature of the lens.
The lensing event time scale is a combination of the lens mass, the transverse velocity, and the distances between the lens, the source and the observer.

Applications of gravitational lensing include:
\begin{itemize}
   \item Cosmology (Hubble constant \citep{2010ApJ...711..201S}, 
   compact objects \citep{1973ApJ...185..397P,2007A&A...469..387T}, 
   $\sigma_8$  \citep{2006ApJ...653..954D})
   \item Astrophysics \citep{2012RAA....12..947M}(stellar atmospheres \citep{2004AN....325..247T}, extrasolar planets, galactic structure, mass estimates)
   \item Fundamental physics (post Newtonian parameters\citep{2006PhRvD..74f1501B}) 
\end{itemize}    

This thesis is focusing on two different lensing phenomena. 
The first one is strong gravitational lensing and the other described in the thesis, similar to microlensing, is called femtolensing.    

\section{Theory}

\begin{figure}[ht!]
  \centering
  \includegraphics[width=15cm,bb=0 0 651 365]{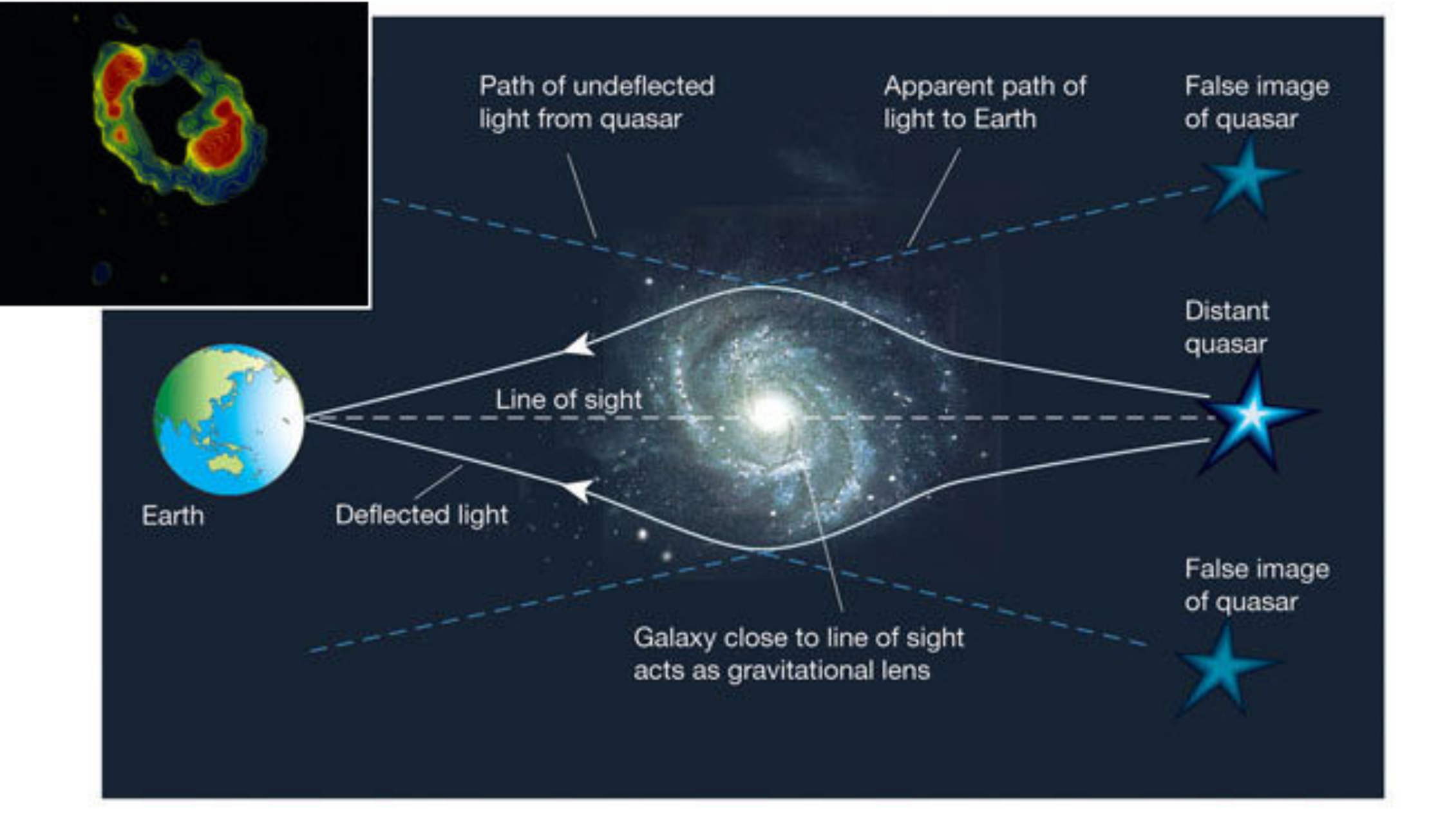}
  \caption{Schematic view of a gravitationally lensed system.}
  \label{fig:GL}
 \end{figure}

The gravitational lensing effect arises when a concentrated mass ("lens") lies in the line of sight from the observer on the Earth to a distant object ("source"), see figure~\ref{fig:GL}.  
The lensing effect magnifies  and distorts the image of the source.
Depending on the geometry of the lens, the resulting image of the lensed object might be an arc, a complete ring, a series of multiple images or a combination of compact images and arcs (see e.g. review by \cite{1992ARA&A..30..311B}). 

The deflection  of photons in the presence of masses is a consequence of the principle of equivalence. 
The first correct formula for the deflection angle $ \alpha$ was derived by Einstein.
The deflection angle $ \alpha$  of  light passing at the distance $r$ from an object of mass $M$ is given by equation:
\begin{equation} 
\alpha = \frac{4GM(r)}{c^2} \frac{1}{r}\,.
\label{eq:alpha}
\end{equation}
Equation~(\ref{eq:alpha})  gives a deflection angle twice larger than the Newtonian deflection for a slow particle.
The light deflection angle by the Sun is: 
\begin{equation} 
\alpha = \frac{4GM_{\odot}}{c^2} \frac{1}{R_\odot}=1.74 \, \mbox{arcsec}\,.
\end{equation}
The deflection angle was first measured by Eddington during a solar total eclipse. 
The Eddington experiment in 1919 brought  the first experimental confirmation of the Einstein predictions 
concerning light deflection.

However, observing the lensing of stars by stars was at that time considered as technologically 
impossible due to the very small image separation of the microlensed event. 
Fritz \cite{1937PhRv...51..679Z,1937PhRv...51..290Z} was the first to point out that galaxies are  likely to be gravitationally lensed 
and the image separation would be detectable.  

The angular separation of images in galaxy or galaxy cluster lensing  
is usually of the order of a few arcseconds. 
A typical galaxy with a mass of  $1.25\times 10^9\,$ M$_{\odot}$ 
acting as a lens will produce an Einstein ring with an angular size of 
$\sim 0.5\,$arcsecond.
A galaxy cluster at $z_L$=0.7 lensing a source at $z_S$=2 
will lead to image separation of the order of 50~arcseconds. 
 
\subsection{Lens Equation \label{sec:le}} 

\begin{figure}
  \centering
  \includegraphics[width=17cm,bb=0 0 792 612]{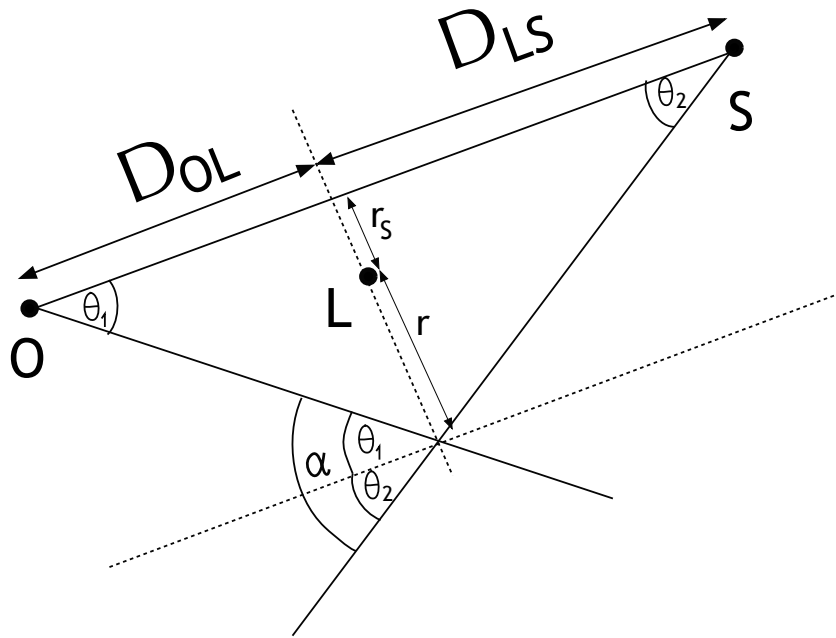}
  \caption{Geometry of a lensing event. The angles and angular diameter distances are shown. The light ray propagates from the source to the observer and passes the lens at distance $r_{S}$ in the lens plane. 
                   The angular diameter  distances between the observer and the source, the observer and the lens
                    are $D_{OL}$, and $D_{LS}$, respectively. }
  \label{fig:lens}
 \end{figure}

Gravitationally lensed systems involve a source, a lens and an observer, as shown on figure~\ref{fig:lens}.
A light ray from the source is deflected by an angle $\alpha$ by the lens and reaches the observer.
Figure \ref{fig:lens} shows the corresponding angular and linear distances in a typical gravitationally lensed system.  
The projected distance from the true source position to the lens is $r_S$ (in the lens plane).
$D_{OL}$ and $D_{LS}$ are respectively the angular diameter  distances between the observer and the source, and the observer and the lens (see section \ref{sec:DistanceFormula}).
The lens is assumed to be point like,  
 thus the light rays emitted by the source are 
splitted  into two images $r_+$ and $r_-$.
For  $\theta_1$,$\theta_2$,$\alpha \, \ll$ 1 and $r_{S}<0$, figure \ref{fig:lens} shows the following relations: 
\begin{equation}
\theta_1=\frac{r-r_S}{D_{OL}}\,,
\end{equation}
\begin{equation}
\theta_2=\frac{r-r_S}{D_{LS}}\,,
\end{equation}
\begin{equation}
\theta_1+\theta_2=\alpha\,.
\end{equation}
The angle $\alpha$ is known from equation \ref{eq:alpha}. 
After simple transformations: 
\begin{equation}
(r-r_S)\left(\frac{1}{D_{OL}} + \frac{1}{D_{LS}} \right)= \frac{4GM}{c^2} \frac{1}{r}\,.
\label{eq:lenseq}
\end{equation}
Equation (\ref{eq:lenseq}) has one solution when the source and the lens are perfectly aligned ($r_S=0$).
The image in the source plane is a ring around the lens with radius: 
\begin{equation}
r_E^2=\frac{4GM}{c^2} \frac{D_{LS}D_{OL}}{D_{OS}}\,,
\label{eq:rE}
\end{equation}
$r_E$ (equation (\ref{eq:rE})) is called the Einstein Radius. 

For a point lens and $r_S \neq 0$, the lens equation (\ref{eq:lenseq}) has two solutions.
The first solution gives  a distance between the first image and the lens:
\begin{equation}
r_{+}=  \frac{1}{2}(r_S+\sqrt{r_S^2+4r_E^2})\,,
 \label{rplus}
\end{equation}
and accordingly a distance between the second image and the lens of:
\begin{equation}
r_{-}=  \frac{1}{2}(r_S-\sqrt{r_S^2+4r_E^2})\,.
 \label{rminus}
\end{equation}
 For an isolated point source, the solution of the lens equation  always  gives 
two images of a background source, with corresponding positions $r_+$ and $r_-$. 

The lensing regime is based on the size of the image separation.  
The angular  radius  of the Einstein ring is given by: 
\begin{equation}
\theta_E = \frac{r_E}{D_{OL}} = \sqrt{\frac{4GM}{c^2} \frac{D_{LS}}{D_{OL} D_{OS}} }\,.
\label{eq:thetaE}
\end{equation} 
If one considers the lensing of  a source by a galaxy at a cosmological distance of $D \sim $~1~Gpc 
and with mass $M\sim 10^{11}$ M$_\odot$ the corresponding Einstein angle is 
\begin{equation} 
\theta_E = 0.9 \left(\frac{M}{10^{11}M_{\bigodot}}\right)^{\frac{1}{2}} \left(\frac{D}{\mbox{Gpc}}\right)^{-\frac{1}{2}} \mbox{arcsec}\,.
\end{equation} 
Image separations of the order of arcseconds can be easily spatially  resolved with  optical and radio telescopes. 
The situation is much different when the lens is an~object with a stellar mass or smaller. 
To illustrate this, the lensing by a star (with a mass M$_{\odot}$),  
 in the Galaxy at a distance of D~$\sim $~10~kpc will produce an Einstein angle of the order of a miliarcsecond: 
\begin{equation} 
\theta_E = 0.9\, \left(\frac{M}{M_{\bigodot}} \right)^{\frac{1}{2}} \left(\frac{D}{10\, \mbox{kpc}}\right)^{-\frac{1}{2}} \mbox{milliarcsec}.
\end{equation}     
When the image separation is of the order of one milliarcsecond, 
the lensing event is called a  microlens. 
When the image separation is of the order of a femtoarcsecond, 
the effect is called femtolensing.
For instance, when the distance of the lens is cosmological and the mass of the lens is in the range $10^{17}$ - $10^{20}\,$g, one gets:
\begin{equation} 
\theta_E = 0.3\,  \left(\frac{M}{10^{17}\,\mbox{g}}\right)^{\frac{1}{2}} \left(\frac{D}{\mbox{Gpc}}\right)^{-\frac{1}{2}} \mbox{femtoarcsec\,}.
\end{equation} 
Objects like primordial black holes or axions clusters are in the mass range relevant to femtolensing.
 
\subsection{Time delay}

\begin{figure}
  \centering
  \includegraphics[width=13cm,bb=0 0 792 612]{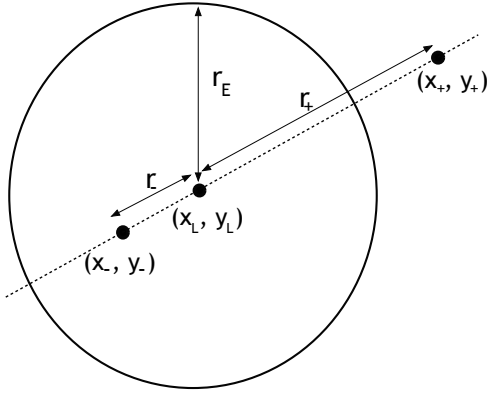}
  \caption{Definition of  distances in the lens plane. }
  \label{fig:lensplane}
 \end{figure}

When multiple images of a source are resolved, their light curves are similar except for a shift in time.
The shift in time comes from the differences in the geometrical path of light  
and the gravitational potential felt by the photon for each individual image. 
Estimation of this time delay, together with the measurements of the brightness ratio of 
individual images are crucial for determining the parameters of the lens, 
such as the lens mass and  geometry.  
The time delay $\delta t$ between the two images (see figure~\ref{fig:lensplane}) is given by
\begin{equation}
c \delta t = V(r_{+};r_S) - V(r_{-};r_S) \,,
 \label{dt2}
\end{equation}
where $V(r;r_{S})$ is the Fermat potential at the position $r$ in the lens plane. 
\begin{equation}
V(r;r_{S}) = \frac{1}{2} \frac{D_{OS}}{D_{LS}D_{OL}}(r-r_{s})^2 - \frac{4GM}{c^2} ln(r)\,. 
\label{eq:Fermat}
\end{equation}
One finds:
\begin{equation}
\frac{c \delta t}{(z_L+1)} =\frac{c (\delta t_{geom}+\delta t_{grav})}{(z_L+1)} = \frac{1}{2} \left( \frac{D_{OS}}{D_{OL}D_{LS}}
\right) (r_{-}^2 - r_{+}^2) - \frac{4GM}{c^2}
\ln\left(\frac{r_{+}}{r_{-}}\right) \,.
 \label{dt3}
\end{equation}
The time delay $ \delta t$ is a function of the image geometry, 
the distances $D_{OL}$, $D_{LS}$ and the gravitational potential. 
The geometrical time delay $\delta t_{geom}$ is caused by 
the extra path length compared to the direct 
line between the observer and the source. 
The gravitational time delay $\delta t_{grav}$  called the Shapiro delay (Shapiro 1964), is  induced by the gravitational potential of the lens. 
The Shapiro effect is due to clocks slowing down in gravitational fields.
Light rays are thus delayed relative to their travel time  in vacuum.

The Fermat potential (equation (\ref{eq:Fermat})) can also be  used to derive the lens equation 
by searching for extrema of the travel time: 
\begin{equation}
\frac{dV}{dr}=0\, ,
\end{equation}
which gives: 
\begin{equation}
(r-r_S)\left(\frac{1}{D_{OL}} + \frac{1}{D_{LS}} \right)= \frac{4GM}{c^2} \frac{1}{r}\, .
\label{eq:lensequationFermat}
\end{equation}
Equation (\ref{eq:lensequationFermat}) is similar to (\ref{eq:lenseq}).

 \subsection{Magnification \label{sec:Magnification}}
 The magnification of a single image of a lensed system is the ratio of the flux of the image to the flux of the unlensed source. If there are several images, the magnifications add up.

In the case of a point mass lens, the amplitude contributed by the
$r_{\pm}$ images is
\begin{equation}
A_{\pm} \propto \frac{\exp(i\phi_{\pm}) }{\sqrt{ |
    1-\frac{r_E^4}{r_{\pm}^4} |}} \,.
\label{A}
\end{equation}
\noindent
The magnification $|A|^{2}$ is obtained by summing the
amplitudes~(\ref{A}) and squaring, which gives
\begin{eqnarray}
|A|^2 &=& |A_{+}+A_-|^2 = \nonumber \\
&=& \frac{1}{ 1-\frac{r_E^4}{r_{+}^4} }
+\frac{1}{1-\frac{r_E^4}{r_{-}^4}} + \frac{2\cos(\Delta
  \phi) }{\sqrt{| 1-\frac{r_E^4}{r_{+}^4} |} \sqrt{|
    1-\frac{r_E^4}{r_{-}^4} |}} \,.
 \label{eq:Mag}
\end{eqnarray}
The phase difference $\Delta \phi$ is given by equation: 
\begin{equation}
\Delta \phi = \frac{E\delta t}{\hbar} \, .
\end{equation}

When light is not coherent ($\Delta \phi \gg 1$), the equation (\ref{eq:Mag}) is reduced to the two first terms. 
If one considers a "traditional"  lensing event (such as microlensing),  
the two light paths are not coherent 
because of large time difference between the images: $\lambda/c \ll \delta t$.
This is essentially the geometric optics  approximation. 
 However,  the condition of light coherence can be fulfilled,
when the lensing object is very small and compact.
In that case, the induced 
time difference between images is comparable to the light wavelength. 
When light is coherent, fringes can be observed in the energy spectrum. 
This interferometry pattern may be observed in GRB spectra 
if a compact object is on the line of sight with mass in the range
 $10^{14} \, g\, -\, 10^{19}\,g$. 
The GRBs emits mostly in the energy range from ${\rm keV}$ to ${\rm MeV}$. 
The condition of coherence ($\Delta \phi~\sim~\left(E/1{\rm MeV}\right)\left(M/1.5\times 10^{17}\,{\rm g}\right)$ close to 1) can thus be satisfied. 
This idea was proposed first by \cite{1992ApJ...386L...5G}, who also invented the word "femtolensing".

\subsection{Projected size of a source}
%
In principle, the finite size of the source and the relative motion of
the observer, the lens and the source have to be taken into account.
In this thesis this effects are important only for GRBs femtolensing. 
The projected size effect is negligible provided $s_{GRB}/r_E$ is~$\ll$~1,
where $s_{GRB}$ is the  size of GRB emission region projected on the lens plane. 
If the GRB is observed at a time $t_{expl}$ after the beginning of the
burst, its size projected onto the lens plane is
\begin{equation}
s_{GRB} \approx \frac{D_{OL}}{D_{OS}} \frac{c t_{expl}}{\Gamma}
\\ \approx \left( c \times 0.01\,\mbox{s} \right) \, \left(
\frac{t_{expl}}{1\,\mbox{s}} \right) \left( \frac{\Gamma}{100} \right)^{-1}
\left( \frac{D_{OL}}{D_{OS}} \right) \,,
\label{sgrb}  
\end{equation}
where $\Gamma$ is the Lorentz factor of the burst. Note that the
Lorentz factor of GRBs is estimated to be in excess of $100$, so that
$s_{GRB}$ given in equation~(\ref{sgrb}) is overestimated.

The Einstein radius $r_{E}$, image position and time delay have been  introduced in section \ref{sec:le}. 
The Einstein radius for the femtolensing event  is of the order of:
\begin{equation}
r_E^2 = \frac{4 G M}{c^2} \frac{D_{OL}D_{LS}}{D_{OS}}
 \\ \approx {\left( c \times 0.3 \,\mbox{s} \right)}^2
\left(
\frac{D_{OL}D_{LS}}{D_{OS}\,1\,\mbox{Gpc}} \right) \left( \frac{M}{10^{19}
  \,\mbox{g}} \right)\,,
 \label{rE}
\end{equation}
The ratio of $s_{GRB}$ to $r_{E}$ is therefore
\begin{eqnarray}
\frac{s_{GRB}}{r_{E}} & \approx & \frac{c^2}{2G^{1/2}} \frac{t_{expl}}{M^{1/2}\Gamma} 
\left( \frac{D_{OS}D_{LS}}{D_{OL}} \right)^{-1/2} \nonumber \\ &\approx& 0.03
 \left( \frac{D_{OS}D_{LS}}{D_{OL}\,1 \,\mbox{Gpc}}\right)^{-\frac{1}{2}} 
  \left( \frac{t_{expl}}{1\,\mbox{s}} \right)  \times  \left( \frac{\Gamma}{100} \right)^{-1}
 \left( \frac{M}{10^{19} \,\mbox{g}}\right)^{-\frac{1}{2}} \,.
\label{finitegrb}
\end{eqnarray}
Equation~(\ref{finitegrb}) shows that the finite size of the GRB can
be in general safely neglected if $t_{expl} < 10\,\mbox{s}$.

\subsection{Time scale of  femtolensing events}
The expected time scale femtolensing-induced event is given in terms of the typical 
Einstein radius and relative velocity $v$ between source and lens.
The Einstein radius crossing time $t_{E}$ is then:
\begin{equation}
t_{E} = \frac{r_E}{v}  \\ \approx 300 \,\mbox{s} \,
\left( \frac{r_{E}}{c \times 0.3 \,\mbox{s}} \right)
  \left(\frac{v}{300\,\mbox{km/s}} \right)^{-1} \,.
\label{einsteincrossingtime}
\end{equation}
Equation (\ref{einsteincrossingtime}) shows that $t_{E} \gg t_{expl}$
under reasonable assumptions on the velocities.  If so, the motion of
the source in the lens plane can also be neglected.  In the analysis
of GRB spectra, it is thus assumed that the point source -- point lens
assumption is valid and that the source stays at a fixed position in
the lens plane.

\subsection{Singular Isothermal Sphere} 
\label{sec:SIS}
Primordial black holes can be modeled as  point lenses.
However, galaxies which are responsible for large time delays between the images are 
extended lenses. 
A simple model of extended lenses is a singular isothermal sphere model (SIS).
In this model, the mass increases proportionally to the radius $r$ and the force is proportional to $1/r$. 
Thus, the isothermal sphere is a first approximation  model for the gravitational field of galaxies and  cluster of galaxies \citep{1988ApJ...333..522R}. 
The three-dimensional density distribution of SIS is given by: 
\begin{equation}
\sigma = \frac{\nu^2}{2\pi G}\frac{1}{r^2} \, ,
\end{equation}
where $\nu$ is the one-dimensional velocity dispersion of stars in the galaxy.

The circularly-symmetric surface mass distribution is obtained by projecting the matter in the lens plane:
 \begin{equation}
\Sigma (r) = \frac{\nu}{2 G}\frac{1}{r} \, .
\end{equation}

The mass inside a sphere of radius  $r$ is given by

\begin{equation}
M(r)=\int_{0}^{r}\Sigma (r')2 \pi r' dr' \, .
\label{eq:M}
\end{equation} 

Using equation (\ref{eq:alpha})  and (\ref{eq:M}) one obtains: 

\begin{equation}
\alpha(r)=4 \pi \frac{\nu^2}{c} \, .
\label{eq:ar}
\end{equation} 
Equation (\ref{eq:ar}) shows that the deflection angle for an isothermal sphere is independent of $r$. 
Equation (\ref{eq:ar}) can be simplified to: 
\begin{equation}
\alpha=1.15 \left(\frac{\sigma_{\nu}}{200\,km\,s^{-1}}\right)^2 \, {\rm arcsec}\, .
\end{equation} 
The Einstein angle calculated in SIS model is: 
\begin{equation}
\theta_E = 4 \pi \frac{\nu^2}{c^2} \frac{D_{LS}}{D_{OS}} = \alpha\frac{D_{LS}}{D_{OS}}  \,.
\end{equation}
SIS  creates two images of the source if it lies inside the Einstein ring, else just one image.
A third image is hidden by the central singularity of the $\ln(r)$ potential.   

\section{Lensing Probability \label{sec:LensingProbability}}
The chance of seeing a lensing event is usually expressed in terms of optical depth. 
This assumes that the optical depth is smaller than 1,   
 so that  it can be understood as a probability.
The concept of optical depth was introduced by \cite{1983ApJ...267..488V}, and is the standard way of determining the probability of lensing.  
The optical depth $\tau$ is  defined, in the context of gravitational lensing, as  a measure of the number of lenses per Einstein ring along the line of sight from the observer to a given source \citep{1989ApJ...341..579N}.  
That is:
\begin{equation}
\tau=\int{\rho_{lens}\pi r_E^2 dl} \, ,
\end{equation}
where $\rho_{lens}$ is the density of lenses along the line of sight.

The connection between the optical depth and the lensing probability depends on 
the observations and on the performances of the instrument. 
For instance in microlensing observations lenses are observed when the magnification is larger than 
1.34, corresponding to one Einstein radius. 
Some of the current instruments like Kepler, can detect  tiny magnitude variations, 
when  sources are bright enough. 
In that case the lensing probability is obtained from the
 maximum area in the lens plane, where a passing source gives  detectable lensing effects 
 \citep{2011PhRvL.107w1101G}. 
  In general it is convinient to define a lensing "cross-section", which reflects the maximum detectable 
 area in the lens plane and depends on the lens observation method (see \cite{1999grle.book.....S} section 11.1), 
 
The microlensing event detection is based on the observation of magnification changes of sources. 
In contrast, the detection of femtolensing is based on the observation of fringes in the energy spectrum 
(see figure \ref{fig:magnification}). 
With GBM, the fringes can be detected  when a lens  is distant of more than 3 $r_E$ from the source.
At this distance the magnification change is not detectable any more. 
In the case of femtolensing the cross-section has been defined as the area in the lens  plane 
where the spectral fringes give a more than 3 sigma detection. 
The femtolensing cross-section depends on  the instrument performance and source brightness, 
so that it was calculated event by event (see part \ref{chap:femto}).

The optical depth is sensitive to  the cosmological model. 
The Universe is  homogeneous and isotropic at large scales and
is well described  by  the Friedmann-Lemaitre-Robertson-Walker (FLRW) geometry. 
The FLRW is characterized by just a few parameters including
the mean mass density $\Omega_M$ and the normalized cosmological constant $\Omega_{\Lambda}$.
However, the local clumpiness has an important effect on photon propagation.
This leads to different formula for the angular diameter distance. 

In the next section  \ref{sec:DistanceFormula} the various  distance formula are described, 
then in sections \ref{sec:LPPM} and \ref{sec:LPSIS} the  formalism of the optical depth calculation for the point mass and SIS models is presented.  
I have followed  the analysis of \cite{1992ApJ...393....3F}.

\subsection{The Distance Formula} 
\label{sec:DistanceFormula}

Based on observations,  the Universe is homogenous on large scales. 
Therefore, on large scales it is well approximated by the FLRW geometry. 
However, at smaller distances, the light propagates through an inhomogeneous space-time 
rather than the averaged smooth space-time. 
The light ray feels the local metric which deviates from the smoothed Robertson-Walker metric.

Even if global parameters  like the mean mass density $\Omega_M$ and the normalized cosmological constant $\Omega_{\Lambda}$  are fixed, the propagation of  light rays, and hence the distance formula is not uniquely determined \citep{1964SvA.....8...13Z,1973ApJ...180L..31D,1992ApJ...393....3F}.  

The distances  calculations are hence made for the 2 cases of the homogenous and isotropic FLRW model and the \cite{1973ApJ...180L..31D} inhomogeneous and clumpy   Universe.

The angular diameter distances for the FLRW model is given by: 
\begin{equation}
d_S(z_1,z_2)=\frac{R_0}{1+z_2} \int_{z_1}^{z_2} \frac{dz}{\sqrt{\Omega_M (1+z)^3 + (1-\Omega_M)}}\, .
\label{eq:ds}
\end{equation}
The angular diameter distances for the Dyer\&Roeder model cosmology is given by: 
\begin{equation}
d_{DR} (z_1,z_2)=R_0(1+z_1)\int_{z_1}^{z_2} \frac{dz}{(1+z)^2 \sqrt{\Omega_M (1+z)^3+(1-\Omega_M-
\Omega_\Lambda)(1+z)^2+\Omega_\Lambda}} \, .
\label{eq:dDRdistance}
\end{equation}
In my analysis I have used  $\Omega_M +\Omega_\Lambda = 1$, so that the $2^{nd}$ term 
in the square root vanishes.  

\subsection{Probability of lensing by point masses}
\label{sec:LPPM}
The effective radius of the lens is characterized by the length:
\begin{equation}
r_{eff}^2=\frac{4GM}{c^2} \frac{D_{OL}D_{LS}}{D_{OS}}\, ,
\end{equation}
and  assuming detection for $r<r_{eff}$ the cross-section $\sigma$ is 
\begin{equation}
\sigma=\pi r^2_{eff} \, .
\end{equation}
The differential probability $d\tau$ of lensing in the path  $dz_L$ is given by
\begin{equation}
d\tau=n_L(0)(1+z_L)^3 \sigma \frac{c\,dt}{dz_L} dz_L=\frac{3}{2} \Omega_L(1+z_L)^3 
\frac{D_{OL}D_{LS}}{R_{0}D_{OS}}\frac{1}{R_0}\frac{cdt}{dz_L}dz_L \, ,
\end{equation}
where $\Omega_L=8\pi  GM n_L(0)/3H_0^2$ is the lens density parameter defined as 
a ratio of the local lens density to the critical density. 
$R_0=c/H_0$ is the Hubble distance and $H_0$ is the Hubble constant. 

The total lensing probability  is calculated by integrating the differential probability  
along the line of sight  to the source:
\begin{equation}
\tau (z_S)=\int_{0}^{z_S} \frac{dz(z_S,z_L)}{dz_L} dz_L .
\label{eq:TotalPro}
\end{equation}
 \begin{figure}[!t]
  \centering
  \includegraphics[width=8cm,angle=-90,bb=0 0 504 720 ]{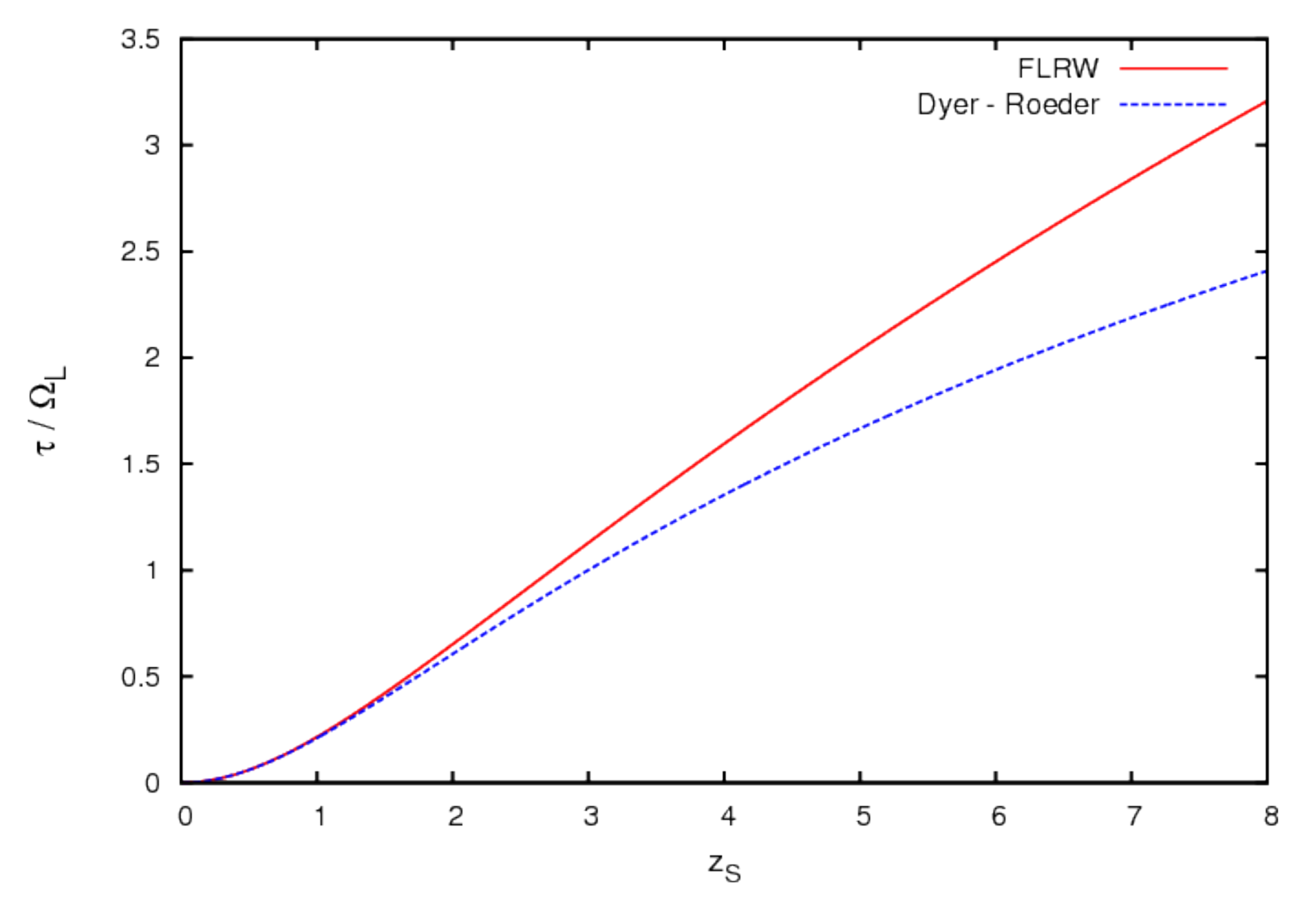}
  \caption{Total lensing optical depth $\tau$ in the FLRW (solid line) 
and Dyer-Roeder (dashed line) formalisms as a function of the redshifts $z$ of sources, assuming point mass lenses.}
  \label{fig:tauPM}
 \end{figure}
Figure \ref{fig:tauPM} shows the total lensing optical depth assuming point mass lenses. 
The optical depth is presented as a function of the redshift $z_S$ of the sources.
 \begin{figure}[!t]
  \centering
  \includegraphics[width=8cm,angle=-90,bb=0 0 504 720 ]{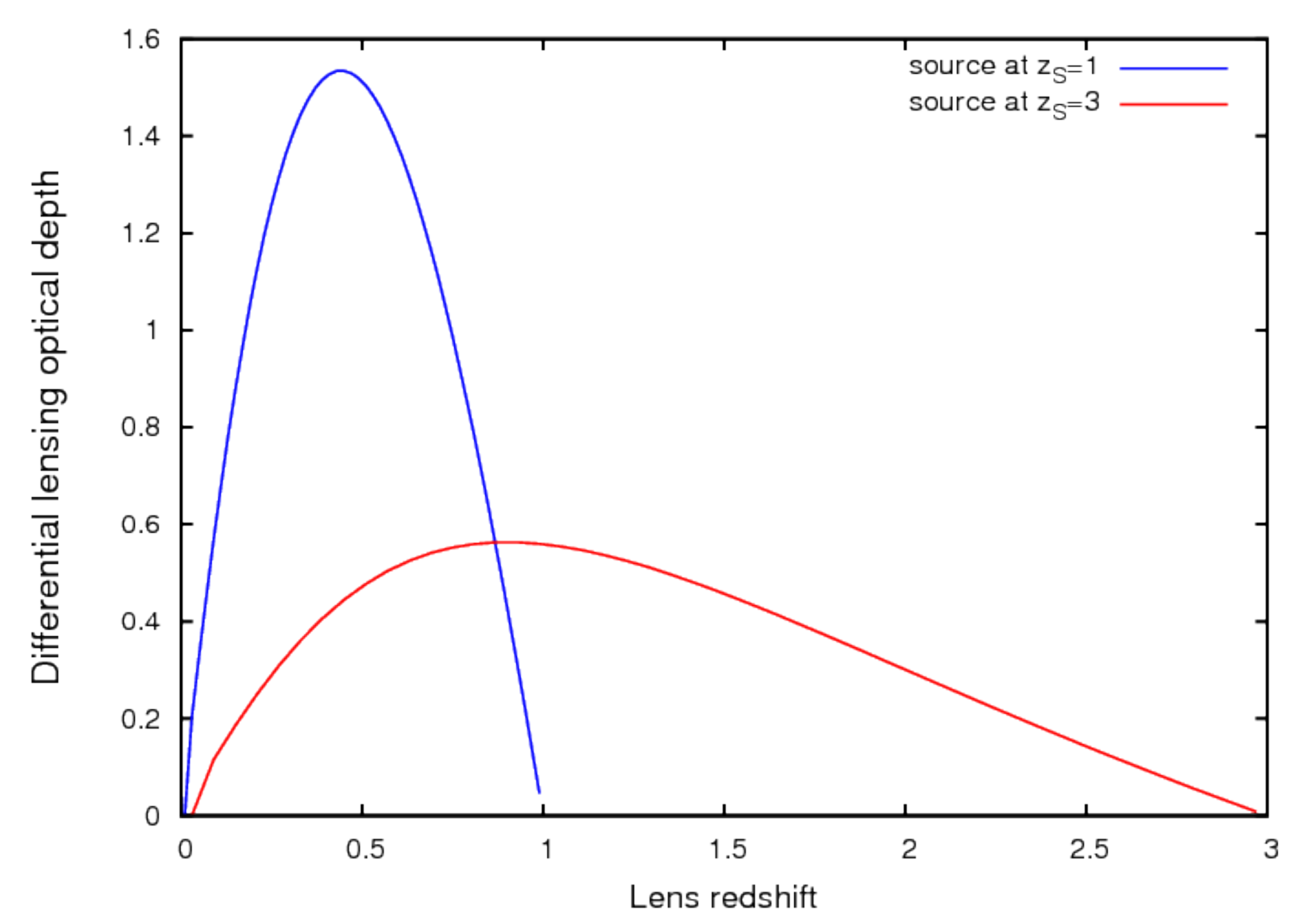}
  \caption{Differential  lensing optical depth $d\tau/dz$  normalized to 1 in the FRLW formalism as a function of redshifts $z$ of the lens. The blue line indicates the case with a source at redshift 1 and the red line a source at redshift 3.  The differential probability has been calculated using a point lens mass model.}
  \label{fig:dtauPM}
 \end{figure}
%
\subsection{Lensing Probability by Singular Isothermal Sphere}
\label{sec:LPSIS}
The SIS model is described in section \ref{sec:SIS}. 
The cross-section  for lensing by the SIS is given by  equation:
\begin{equation}
\sigma=\pi r^2_{eff}=16\pi^3\left(\frac{\nu}{c}\right)^4\left(\frac{D_{OL}D_{LS}}{D_{OS}} \right)^2
\end{equation}
The differential probability of lensing is:
\begin{equation}
d\tau =n_{0}(1+z_{L})^3\sigma \frac{cdt}{dz_{L}}dz_{L} 
= F(1+z_{L})^3\left(\frac{D_{OL}D_{LS}}{R_{0}D_{OS}} \right)^2\frac{1}{R_0}\frac{cdt}{dz_L}dz_L \, ,
\end{equation}
where:
\begin{equation}
\frac{cdt}{dz_{L}} = \frac{R_0}{1+z_L}\frac{1}{\sqrt{\Omega_\lambda(1+z_L)^3+(1-\Omega_\lambda-\lambda_0)(1+z_L)^2+\lambda_0)}} \, ,
\end{equation}
and $F$ is a quantity which measures the effectiveness of matter in producing double images \citep{1984ApJ...284....1T} and is given by equation: 
\begin{equation}
F=16\pi^3 n_0 \left(\frac{\nu}{c}\right)^4 R_0^3.  
\end{equation}
The value of $F$ used by \cite{1992ApJ...393....3F} was 0.047. 
The total probability of lensing is calculated as in equation (\ref{eq:TotalPro}).

 \begin{figure}[!t]
  \centering
  \includegraphics[width=8cm,angle=-90,bb=0 0 504 720 ]{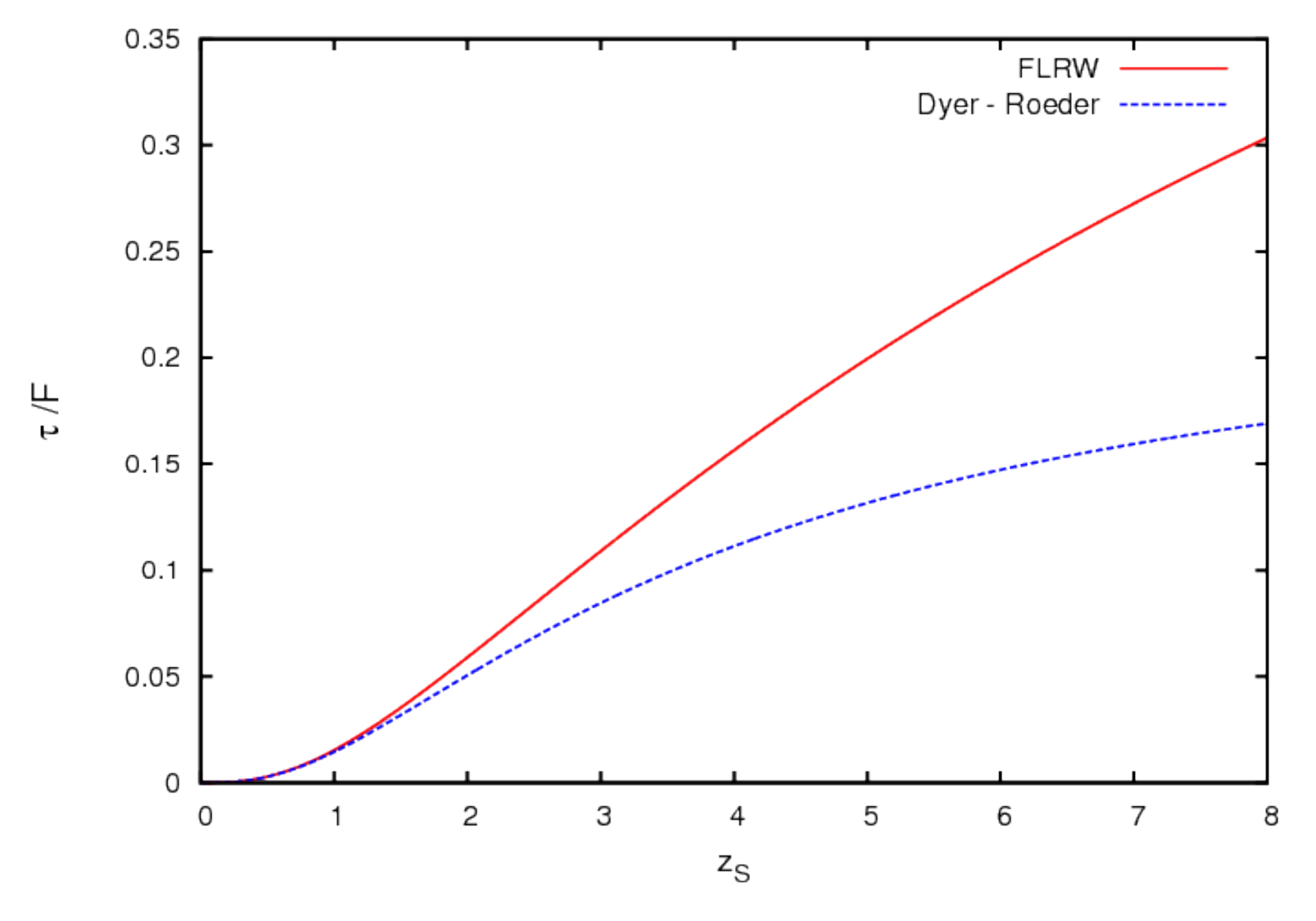}
  \caption{Total lensing optical depth $\tau$ in the FRLW (solid line) 
and Dyer-Roeder (dashed line) formalisms as a function of redshifts $z$ of the source.}
  \label{lp_fig}
 \end{figure}
 \begin{figure}[!t]
  \centering
  \includegraphics[width=8cm,angle=-90,bb=0 0 504 720 ]{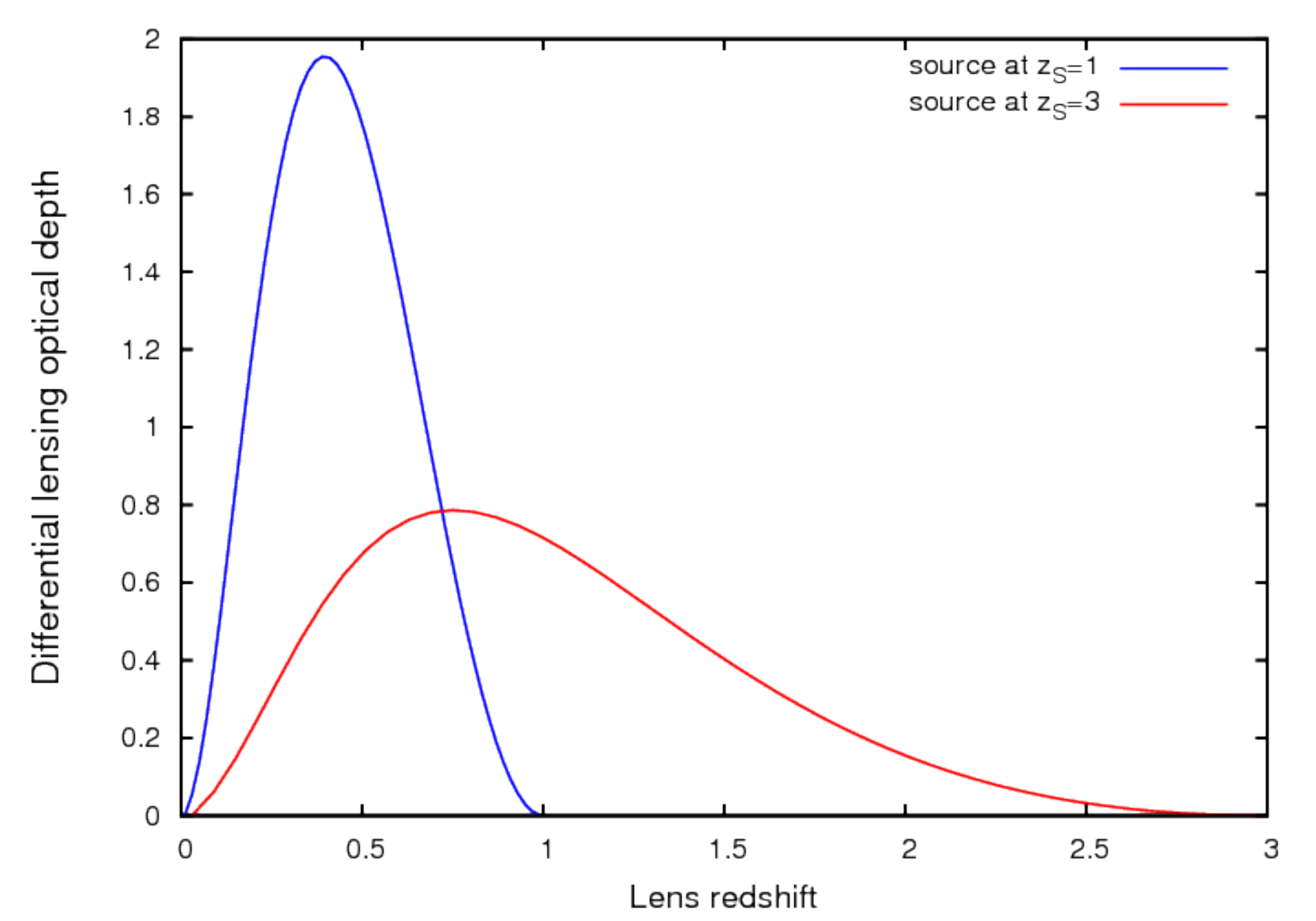}
  \caption{Differential  lensing optical depth $d\tau/dz$  normalized to 1 in the FRLW formalism as a function of redshift $z$ of the lens. The blue line indicates the case with a source at redshift 1 and the red line a source at redshift 3.  The differential probability has been calculated using SIS as a lens model.}
  \label{fid:dtauSIS}
 \end{figure}

\setcounter{chapter}{4}
\setcounter{section}{0}
\setcounter{equation}{0}
\setcounter{figure}{0}
\part{Gravitational lensing in high energy   \label{chapter:helens}}  
\section{Introduction}

The precise estimation of the time delay between components of lensed Active Galactic Nuclei (AGN) is crucial
for modeling the lensing objects. In turn, more accurate lens models give better constraints on the Hubble constant.
 Recent years, more than 200 strong lens systems have been discovered, most of  them by dedicated surveys such as
the Cosmic Lens All-Sky Survey \citep{2003Myers,2003Browne}
and the Sloan Lens ACS Survey \citep{2004Bolton}.

The launch of the FERMI satellite \citep{2009Atwood} in 2008 gives the opportunity
to investigate gravitational lensing phenomena with high energy gamma rays.
The observation strategy of FERMI-LAT, which surveys
the whole sky in 190 minutes, allows a regular sampling of quasar light curves with
a period of a few hours.
 
The multiple images of a gravitational lensed AGN cannot be  
directly observed with high energy gamma-ray instruments such as  
 FERMI-LAT, Swift or ground based Cherenkov telescopes,
due to their limited angular resolutions. The angular resolution of
these instruments is at best a few arcminutes (in the case of the H.E.S.S.),
when the typical separation of the images for quasar lensed by galaxies is a few arcseconds.

The  analysis described in this part of my thesis is concerned on the time delay estimation in strong lenses  
when only spatially unresolved data are available. 
I have developed a method that handles the problem of poor instrument angular resolution \citep{2011A&A...528L...3B}.
It is based on the so-called "double power spectrum" analysis.    
 
The method of time delay estimation have been tested on simulated light curves and on FERMI LAT observations
of the very bright radio quasar PKS 1830-211, for which the
time delay was previously estimated 
by \cite{1998Lovell} using radio observations.   

This part of the thesis  is organized as follows: 
first, I discus the lensing probability of the FERMI AGNs (section \ref{sec:FermiProb}), 
then I  give a very brief summary of the properties of PKS 1830-211(section \ref{sec:pks1830}), 
of  the FERMI LAT satellite (section \ref{sec:FERMIsat}) 
and the data  towards this AGN (section \ref{sec:PKS1830data}).
In section \ref{sec:TimeDelayMethods}, I discus the methods for time delay estimation. 
Section  \ref{sec:DPSM} describes  the measurement of the time delay
between the two compact components of PKS 1830-211.
Section  \ref{sec:resultsGL} contains the summary of results.
The last section  (\ref{sec:Hubble}),  is devoted to gravitational lens time delay and the Hubble constant. 
%
\section{Probability of lensing of FERMI AGNs} 
\label{sec:FermiProb}
The 2st FERMI catalogue \citep{2012ApJS..199...31N} 
 contains 1873 sources detected  in the 100 MeV to 100 GeV range.
Among the AGN associations the 2st FERMI catalogue lists  1064 blazars, 
consisting of 432 BL Lacertae objects (BL Lacs), 
370 flat-spectrum radio quasars (FSRQs), 
 and 262 of unknown type. 
From the formalism described in part \ref{chapter:TheoryGL}  
the expected numbers of lenses in  the FSRQ class  is 9.4$\times F$ 
- assuming the homogeneous FLRW model of the universe, 
and 8.25$\times F$ for the Dyer-Roeder model. 

The probability of lensing was also calculated for BL Lacs objects. 
The expected number of lenses is 0.61$\times F$ for the FLRW model 
and 0.58$\times F$ for the Dyer-Roeder model. 
The AGN are lensed by galaxies which are modeled  as SIS lenses (chapter \ref{chapter:TheoryGL}).
The total probability of lensing by SIS lenses is shown on figure \ref{lp_fig}.

The large difference in number of expected lensed objects in FSRQ and BL Lac groups 
is arising from the redshift distribution of both groups. 
The BL Lac objects have been detected only at small redshifts below 1.5.
Most of them have a z $<$ 0.5. with majority below 0.5. 
In addition, only half of BL Lac objects have measured redshift. 
It is very difficult to measured redshift of BL Lac object since this objects are lineless.

\begin{figure}[!t]
  \centering
  \includegraphics[width=12cm,angle=0,bb=0 0 360 216 ]{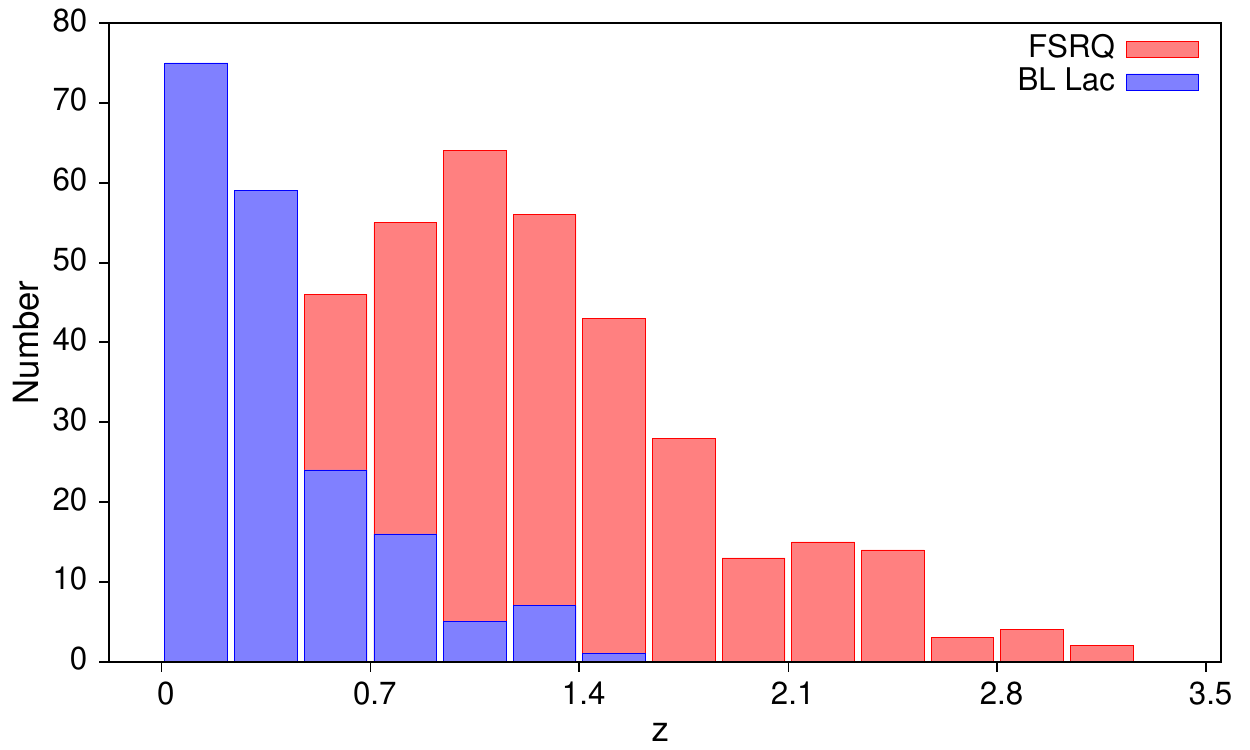}
  \caption{The redshift distribution of FSRQ and BL Lac objects included in second Fermi catalogue.}
  \label{fig:reddis}
 \end{figure}

The FSRQ have been detected at much larger distances. 
The most distant FSRQ in FERMI catalog was detected at a redshift $>$ 3. 
The redshift distribution for BL Lac and FSRQ objects from the second FERMI catalogue is shown on figure~\ref{fig:reddis}

The lensing probability calculated above does not take into account the magnification bias. 
The magnification bias  
 can substantially increase the probability of lensing for bright optical quasars
 \citep{1980ApJ...242L.135T,1984ApJ...284....1T,1993LIACo..31..217N}.

\section{The PKS 1830-211 gravitational lens system}
\label{sec:pks1830}
PKS 1830-211 is one of the object observed in the high energy domain by the FERMI satellite. 
PKS 1830-211 is a variable, bright radio source and an X-ray blazar.
Its redshift was measured to be z=2.507 \citep{1999Lidman}.

The blazar was detected in $\gamma$-rays with EGRET.
The association of the EGRET source
with the radio source was done by \cite{1997Mattox}.
The classification of PKS 1830-211 as a gravitational lensed quasi-stellar object
was first proposed by \cite{1988Rao}.

The lensing galaxy is a face-on spiral galaxy, identified by \cite{2002Winn} and \cite{2002Courbin}, and 
located at redshift z=0.89 \citep{1996Wiklind}. 

PKS 1830-211 is observed in radio as an elliptical ring-like structure connecting two bright sources distant of roughly one arcsecond \citep{1991Jauncey}, see figure \ref{fig:image_radio}.
The compact components were separately observed
by the Australia Telescope Compact Array at 8.6 GHz for 18 months.
These observations and the subsequent analysis made by \cite{1998Lovell}
gave a magnification ratio between the 2 images of $1.52\pm 0.05$
and a time delay of $26^{+4}_{-5}$ days. 
A separate measurement of a time delay of $24^{+5}_{-4}$ days was made by \cite{2001Wiklind}
 using molecular absorption lines.

\section{The FERMI Satellite} 
\label{sec:FERMIsat}
The FERMI Gamma-ray Space Telescope is a space observatory aimed at performing gamma-ray astronomy observations. 
Fermi was launched on 11 June 2008.
The key scientific goals of the FERMI mission are then studies
of active galaxy nuclei, supernova remnants, gamma-ray bursts and dark matter. 

The observatory includes two scientific instruments. 
One is a calorimeter, the Large Area Telescope (LAT) sensitive to photons 
in energy range from 30 MeV to 300 GeV.
The LAT  field of view is about 20\% (2 sr) of the sky.
The second instrument is a Gamma-ray Burst Monitor (GBM) used to detect gamma-ray bursts 
in the energy range from a few keV to 30 MeV.

\section{FERMI LAT data on PKS 1830-211}
\label{sec:PKS1830data}
  \begin{figure*}
  \centering
\includegraphics[width=15cm,angle=0,bb=0 0 567 183]{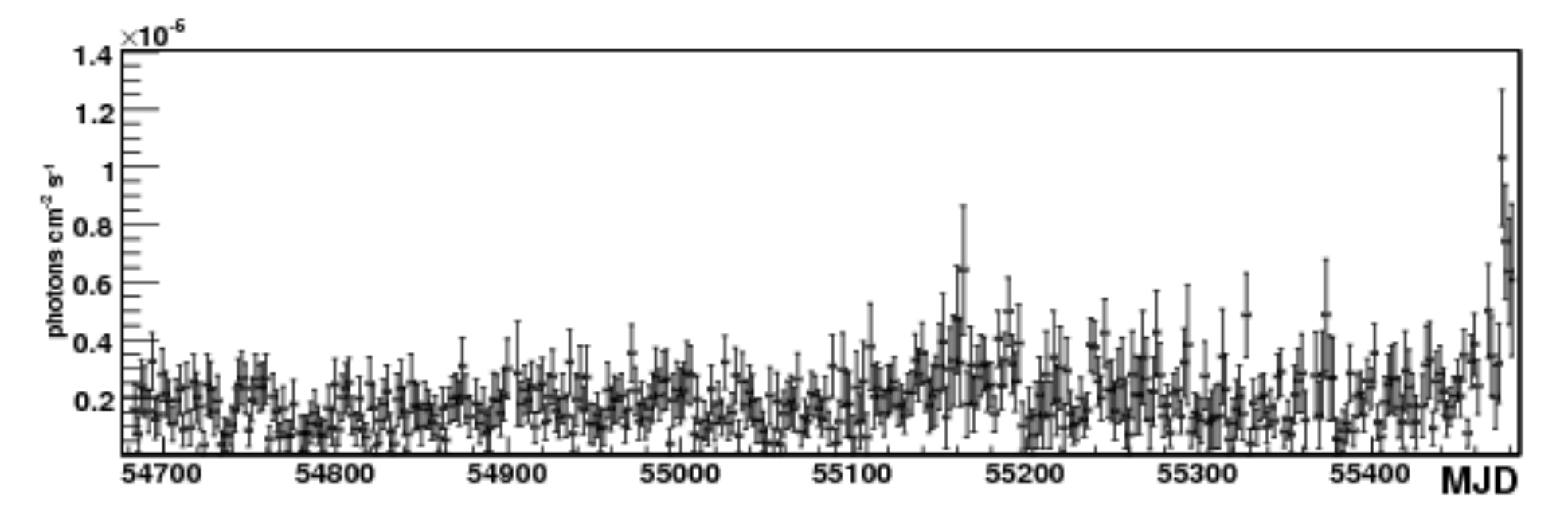} \\
\caption{Fermi LAT light curve of PKS 1830-211,  with a 2 days binning. Photons with energy between  300 MeV and 300 GeV were selected.}
            \label{lc}
  \end{figure*}

\begin{figure*}
  \centering
\includegraphics[width=15cm,angle=0,bb=0 0 77 29]{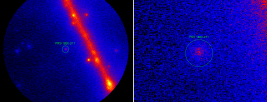} \\
\caption{Fermi count map centered on PKS 1830-211. Photons in the energy range from 300 MeV to 300 GeV were selected.}
            \label{fig:countsmap}
  \end{figure*}

PKS 1830-211 has been detected by the FERMI LAT instrument with 
a detection significance above 41 FERMI Test Statistic (TS), equivalent to a 6 $\sigma$ effect \citep{2010Abdo}. 
The long-term light curve is presented on figure \ref{lc} with a two days binning and the counts map centered at position of PKS 1830-211 (figure \ref{fig:countsmap}).

The data analysis was performed using a two days binning, 
which provides a sufficient photon statistic per bin with a time span per bin much shorter than 28 days. 
The data analysis was cross-checked by binning the light-curve into 1 day and 23 hours bins, 
with similar results.  

The data were taken between August 4 2008 and October 13 2010, 
and processed by the publicly available Fermi Science Tools version 9.
The v9r15p2 software version and the P6\_V3\_DIFFUSE instrument response functions have been used. 
The light curve has been produced by aperture photometry, selecting photons from a region with radius 0.5 deg around the nominal position of PKS~1830-211 and energies between 300 MeV and 300 GeV.  
 
\section{Methods of Time Delay estimation}
\label{sec:TimeDelayMethods}

The most popular  methods of time delay estimation, in the context of gravitational lensing, 
are  the cross-correlation method \citep{1997ApJ...482...75K}, and the dispersion spectra method \citep{1996A&A...305...97P}. 

\begin{figure}[ht]
\begin{minipage}[b]{0.63\linewidth}
\centering
\includegraphics[width=\textwidth,angle=-90,bb=0 0 501 338]{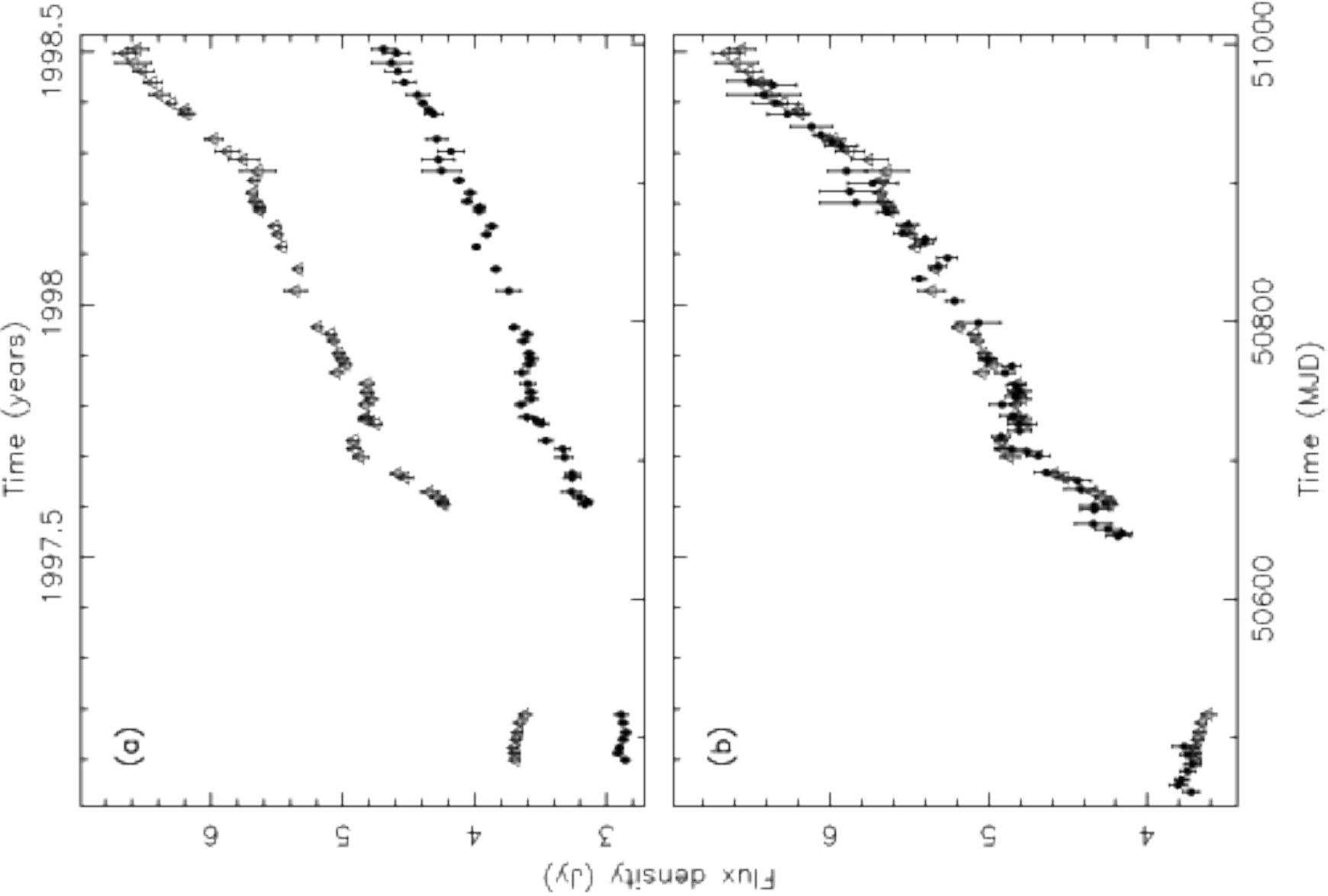}
\caption{(a) The 8.6 GHz light curve for both components of PKS 1830-211. 
                (b) Light curves superimposed  applying the dispersion analysis with a time delay of -23 days             
                and magnification ratio 1.52. Figure is taken  from  \cite{1998Lovell}.}
\label{fig:lc_radio}
\end{minipage}
\hspace{0.01cm}
\begin{minipage}[b]{0.37\linewidth}
\centering
\includegraphics[width=\textwidth,bb=0 0 585.133 713.162]{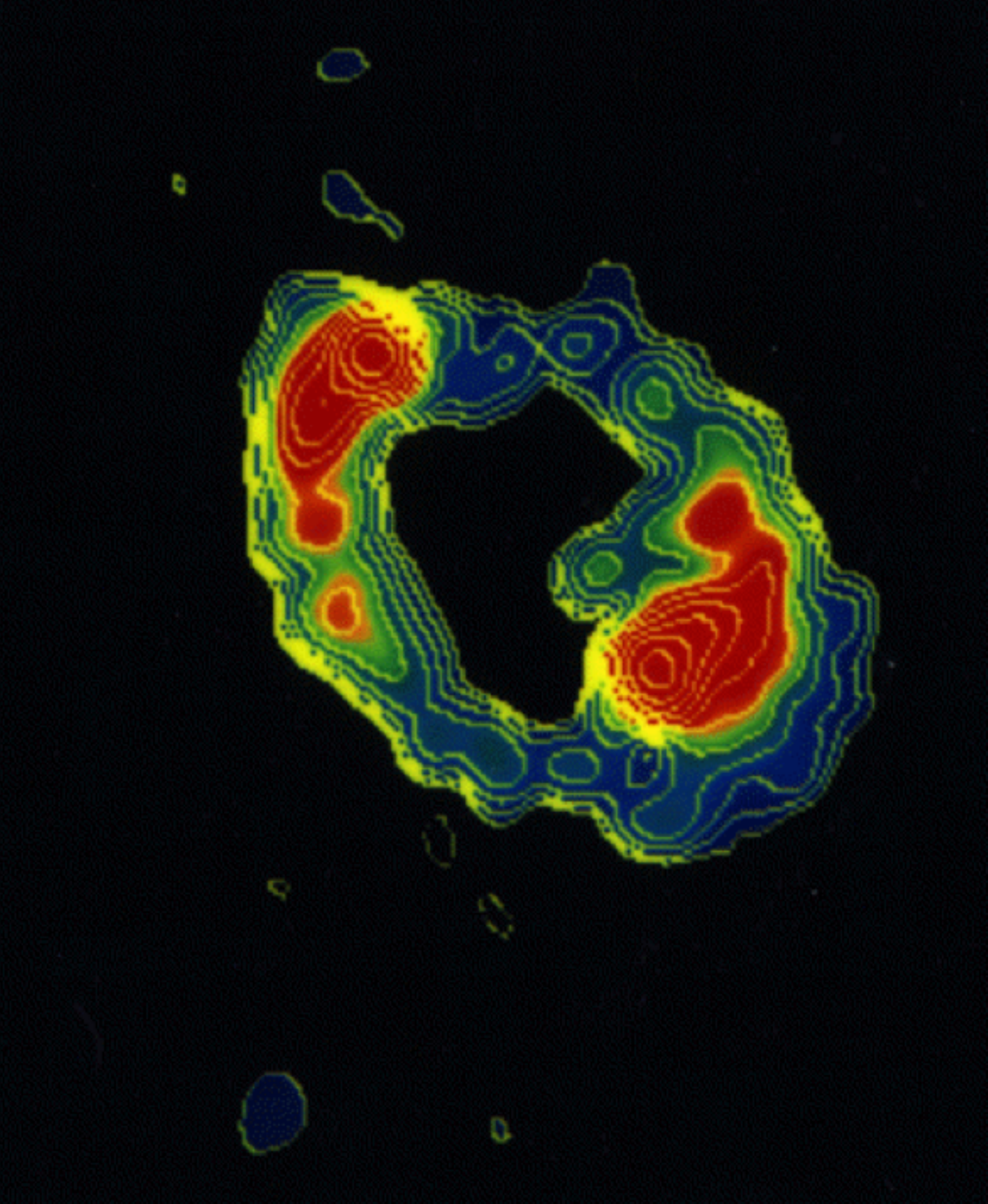}
\caption{High-resolution radio image of PKS1830--211 obtained from the VLBI interferometric radiotelescope networks. The image shows  an  elliptical ring-like structure connecting the two brighter components separated by distance of one arcsecond.
Image credited from \cite{1991Jauncey}}
\label{fig:image_radio}
\end{minipage}
\end{figure}

In the case of PKS 1830-211, the previous time delay estimation given by \cite{1998Lovell} based on
the dispersion analysis method \citep{1994A&A...286..775P,1996A&A...305...97P}. 
The light curve analysis has been performed using the data taken from 1997 January to 1998 July. 
The observation was done with the Australian Telescope Compact Array operating at 8.6 GHz. 
The light curves taken from \cite{1998Lovell}  is shown on figure \ref{fig:lc_radio}. 
 
The radio image (figure \ref{fig:image_radio}) shows two bright,  well resolved components connected by 
an elliptical ring like structure. 
The analysis presented in \cite{1998Lovell} decomposes the observed system  into two lensed images $S_1(t)$
and $S_2(t)$ and  an additional component  from the ring like structure. 
The total observation flux density  was thus defined as: 

\begin{equation}
S_m(t) = S_1(t) + S_2(t) + S_{ring} \,.
\end{equation} 

The intensity of the lensed images changes in time in similar fashion, 
but the light curves  are shifted in time by $\Delta \tau$. 
The flux of the two magnifier images change with time  
but the  magnification ratio  $\mu$ is constant. 
Thus, the flux of the second image can be define as: 

\begin{equation}
S_{2} (t) = \frac{1}{\mu} S_{1} (t+\Delta \tau ) \,.
\end{equation}
 
\cite{1998Lovell} followed the analysis of \cite{1994A&A...286..775P,1996A&A...305...97P}. 
They obtain two light curved datasets 
$a_i$ and $b_i$ ($i=1,...,N$ where $N$ is a number of observations) for every data value of $\mu$ and $\Delta \tau$. 

The dispersion $D^2$ of combined light curves is given by:

\begin{equation}
D^2(\Delta \tau, \mu) = \frac{\sum_{ij}W_{ij}V^{'}_{ij}(a_i - b_j)^2}{2\sum_{ij}W_{ij}V^{'}_{ij}} \,,
\label{eq:dispersion1}
\end{equation}

where $V^{'}_{ij}$ are weights which select only pairs of observations whose observing time t, 
do not differ more than $\delta t$.


and $W_{ij}$ is a combined statistical weight defined as

\begin{equation}
W_{ij}=\frac{W_iW_j}{W_i+W_j} \,,
\end{equation}

where  $W_i=\frac{1}{\sigma^2_i}$ and $W_j=\frac{1}{\sigma^2_j}$ are the errors from the observations. 
Equation (\ref{eq:dispersion1}) 
 represents the weighted sum 
of the squared difference between $a_i$, $b_i$ pairs over the entire light curve.  
$D^2$ is minimized with respect to $\Delta \tau$ and $\mu$.
 \cite{1998Lovell}  obtain a time delay of  $26^{+4}_{-5}$ days and a magnification ratio $\mu=1.52\pm0.05$.

\section{Double power spectrum method}
\label{sec:DPSM}

At the high  energy (HE $>$ 100 MeV) domain, the multiple images of a lensed AGN cannot be directly observed due to the limited angular resolution of the existing detectors.
This section presents a new method for estimating the time delay of the images of a lensed quasar \citep{2011A&A...528L...3B}. 
This method is usable when the images are unresolved. 

\subsection{Idea}
\label{idea}
As it was discussed in part \ref{chapter:TheoryGL}, if a distant source (in our case an AGN) is gravitationally lensed,
the light reaches the observer through at least two different paths.
For the moment only  two light paths are assumed.
In reality,  the light curves of the two images are not totally identical since 
(in addition to differences due to photon noise)
the source can be microlensed in one of the two paths. 
The microlensing is cased by a star from the lensing galaxy crossing one of the paths.

Neglecting for the moment the background light  and the differences due to microlensing, 
the observed flux can be decomposed into two components.
One of the components is the intrinsic AGN light curve, given by $f(t),$ with Fourier transform $\tilde{f}(\nu ).$
The other component has a similar time evolution than the first one, but is shifted in time with a delay $a.$
In addition, the brightness of the second component  differs by a factor $b$
from that of the first component,
so that it can be written as $b f(t+a)$ and its transform to the Fourier space gives  $b\tilde{f}(\nu ) e^{-2\pi i\nu a}$.

The sum of two component gives

\begin{equation}
g(t) =  f(t) + bf(t + a) \,,
\end{equation}

which transforms into
\begin{equation}
       \tilde{g}(\nu ) = \tilde{f}(\nu ) (1 + b e^{-2\pi i\nu a}) \,,
\end{equation}
in Fourier space.

 The power spectrum  $P_{\nu}$ of the source is obtained by computing the square modulus of $\tilde{g}(\nu )$:
 
 \begin{equation}
         P_{\nu} = |\tilde{g}(\nu )|^{2} = |\tilde{f}(\nu )|^{2}(1 + b^{2} + 2b cos(2\pi \nu a)) \,.
\label{Powereq}
 \end{equation}
 
The measured  $P_{\nu}$ is the product of the ``true'' power spectrum of the source times a periodic component
with a period (in the frequency domain) equal to the inverse of the relative time delay $a$. 
The microlensing of one of the components, when taken into account, 
 gives a modulation of the amplitude of the oscillatory pattern at low frequencies. 
 
 The typical time interval for a quasar to cross its own  diameter $R_{source}$ is 
 
 \begin{equation}
 t_{cross} = R_{source}/v_{\perp} \approx 4R_{15}v^{-1}_{600} \, {\rm months} \,,
 \end{equation}
 
 where $R_{15}$ is the quasar size in units of $10^{15}\,$cm, and $v_{\perp}$ is a transverse 
 velocity, perpendicular to the line of sight, here $v_{600}$ in 600 km/s. 
The typical time of a caustic crossing microlensing event is thus a few months \citep{2001ASPC..237..185W}.
Therefore,  only frequencies under $3 10^{-7}$ Hz will be affected by microlensing event.

The usual way of measuring the time delay $a$ is
to calculate the autocorrelation function of f(t). 
This method was investigated by \cite{1996Geiger}.
The computation of the autocorrelation of a light curve with
uneven sampled data is described in \cite{1988Edelson}. 

The Fermi light curve of PKS 1830-211 has very few gaps, and only one  notable four-day gap. 
The missing data have been linearly interpolated.
However, simulations with an artificial gap have shown that the results  are little affected by this gap.
The autocorrelation function can be written as the sum of three terms.

One of these terms models the "intrinsic" autocorrelation of the AGN,
decreasing with a time constant $\lambda$.
If $\lambda$ is larger than $a$, the autocorrelation method fails,
because the time delay peak merges with the intrinsic component of the AGN.
Another potential problem with the autocorrelation method is the sensitivity
to spurious periodicities such as the one coming from the motion and
rotation of the Fermi satellite.
 
The periodic modulation of $P_{\nu}$ suggests the use of another method,
based on the computation of the power spectrum of $ P_{\nu} $, noted $D_{a}$.
This method is similar in spirit to the cepstrum analysis \citep{Bogert1963} used in seismology and speech processing.

If $|\tilde{f}(\nu )|^{2}$ were a constant function of $\nu$, $D_{a}$ would have a peak at the time delay $a$.
In the general case, $D_{a}$ is obtained by the convolution of a Dirac function, 
coming from the cosine modulation, by the Fourier transform of the function:
 
\begin{equation}
            \tilde{h}(\nu) = |\tilde{f}(\nu )|^{2}  W(0,\nu_{\mbox{max}})\,,
\end{equation}
                                              
where W(0,$\nu_{\mbox{max}}$) is the window in frequency of $P_{\nu}$ and $\nu_{\mbox{max}}$ is maximum available frequency.
The Fourier transform h(a) of $ \tilde{h}(\nu)$ defines the width of the time delay peak
in the double power spectrum $D_{a}$.
 For instance if $|\tilde{f}(\nu)|^{2} = e^{-\lambda\nu}$ and $\lambda W >> 1 ,$
then the time delay peak in $D_{a}$ has a Lorentzian shape  with a FWHM of ${\lambda}/{\pi}.$
For a typical value of $\lambda = 10$ days, one has a FWHM of 3 days.

In the next section we describe the calculation of $P_{\nu}$ and $D_{a}$ and
illustrate the procedure with Monte Carlo simulations.

\subsection{Power spectrum calculation}
\label{FPSsubs}

The Fourier transform is a powerful technique to analyze astronomical data,
but it requires a proper preparation of the dataset.
To avoid problems arising from the finite length of measurements,
sampling and aliasing, we use the procedure for data reduction described 
by \cite{1971Brault} and \cite{2007Press}.
 
First the whole light curve is divided  into several segments of equal length. 
The data have to be evenly spaced
and the number of points per segment needs to be equal to a power of 2.

The choice of the segment length is a compromise between the spectral resolution 
and the size of error bars on points in the final spectrum. 
The resolution of lines increases with the number of points in the segment,
but the error on each point in the spectrum decreases as the number of segments.  
 
As suggested by \cite{1971Brault},  
the segments are overlapped to obtain a larger number of segments
with a sufficient number of  points.
Then the data in each segment are transformed with the following procedure:
\begin{enumerate}
\item Data gaps are removed by linear interpolation
\item The mean is subtracted from the series to avoid having a large value in the first bin of the transform
\item The data are oversampled to remove aliasing.
\item The light curve is multiply by the window
\item Zeros are added to the end of the series to reduce power at high frequencies
\end{enumerate}

After the interpolation of missing data, and standard operations (oversampling, zero-padding), the power spectrum is calculated in every segment. 
Figures \ref{fig:mc_alias} and \ref{fig:mc_alias} illustrate with the artificial light curve 
how step 3 of the data processing procedure removes the aliasing. 
During the procedure evoluation a lot of different window functions   
have been tested.  
Monte Carlo simulations show, that for the first power spectrum calculation, most efficient was a rectangular window function. 
The rectangular windows function had the smallest smoothing effect on the oscillations expected when the signal is delayed. 
Finally, the power spectrum is averaged over all segments.

\begin{landscape}
  \begin{figure}
  \centering
\includegraphics[width=18cm,angle=0,bb=0 0 1024 768]{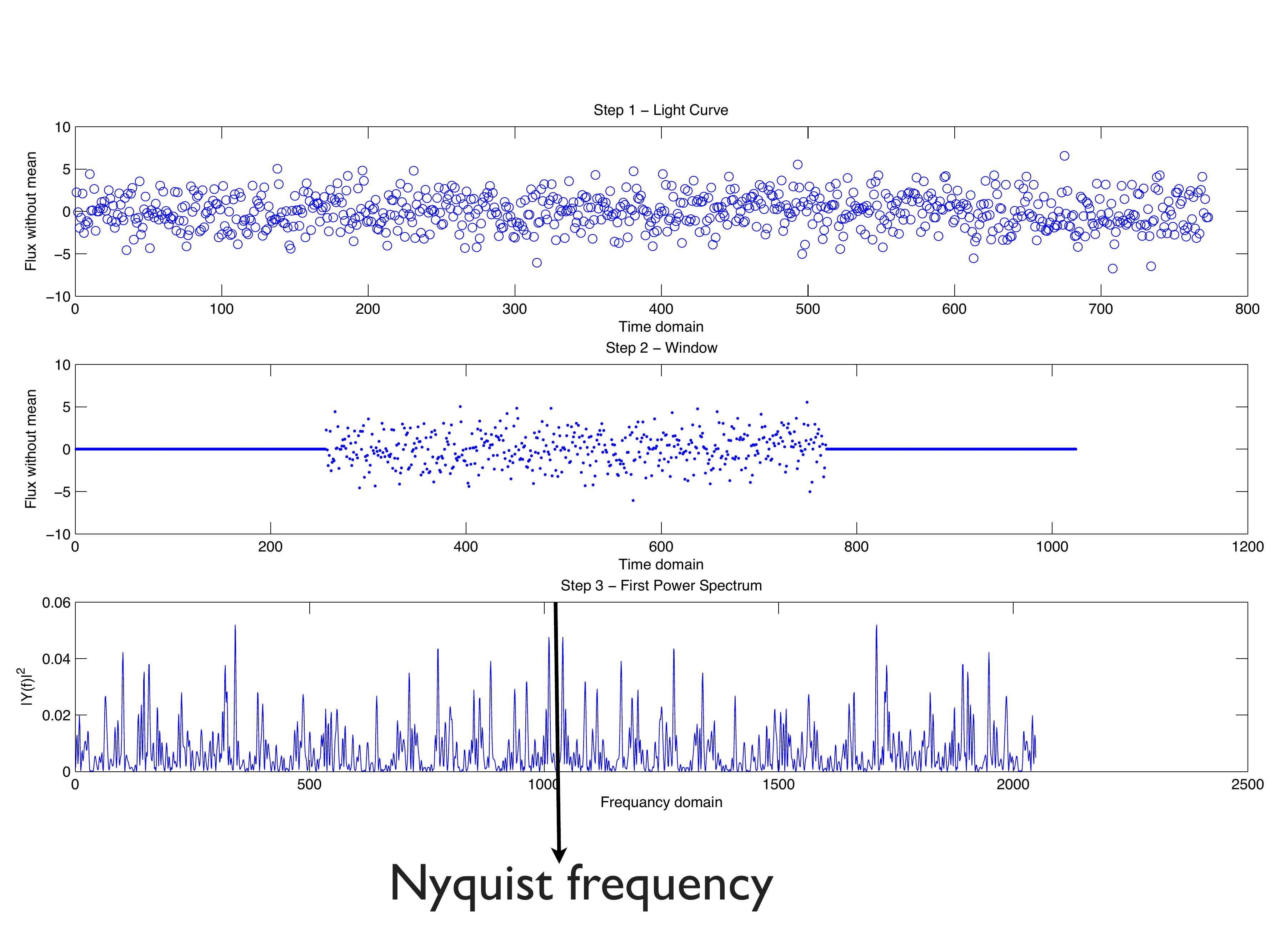} \\
\caption{Plot illustrating processing of an artificial light curve. Top plot shows an artificial light curve. Middle one shows the artificial light curve multiply by rectangular window and with added zeros to the end and the beginning of dataset. Bottom plot shows  the power spectrum of light curve presented on middle plot. The arrow indicates the Nyquist frequency. The power spectrum is strongly aliased (the amplitudes are not going to zero when frequencies are  going to Nyquist frequency).  Units of x-axises are arbitrary.}
            \label{fig:mc_alias}
  \end{figure}
  
    \begin{figure}
  \centering
\includegraphics[width=18cm,angle=0,bb=0 0 1024 768]{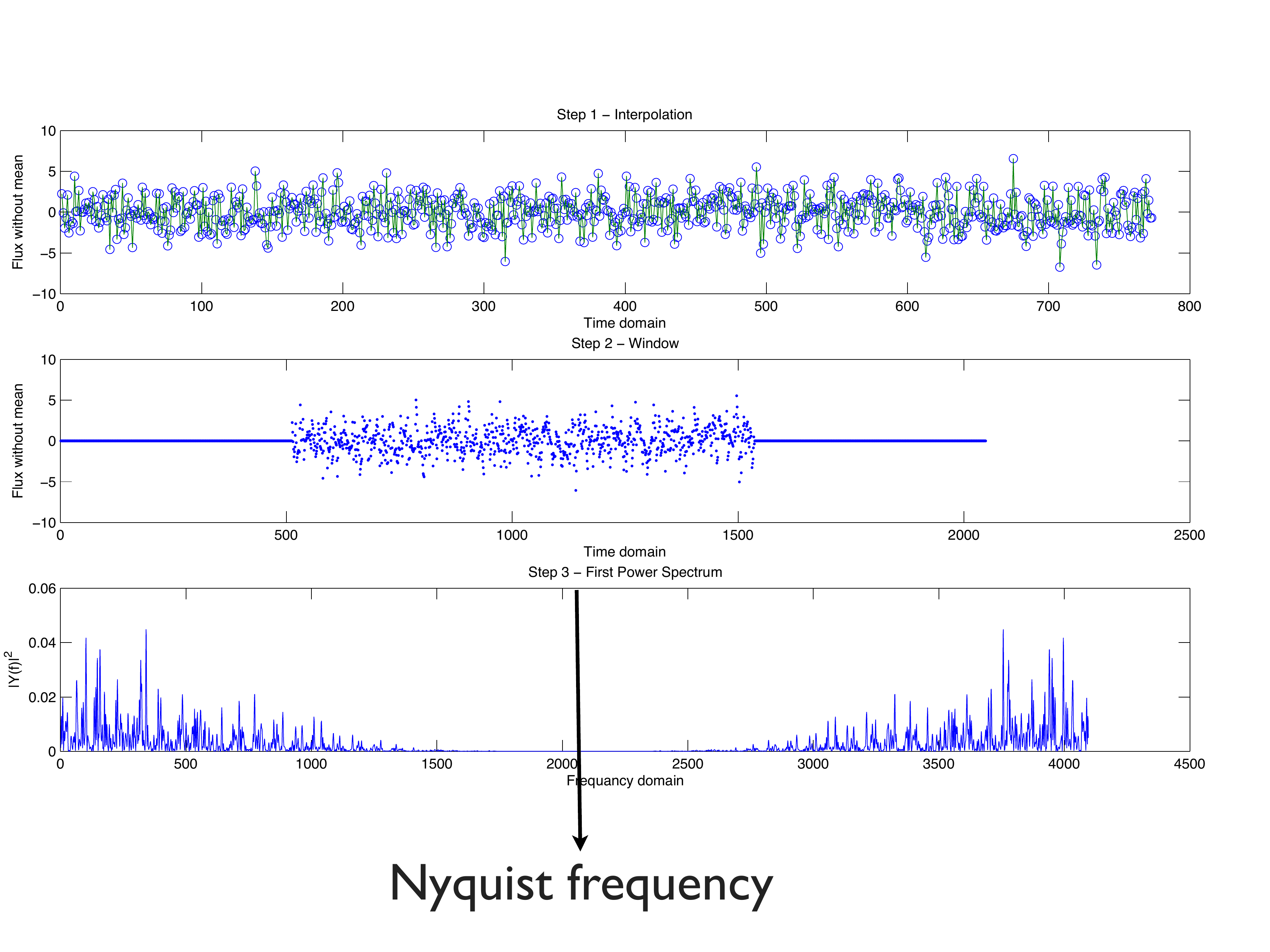} \\
\caption{Top plot shows the artificial light curve with doubled points. Middle plot shows this artificial light curve multiply by rectangular window and with zeros added to both ends of the series. Bottom plot shows  the power spectrum of light curve presented on middle plot. The arrow indicates the Nyquist frequency. 
After doubling the points the power spectrum is not aliased any more (the amplitude goes to zero when frequency goes to Nyquist frequency).  Units of x-axises are arbitrary.   Bottom figure presents the power spectrum calculated with just one data slice. }
            \label{fig:mc_noalias}
  \end{figure}
\end{landscape}

\subsection{Monte Carlo simulations}

Artificial light curves were produced by summing three simulated components.
The light curve of PKS 1830-211 shown on figure \ref{lc}
does not exhibit any easily recognizable features, but  has a rather random-like aspect.
The first component was thus simulated as a white noise, with a Poisson distribution.
It would be more realistic to use red noise instead of white noise 
but the latter is sufficient for most of our purposes, such as computing $D_{a}.$ 
The second component is obtained from the first by shifting the light curve with a 28 days time lag. 
The effect of differential magnification of the images has also been included.
The background photon noise was taken into account by adding a third component
with a Poisson distribution.
 
The mean number of counts per 2 day bin for PKS 1830-211 is 5.42.
This value was used  to generate artificial light curves.
The first and second components account for 80\% of the simulated count rate
and the rest is contributed by the Poisson noise.
 
\begin{figure}
  \centering
\includegraphics[width=13cm,angle=0,bb=0 0 567 358]{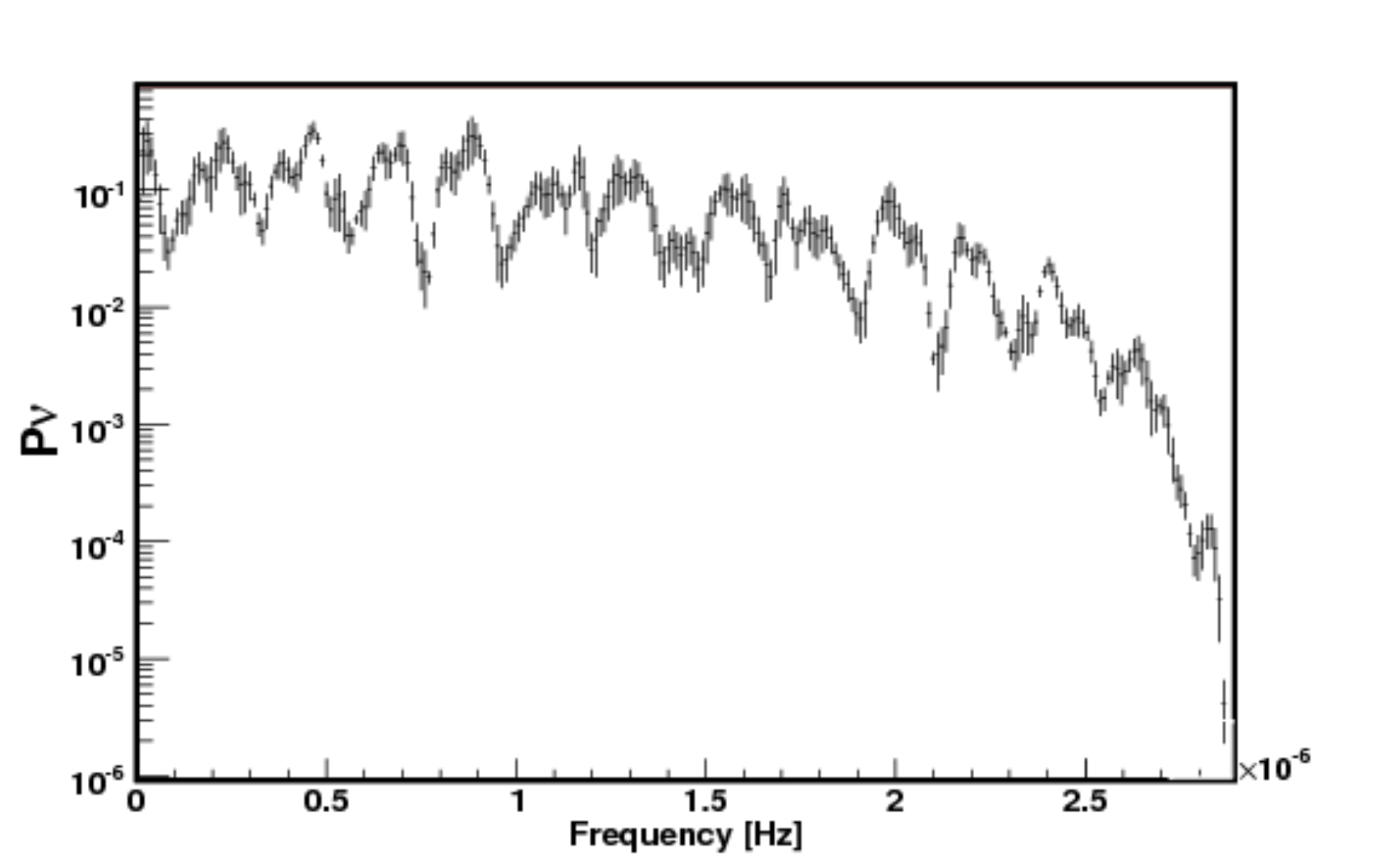} \\
\caption{Simulated power spectrum of a lensed AGN with a time delay between images of 28 days. The power $P_{\nu}$ is in arbitrary units. }
            \label{mc_fps}
  \end{figure}

The power spectrum obtained with an artificial light curve is shown on figure \ref{mc_fps}.

\subsection{Time delay determination}
\label{SubTDD}
The methods of time delay determination use the power spectrum $P_{\nu}$ obtained 
as described in the previous section.
The simulated $P_{\nu}$ presented on figure \ref{mc_fps} shows a very clear periodic pattern.
From equation (\ref{Powereq}), it is known that the period of the observed oscillations is equal to the inverse of the time delay between the images.
 
The preferred approach here was to calculate a so-called double power spectrum $D_{a}$.  
As in section \ref{FPSsubs}, the power spectrum $P_{\nu}$ has to be prepared
before undergoing a Fourier transform to the ``time delay'' domain.
The low frequency part ($\nu < 1/55 \mbox{day}^{-1}$) of $P_{\nu}$ was cut off. 
This cut arises because of the large power observed at low frequencies in the power spectrum of PKS~1830-211.  

The high frequency part  of the spectrum $P_{\nu}$ was also removed because
the power at high frequency is small (it goes to 0 at the Nyquist frequency).          
The calculation of $D_{a}$ proceeds like in section \ref{FPSsubs},
except that the $P_{\nu}$ data are bent to zero by multiplication with a cosine bell.
This eliminates spurious high frequencies when zeros are added to the $P_{\nu}$ series.

\begin{figure}
  \centering
\includegraphics[width=13cm,angle=0,bb=0 0 567 358]{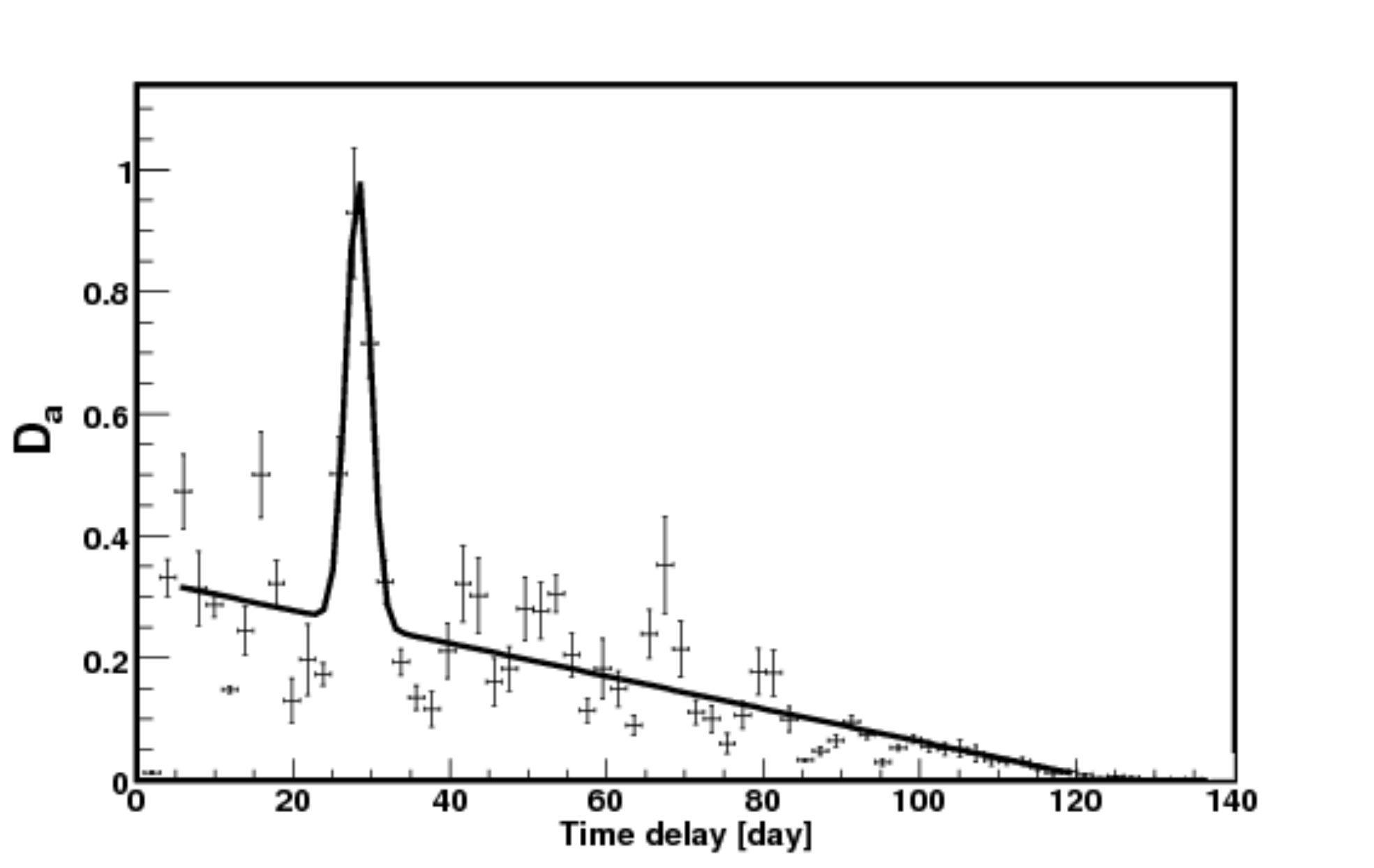} \\
\caption{Double power spectrum $D_{a}$ for the simulated lensed AGN of figure \ref{mc_fps}. $D_{a}$ is plotted in arbitrary units.
             The solid line is a fit to a linear plus Gaussian profile. }
            \label{mc_sps}
  \end{figure}

The $D_{a}$ distribution is estimated from 5 segments of the light curve.
In every bin of the  $D_{a}$ distribution, the estimated double power spectrum
is given by the average over the 5 segments.
The errors bars on $D_{a}$ are estimated from the dispersion of bin values divided by 2 (since there are 5 segments).
Due to the random nature of the sampling process, some of the error bars obtained
are much smaller than the typical dispersion in the  $D_{a}$ points.
To take this into account, a small systematic error bar was added quadratically to all points.
The result (with statistical error bars only) is presented on
figure \ref{mc_sps}.

As described in section \ref{FPSsubs}, we simulated light curves with a time delay of 28 days.
A peak is apparent near a time delay of 28 days on the $D_{a}$ distribution
shown on figure \ref{mc_sps}.
The points just outside the peak are compatible with a flat distribution.
Including also the points in the peak gives a distribution which is incompatible with a flat distribution at the 12.9 sigma level.
The parameters of the peak were determined by fitting the sum of a linear function for the background plus a Gaussian function for the signal.  
In the case shown on figure \ref{mc_sps}, the time delay estimated from $D_{a}$ is $28.35\pm0.56$ days.
 
\begin{figure}
  \centering
\includegraphics[width=13cm,angle=0,bb=0 0 567 358]{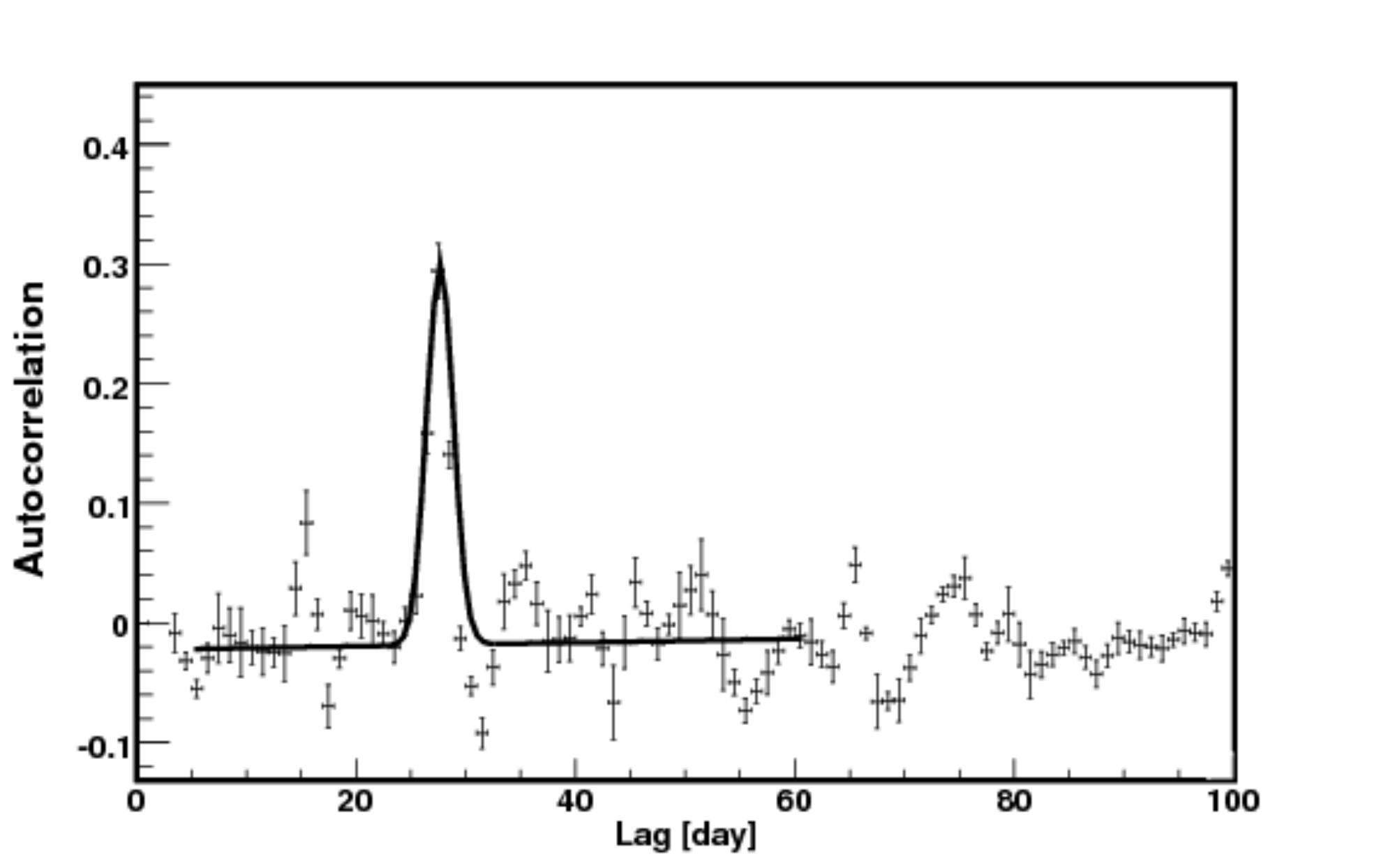} \\
\caption{Autocorrelation function of the simulated lensed AGN of figure \ref{mc_fps}.
             The solid line is a fit to a linear plus Gaussian profile.}
            \label{mc_ac}
  \end{figure}

As mentioned in section \ref{idea}, the usual approach for the time delay estimation
is to compute the autocorrelation of the light curve.
The auto-covariance is obtained by taking the real part of the inverse Fourier transform of $P_{\nu}$.
The auto-covariance is normalized (divided by the value at zero time lag) to get the autocorrelation.
The autocorrelation function of an artificial light curve simulated as in section \ref{FPSsubs} is presented on figure \ref{mc_ac}.
 
A peak with a significance of roughly 16 $\sigma$ is present at  $27.85\pm0.14$ days.
However the significance of this peak is overestimated since
light curves are simulated with white noise instead of red noise.
The autocorrelation function of a light curve driven by red noise is given by  $e^{-a/ \lambda}.$
In the case of our simulated light curves, $\lambda = 0$, so that the peak is a little affected by the background
of the AGN.
 
For the simulated light curves, both approaches of time delay determination give reasonable and compatible results.
 
\section{Results}
\label{sec:resultsGL}
 
\begin{figure}
  \centering
\includegraphics[width=13cm,angle=0,bb=0 0 567 358]{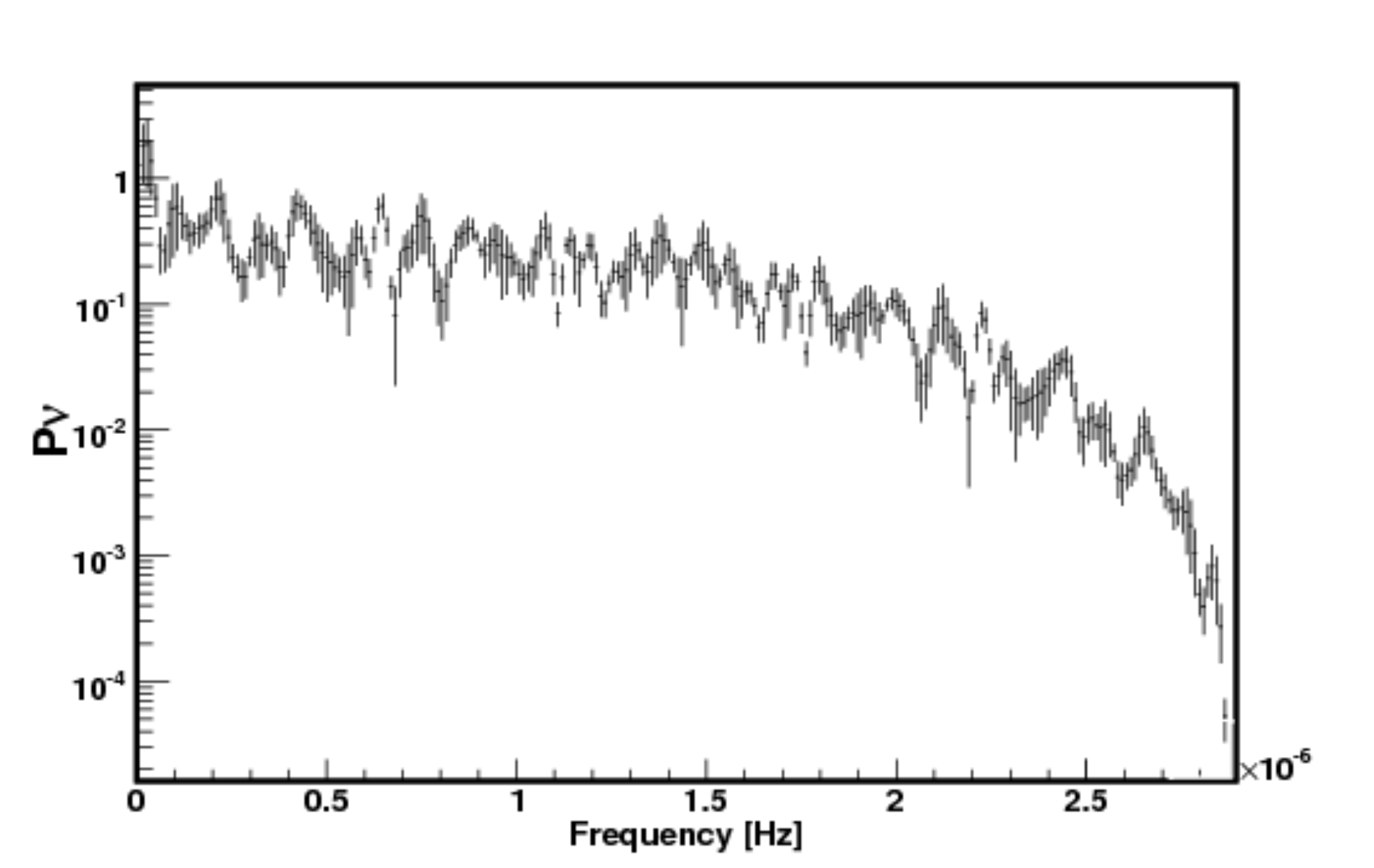} \\
\caption{ Measured power spectrum of PKS 1830-211, plotted in arbitrary units.  }
            \label{pks1830_fps}
  \end{figure}

The results for real data were obtained with the same procedure as was presented for the simulated light curves.
Figure \ref{pks1830_fps} shows the power spectrum  $P_{\nu}$ calculated from the light curve of PKS 1830-211. 
 
An oscillatory pattern is clearly visible in the power spectrum. 
It is similar to the pattern expected from the simulations shown on figure \ref{mc_fps}.
 
  \begin{figure}
  \centering
\includegraphics[width=13cm,angle=0,bb=0 0 567 358]{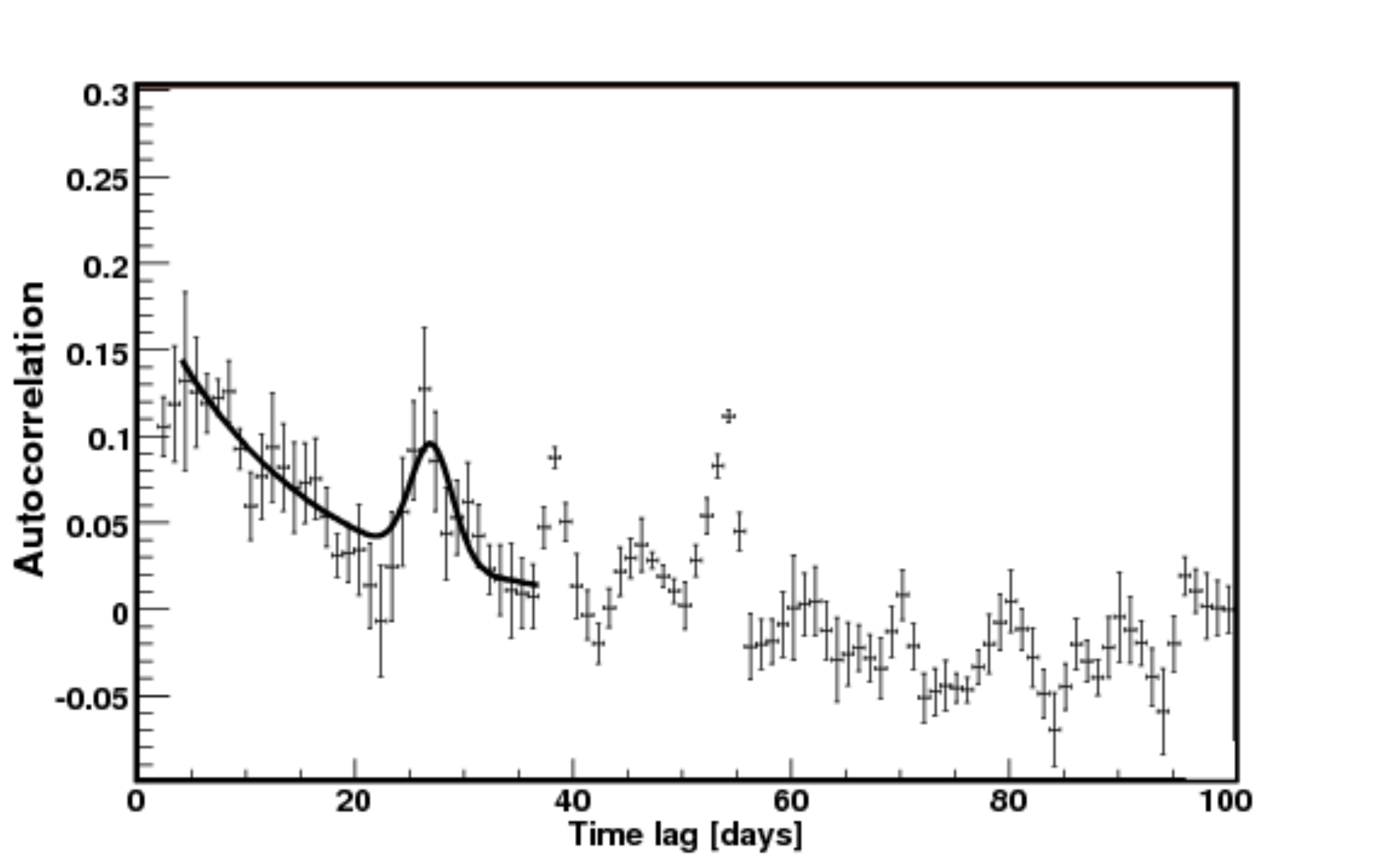} \\
\caption{Measured autocorrelation function of PKS 1830-211. The solid line is a fit to an exponential plus Gaussian profile.}
            \label{pks1830_ac}
  \end{figure}
\begin{figure}
  \centering
\includegraphics[width=13cm,angle=0,bb=0 0 567 358]{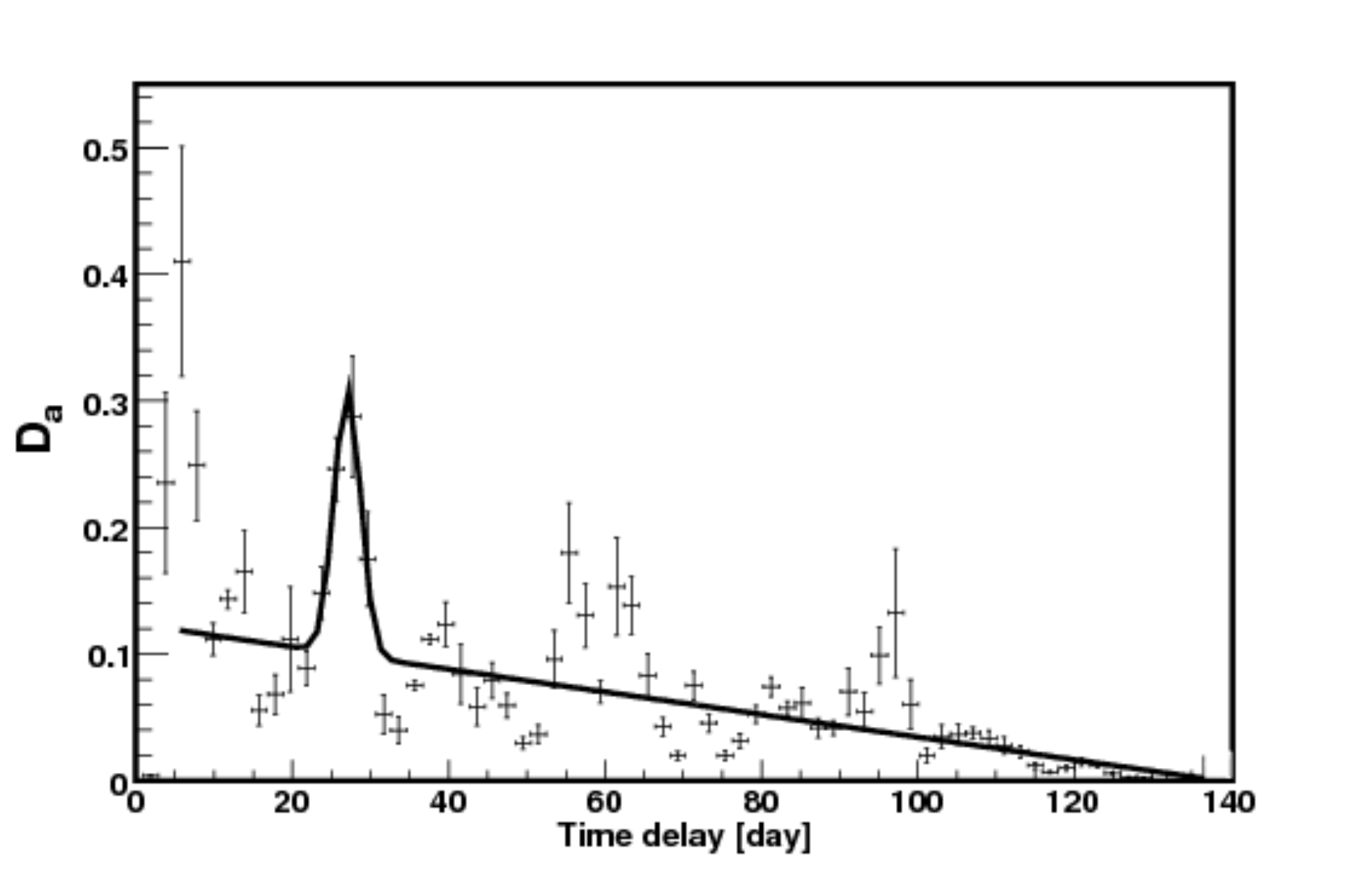} \\
\caption{Double power spectrum of PKS 1830-211 plotted in arbitrary units.
             The solid line is a fit to a linear plus Gaussian profile. }
            \label{pks1830_sps}
  \end{figure}

The autocorrelation function and the $D_{a}$ distribution calculated for real data
are shown on figures \ref{pks1830_ac} and \ref{pks1830_sps}.
 
A peak around 27 days is seen in both distributions.
Several other peaks are present on the autocorrelation function as was already noted by \cite{1996Geiger} (see their figure 1).

The peak around 5 days in the $D_{a}$ distribution is likely to be an artefact connected 
to the time variation of the exposure of the LAT instrument on PKS 1830-211.
Using the method described in section \ref{SubTDD}, the significance of the peak around 27 days is found to be 1.1 $\sigma$
in the autocorrelation function and 4.2 $\sigma$ in the double power spectrum $D_{a}$.
Fitting the position of the peak gives a time delay of $a= 27.1\pm0.6$ days for the $D_{a}$ distribution.
The fit of the autocorrelation function to a Gaussian peak over an exponential background gives $a =  27.1\pm0.45$ days.
 In both cases, the quoted error is derived from the fit. 
 
The double power spectrum distribution obtained for PKS 1830-211 provides the first evidence for gravitational lensing
phenomena in high energy gamma rays.
The evidence is still at the  4.2 $\sigma$ level but will likely improve by a factor of 2 over the lifetime of the FERMI satellite.   
Thanks to the uniform light curve sampling provided by FERMI LAT instrument,
it is not necessary to identify features on the light curve to apply Fourier transform methods.
The example of PKS 1830-211 shows that the method works in spite of the low photon statistic.
Possible extensions of the present work are finding multiple delays in complicated lens systems
or looking for unknown lensing systems in the FERMI catalog of AGNs.
 
\section{ Gravitational lens time delay and the Hubble constant}
\label{sec:Hubble}
The Hubble constant ($H_0$) estimation bases on the distance determination in the Universe. 
The most common methods of the distances determination  in the nearby Universe use the Cepheids, the tip of the red giant branch or maser galaxies. 
Larger distances are estimated using Tully-Fisher relation for spiral galaxies, the surface brightness fluctuation  method or the maximum luminosity of Type Ia supernovae (for full review see \cite{2010ARA&A..48..673F}). 
This methods give direct way of $H_0$ estimation.
However, there are also indirect techniques  to estimate $H_0$, 
for example with using the Sunyaev-Zel'dovich effect, 
the anisotropy in the cosmic microwave effect or the gravitation lensing. 

Recently, the most accurate $H_0$ estimation is $70.4^{+1.3}_{-1.4}$(km/s)/Mpc.
This result was provided with the seven years Wilkinson Microwave Anisotropy Probe (WMAP) observations combined with  BAO and $H_0$ data \citep{2011ApJS..192...18K}. 
The $H_0$ estimated with alternative and independent methods gives consistent value.
For example, in the case of gravitational lensing method, the best estimation of $H_0$ was obtained 
with a Bayesian analysis of the strong gravitational lens system B1608+656.
This gave $H_0=70.6^{+3.1}_{-3.1}\,$(km/s)/Mpc  \citep{2010ApJ...711..201S}.

The $H_0$ estimation, based on the lensed quasar  PKS 1830-211, 
has provided rather weak estimation as yet. 
The value given by \cite{2002Winn} was based on previous time delay measurement and
assumption that the lens galaxy has an isothermal 
mass distribution, they estimate $H_0=44 \pm 9$ (km/s)/Mpc.

The gravitational lens PKS 1830-211 observed in radio band, 
consist of two bright components 
separated by 1~arcsec and connected by ring-like structure. 
The mass of lensed galaxy can be estimated using
 the Einstein ring radius $\theta_E$ from equation~(\ref{eq:thetaE}):
 
 \begin{equation}
 M_L=\frac{\theta^2_Ec^2}{4G}\frac{D_{OL}D_{OS}}{D_{LS}} \,.
 \label{eq:mass}
 \end{equation}

\begin{figure}
  \centering
\includegraphics[width=13cm,angle=0,bb=0 0 1024 768]{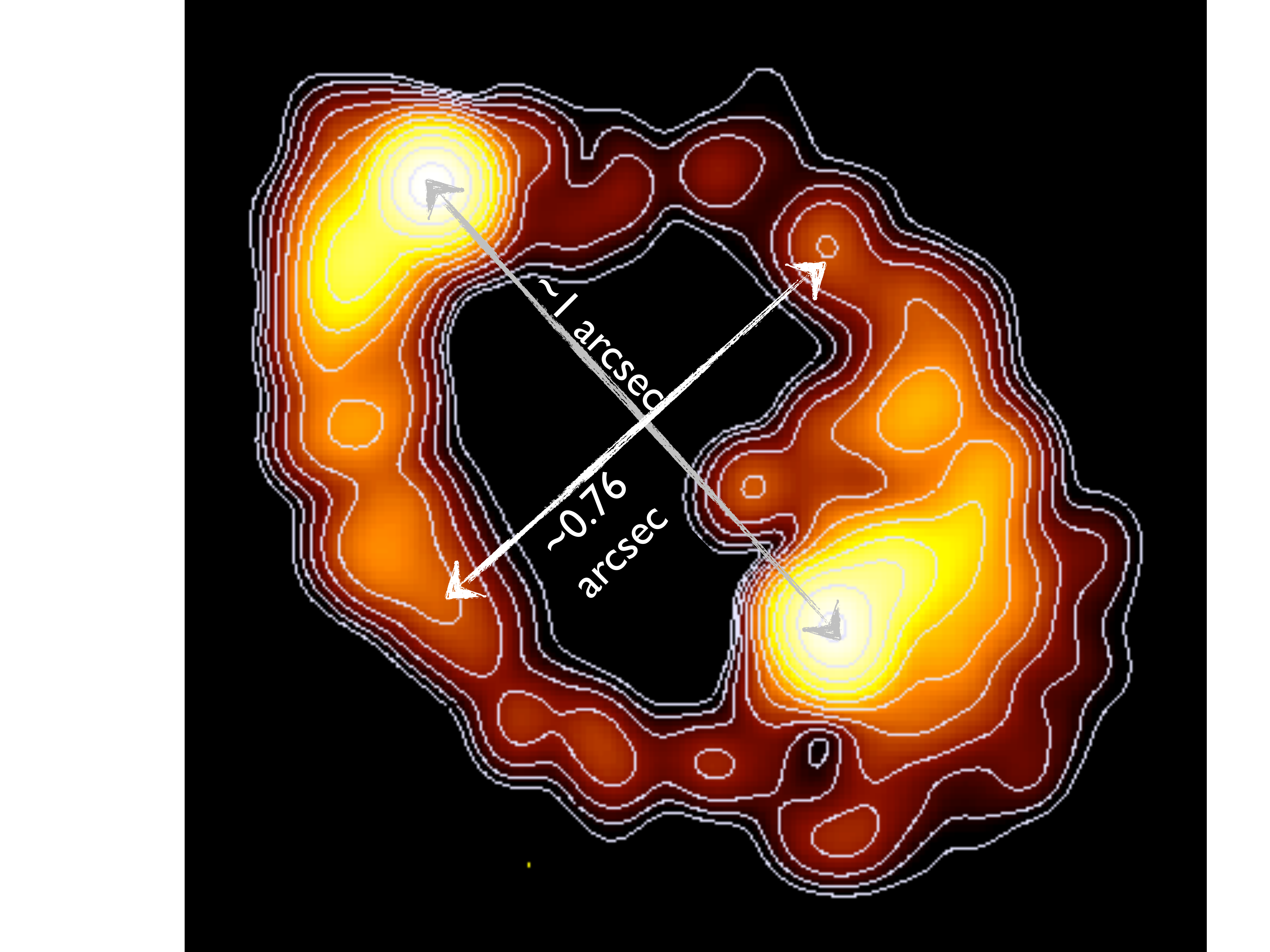} \\
\caption{The extragalactic gravitational lens PKS 1830-211 with indicated "Einstein diameters" (Image credit: ATCA). }
            \label{fig:pks1830mass}
  \end{figure}

 I will use the $\theta_E$ measured from radio ring like structure. 
The lensed image structure has rather elliptical shape. 
 Therefore, to have more realistic mass estimation in this simplify  case, 
 the Einstein radius is measured for two different distances in the elliptical structure 
 (see figure \ref{fig:pks1830mass}).
This approach  gives two values of $\theta_E$: 0.5 arcsec and 0.38 arcsec. 
The mass estimated using this values is $6.46\times10^{10}{\rm h^{-1} M_{\odot}}$
and  $3.69\times10^{10}{\rm h^{-1} M_{\odot}}$, respectively. 
The distances were calculated using FLRW formalism (see section \ref{sec:DistanceFormula}),
 $\Omega_M=0.3$, $\Omega_\Lambda=0.7$ and $H_0=100{\rm h (km/s)/Mpc}$.
For simplicity, I assumed a point mass lens. 
Using the time delay between two images separated by 2$\theta_E$ given by \cite{1991PhDT........77L} and mass of the lens expressed in units of $h^{-1}$ one can get:

\begin{equation}
h=\frac{1+z_L}{\delta t}\frac{4GM}{c^3}\left[\frac{\theta_S \sqrt{\theta_S^2+4\theta_E^2}}{2\theta_E^2}  
+ \ln{\frac{\theta_S+\sqrt{\theta_S^2+\theta_E^2}}{|\theta_S-\sqrt{\theta_S^2+\theta_E^2}|}} \right] \,.
\end{equation}

where $\theta_S$ is the angle between the position of the lensed galaxy and the true (undeflected) position of the background source. 
Usually, $\theta_S$ is obtain from detail lens modeling. 
In my analysis I used magnification $\mu=1.52\pm 0.05$ ratio provided by \cite{1998Lovell}.
The magnification factor $\mu_{\pm}$ for a point mass lens is:

\begin{equation}
\mu_{\pm} = \left[ 1 - \left(\frac{\theta_E}{\theta_{\pm}} \right)^4\right]^{-1} = \frac{u^2+2}{2u\sqrt{u^2+4}}\pm \frac{1}{2} \,,
\end{equation} 

where $u$ is the angular separation of the source from the point mass lens 
$\theta_S = r_S/D_{OL}$ in units of the Einstein angle.  
The magnification ratio of the two images approximately is: 

\begin{equation}
\frac{\mu_+}{\mu_-} \approx \frac{u+1}{u-1} \,.
\label{eq:mu}
\end{equation} 

Using equation (\ref{eq:mu}) and the magnification ratio  $\mu=1.52\pm 0.05$ one can get $\theta_S=0.2\theta_E$. 
The Hubble constant calculated using lens mass and $r_S$ obtained above is 
$H_0=61.1\pm13$ using new time delay and Einstein radius of 0.5 arcsecond, and 
$H_0=35.6\pm 0.5$ for 0.38 arcsec Einstein radius.
The most probable value is  then $H_0=48.7\pm 13.7$.

The strong influence on the accuracy of $H_0$  have the systematics errors. 
Basically, the point mass lens model assumption and value of $\theta_{S}$. 
The further improvement require reduction of systematics errors by 
more realistic lens modelling, measurement of the magnification ratio using FERMI data 
and improvement in the double power spectrum  method. 
The time delay estimation method  can be improved when the non-stationary processes like flares 
or microlensing effect will be taken into account. 

\setcounter{chapter}{5}
\setcounter{section}{0}
\setcounter{equation}{0}
\setcounter{figure}{0}
\part{Femtolensing of GRBs \label{chap:femto}}  
\section{Introduction}
%
%
Dark matter is one of the most challenging open problems in
cosmology or particle physics. 
A number of candidates for particle dark matter has been proposed over the years~\citep{feng10}.

An alternative idea that the missing matter consists of compact
astrophysical objects was first proposed in the
1970s~\citep{1974MNRAS.168..399C,1974Natur.248...30H,1971MNRAS.152...75H}.
An example of such compact objects are primordial black holes (PBHs)
created in the very early Universe from matter density perturbations.
PBHs would form during the radiation-dominated era, 
and  would be non-baryonic.
That  satisfy the big bang nucelosynthesis limits on baryons, 
and PBHs would be thus  classified as cold dark matter in agreement with the current paradigm 
\citep{2012arXiv1210.7729C}.
Another famous example is "brown dwarfs", 
excluded by the EROS, MACHO and OGLE searches. 

The abundance of PBH above $10^{15}\,$g is a probe of gravitational collapse and large scale structure theory \citep{2005astro.ph.11743C}. 
In particular, it  constrains 
the gravitational wave background produced from primordial scalar perturbations
in the radiation era of the early Universe  \citep{PhysRevD.83.083521}.

Recent advances in experimental astrophysics, especially the
launch of the FERMI satellite with its unprecedented sensitivity,
has revived the interest in PBH
physics~\citep{2010PhRvD..81j4019C,2011PhRvL.107w1101G}. 

In this part, I present the results of a femtolensing search performed on 
the spectra of GRBs with known redshifts detected by the Gamma-ray Burst Monitor
(GBM) on board the FERMI satellite \citep{2012PhRvD..86d3001B}.  
The non observation of femtolensing on these bursts provides new constraints
on the PBHs fraction in the mass range $10^{17} -
10^{20}\,$g.  I describe in details the optical depth
derivation based on simulations applied to each burst
individually.  The sensitivity of the GBM to the femtolensing
detection is also calculated.

This chapter is organized as follows: 
The first section introduces Primordial Black Holes (PBH).
In section ~\ref{sec:estimate} the basics of femtolensing are given. 
Section~\ref{sec:data} describes the data sample and simulations.  
In section~\ref{sec:results} the results are presented, while
section~\ref{sec:conclusions} is devoted to discussion and conclusions.

\section{Primordial Black Holes} 

As I mentioned in the introduction section, PBHs could have formed in the early Universe. 
PBHs are not associated with the collapse of a massive star,
so that they could have formed with a wide range of masses.
In this scenario,  the mass of PBH would depend on their formation time $t$:

\begin{equation} 
M_H(t)\approx \frac{c^3t}{G}\approx 10^{15} \left( \frac{t}{10^{-23}\,s}\right) \, \mbox{g}\,.
\end{equation}

PBH formed at the Planck time ($10^{-43}\,s$) just after the Big Bang would have a mass equal to the Planck mass ($10^{-5}\,$g). 
However, those formed 1 s after the Big Bang would have a mass of $10^5M_{\odot}$. 

\cite{1973PhRvD...7.2333B} and \cite{1974Natur.248...30H} discovered 
that black holes  produce  a thermal radiation with a temperature: 

\begin{equation}
T=\frac{\hbar c^3}{8\pi GMk} \approx 10^{-7} \left(\frac{M}{M_{\odot}} \right)^{-1} \, \mbox{K} \,.
\end{equation}

The radiation has a black body spectrum and is inversely proportional to black hole mass. 
Black holes of mass M should evaporate on a timescale: 

\begin{equation}
\tau (M) \approx \frac{\hbar c^4}{G^{2}M^3} \approx 10^{64} \left(\frac{M}{M_{\odot}} \right)^{3} \, \mbox{years}\,.
\label{eq:evaporation}
\end{equation}

Equation (\ref{eq:evaporation}) suggests that PBHs with a  mass smaller than $10^{15}\,$g would have evaporated thus far. 
The $10^{15}\,$g PBHs have a $\sim$ 100 MeV  temperature at the present epoch.
Observations of the  $\gamma$-ray background  constrains  the density of PBH with masses of less than  $10^{15}\,$g. 
A recent analysis of EGRET data  \citep{2009A&A...502...37L} shows that the   PBH density does not exceed $10^{-8}$ times the critical density.

\section{Femtolensing\label{sec:estimate}}
 One of the
most promising ways to search for PBHs is to look for lensing
effects caused by these compact objects.  
Since the Schwarzschild radius of PBH is comparable to the photon wavelength, 
the wave nature of electromagnetic radiation has to be taken into account.  
In such a case, the lensing caused by PBHs introduces an interferometry pattern in
the energy spectrum of the lensed object~\citep{1981SvAL....7..213M}.
This effect is called 'femtolensing'~\citep{1992ApJ...386L...5G} due to
 the  $\sim 10^{-15}$ arcseconds angular distance between the images of a source lensed by a $10^{18}\,$g lens.
The phenomenon
has been a matter of extensive studies in the
past~\citep{2001ASPC..239....3G},
but the research was
almost entirely theoretical since no case of femtolensing has been
detected as yet.  \cite{1992ApJ...386L...5G} first suggested
that the femtolensing of gamma-ray bursts (GRBs) at cosmological distances
could be used to search for dark matter objects in the 
$10^{17} - 10^{20}\,$g mass range.
Femtolensing could also be a signature of another
dark matter candidate: clustered axions \citep{1996ApJ...460L..25K}.

\subsection{Magnification and spectral pattern}
%
Consider the lensing  of a GRB event by a compact object.  
The magnification of a point like source has been introduced in section~\ref{sec:Magnification}. 
Equation~(\ref{eq:Mag}) indicates that the magnification depends on the phase.
When the two lights paths are not temporally coherent equation~(\ref{eq:Mag})  is reduced to the  two first components.

\begin{landscape}
  \begin{figure}
 \includegraphics[width=23cm,angle=0,bb=0 0 1062 563]{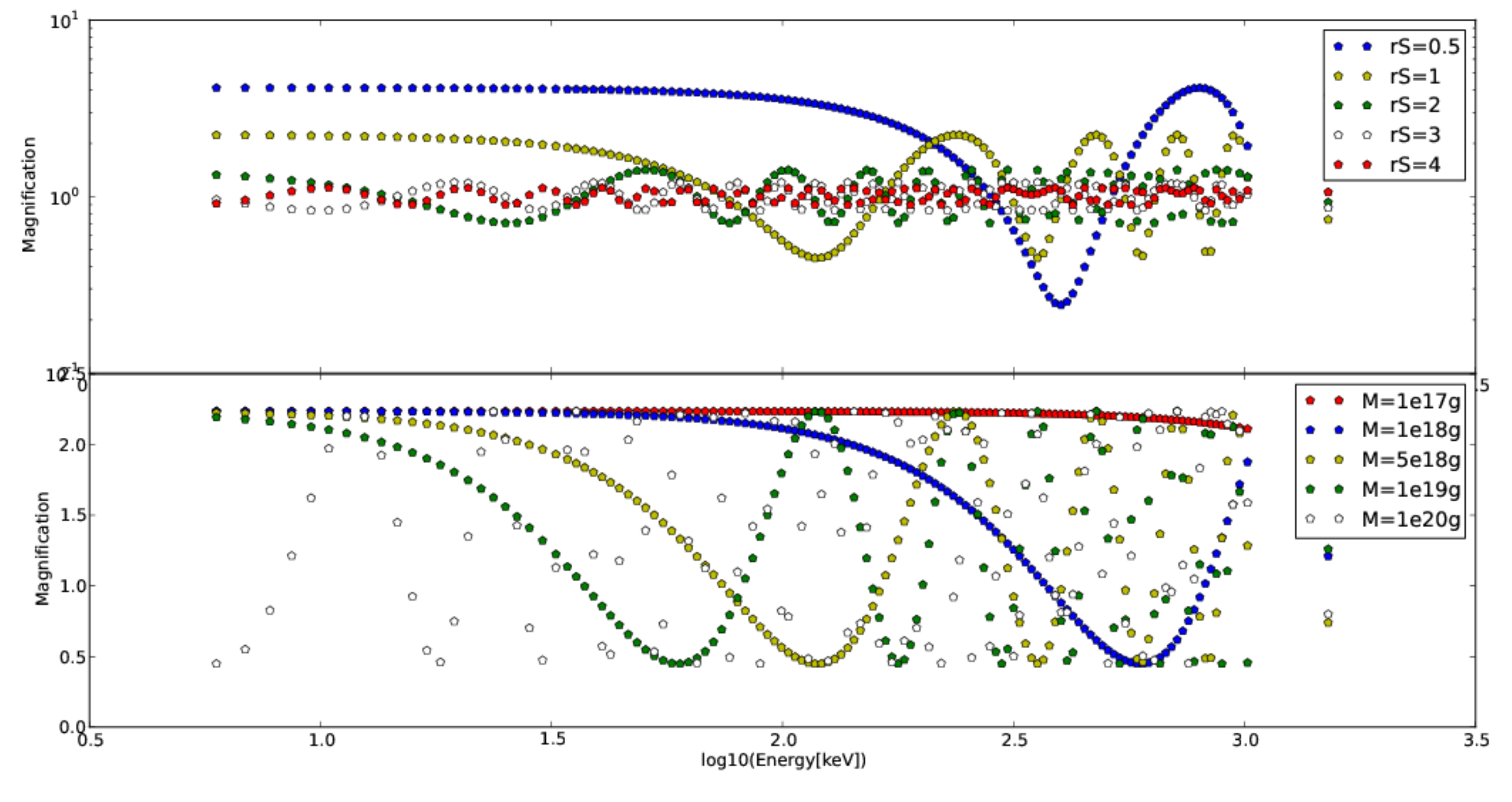}%
 \caption{\label{fig:magnification} Magnification pattern for different lens masses $M$ and projected distances of the source to the lens  ($r_S$) in the lens plane.  The points have been calculated at the values of the energy channels of the GBM detector. }
 \end{figure}
 \end{landscape}

In the case of femtolensing, the phase shift between the two images is:
\begin{equation} 
\Delta \phi~=~\frac{E~\delta t}{\hbar}, 
\end{equation}
where $E$ is the energy of the photon.
The energy dependent magnification produces fringes in the energy spectrum of the lensed object.   
The magnification pattern for different configurations  is presented on figure 
\ref{fig:magnification} as a function of the photon energy. 

\subsection{Lensing probability\label{sec:probability}}
%

The lensing probability of gamma ray burst events is calculated in two steps. 
First, the optical depth $\tau$ for lensing by compact objects is calculated 
according to the formalism described in section \ref{sec:LensingProbability}. 
The cosmological parameters used in the calculation are:
a mean mass density $\Omega_{M}=0.3$ and a normalized cosmological constant $\Omega_{\Lambda}=0.7.$ 
The calculations are made for both the FLRW 
and the  \cite{1973ApJ...180L..31D} cosmology. 
In the sample, the GRB redshift $z_{s}$ is known. The lens redshift $z_{L}$ is 
assumed to be given by the maximum of the $d\tau/dz_{L} (z_{S})$ distribution
(see figure \ref{fig:dtauPM}). 
When $\tau \ll 1,$ the lensing probability $p$ is given by 
$p = \tau \sigma$ where $\sigma$ is the ``lensing cross-section'' (see Chap. 11 of \cite{1992grle.book.....S}). 
    
In this analysis, the cross-section is defined in the following way. Fringes are 
searched in the spectra of GRBs. These fringes are detectable only for certain
positions $r_{S}$ of the source. The exact criteria for detectability will be 
given in section \ref{sec:simulations}. The maximum and minimum position  of $r_{S},$ in units 
of $r_{E}$ are noted $r_{S,min}$ and $r_{S,max}.$ They are found by simulation 
and depend on the GRB redshift and luminosity.
 A minimum value of $r_S$ occurs because the period of the spectral fringes becomes larger than the GBM energy range at small $r_{S}$. 

The femtolensing ``cross-section'' is simply 
\begin{equation}
\sigma = r^{2}_{S,max}-r^{2}_{S,min} \,.
 \label{sigma}
\end{equation}

The lensing probability does not depend on the individual masses of lenses, but only on the density of compact objects $\Omega_{CO}$. 
In the optical depth calculation, an increase in the mass of the lenses is compensated by a decrease in the number of scatterers.
Therefore, the constraints for a given mass depend only on the cross section $\sigma$. 

\section{FERMI Gamma-ray Bursts Monitor GBM} 
    \begin{figure}[ht!]
    \center
 \includegraphics[width=15cm,angle=0,bb=0 0 1037 491]{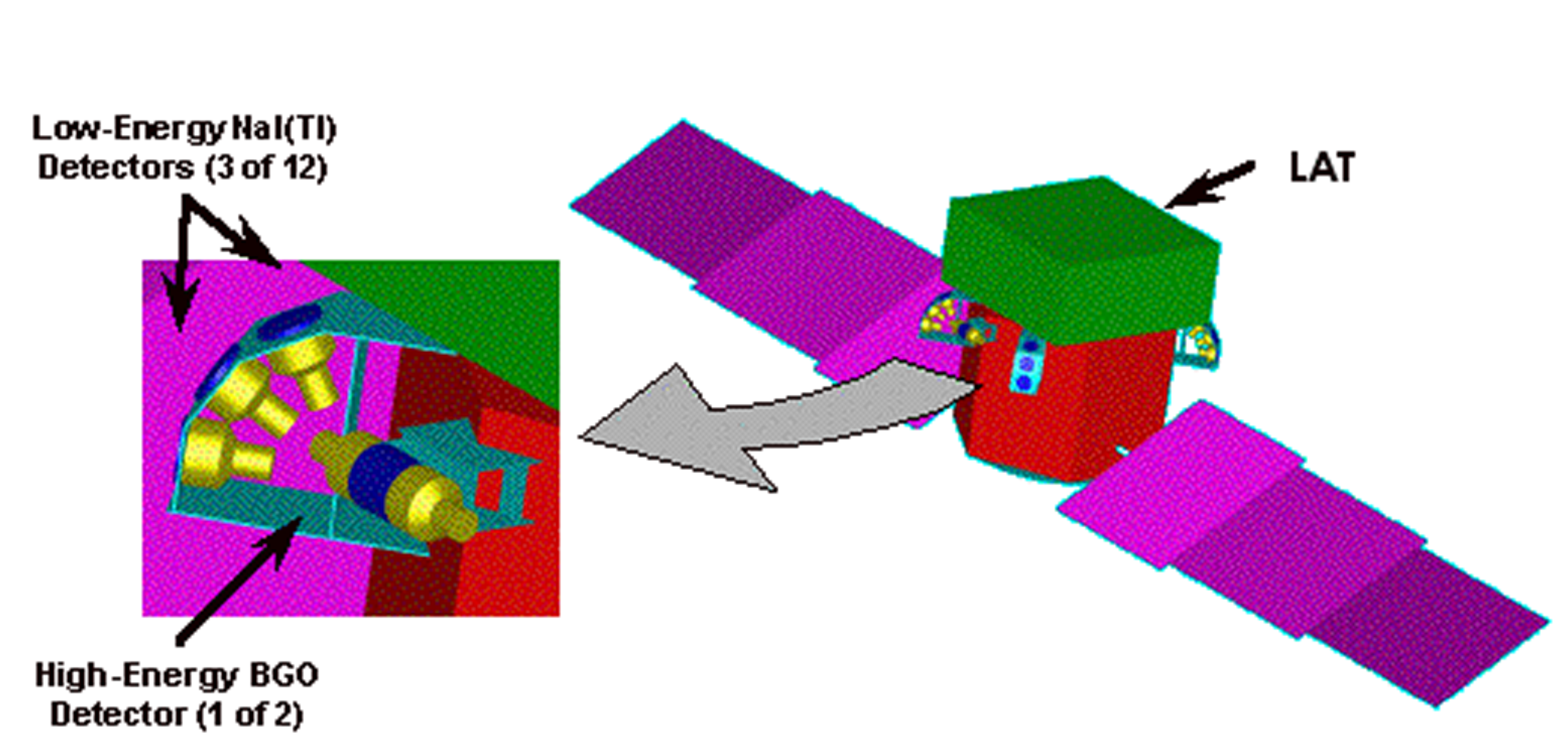}%
 \caption{\label{fig:GBMDetector}  The GBM detector on-board of FERMI sattelite. Image credited from http://gammaray.msfc.nasa.gov/gbm/instrument/description/}
 \end{figure}
 
The Gamma-Ray Burst (GBM) detector \citep{2009ApJ...702..791M} on-board the Fermi satellite 
consists of 12 NaI and 2 BGO scintillators 
which cover the energy range from 8 keV up to 40 MeV in 128 energy bins. 
The energy resolution at 100 keV reaches $\sim$15\% and $\sim$10\% at 1~MeV.
These detectors monitor the entire sky. 
The GBM can locate a burst  with an accuracy of $<$ 15 deg.
Currently, the average burst trigger rate is $\sim$ 260 bursts per year.
In the first two years of operation, the GBM triggered on roughly 500 GRBs.

\section{Data analysis\label{sec:data}} 
In our analysis, we use a sample of GRBs with known redshifts. 
The selection of these bursts is described in section~\ref{sec:selection}.
Each burst is then fitted to a standard  spectral model, as explained in section~\ref{sec:analysis}.
Finally, the sensitivity of each burst to femtolensing is studied with simulated data.
The simulation is described in section~\ref{sec:simulations}.

\subsection{Data selection\label{sec:selection}}
%
In this analysis, only the bursts with known redshifts  have been investigated. 
The initial  sample consisted of 32 bursts taken from 
\cite{2011A&A...531A..20G} 
and 5 additional bursts from the GRB Coordinates Network (GCN) circulars\footnote{http://gcn.gsfc.nasa.gov}. 
For 17 bursts, the amount  of available data was not sufficient to obtain good quality spectra. 
The final sample thus  consists of 20 bursts, which are listed in table~\ref{tab:table}.
 %
\subsection{Data processing and spectral analysis\label{sec:analysis}}
\begin{landscape}
    \begin{figure}[ht!]
    \center
 \includegraphics[width=22cm,angle=0,bb=0 0 1245 309]{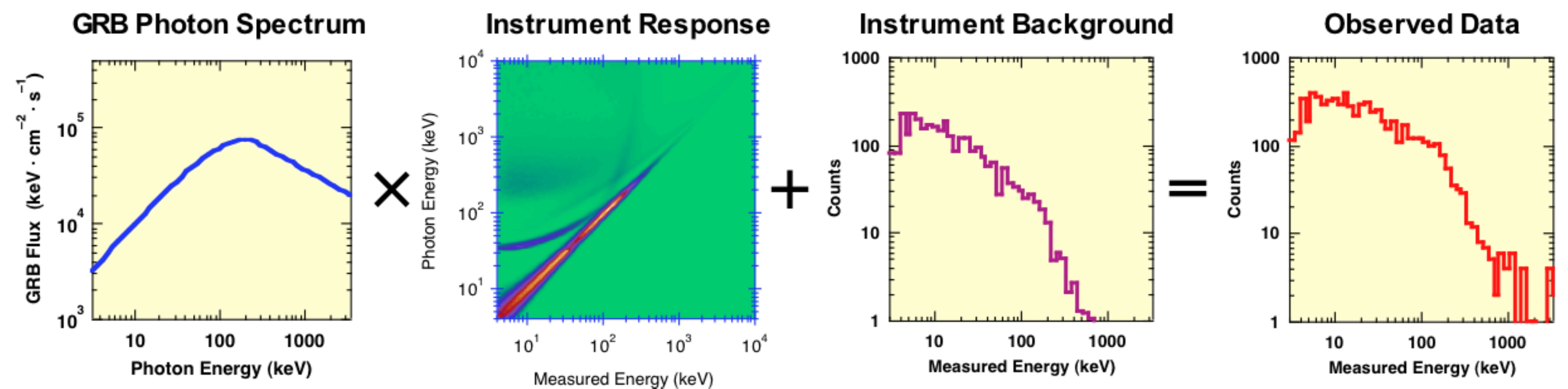}%
 \caption{\label{fig:ObdrvedData}  Decomposition of  observed GBM data. Figure credited from R.M. Kippen (LANL).}
 \end{figure}
\end{landscape}

The GBM data are publicly available in the CSPEC format and were downloaded from the FERMI FSSC website \footnote{http://fermi.gsfc.nasa.gov/ssc/}.
The CSPEC files contain the counts in 128 energy channels with 1.024 s bins for all detectors. 
Only detectors with a minimal signal to noise ratio of 5.5 in each bin were selected for the analysis.   
Data were analyzed with the {\tt RMfit}  version 33pr7 program.   
The {\tt RMfit} software package was originally developed for the time-resolved 
analysis of BATSE GRB data but has been adapted to GBM and other instruments.
The observed data is a convolution of the GRB photon spectrum with an instrument response function.
This is illustrated on figure \ref{fig:ObdrvedData}.
For each detector with sufficient data, the background was subtracted and the 
counts spectrum of the first ten seconds of the burst (or less if the burst was shorter) was extracted. 

The energy spectrum was obtained with a standard forward-folding algorithm.
Several  GRB spectral models such as  a broken power law (BKN),  Band's model (BAND) or a smoothly broken power law (SBKN) where considered. 
The femtolensing effect was added as a separate model. 
The magnification and the oscillating fringes were calculated according to equation~(\ref{eq:Mag}), 
then multiplied with the BKN or BAND functions.
\subsection{Simulations\label{sec:simulations}}

 The detectability of spectral fringes has been studied with  simulated signals. 
 The detectability first depends on the luminosity and  redshift of the bursts, 
and second on the detector's energy resolution and data quality. 
The sensitivity of the GBM to the lens mass $M$ depends strongly also on the energy range and
resolution of the GBM detectors. When small masses are considered, 
the pattern of spectral fringes appears outside of the energy range. 
The large masses produce fringes with hardly detectable amplitudes and periods smaller than the 
energy bin size. 

Because the data quality and the background are not easily simulated, 
the detectability estimation is performed on real data.
Namely, GRB events with known redshifts are selected. 
Since the source redshift is known,
the lens redshift is assumed to be the maximum value of $d\tau/dz_{L} (z_{S})$  as explained 
in section~\ref{sec:LPPM}. 
For a given observed GRB,  the femtolensing signal depends thus only on 2 parameters: 
the lens mass $M$ and the source position in the lens plane $r_{S}.$ 
The data are then processed  as follows: 
\begin{enumerate}
\item The magnification (equation~\ref{eq:Mag}) as a function of the energy is calculated for the given
lens mass $M$ and position of the source $r_{S}.$ 
\item This magnification is then convolved  with the instrumental resolution matrix to obtain magnification factors for each channel of the detector.  
\item The spectral signal is extracted from the data by subtracting the background. It is then
multiplied by the corrected magnification. 
\item The background is added back.
\end{enumerate}

I now illustrate the detectability calculation  with the luminous  burst GRB090424592. 
The spectral data of this burst were first fitted with standard spectral models: BKN, SBKN and BAND.  
The GRB090424952 burst is best fitted with the BAND model. 
The BAND model has 4 free parameters: the amplitude A, 
the low energy spectral index $\alpha$, the high energy spectral index $\beta$ and the peak energy $E_{peak}$ \citep{2012arXiv1201.2981G}. The fit has $\chi^{2} = 78$ for 67 degrees of freedom (d.o.f). 

\begin{figure}[ht!]
      \centering
 \includegraphics[width=9cm,angle=-90,bb=0 0 504 720]{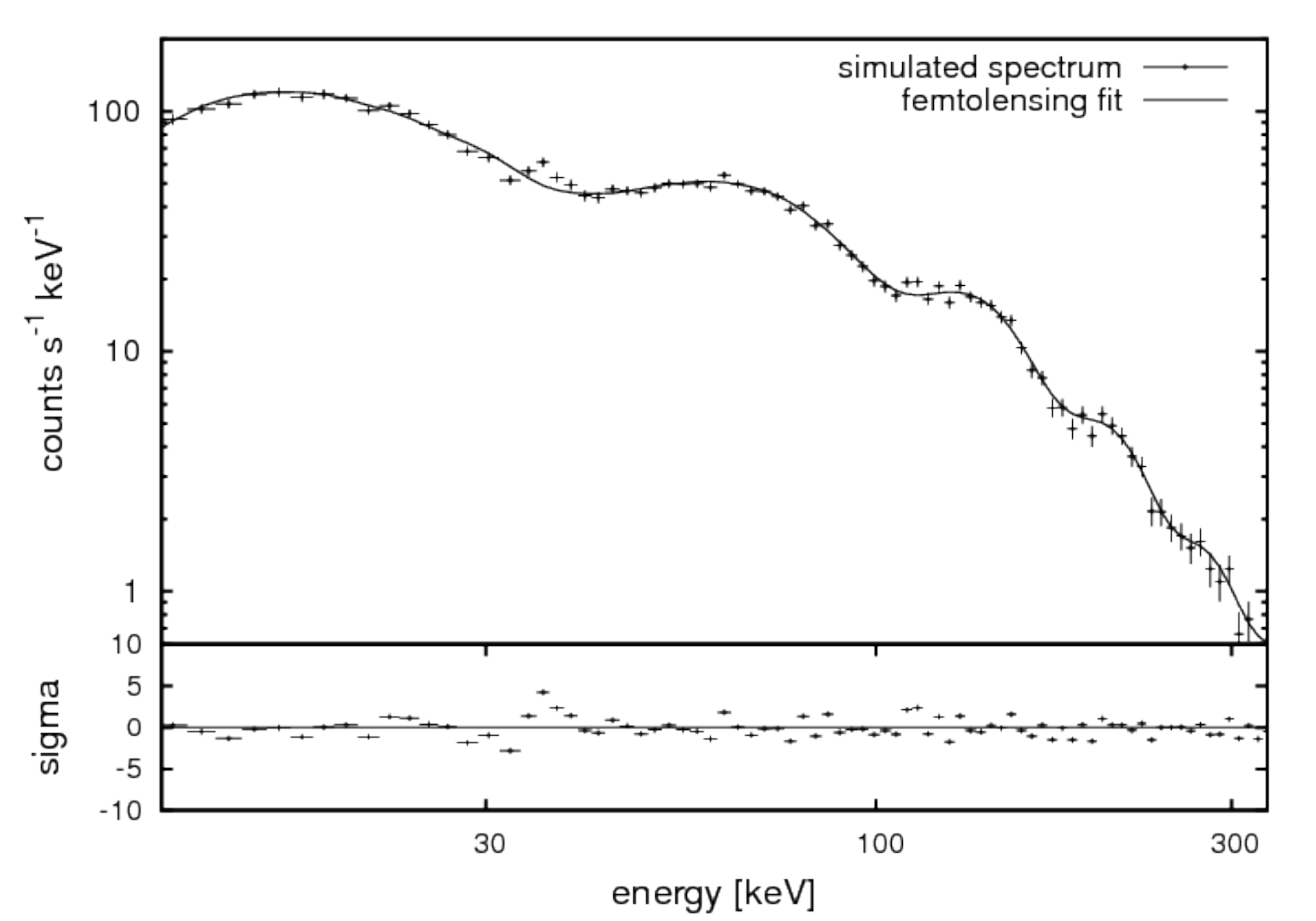}%
 \caption{\label{MD_femto} Simulated spectrum obtained with GRB 090424592.  
The spectrum  was fitted with femtolensing$+$BAND model.
The fit has $\chi^2$ = 79 for 73 d.o.f. 
The fit parameters are: $A=0.32\pm0.01\,$ph$\,$s$^{-1}\,$cm$^{-2}\,$keV$^{-1}$, $E_{peak}=179\pm3\,$keV, 
$\alpha=-0.87\pm0.02$ and $\beta=-3.9\pm7.5$. 
The simulated femtolensing effect is caused by a lens at redshift $z_L=0.256$ acting on a source at $z_S=0.544$.
The simulated mass is $M = 1 \times 10^{18} \,$ g and the mass reconstructed from the fit is $1.01 \times 10^{18}\,$g.
 The source is  simulated at position $r_S=2$. The position reconstructed from the fit is $r_S =1.9$.}  
 \end{figure}

The data are then modified by incorporating the spectral 
fringe patterns for a range of lens masses $M$ and source positions $r_{S}.$  
The simulated data and the corresponding femtolensing fit are presented in figure~\ref{MD_femto}.

    \begin{figure}[ht!]
       \centering
 \includegraphics[width=9cm,angle=-90,bb=0 0 504 720]{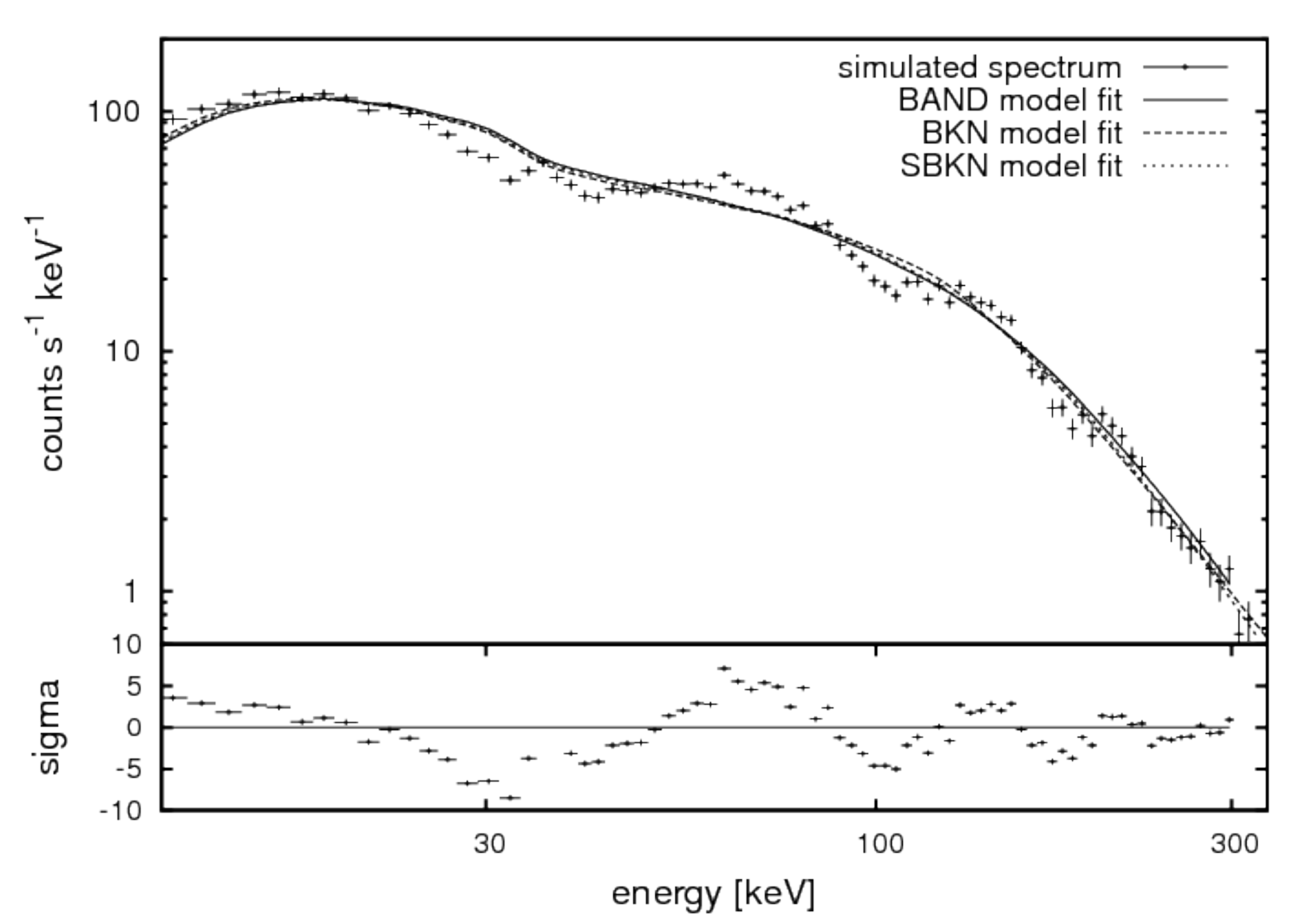}%
 \caption{\label{MD_BAND}  Simulated femtolensed spectrum fitted with the BAND model. 
 The fit has $\chi^2 = 752$ for $75$ d.o.f. 
 The fit parameters are: 
 $A= 0.36\pm0.01\,$ph$\,$s$^{-1}\,$cm$^{-2}\,$keV$^{-1}$,  $E_{peak}=174\pm5\,$keV,
 $\alpha=-0.8\pm0.02$ and $\beta=-2.4\pm0.1$.
 The SBKN model fit is almost indistinguishable from the BKN model fit.}
 \end{figure}
\begin{figure}[ht!]
   \centering
   \includegraphics[width=9cm,angle=-90,bb=0 0 504 720]{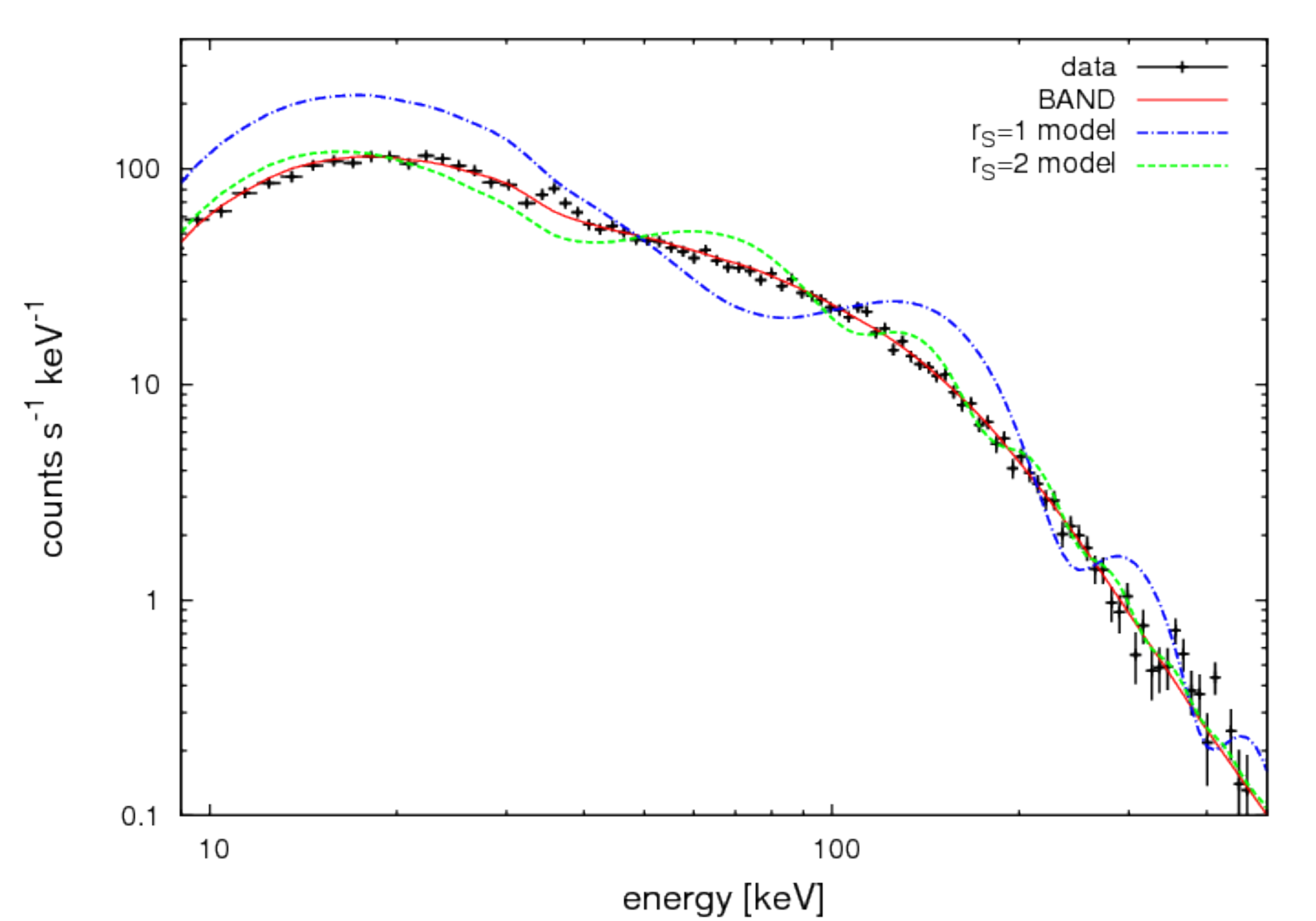}%
   \caption{\label{rS12} The spectrum of GRB 090424592 using NaI
     detector n7, with the BAND and femtolensing fits superimposed.
      The parameters are $r_{S} = 1$, $2$, and lens mass $1\times10^{18} \,$g.
     The models are convolved with the response matrix. }
 \end{figure}
\begin{figure}[ht!]
    \centering
   \includegraphics[width=9cm,angle=-90,bb=0 0 504 720]{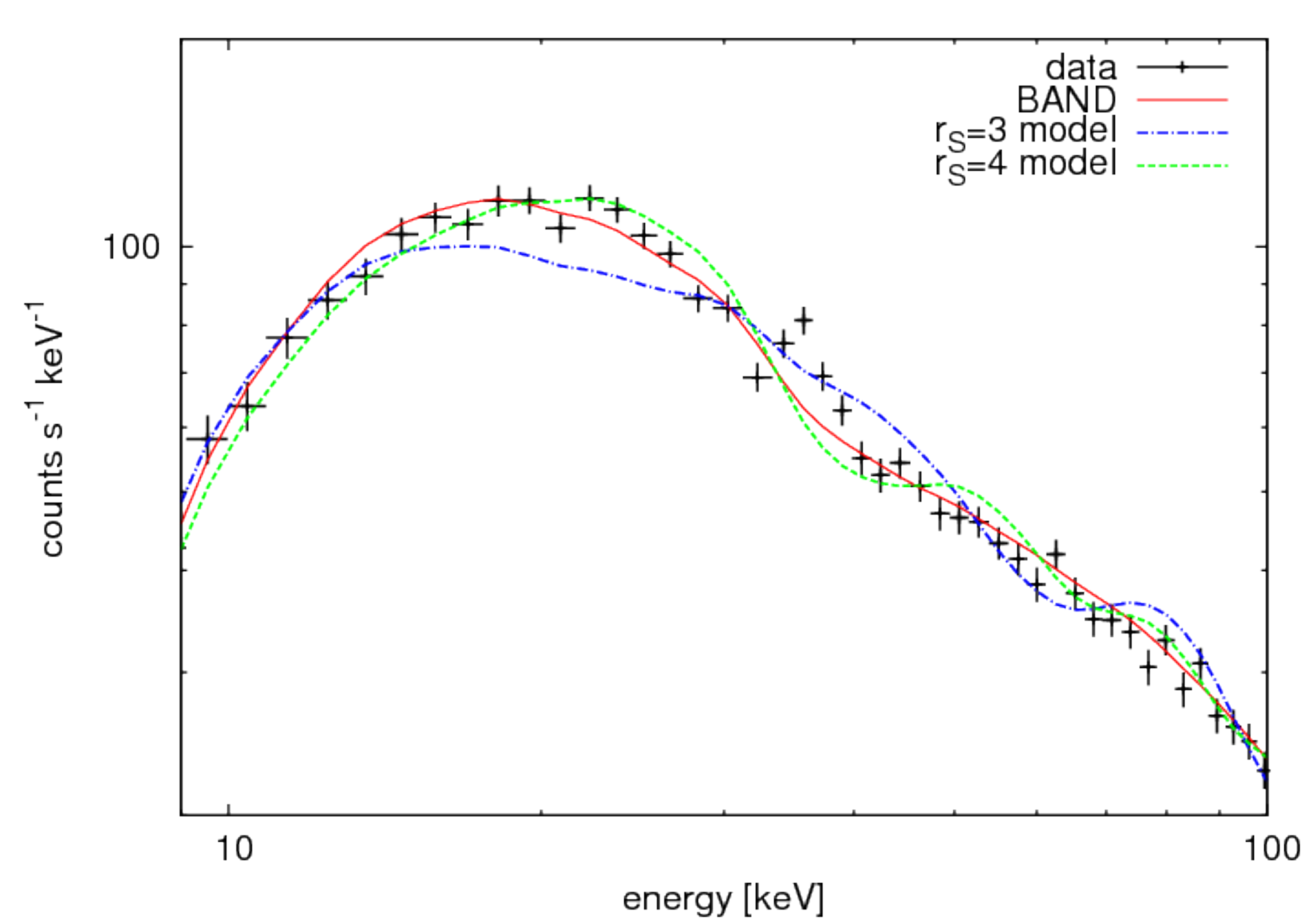}%
   \caption{\label{rS34} The spectrum of GRB 090424592 using NaI
     detector n7. The BAND and femtolensing fits are superimposed. 
     The parameters are $r_{S} = 3$, $4$, and lens mass $1\times 10^{18} \,$g.
     The excess at $33\,$keV (K-edge) is an instrumental effect
     seen on many bright bursts. }
 \end{figure}

Neither  BKN nor BAND models are able to fit the simulated data (see figure~\ref{MD_BAND}). 
The values of $r_{S}$ are then changed until the $\chi^2$ of the fit obtained is not
significantly different from the $\chi^2$ of the unmodified data.  
More precisely, the $\chi^2$ difference $\Delta \chi^2$ should be distributed
in the large sample limit as a $\chi^2$ distribution with 2 degrees of freedom
according to  Wilk's theorem \citep{1996ApJ...461..396M}. 
The value $\Delta \chi^2 = 5.99,$ which corresponds to a $\chi^2$ probability of 5\% for 2 d.o.f, 
was taken as the cut value. 
The effect of changing $r_{S}$  on the femtolensing model is illustrated on figures~\ref{rS12} and \ref{rS34}. 
 
The pattern in energy is visible when  the phase shift between the two images 
$\Delta \phi~\sim~\left(E/1MeV\right)\left(M/1.5\times 10^{17}\,g\right)$ is close to 1.

The GBM detector can detect photons with energy from few keV to $\sim$ MeV. 
Lens masses from $10^{17}\,$g to $10^{20}\,$g are thus detectable with GBM.
The femtolensing pattern can be  detected when the period of the fringes  is larger than the detector energy resolution
and smaller than the detector energy range. 
The value of $r_{S,max}$ comes from the comparison of the period of the oscillating pattern to the detector energy resolution. 
The value of $r_{S,min}$ arises from the comparison of the period of the fringes  to the detector energy range.
Because of these constraints, the most sensitive mass range is $10^{18} \,$g to  $10^{19} \,$g. 

    \begin{figure}[ht!]
    \center
 \includegraphics[width=13cm,angle=0,bb=0 0 595 842]{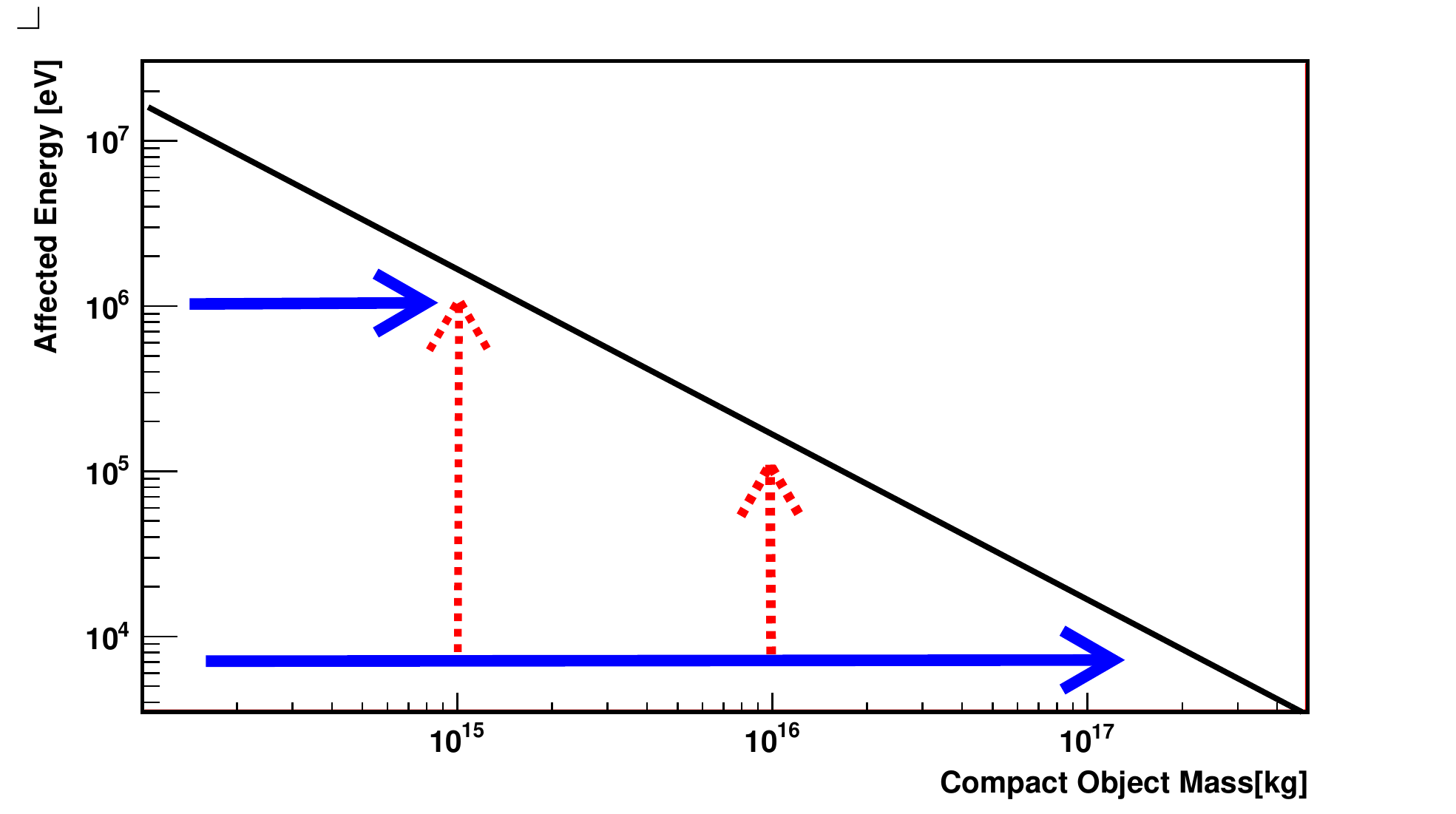}%
 \caption{\label{sensitivity}  Energy of photons affected by femtolensing as a function of the compact lens mass. The dotted arrows indicate the GBM energy range and the solid arrows show the masses which can be best detected by GBM.}
 \end{figure}

In figure~\ref{effR} I show the maximum and minimum detectable $r_S$ for different lens masses. 
The maximum difference between $r_{S,max}$ and $r_{S,min}$ is at $M = 1 \times 10^{18}\,$g.
 The largest femtolensing cross-section occurs for this mass.


\section{Results\label{sec:results}}
The 20 burst sample from table~\ref{tab:table} have been fitted with the standard BKN, BAND and SBKN models. 
The models with the best $\chi^2$ probability were selected and are shown on table~\ref{tab:table}. 
The bursts are well fitted by these standard models, so there is no evidence for femtolensing in the data. 

\begin{figure}[ht!]
     \centering
 \includegraphics[width=8cm,angle=-90,bb=0 0 504 720]{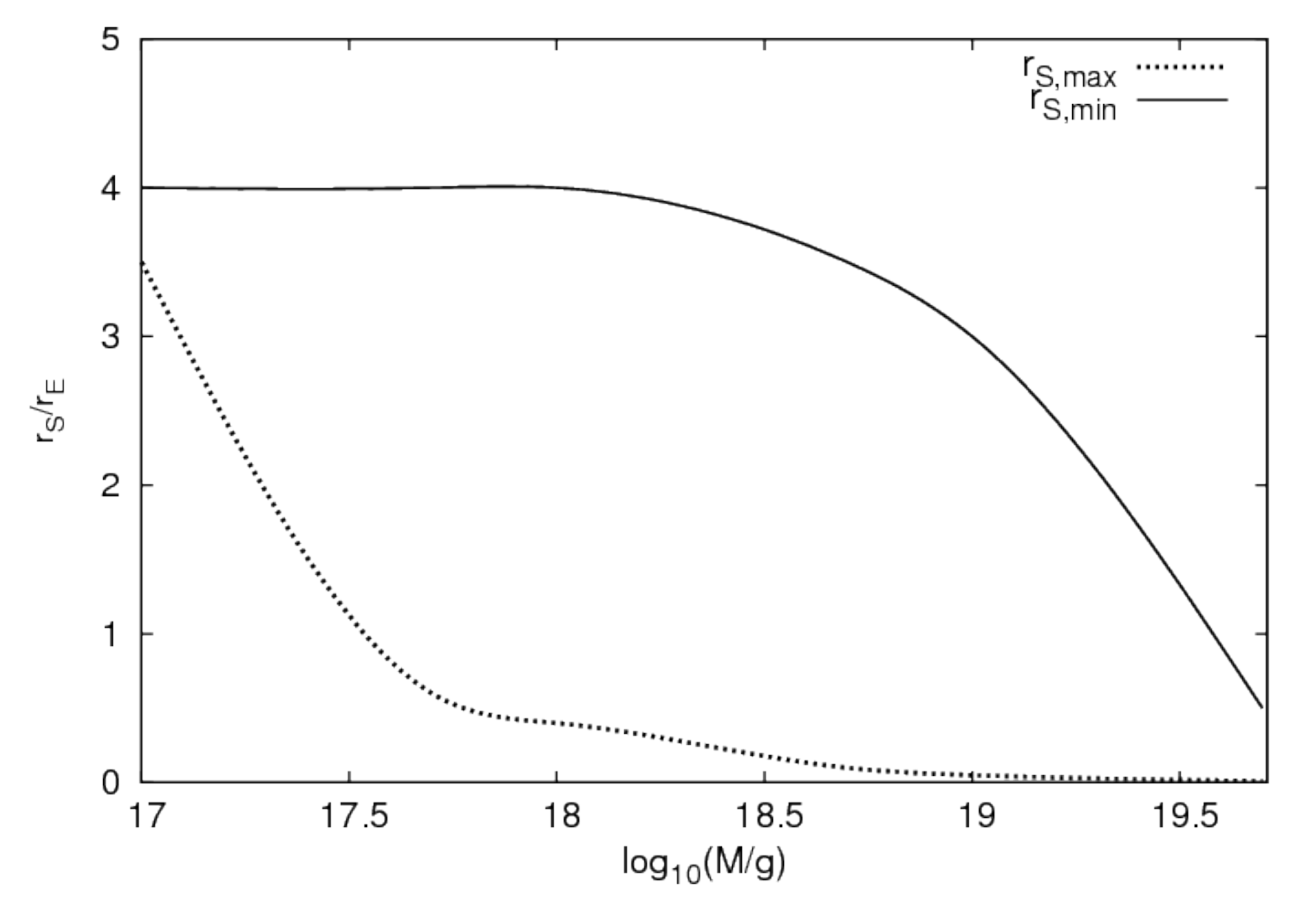}%
 \caption{\label{effR} Minimum and maximum detectable $r_S/r_E$ as a function of lens mass for GRB 090424592.  }
 \end{figure}

As explained in section \ref{sec:probability}, the lensing probability 
for each burst depends on the lens mass and on the $r_{S,min}$ and $r_{S,max}$ values. Since the sensitivity
of GBM to femtolensing is maximal for lens masses of $\sim 1  \times 10^{18}\, $ g (see figure~\ref{effR}), 
the values of $ r_{S,min}$ and $r_{S,max}$ for each event were first determined at a mass $M = 1  \times 10^{18}\, $ g by simulation.  
As explained in section~\ref{sec:probability}, the value of $r_{S,min}$ is set by the period of the spectral fringes 
so that it is independent of the burst luminosity. 
The values of  $r_{S,max}$ obtained are listed  in table~\ref{tab:table}.
The lensing probability is  then calculated for both  the FRLW and Dyer \& Roeder 
cosmological models using the redshift of each burst, the most probable lens position 
and the values of $r_{S,min}$ and $r_{S,max}$ for the mass $M = 1  \times 10^{18}\,$g. 
The number of expected lensed bursts in the sample is the sum of the lensing probabilities. 
It depends linearly on $\Omega_{CO}.$

\begin{figure}[ht!]
     \centering
 \includegraphics[width=8cm,angle=-90,bb=0 0 504 720]{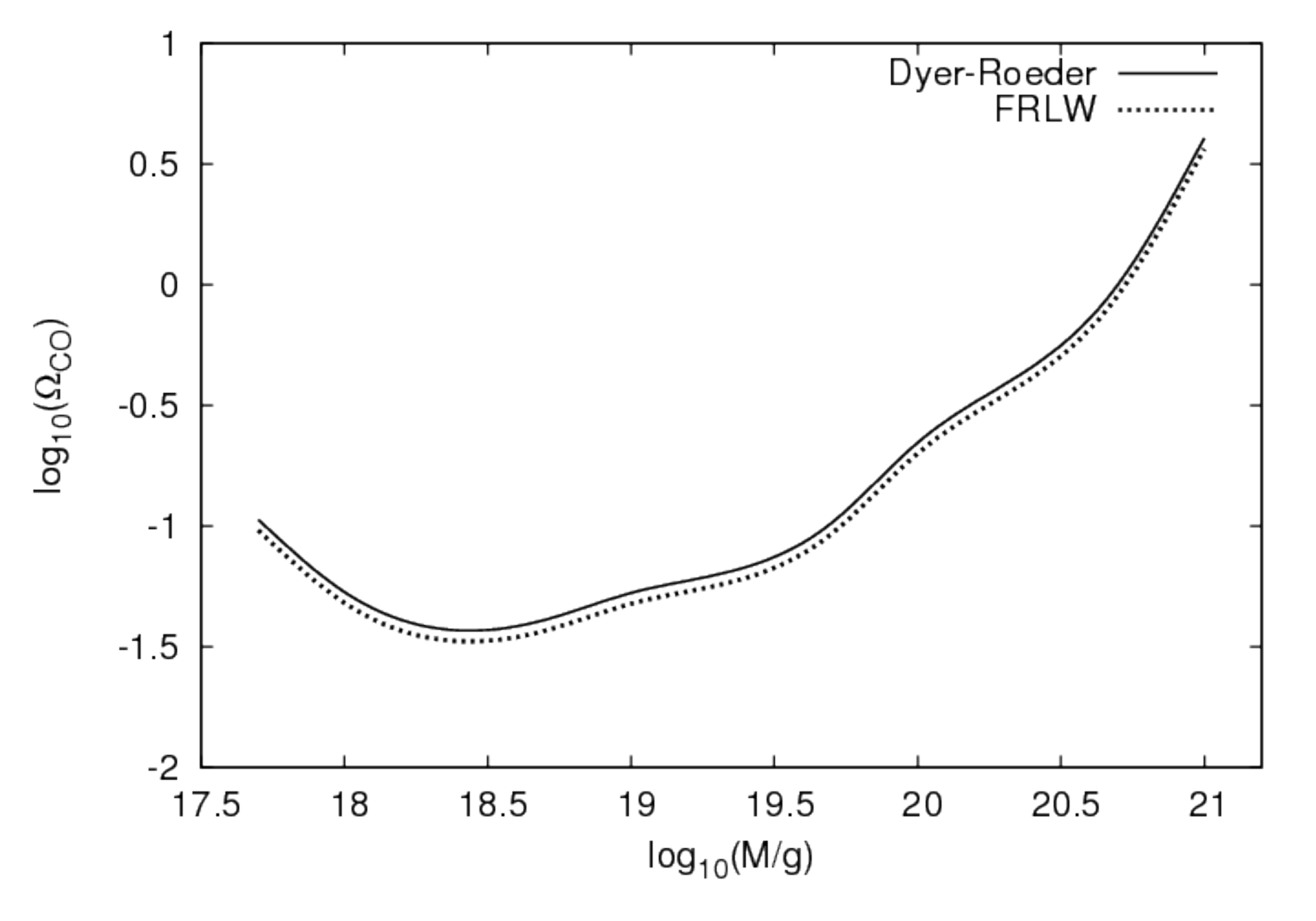}%
 \caption{\label{fraction} Constraints on the fraction (or normalized density) of compact objects. 
 The zones above the curves are excluded at the 95\% confidence level.}
 \end{figure}

Since no femtolensing is observed, the number of expected events 
should be less than 3 at 95\% confidence level (C.L.).
The constraints on  the density of compact objects $\Omega_{CO} $  
is derived to be less than 4\% at 95\% C.L for both cosmological models.
The values of the lensing probabilities for all the bursts in our sample 
assuming the constrained density of compact objects are shown in table~\ref{tab:table}. 
The limits at other lens masses are obtained by normalizing the $\Omega_{CO} $ at $M = 1\times 10^{18} \mbox{g}$ by the cross section $\sigma$. 
The cross section  is calculated 
using equation.~(\ref{sigma}) and the values of $r_{S,min}$ and $r_{S,max}$  from figure~\ref{effR}.  
The limits on $\Omega_{CO}$   at $95\%\,$C.L. are plotted in figure~\ref{fraction}.

   
\begin{landscape}
\tiny

\begin{table} 
\label{tab:table}
 \caption{ The sample of 20 GBM GRBs used in the analysis.} 
 \begin{center}
 \begin{tabular}{cccllccccc}
          &                          &               &   
                                                                           &  \multicolumn{2}{c}{Fit to simulated data}  
                                                                                                                        &                 &       &                & \\ 
 Name          
                                       & $z_S$ &   \multicolumn{2}{c}{Fit to the data \footnote{Fit has been performed using only the photons arrived in less than 10s from the beginning of the burst.}}             
                                                                          & Model   &Femtolensing  
                                                                                                                         & $r_{S,max}$&$z_L$&\multicolumn{2}{c}{ Lensing Probability }  \\
                                     &                 & Model     & $\chi^2$/d.o.f  
                                                                                             &  $\chi^2$/d.o.f &$\chi^2$/d.o.f
                                                                                                                                   	& 
	                                                                                                                                           &     & FRLW \footnote{for assumed $\Omega_{CO}=0.0310$}  & Dyer-Roeder \footnote{for assumed $\Omega_{CO}=0.0336$}\\
 \hline
GBM 080804972	&2.2045   &BAND	&68/74	&129/74		&80/72	&2.5  &0.770	& 0.145 & 0.145 \\
GBM 080916009C   &4.3500	&BKN	&75/74	&115/74		&93/72	&3   &1.087	& 0.489 & 0.444 \\
GBM 080916406A   &0.6890	&BKN	&58/57	&91/55		&67/53	&3   &0.324	& 0.031 & 0.033 \\
GBM 081121858	&2.5120	&BKN	&39/49	&63/52		&50/50	&3   &0.829	& 0.250 & 0.245 \\
GRB 081222204 	&2.7000   &BKN	&73/66	&89/62		&68/60      &3.5   &0.859	& 0.374 & 0.364 \\ 
GRB 090102122	&1.5470	&BAND	&81/85	&124/85		&92/83	&3   &0.603	& 0.134 & 0.127 \\
GRB 090323002	&3.5700	&BAND	&77/77	&106/64		&95/62	&2   &0.964	& 0.173 & 0.162 \\
GRB 090328401	&0.7360	&BKN	&105/70	&124/70		&64/68	&2.5   &0.346	& 0.024 & 0.026 \\
GRB 090424592	&0.5440     &BAND	&78/67	&215/78		&104/76	&4   &0.256	& 0.035 & 0.038 \\
GRB 090510016	&0.9030	&BKN	&62/66	&108/98		&94/96	&1.5&0.406	& 0.012 & 0.013 \\
GRB 090618353	&0.5400	&BAND	&59/72	&158/69		&93/67	&3    &0.254	& 0.019 & 0.021 \\
GRB 090926181	&2.1062	&BAND	&87/81	&247/81		&123/79	&4    &0.737	& 0.348 & 0.349 \\
GRB 091003191	&0.8969	&BKN	&93/94	&140/94		&96/92	&3    &0.400  	& 0.049 & 0.053 \\
GRB 091020900	&1.7100	&BKN	&74/69	&100/69		&77/67	&3   &0.667	& 0.144 & 0.147 \\
GRB 091127976	&0.4900	&BAND	&78/74	&84/74		&71/72	&4    &0.240	& 0.029 & 0.031 \\
GRB 091208410	&1.0630	&BAND	&55/55	&101/55		&54/53	&3  &0.457	& 0.066 & 0.070 \\
GRB 100414097	&1.3680	&BKN	&65/61	&120/68		&92/66	&2.5 &0.560	& 0.070 & 0.073 \\
GRB 100814160A	&1.4400	&BKN	&86/70	&181/70		&110/68 	&2    &0.590	&  0.049 & 0.051 \\
GRB 100816009	&0.8049	&BKN	&67/56	&83/56		&68/54	&3 &0.360	&  0.041 & 0.043 \\
GRB 110731465	&2.8300	&SBKN	&72/64	&96/53		&78/51	&3    &0.877	&  0.292 & 0.283 \\
 \end{tabular}
 \end{center}
\end{table}
\end{landscape}

\section{Discussion and conclusions\label{sec:conclusions}}

\begin{figure}[ht!]
     \centering
 \includegraphics[width=15cm,angle=0,bb=0 0 954 694]{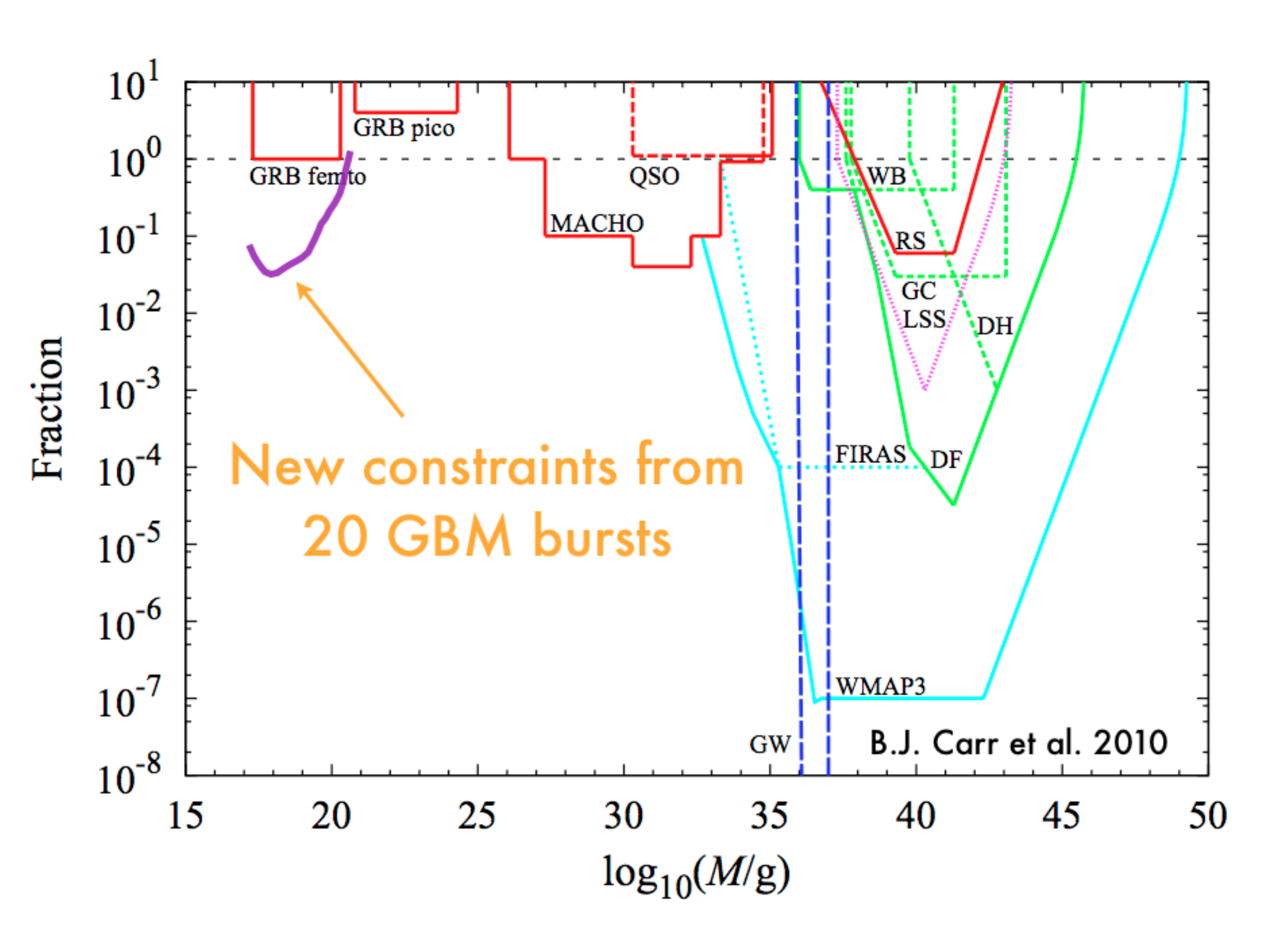}%
 \caption{\label{fig:Carr} Constraints on the fraction of compact objects for the various effects. 
 Figure credited from \cite{2010PhRvD..81j4019C} }
 \end{figure}

Cosmological constraints on the PBH abundance are reviewed by \cite{2010PhRvD..81j4019C}. 
One way to obtain the abundance of PBH is to constrain the density of compact objects $\Omega_{CO}$.  
Note that the limits on the compact object abundance  in the $10^{26} - 10^{34}\,$g range obtained with microlensing are at the 1$\%$~level (MACHO in figure \ref{fig:Carr}).

It is stated in \cite{2009ApJ...705..659A}
that the mass range $10^{16}\,$g $< M_{BH} < 10^{26}\,$g is virtually unconstrained. 
Indeed, constraints in this  mass range were given by just one group (\cite{1999ApJ...512L..13M}). The limits are shown on figure  \ref{fig:Carr} as GRB femto and pico.
Their results is based on a sample of 117 bright bursts detected by the BATSE satellite. 
The bursts were searched for spectral features by \cite{1998AIPC..428..299B}. 
The constraints  reported by \cite{1999ApJ...512L..13M} are 
$\Omega_{CO} < 0.2$ if the average distance to the GRBs is $z_{GRB} \sim 1$ or  $\Omega_{CO} < 0.1$ if $z_{GRB} \sim 2$.  

Under the mass $5 \times 10^{14}\,$g, $\Omega_{CO}$ is constrained by PBH evaporation. 
Above the femtolensing range, the constraints come from microlensing. 
The new idea by \cite{2011PhRvL.107w1101G} shows that the microlensing limit could be improved and
get constraints down to $10^{20}\,$g with the Kepler satellite observations.Ó

The FERMI satellite  was launched three and a half years ago.
Since then, almost 1000 of GRB were observed with the GBM detector. 
In many cases data quality is good enough to reconstruct  time-resolved spectra. 
This unique feature is exploited in our femtolensing search by selecting the first few seconds of a burst in data analysis.

Limits from the present thesis were obtained by selecting only those bursts with known redshifts in the GBM data. 
This reduces the data sample from the 500 bursts detected in the first 2 years to only 20.
The constraints on $\Omega_{CO}$   obtained at  the $95\%\,$ C.L. are shown on figure.~\ref{fraction}.
These constraints improve the existing constraints by a factor of 4 in the mass range $1 \times 10^{17} $ -- $10^{20}\,$g.

After ten years of operation, the GBM detector should collect over 2500 bursts. 
Only a few of the bursts, say 100, will have a measured redshift and sufficient spectral coverage.
By applying the methods described in this thesis, the limits will improve by a factor of 5 reaching 
a sensitivity to density of compact objects down to the 1\% level.  



\part{Conclusions \label{chap:conclusions}}  

In this thesis I have presented my work on the Level 2 trigger system for the H.E.S.S. II telescope 
developed in cooperation with IRFU CEA-Saclay in France. 
The H.E.S.S. II telescope was built to enlarge the current energy range 
of the existing H.E.S.S. system down to tens of GeV. 
In the low energy part of the energy range, the H.E.S.S. II telescope has to 
carry the  observations without the support of the smaller telescopes - in the, so called, monomode. 
In the monomode, very high trigger rates are expected due to a large flux of single muons. 
To reduce the trigger rate, the Level 2 trigger system has been implemented in the H.E.S.S. II telescope. 
The system consists of both hardware and software solutions. 
My work on the project focused on the algorithm development 
and the Monte Carlo simulations of the trigger system and overall instrument. 
I have developed and successfully tested the algorithm suitable to reject a major fraction 
of the background events and to reduce the trigger rate
 to the level possible to process by the data acquisition system 
\citep*{2011APh....34..568M,2011ITNS...58.1685M}.
 
 I have also been analyzing the H.E.S.S. data of the particular blazar PKS 1510-089.
The work on this blazar concerned  the data analysis and modeling of broad-band emission  of PKS 1510-089 observed in a flaring state in very high energy (VHE) range by the H.E.S.S. observatory. 
PKS 1510-089 is an example of the, so-called, flat spectrum radio quasars (FSRQs) for which no VHE emission is expected due to the Klein-Nishina effects and strong absorption in the broad line region \citep{2005MNRAS.363..954M}. Recent detection of at least three FSRQs by Cherenkov telescopes has forced 
a revision of our understanding of these objects. 
In this thesis I have presented the analysis of the H.E.S.S. data: 
a light curve and a spectrum of the PKS 1510-089, 
together with the FERMI data and a collection of multi-wavelength data obtained by various instruments. 
I have successfully modeled  PKS 1510-089 by applying the single zone internal shock scenario. 
In this scenario, the highest energy emission is the result of the Comptonization of the infrared photons from the dusty torus (thus avoiding Klein-Nishina limit), while the bulk of the emission in the MeV-GeV range is still dominated by the Comptonization of the radiation coming from the broad line region.
 
The strategy of observation of FERMI-LAT, which surveys the whole sky in 190 minutes, 
allows a regular sampling of quasar light curves with a period of a few hours. 
This gives the opportunity to investigate lensing phenomena with high energy (HE) gamma-rays. 
However, the multiple images of a gravitational lensed AGN cannot be directly observed 
with HE gamma-ray instruments due to their limited angular resolutions. 
I have developed a method of time delay estimation that handles 
the problem of the limited instrument angular resolution  \citep{2011A&A...528L...3B}. 
It is called a  "double power spectrum" analysis and relies on the double Fourier transform of the observed light curves. 
The method has been tested on simulated light curves and on FERMI LAT observations of the very bright radio quasar PKS 1830-211, 
for which the time delay of $26^{+4}_{-5}$ days was previously estimated by \cite{1998Lovell} using radio observations. 
The double power spectrum analysis has resulted in an estimation of the delay of 27.1$\pm$0.6 days, with a statistical uncertainty 
an order of magnitude better as compared to previous derivations. 
PKS 1830-211 has thus become the first gravitationally lensed object with its echo detected using the HE instrument.

Dark matter is one of the most challenging open problems in cosmology and particle physics. 
Although currently the weakly interacting massive particles (WIMPs) seems to be favored 
as a possible constituent of the dark matter, 
the alternative idea that the missing matter consists of compact astrophysical objects was also proposed 
as early as in the 1970s~\citep{1974MNRAS.168..399C,1974Natur.248...30H,1971MNRAS.152...75H}. 
An example of such compact objects might be primordial black holes (PBHs) 
created in the very early Universe from matter density perturbations. 
To derive the constraints on the compact object abundance I have search 
for  femtolensing effects in the spectra of gamma-ray bursts with known redshifts 
detected by the Gamma-ray Burst Monitor (GBM) on board the FERMI satellite. 
The search involved the analysis of the FERMI data, which have been used to estimate the femtolensing effect detectability. 
The lack of detection of the femtolensing effect has provided new constraints on the PBHs fraction in the mass range $10^{17}-10^{20}\,$g.
 Less than 3\% of critical density are contributed by PBHs (or any compact objects) at 95\% confidence level \citep{2012PhRvD..86d3001B}.


\bibliographystyle{ab12}
\bibliography{thesis}

\end{document}